\documentclass[superscriptaddress, twocolumn,amsmath,amssymb,prb]{revtex4-1}
\pdfoutput=1
\usepackage{amssymb}
\usepackage{txfonts}
\usepackage{amsmath}
\usepackage{mathtools}
\usepackage{amsmath}
\usepackage{graphicx}
\usepackage{dcolumn}
\usepackage{multirow}
\usepackage{natbib}
\usepackage{booktabs}
\usepackage{bm,color}
\usepackage{graphicx}
\usepackage{graphicx, array, blindtext}
\usepackage{subcaption}
\usepackage{adjustbox}

\begin{document}

\title{Symmetry Demanded Topological Nodal-line Materials}

\author{Shuo-Ying Yang}
\affiliation{Max Planck Institute of Microstructure Physics, Weinberg 2, 06120 Halle, Germany}

\author{Hao Yang}
\affiliation{Max Planck Institute of Microstructure Physics, Weinberg 2, 06120 Halle, Germany}
\affiliation{Max Planck Institute for Chemical Physics of Solids, 01187 Dresden, Germany}

\author{Elena Derunova}
\affiliation{Max Planck Institute of Microstructure Physics, Weinberg 2, 06120 Halle, Germany}

\author{Stuart S. P. Parkin}
\affiliation{Max Planck Institute of Microstructure Physics, Weinberg 2, 06120 Halle, Germany}

\author{Binghai Yan}
\affiliation{Max Planck Institute for Chemical Physics of Solids, 01187 Dresden, Germany}
\affiliation{Department of Condensed Matter Physics, Weizmann Institute of Science, Rehovot 76100, Israel}

\author{Mazhar N. Ali}
\email{maz@berkeley.edu}
\affiliation{Max Planck Institute of Microstructure Physics, Weinberg 2, 06120 Halle, Germany}

\begin{abstract}

The realization of Dirac and Weyl physics in solids has made topological materials one of the main focuses of condensed matter physics. Recently, the topic of topological nodal line semimetals, materials in which Dirac or Weyl-like crossings along special lines in momentum space create either a closed ring or line of degeneracies, rather than discrete points, has become a hot topic in topological quantum matter. Here we review the experimentally confirmed and theoretically predicted topological nodal line semimetals, focusing in particular on the symmetry protection mechanisms of the nodal lines in various materials. Three different mechanisms: a combination of inversion and time-reversal symmetry, mirror reflection symmetry, and non-symmorphic symmetry, and their robustness under the effect of spin orbit coupling are discussed. We also present a new Weyl nodal line material, the Te-square net compound KCu$_2$EuTe$_4$, which has several Weyl nodal lines including one extremely close to the Fermi level ($<$30 meV below E$_F$). Finally, we discuss potential experimental signatures for observing exotic properties of nodal line physics.

\end{abstract}

\maketitle

\section{Introduction}

Topologically nontrivial states of matter have been of great interest in the field of topological quantum materials during the past decade for their rich and novel physics. The field was ignited after the discovery of topological insulators (TIs) - 
materials that exhibit robust metallic surface states protected by the topology in the insulating bulk\cite{hasan2010colloquium, qi2011topological}. Recently, topological semimetals (TSM) such as Dirac and Weyl semimetals were discovered which can also host metallic surface states with a semi-metallic bulk\cite{armitage2017weyl}. \par

Three types of TSMs, Dirac (DSM), Weyl (WSM), and nodal-line semimetals (NLS) have been discovered. A DSM, shown in Figure 1a, is the result of two doubly degenerate bands crossing near the Fermi level (E$_F$) at a discrete point in $k$-space, known as a Dirac point. Demanded by both inversion and time-reversal symmetry, the four-fold degenerate Dirac point has a band dispersion that is linearly dependent with $k$. Such a linearly dispersed band structure causes the low energy excitations near the crossing point to be Dirac fermions. The existence of a 3D DSM was first confirmed by angular-resolved photoemission spectroscopy (ARPES) in Na$_3$Bi \cite{wang2012dirac, liu2014discovery} and Cd$_3$As$_2$\cite{neupane2014observation, liu2014stable, borisenko2014experimental}, where the Dirac points are topologically protected by C$_3$ and C$_4$ rotation symmetry respectively\cite{wang2012dirac, wang2013three, ali2014crystal}. These materials exhibit many exotic transport properties, such as ultrahigh mobility and titanic MR (MR) \cite{wang2013three, jeon2014landau, feng2015large, liang2015ultrahigh, wang2016aharonov}. \par

When either inversion symmetry or time reversal symmetry is broken, doubly degenerate bands become spin split when considering SOC, separating into two singly degenerate band crossings called Weyl points, as shown in Figure 1b. According to the `no-go theorem'\cite{nielsen1981absence, nielsen1981absence1}, they can only appear in pairs of opposite chirality \cite{fang2003anomalous, wan2011topological, balents2011viewpoint, xu2011chern}. The existence of Weyl points near the E$_F$ leads to several special characteristics, including Fermi arcs on the surface \cite{weng2015weyl, wan2011topological, huang2015weyl, balents2011viewpoint} and the chiral anomaly in the bulk \cite{nielsen1983adler, hosur2012charge, son2013chiral, kim2013dirac, parameswaran2014probing, xiong2015evidence, li2015giant, arnold2016negative, zhang2017room, hirschberger2016chiral}, among many others \cite{hosur2013recent, volovik2015standard}. The Weyl points can be seen as singularity points of Berry curvature (source or sink), or `magnetic monopoles' in $k$-space. The Fermi arcs, unlike closed Fermi surfaces in 2$d$ or 3$d$ metals, are open `arcs' that connect the two opposite Weyl points on the boundary. Weyl-like states have first been observed in inversion symmetry-breaking compounds TaAs\cite{xu2015discovery, lv2015experimental, yang2015weyl}, NbAs \cite{xu2015discovery1}, TaP, NbP\cite{souma2016direct, liu2016evolution} as well as in photonic crystals \cite{lu2015experimental}. They have also been observed in other inversion symmetry-breaking materials such as MoTe$_2$, WTe$_2$ and other Ta or Nb monopnictides \cite{soluyanov2015type, weng2015weyl, sun2015prediction, huang2016spectroscopic, deng2016experimental, jiang2017signature, tamai2016fermi}. These Weyl particles are associated with a number of novel transport phenomena, such as ultra-high mobility, titanic MR and the chiral anomaly \cite{son2013chiral, huang2015observation, arnold2016negative, du2016large, yang2015chiral, ali2014large, shekhar2015extremely}. \par

In contrast to DSMs and WSMs, which have zero dimensional band crossings, NLSs have extended band crossings along special lines in $k$-space. Nodal lines can cross the Brillouin zone (BZ) in the shape of a closed ring or a line \cite{bzdusek2016nodal}, as shown in Figure 1c. Similar to the characteristic 1D Fermi arc surface state of WSMs, nodal ring semimetals are characterized by the 2D topological `drumhead' surface state\cite{weng2015topological}. The distinguishing characteristic of these `drumhead' surface states is that they are embedded inside the `direct gap' between conduction and valence bands in the 2D projection of the nodal ring\cite{burkov2011topological, heikkila2011dimensional}. These edge states are approximately dispersionless, analogous to the acoustic vibration on the surface of a drum, and give rise to a large density of states\cite{kopnin2011high}. The flat band surface states could potentially realize high temperature superconductivity, magnetism, or other correlated effects \cite{kopnin2011high, heikkila2011flat}. There have been several predictions of novel phenomena in Nodal line/ ring systems. In nodal ring systems, long range coulomb interactions are expected due to a vanishing density of states at the E$_F$ and partially screened Coulomb interaction \cite{huh2016long}. In addition, the appearance of zero modes due to non-dispersive Landau levels is predicted to exist inside the nodal ring \cite{rhim2015landau}. Nodal line systems are predicted to show a quasi-topological electromagnetic response which is related to charge polarization and orbital magnetization \cite{ramamurthy2015quasi}, as well as a light-induced Floquet effect in which NLS can be driven into a WSM state through a circularly polarized light\cite{ebihara2016chiral, chan2016chiral, taguchi2016photovoltaic}. \par

\begin{figure}
 \begin{center}
\begin{subfigure}[b]{.3\linewidth}
\includegraphics[width=\linewidth,height=2.7cm]{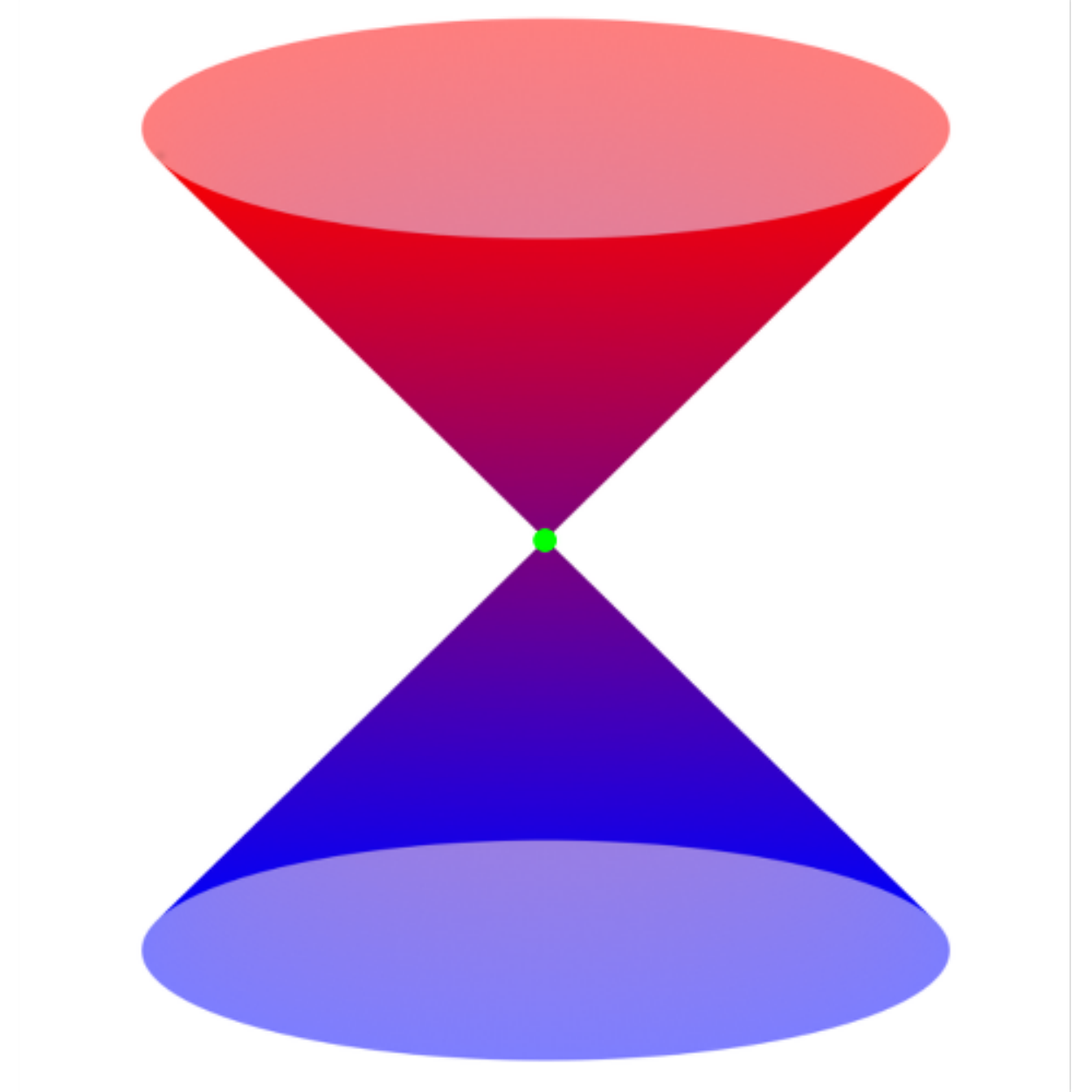}
\caption{Dirac semimetal}\label{fig:mouse}
\end{subfigure}\hspace{10mm}
\begin{subfigure}[b]{.4\linewidth}
\includegraphics[width=\linewidth,height=2.7cm]{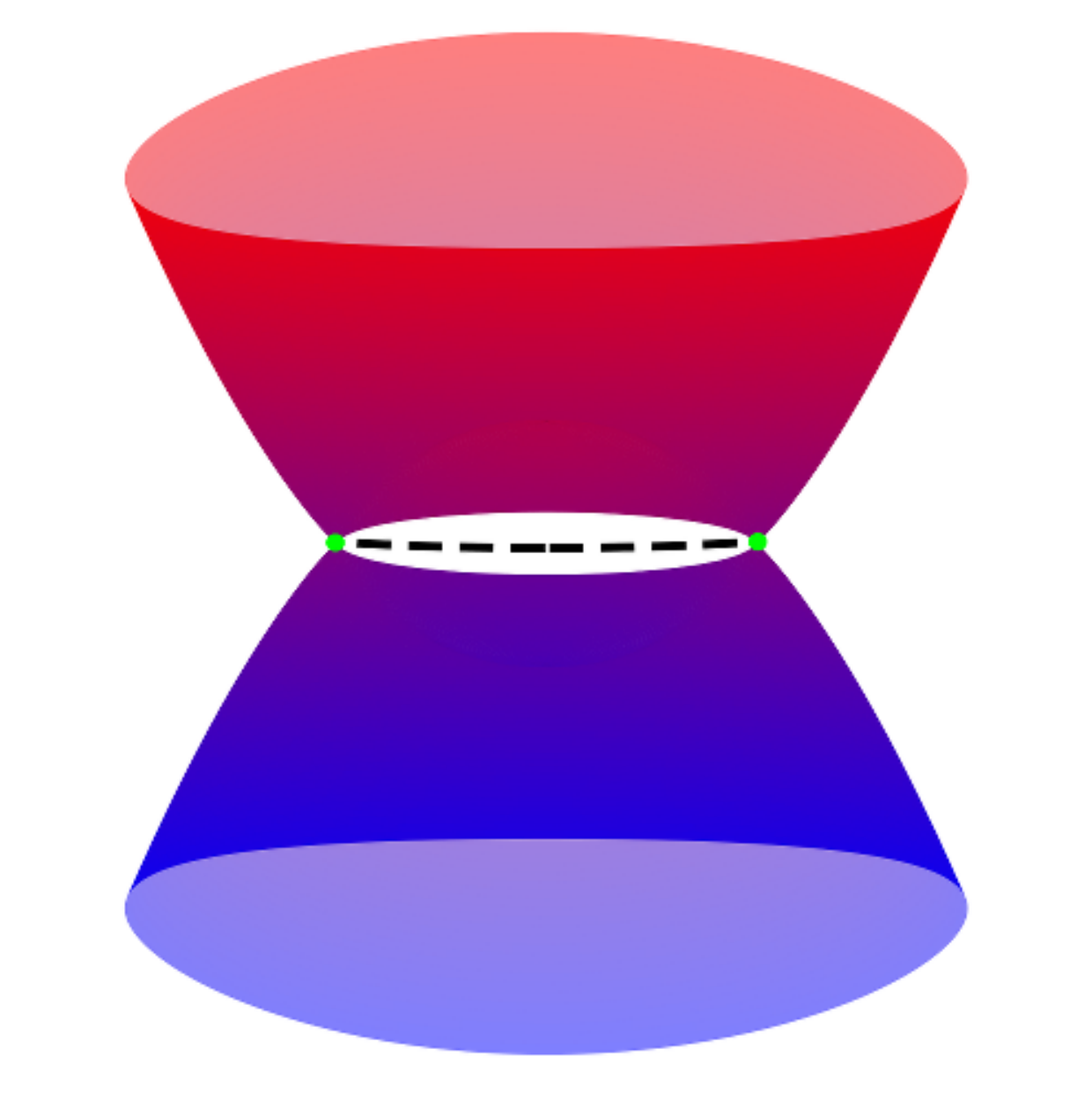}
\caption{Weyl semimetal}\label{fig:gull}
\end{subfigure}

\begin{subfigure}[b]{.75\linewidth}
\includegraphics[width=\linewidth,height=3.5cm]{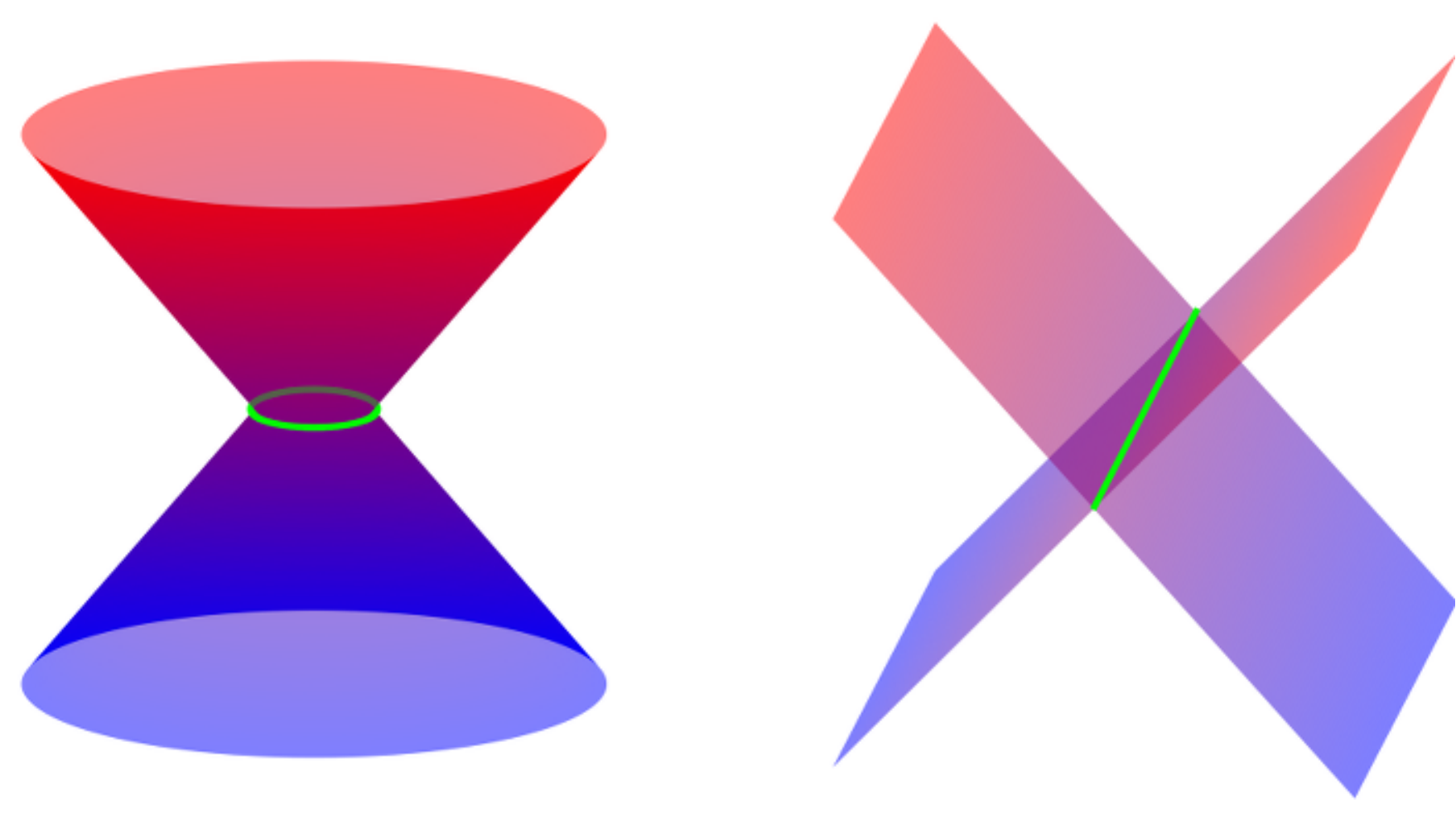}
\caption{Nodal line semimetals}\label{fig:tiger}
\end{subfigure}
\end{center}

\captionsetup{justification   = raggedright,
              singlelinecheck = false}

\caption{Schematic illustration of Dirac node, Weyl node and Nodal line/ ring in momentum space. (a) Schematic of a Dirac semimetal where the bands are linearly dispersed around the Dirac point. The Dirac point is shown by the green dot. (b) Weyl semimetal, in which the Weyl points with opposite chirality are connected by the characteristic Fermi arc. The Weyl points are shown by the green dot and Fermi arc is shown by the black dotted line. (c) Nodal line semimetals where valence and conduction bands cross along special lines in momentum space forming either a ring-shaped line or 1D line, shown by the green circle/ line. }
\label{fig:animals}
\end{figure}

Crystallographic symmetries play an important role in realizing TSMs. Depending on the symmetry protection, NLS can typically be categorized into two types: Dirac Nodal Line Semimetals (DNLS) and Weyl Nodal Line Semimetals (WNLS). The DNLS are found in materials with both inversion symmetry and time reversal symmetry. When SOC is neglected or negligibly small, band inversions happen at one or more high symmetry points along the BZ, resulting in two doubly degenerate bands crossing each other to form a fourfold degenerate nodal line. WNLS lack either inversion or time-reversal symmetry allowing for spin slitting. Therefore, the otherwise four-fold degenerate nodal lines split into two singly degenerate nodal lines which are protected by one additional symmetry. Although it is known that certain crystalline symmetries play an important role in protecting band degeneracy\cite{takahashi2017topological}, finding materials with stability of the degeneracy near the E$_F$, especially in the presence of SOC, is still one of the main goals of the field\cite{fang2015topological, fang2016topological}. \par
                            
In the last few years, this field has gone through a lot of development, in regards to the theoretical proposals, the discovery of new materials, and the study of experimental phenomena. A number of comprehensive reviews on related topics have also came up\cite{armitage2017weyl, yan2017topological, jia2016weyl, wang2017quantum, hosur2013recent, fang2016topological, yu2017topological, weng2016topological}. This work will detail the aforementioned protection mechanisms in the context of both predicted as well as the verified NLS materials. Section II reviews three types of crystalline symmetries that generate NLS, including inversion plus time-reversal symmetry, mirror reflection symmetry, and non-symmorphic symmetry. The effect of SOC on these symmetry protections will also be discussed. Section III presents examples of materials which are observed or predicted, by these three protection mechanisms, to be NLS. Out of the currently studied materials only Pb(Tl)TaSe$_2$ and Zr(Hf)SiS have WNL and DNL's, respectively, in a real-life scenario. PbTaSe$_2$ was the only WNLS, in which two Weyl nodal rings are 0.05 eV and 0.15 eV above the E$_F$. It also has one accidental nodal ring created by SOC\cite{bian2016topological}. Zr(Hf)SiS, SrIrO$_3$, and IrO$_2$ have nodal ring/ line protected by non-symmorphic symmetry that are robust against SOC \cite{schoop2016dirac, fang2015topological, sun2017dirac}. SrIrO$_3$, however, can gap out when magnetic ordering is included, and is still being investigated\cite{liu2016direct}. All other DNLS reviewed here have nodal rings without SOC, however, including SOC turns them into Weyl semimetals or TIs. Section IV predicts a new WNLS candidate, KCu$_2$EuTe$_4$, which, like PbTaSe$_2$, has Weyl nodal lines and rings very close to E$_F$ in the presence of SOC. Section V summarizes and discusses the potential future applications of NLS. \par


\section{Crystalline Symmetries}

The DNLS can be considered as a starting point for realizing many other topological states. Figure 2 shows the relationship between these different topological classes. Starting from a spinless DNLS, SOC could lift the degeneracy along the crossing line, leading to 1) a spinful DNLS when the line of degeneracy is protected by a combination of time-reversal, inversion, and non-symmorphic symmetry (i.e. ZrSiS\cite{schoop2016dirac, neupane2016observation, chen2017dirac}); 2) a spinful WNLS (i.e. PbTaSe$_2$ \cite{bian2016topological, ali2014noncentrosymmetric, guan2016superconducting, chang2016topological, bian2016drumhead}) when mirror reflection within high symmetry plane protects the nodal lines, but broken inversion symmetry (or broken time-reversal symmetry) splits the spin component at the nodal ring, giving rise to two Weyl rings; 3) a 3D DSM (i.e. Cu$_3$PdN\cite{kim2015dirac, yu2015topological}) if certain symmetry-invariant points are protected by rotation symmetry as well; 4) a WSM (i.e. TaAs\cite{weng2015weyl, yang2015weyl}) when the system has mirror reflection symmetry, orthogonal orbital component making up the electronic states, and proper strength of SOC; 5) a TI (i.e. Mackay-Terrones crystal\cite{weng2015topological}) if the material has only inversion and time-reversal symmetry. This would give rise to a gapped bulk state but linearly dispersed projected surface states. Starting from DSMs, breaking either inversion or time-reversal symmetry will lead to WSMs. \par

\begin{figure*}
 \begin{center}
 \includegraphics[width=1\textwidth]{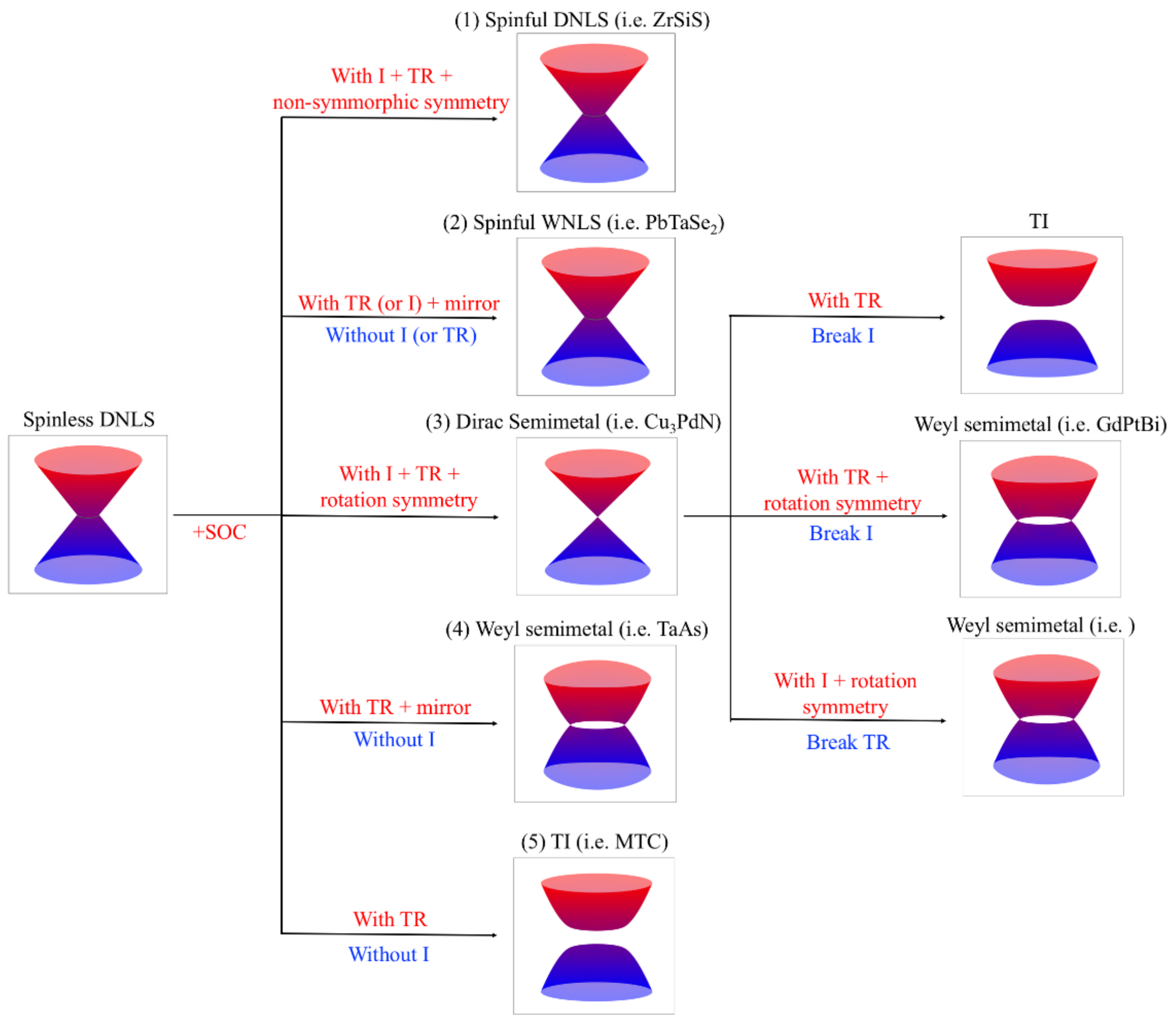}
 \end{center}
 
       \captionsetup{justification   = raggedright,
              singlelinecheck = false}
              
 \caption{Schematic of different topological states and their relationship with each other. The symmetries in red (blue) caption indicate the global symmetry or spatial/ lattice symmetry elements need to be preserved (broken) to realize the new topological phase. \\
*: The evolvement to Weyl semimetal require mirror reflection symmetry, atomic orbital components making up the bands forming the nodal lines, as well as the proper strength of SOC. If the strength of SOC is too strong, the system turns into a fully gapped insulator \cite{yu2017nodal}. }
 
\end{figure*}

The symmetry of a crystal structure and the atomic orbitals making up the electronic states at band crossings determine which topological state is realized under consideration of SOC: the electronic states must be orthogonal to each other in order to not get hybridized with each other and open a gap. This means one has to consider both crystal symmetry and orbital symmetry and, when taking spins of electrons into account, to consider the number of irreducible representations in the double group. For example, the C$_2v$ point group (without spin) has four irreducible representations, but the C$_2v$ double group (including spins) only has one irreducible representation. Since two bands with the same irreducible representation hybridize (mixing because they are not orthogonal), C$_2v$ is gapped in the presence of SOC \cite{gibson2015three}. \par

Global symmetry and spatial/ lattice symmtry are needed to realize NLS. Currently, three different recipes are known to yield a NLS. One is the combination of inversion and time-reversal symmetry. The second recipe is to introduce additional mirror symmetry, where the degeneracy on the mirror plane is protected. The third recipe is to introduce non-symmorphic symmetry such as glide mirror or screw rotation. The non-symmorphic symmetry demands normally singly degenerate points to cross at certain doubly degenerate points at the BZ zone boundary\cite{gibson2015three, young2015dirac}. SOC plays different roles in three scenarios. In this section, these three recipes, as well as the effect of SOC on them will be discussed.\par

\subsection{Inversion and time-reversal symmetry}

\begin{figure*}[t]
 \begin{center}
 \includegraphics[width=0.8\textwidth]{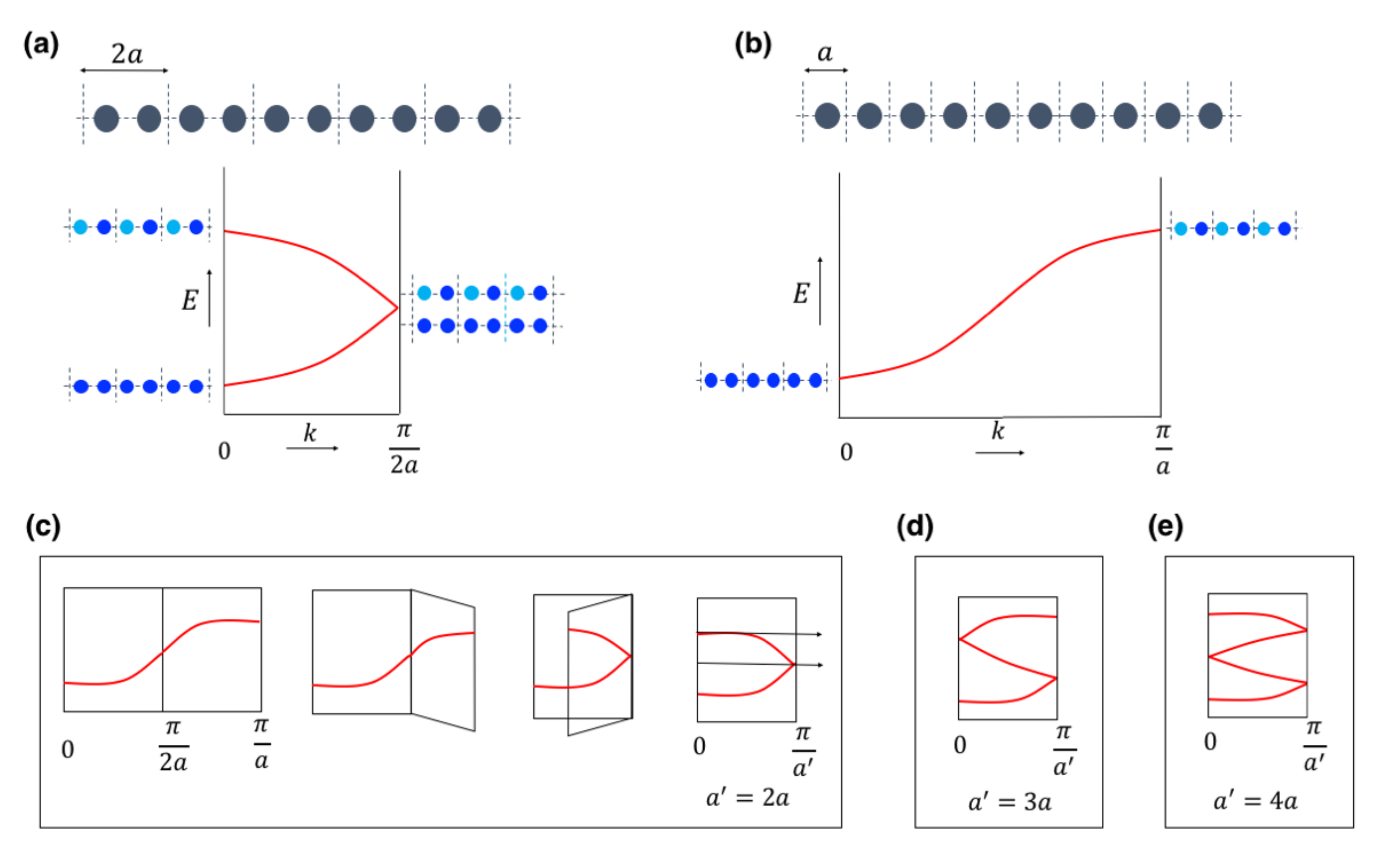}
 \end{center}
              
              \captionsetup{justification   = raggedright,
              singlelinecheck = false}
              
 \caption{Band folding in one-dimensional polymers. (a) Band structure of a polymer in which there are two atoms per unit cell. The bands contain two branches: one from the `running up' bonding bands, and one from the `running down' antibonding bands. The two branches intersect at $k=\pi/(2a)$. (b) Band structure of a polymer in which there is one atom per unit cell. The BZ is doubled because the unit cell is one half as it is in (a). (c) Production of band structure in (a) by folding band structure in (b). (d)-(e) Enlargement of the unit cell causes the multiplicity of bands. Bands are folded three times when the unit cell is tripled, four times when quadrupled.}
\end{figure*}

\begin{figure*}[t]
 \begin{center}
 \includegraphics[width=0.8\textwidth]{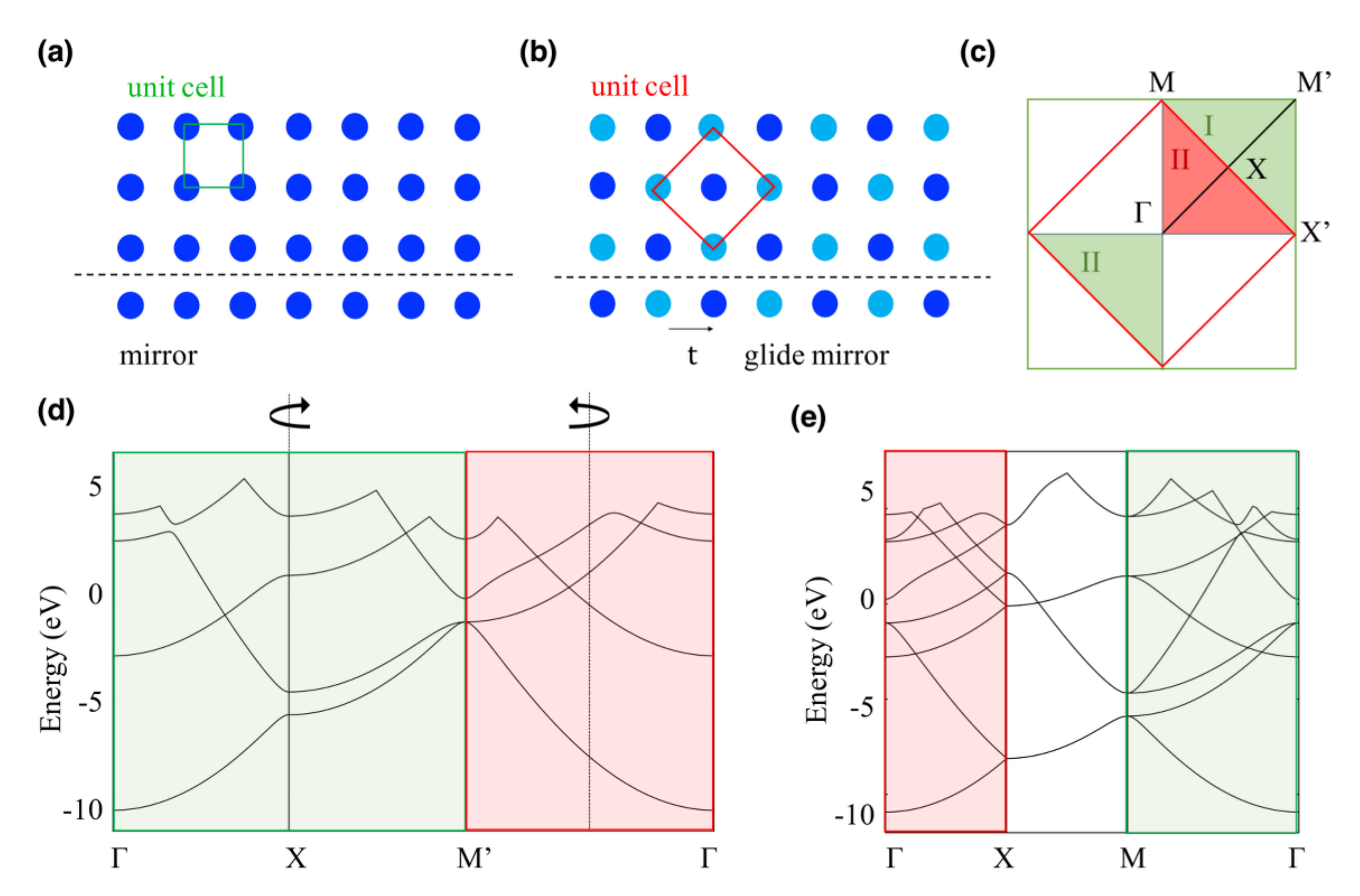}
 \end{center}
 
 \captionsetup{justification   = raggedright,
              singlelinecheck = false}
 
 \caption{Band folding in two-dimensional square lattice. (a) Symmorphic square lattice with mirror reflection symmetry, in which the unit cell is shown in green. (b) Non-symmorphic square lattice with glide symmetry. The unit cell shown in red is larger than the unit cell of symmorphic square lattice. (c) The BZ of the symmorphic and non-symmorphic square lattices, which are enclosed by green and red lines respectively. The green shaded area I is equal to the green shaded area II by translation, which is also equivalent to red shaded area II by time-reversal symmetry. (d)-(e) The band structure of non-symmorphic square lattice (e) can be obtained from the band structure of symmorphic square lattice (d). Folding the square made by $\Gamma$-X'-M'-M-$\Gamma$ along M-X' makes M' coincide with $\Gamma$. Therefore, $\Gamma$-X of the non-symmorphic lattice can be obtained by folding $\Gamma$-M' of the symmorphic square lattice in half. Similarly, the same folding makes X'-M' superimpose onto $\Gamma$-X' which is equivalent to $\Gamma$-M of the non-symmorphic lattice. The dashed lines show folding points of the symmorphic band structure.}
\end{figure*}

In the absence of SOC, a crystalline system with simultaneous presence of time-reversal and inversion symmetries generates a DNL in 3D\cite{kim2015dirac}. In such case when the 3D block Hamiltonian is invariant under inversion and time-reversal symmetry, inversion and time-reversal symmetry constrain the Berry phase ($\Phi$) on a closed loop in momentum space ($C$) to satisfy $\Phi$($C$) = $\Phi$(-$C$) and $\Phi$($C$) = $\Phi$$(-$$C$$)^{\ast}$. A non-trivial loop, where $\Phi$($C$) = -1, means that the loop contains a degeneracy. In 3D, this guarantees a small nodal loop at the band inversion\cite{kim2015dirac}.\par

In the presence of SOC, Dirac crossings generated with this mechanism are, in general, gapped out owing to band repulsion, unless the SOC is negligibly small, turning the system into a TI. However, if the system has an additional C$_n$, in particular, C$_3$, C$_4$, and C$_6$ rotation symmetries, the rotation symmetry can protect the band crossing at certain generic discrete points, turning the system into a DSM in the presence of SOC. Specifically, one has to consider the crystal structure and orbital symmetry in a double group along the rotation axis. If the electronic states along the rotation axis have different irreducible representations, rotation symmetry protects the Dirac points on the axis. Everything else over the BZ is gapped, thus giving rise to a 3D DSM in the SOC regime\cite{gibson2015three}. \par

\subsection{Mirror reflection symmetry}

When the mirror operation and the Hamiltonian commute, nodal rings can be generated and protected by mirror reflection symmetry in the absence of SOC. For example, assuming a system described by Hamiltonian $H$($k$$_x$, $k$$_y$, $k$$_z$) has mirror reflection $M$$_x$$_y$, one has $M$$H$($k$$_x$, $k$$_y$, -$k$$_z$)$M$$^{-1}$ = $H$($k$$_x$, $k$$_y$, $k$$_z$). In two high symmetry planes, k$_z$ = 0 and k$_z$ = $\pi$, the mirror operation and the Hamiltonian commute. This makes them simultaneously diagonalizable, resulting in oppositely signed mirror eigenvalues. Therefore, without SOC, these two bands that cross at the Dirac point are protected by the reflection symmetry from hybridizing with each other \cite{lecturenotes, fang2016topological, fang2015topological, chan20163}. \par

Taking into account of SOC, depending on the atomic orbitals making up the electronic states and the strength of SOC, SOC can turn a spinless Dirac nodal rings into 1) a spinful Weyl nodal rings within the mirror planes (i.e. PbTaSe$_2$ \cite{bian2016topological, ali2014noncentrosymmetric, guan2016superconducting, chang2016topological, bian2016drumhead}). In this case, the WNLs are protected by mirror reflection symmetry; 2) a Weyl semimetal which has Weyl points off the mirror planes (i.e. TaAs \cite{weng2015weyl, yang2015weyl}). These Weyl points are completely accidental, highly depends on the strength of SOC, and are no longer a result of mirror symmetry. 3) a TI when the strength of SOC is strong enough (i.e. CaAgAs \cite{chan20163, yu2017nodal}).   \par

\subsection{Non-symmorhpic symmetry}

To understand how non-symmorphic symmetry generates a NLS, it is important to first understand how more than one electronic unit in a unit cell, or the choice of unit cell, results in band folding. The following explanation was first laid out by Hoffman et al \cite{hoffmann1987chemistry, tremel1987square}. Consider a 1D chain of molecules, one can either see it as a polymer with one atom per unit cell, or twice as many atoms per unit cell. Constructing the orbitals of the polymer with two different unit cells, when the polymer is considered to have two atoms per unit cell, as shown in Figure 3a, it has one branch disperse upward from the bonding orbitals, and the other branch disperse downward from the antibonding orbitals. The bonding and antibonding orbitals are degenerate precisely at $k=\pi/(2a)$. When the polymer is regarded as having one atom per unit cell, the band structure is plotted in Figure 3b with a larger BZ. \par

Since the band structure of a polymer should not depend on the choice of unit cell, the two unit cell construction must result in identical band structures. In other words, the band structure in Figure 3a with two bands and the band structure in Figure 3b with one band must carry the same information about electronic bands. This could be understood in the way that the production of Figure 3a is Figure 3b being `folded back', as the process shown in Figure 3c. This band-folding process can be continued. If the unit cell is tripled, the band will fold as shown in Figure 3d. If it is quadrupled, it folds like shown in Figure 3e, and so on \cite{hoffmann1987chemistry}. \par

This is essentially what happens in non-symmoprhic symmetry crystals. A non-symmorphic symmetry element $G=\left \{ g\mid t \right \}$, shown in Figure 4b, is composed of a point group symmetry operation $g$ and a partial lattice translation $t$. The translation operation do not conserve spatial origin, and causes enlargement of a unit cell in comparison to the symmorphic space group (comparing Figure 4a and 4b). The enlargement of a unit cell is analogous to the double-atom-per-unit-cell idea discussed in Figure 3, which causes the folding of $k$-space, and forces a band degeneracy the BZ boundary. \par

The band structures of non-symmorphic symmetry crystals can simply be generated by the folding back procedure. Given the two unit cells in Figure 4a and 4b, one can construct the corresponding BZ as shown in Figure 4c, where the BZ of the symmorphic square lattice is shown in green and that of non-symmorphic square lattice is shown in red. The green shaded area I can be translated back to the green shaded area II, which is equal to the red shaded area by time-reversal symmetry because time-reversal symmetry implies E$_n$(k)= E$_n$(-k). Since X lies exactly at the midpoint of $\Gamma$-M', and also of M-X', folding the $\Gamma$-X'-M'-M square in half across the M-X' diagonal like a sheet of paper, one can see that X-M' folds directly onto $\Gamma$-X. Similarly, X'-M' folds onto $\Gamma$-X'. In addition, $\Gamma$-X' is equivalent to $\Gamma$-M, thus explaining the band structure in Figure 4e as a simple folding of Figure 4d\cite{tremel1987square}. In particular, $\Gamma$-X of the non-symmorphic square can be constructed by folding $\Gamma$-M' of the symmorphic square lattice in half, and $\Gamma$-M (which is equivalent to $\Gamma$-X') of the non-symmorphic square can be constructed by the superposition (caused by the folding) of $\Gamma$-X' and X'-M' of the symmorphic square lattice\cite{tremel1987square}.\par

What is interesting about the non-symmorphic protected nodal lines is that they cannot be gapped by SOC because they are protected by a lattice translation\cite{young2015dirac}. Take glide mirror as an example, since the translation $t$ is a fraction of a primitive unit vector, in spinless systems for Bloch states at $k$, $G^{2}=e^{-i\mathbf{k}\cdot \mathbf{t}}$. Therefore, the glide eigenvalues are $\pm e^{-i\mathbf{k}\cdot \mathbf{t}/2}$. Considering spins, the glide eigenvalues are $\pm i e^{-i\mathbf{k}\cdot \mathbf{t}/2}$. Either with or without SOC, the non-symmorphic symmetry give two distinct eigenvalues, therefore the bands are protected from being hybridized\cite{lecturenotes}. \par


\section{Material Realization}

\subsection{Time-reversal and inversion symmetry protected nodal line materials}

Centrosymmetric materials have inversion symmetry. If time-reversal symmetry is also presence in the system, four-fold degeneracies along nodal rings can be realized. Nodal rings created in this way generally exist only in the limit of vanishing SOC. Including SOC typically turns the DNLS into either a DSM or a TI. Some predicted centrosymmetric DNL materials include cubic antiperovskite materials Cu$_3$N$X$ \cite{kim2015dirac, yu2015topological}, CaTe\cite{du2016cate}, La$X$ \cite{zeng2015topological}, Ca$_3$P$_2$\cite{xie2015new}, CaP$_3$ family \cite{xu2017topological}, BaSn$_2$ \cite{huang2016topological, nayak2017multiple}, AlB$_2$-type diborides\cite{feng2017topological, kumar2012electronic, zhang2017coexistence}, and 3D carbon allotrope materials with negligible SOC such as Mackay-Terrones crystals\cite{weng2015topological} and hyperhoneycomb lattices\cite{mullen2015line}. In addition, two-dimensional DNL materials have also been proposed in monolayer Cu$_2$Si\cite{feng2016discovery} and honeycomb-kagome lattice\cite{lu2016two}. \par

Table I to III show the crystal structures, electronic band structures without and with SOC, as well as nodal line distributions without SOC within a BZ. Without SOC, the Dirac points responsible for the nodal lines are coded with red square box in the band structure. With SOC, all three compounds shown in Table I turn into DSM. All the other compounds shown in Table II and III are gapped into TIs  (except TiB$_2$). The Dirac points of DSM with SOC are coded with red square box in their corresponding band structures. Nodal lines without SOC within a BZ are shown in red lines, and the high symmetry lines are shown in black lines together with high symmetry points. \par

\begin{table*}[htbp]
\caption{Centrosymmetric nodal line materials}
\centering

\begin{adjustbox}{width=0.9\textwidth}

\newcommand*{\TitleParbox}[1]{\parbox[c]{1.75cm}{\raggedright #1}}%

\makeatletter
\newcommand{\thickhline}{%
    \noalign {\ifnum 0=`}\fi \hrule height 1pt
    \futurelet \reserved@a \@xhline
}
\newcolumntype{"}{@{\hskip\tabcolsep\vrule width 1pt\hskip\tabcolsep}}
\makeatother

\newcolumntype{?}{!{\vrule width 1pt}}

\hspace*{-1cm}
\begin{tabular}{c?c c c } 
 
\toprule

\textbf{Materials} &  \textbf{Cu$_3$NPd} & \textbf{CaTe}  & \textbf{LaN} \\
\thickhline

\midrule

\raisebox{3\totalheight}{\parbox[c|]{2cm}{\raggedright Crystal structure}} &
\includegraphics[width=1.2in]{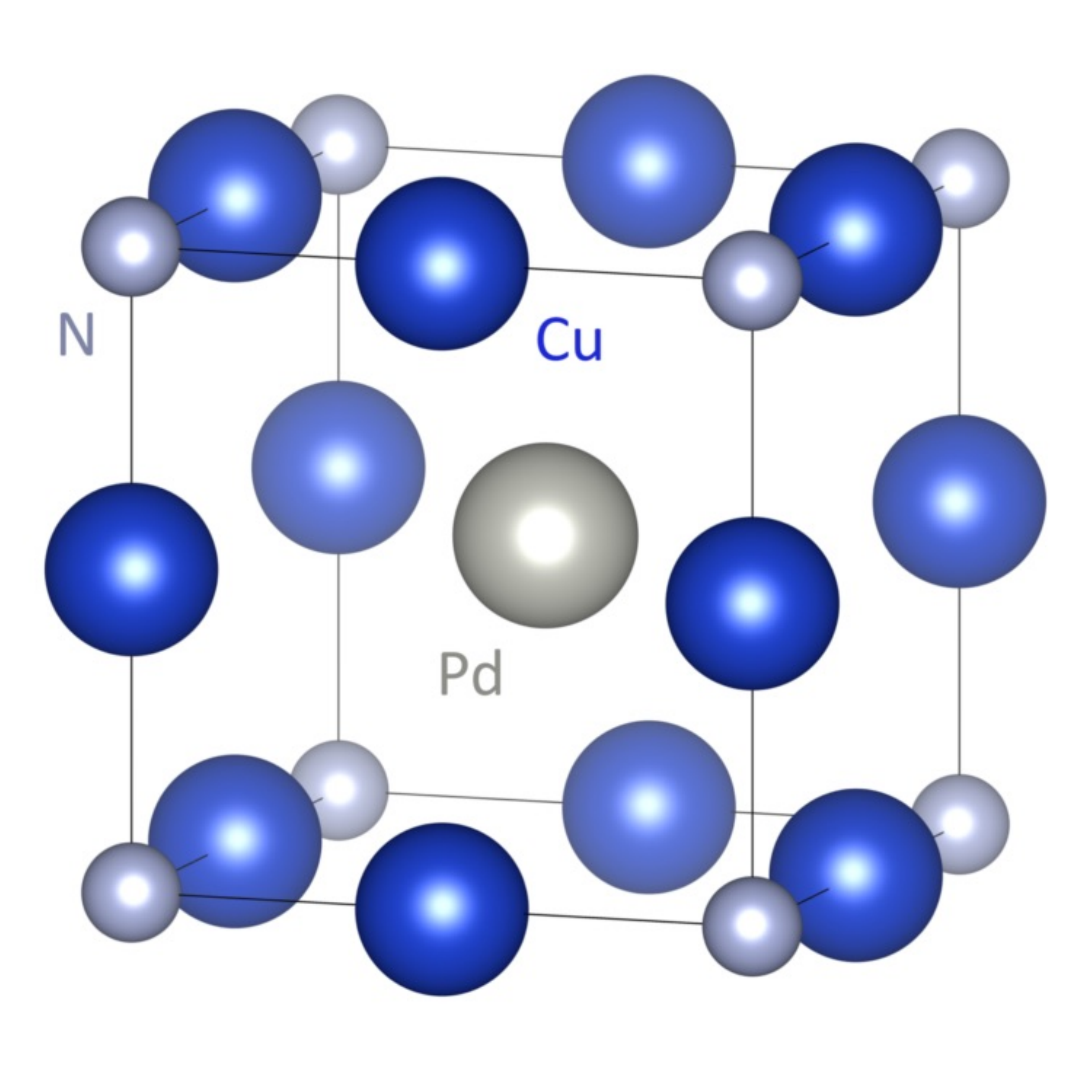} &
\includegraphics[width=1.25in]{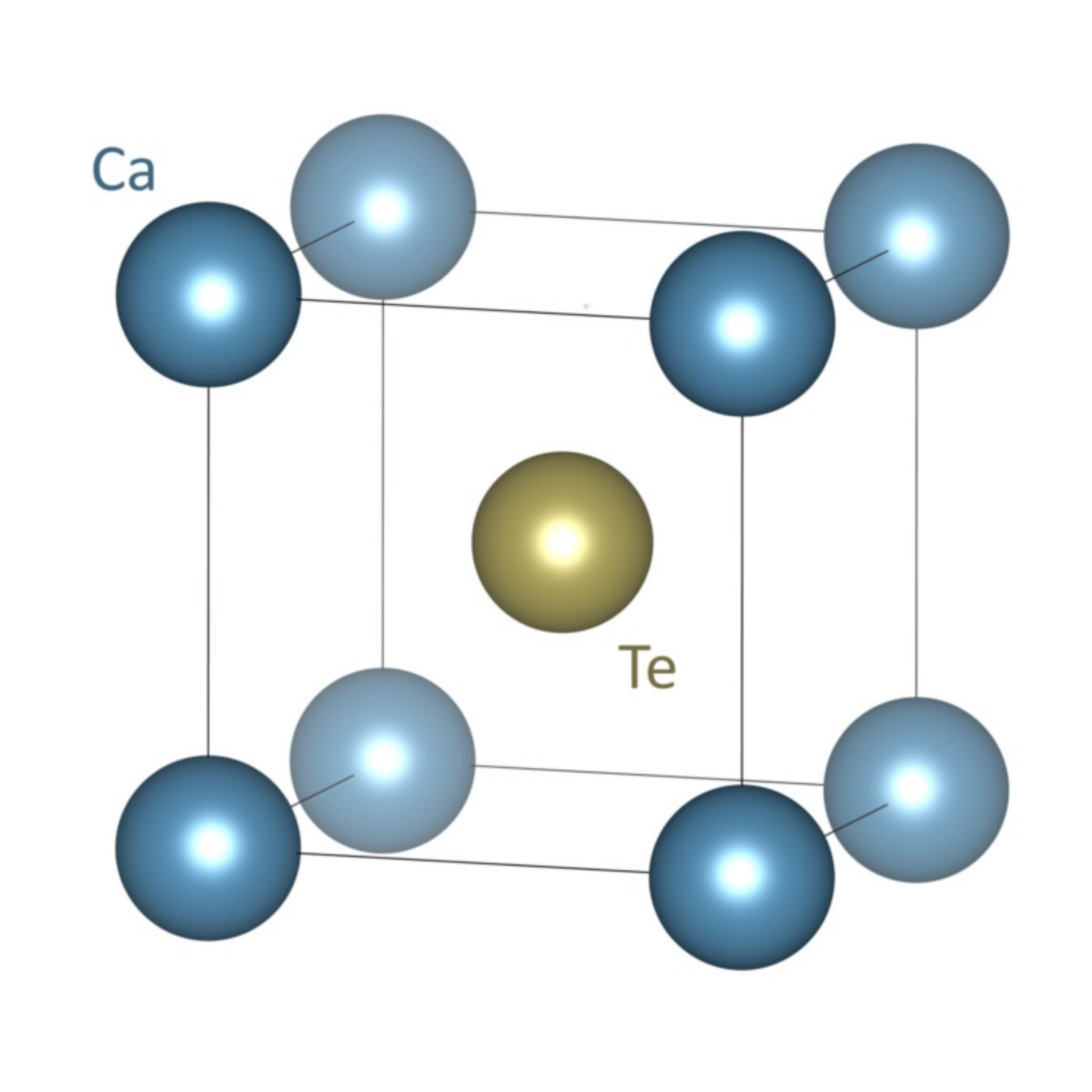} &
\includegraphics[width=1.2in]{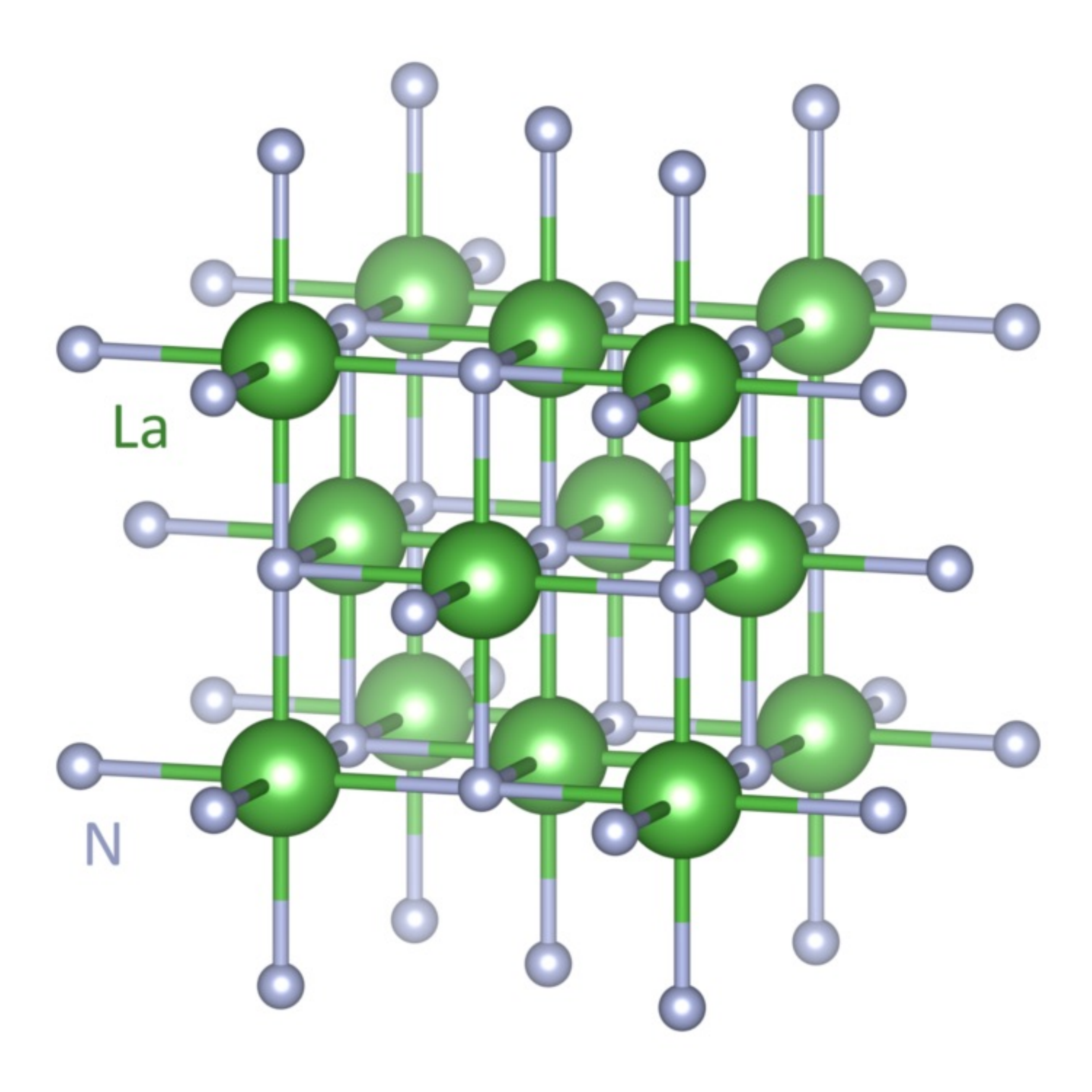} \\
\hline

\raisebox{2\totalheight}{\parbox[c|]{2cm}{\raggedright Band structure without SOC}} &
\includegraphics[width=1.5in]{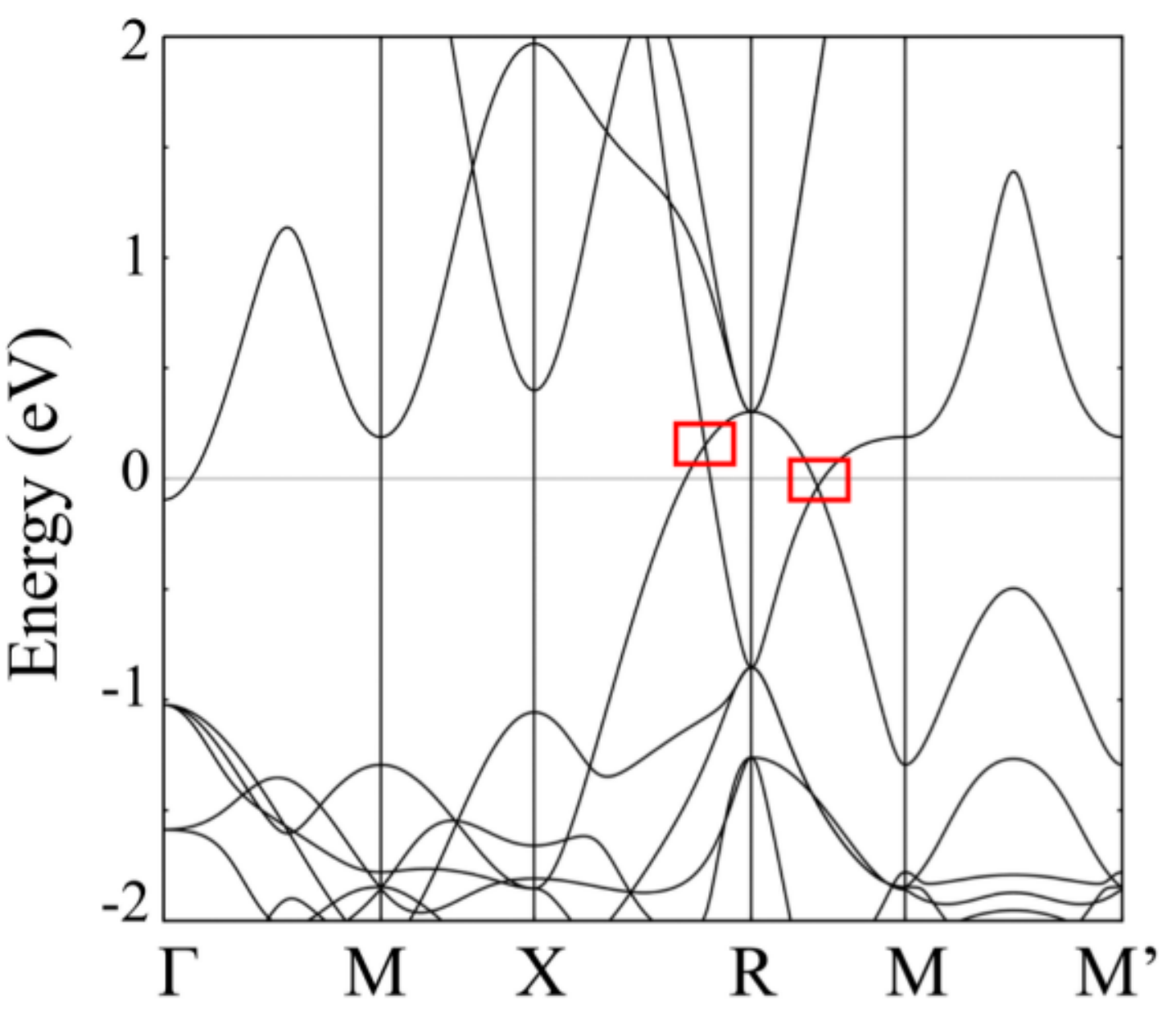} &
\includegraphics[width=1.5in]{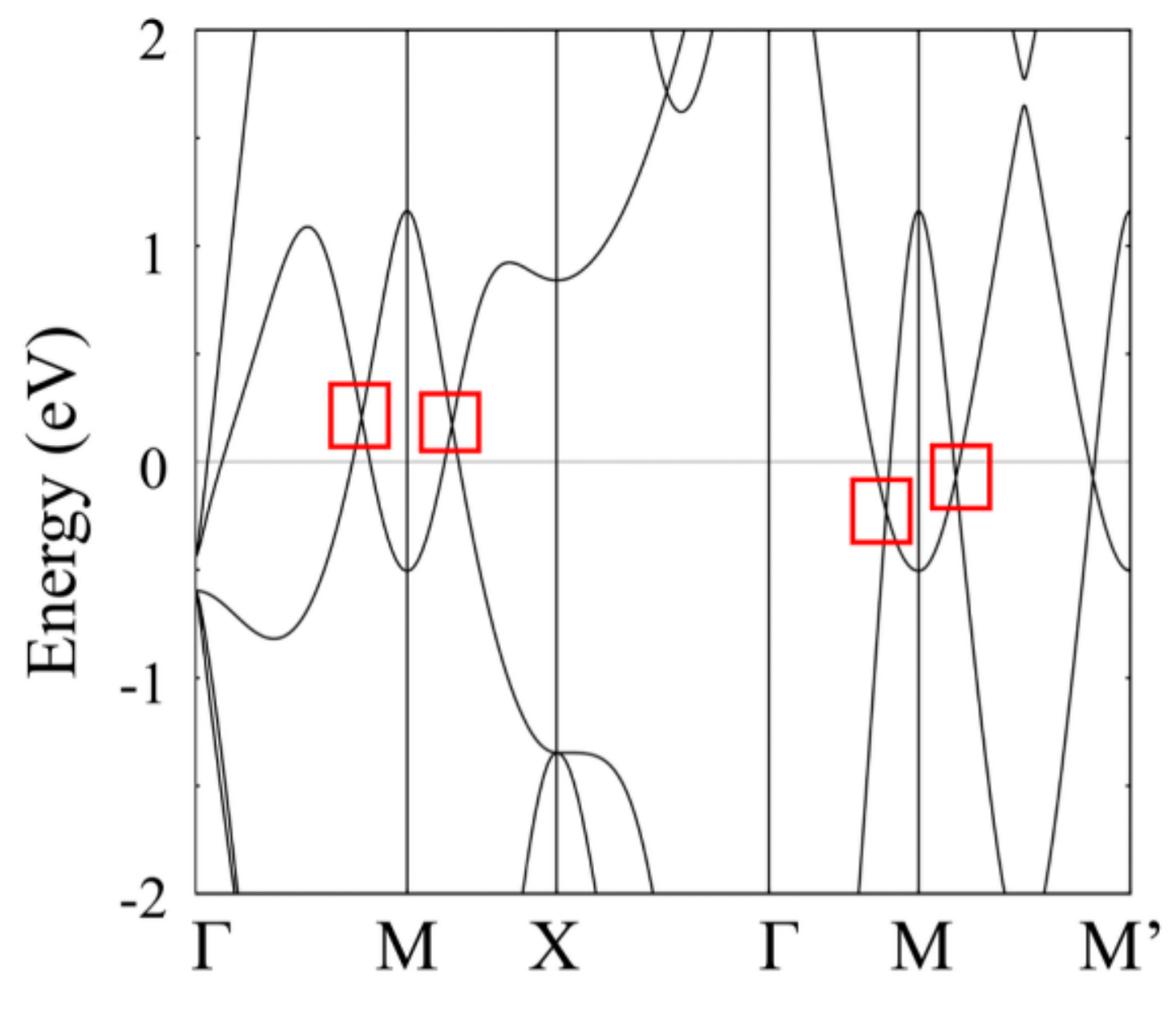} &
\includegraphics[width=1.5in]{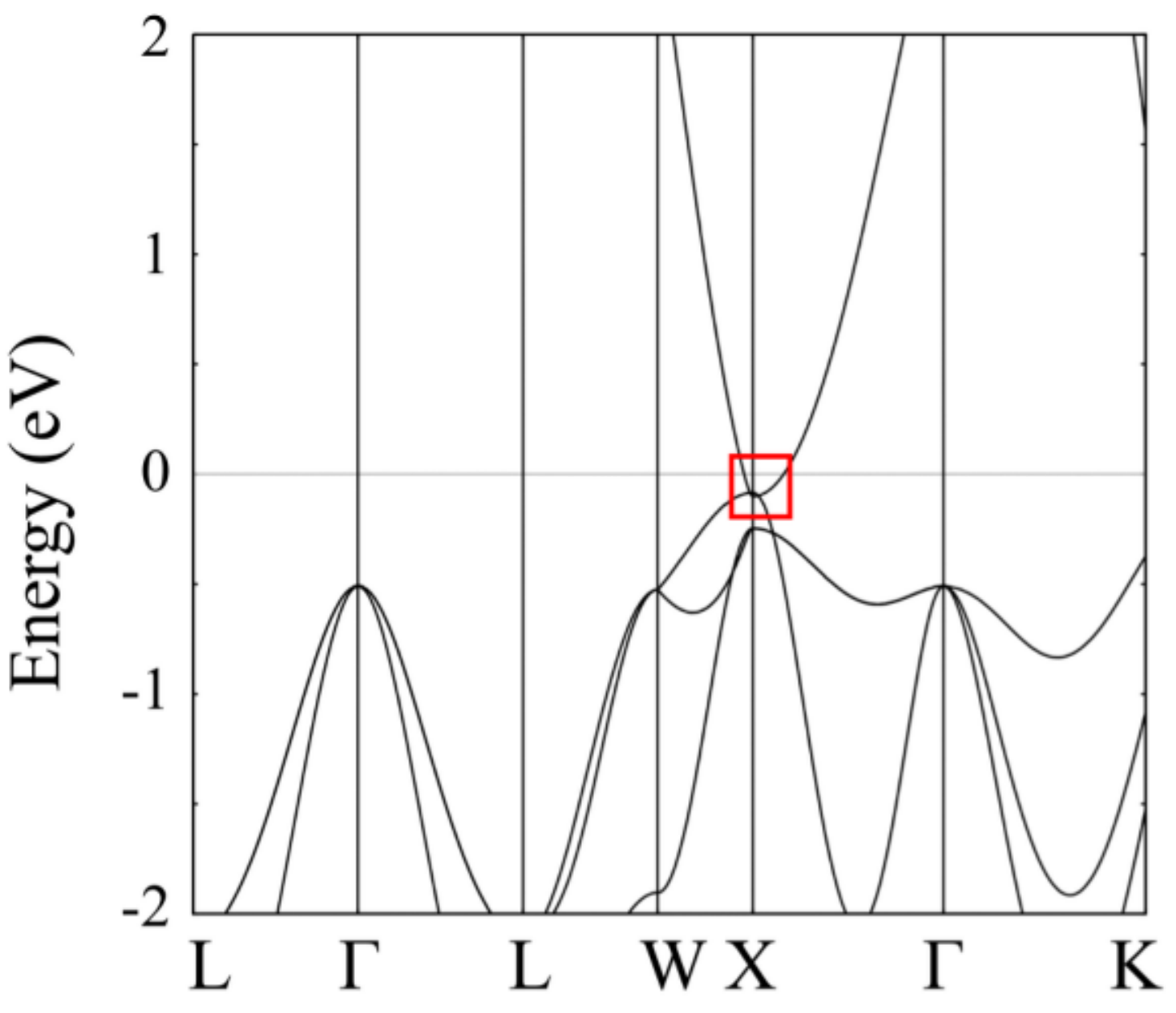} \\
\hline

\raisebox{2.5\totalheight}{\parbox[c|]{2cm}{\raggedright Band structure with SOC}} &
\includegraphics[width=1.5in]{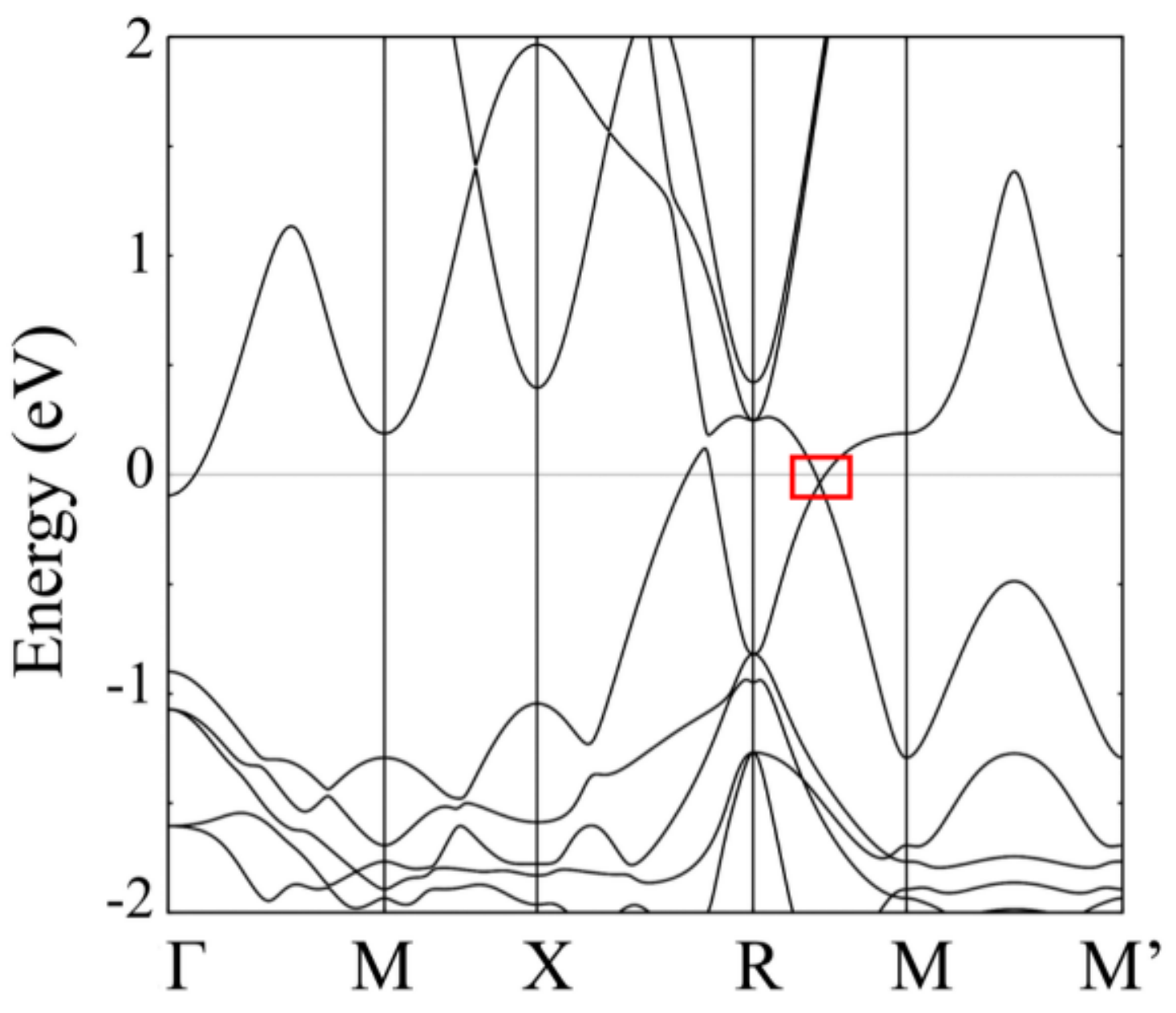} &
\includegraphics[width=1.5in]{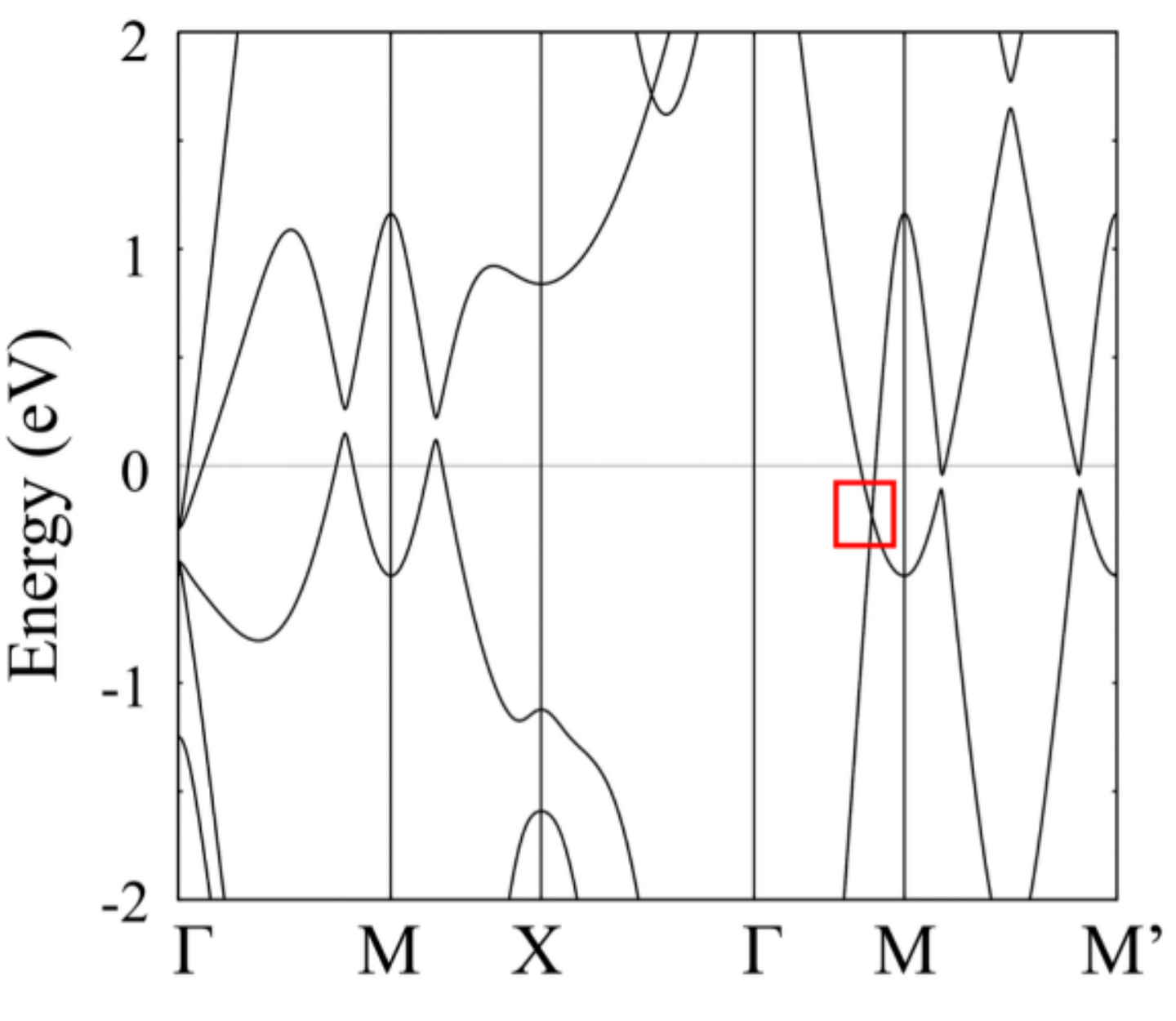} &
\includegraphics[width=1.5in]{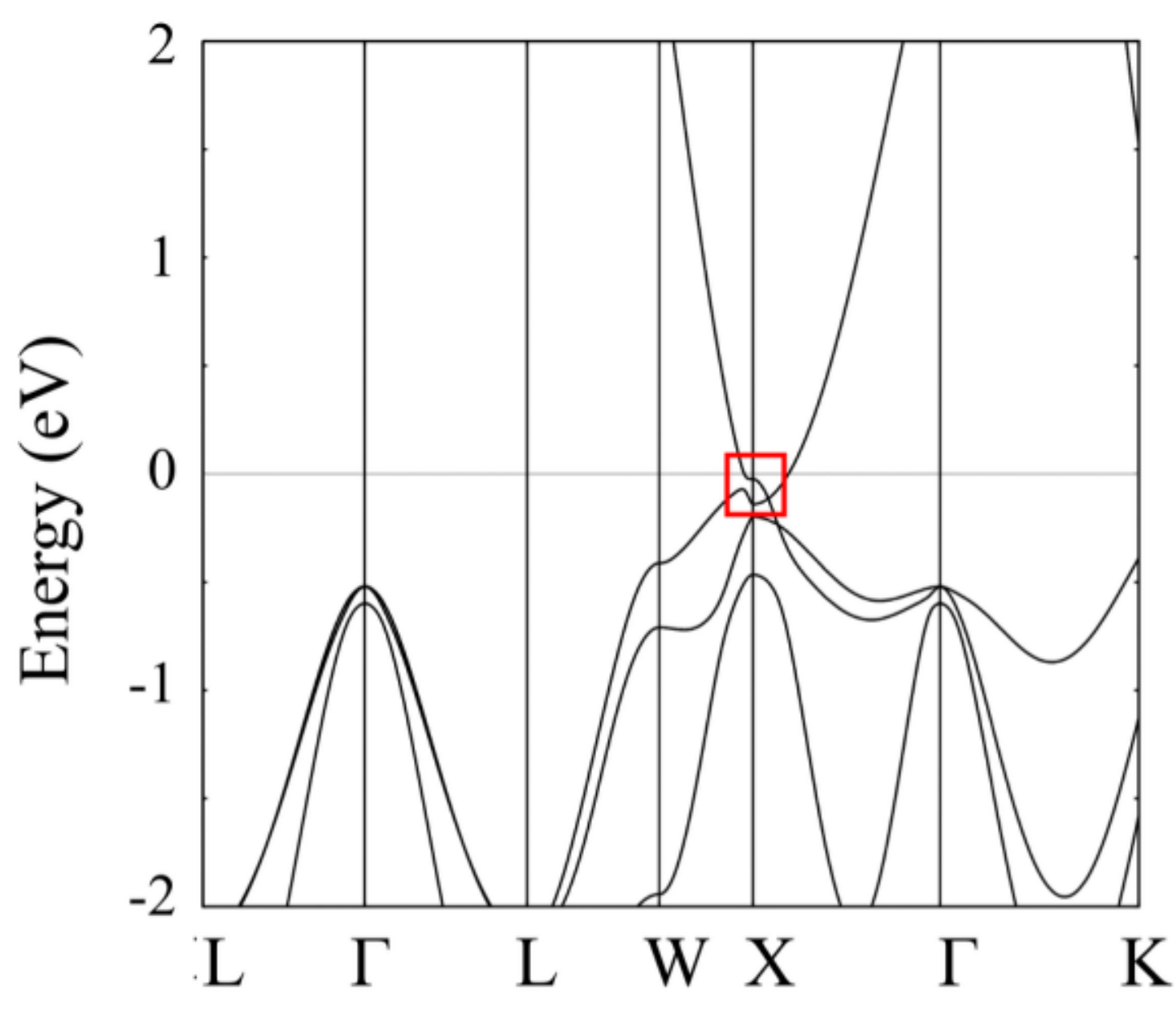} \\
\hline

\raisebox{1.5\totalheight}{\parbox[c|]{2cm}{\raggedright Nodal line distribution without SOC}} &
\includegraphics[width=1.15in]{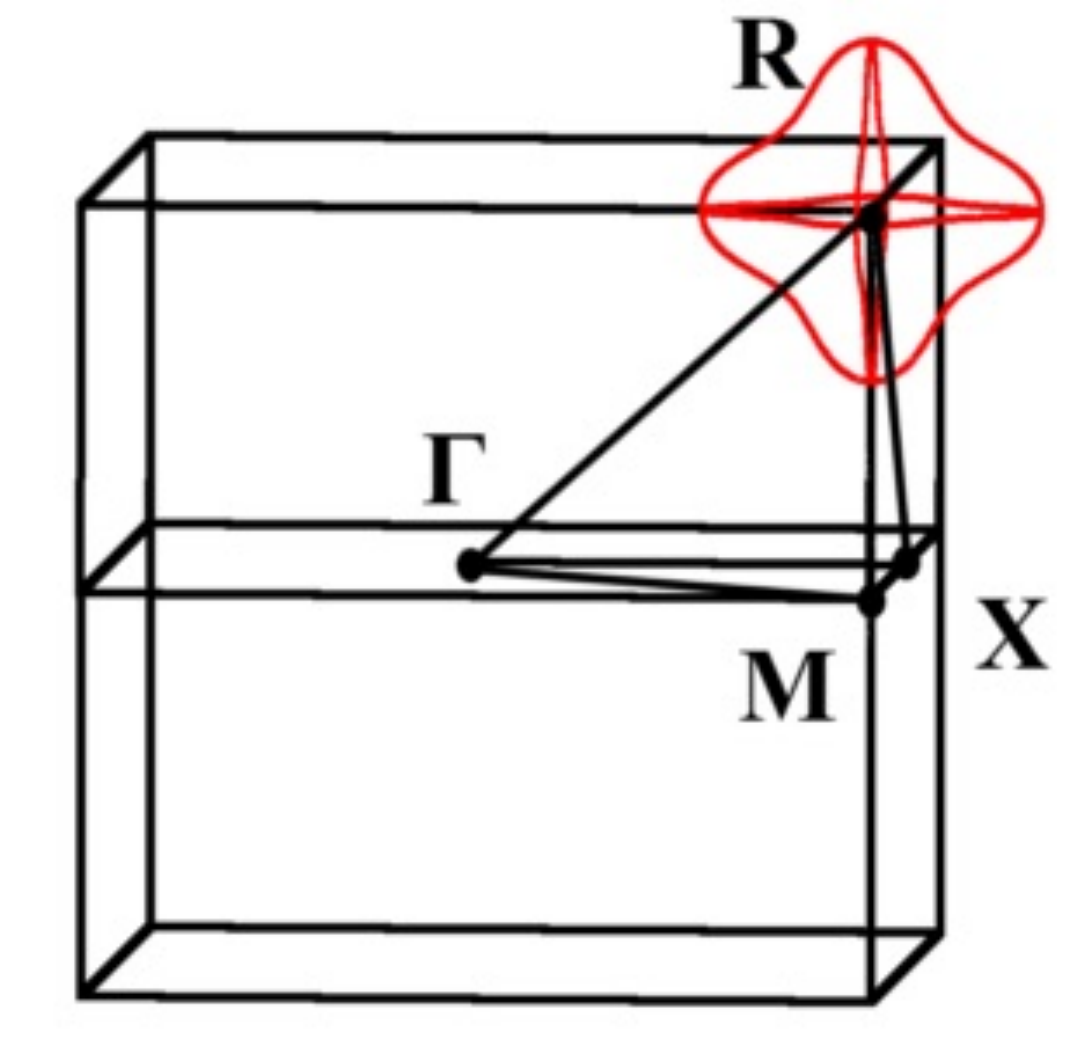} &
\includegraphics[width=1.2in]{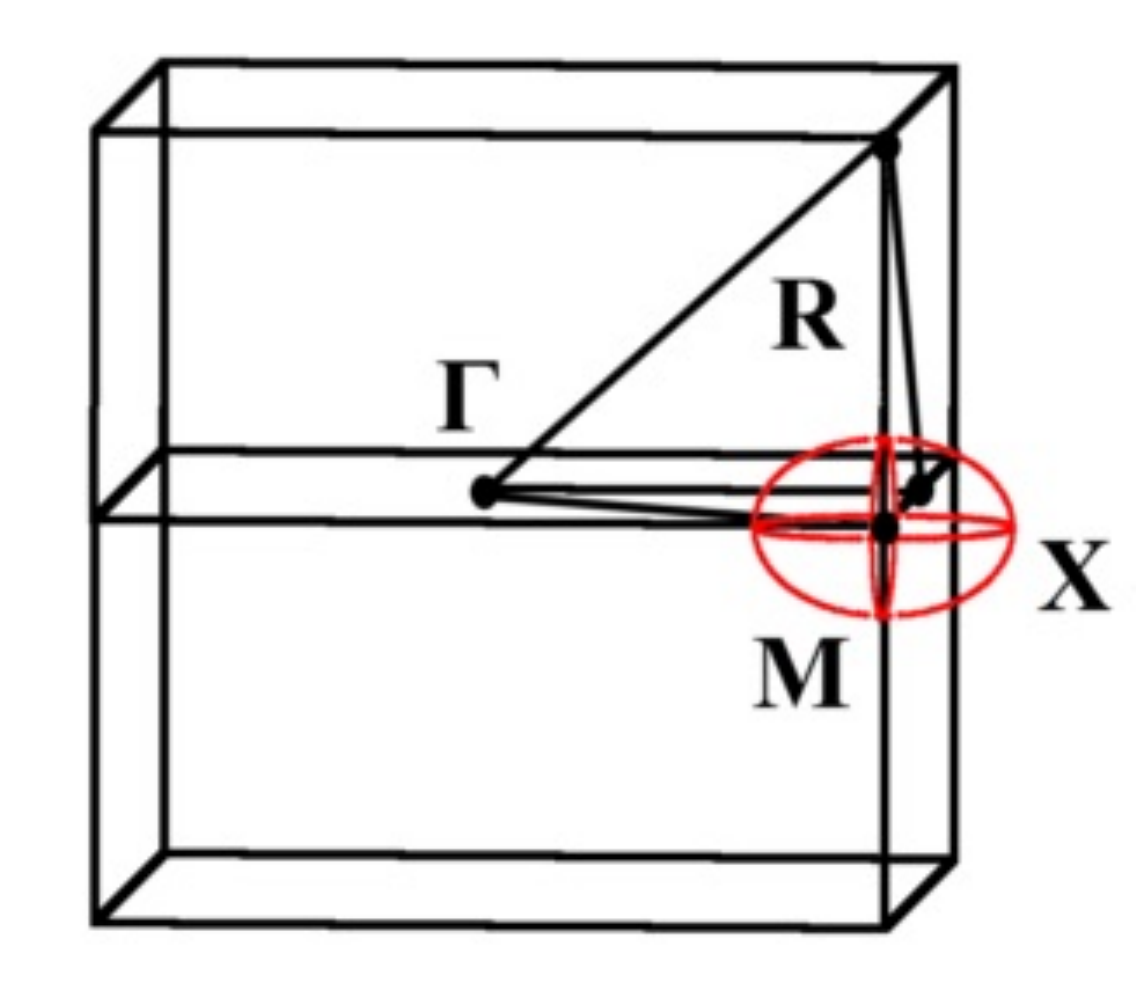} &
\includegraphics[width=1.1in]{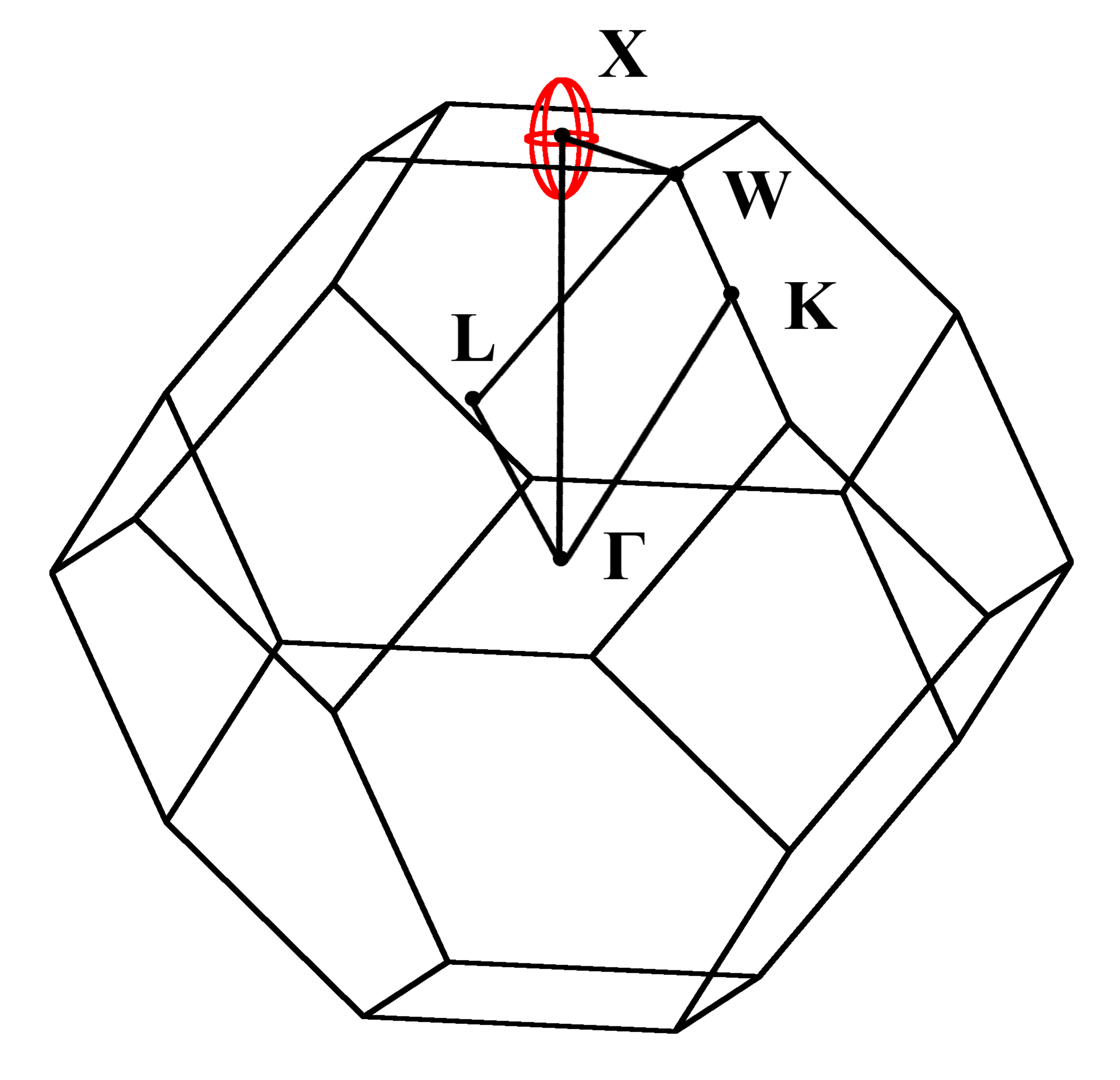} \\
\hline

\raisebox{0.1\totalheight}{\parbox[c|]{2cm}{\raggedright References}} &
{[83], [84]} &
{[92]} &
{[93], [97], [104-107]} \\
\hline

\bottomrule

\end{tabular}
\label{tab:gt}
\end{adjustbox}

\end{table*}

\begin{table*}[htbp]
\caption{Time-reversal and inversion symmetry protected nodal line materials (continued)}
\centering

\begin{adjustbox}{width=0.9\textwidth}

\newcommand*{\TitleParbox}[1]{\parbox[c]{1.75cm}{\raggedright #1}}%

\makeatletter
\newcommand{\thickhline}{%
    \noalign {\ifnum 0=`}\fi \hrule height 1pt
    \futurelet \reserved@a \@xhline
}
\newcolumntype{"}{@{\hskip\tabcolsep\vrule width 1pt\hskip\tabcolsep}}
\makeatother

\newcolumntype{?}{!{\vrule width 1pt}}

\hspace*{-1cm}
\begin{tabular}{c?c c c c} 
 
 \toprule

\textbf{Materials} & \textbf{Ca$_3$P$_2$} & \textbf{BaAs$_3$} & \textbf{BaSn$_2$} & \textbf{TiB$_2$} \\
\thickhline

\midrule

\raisebox{3\totalheight}{\parbox[c|]{2cm}{\raggedright Crystal structure}} &
\includegraphics[width=1.6in]{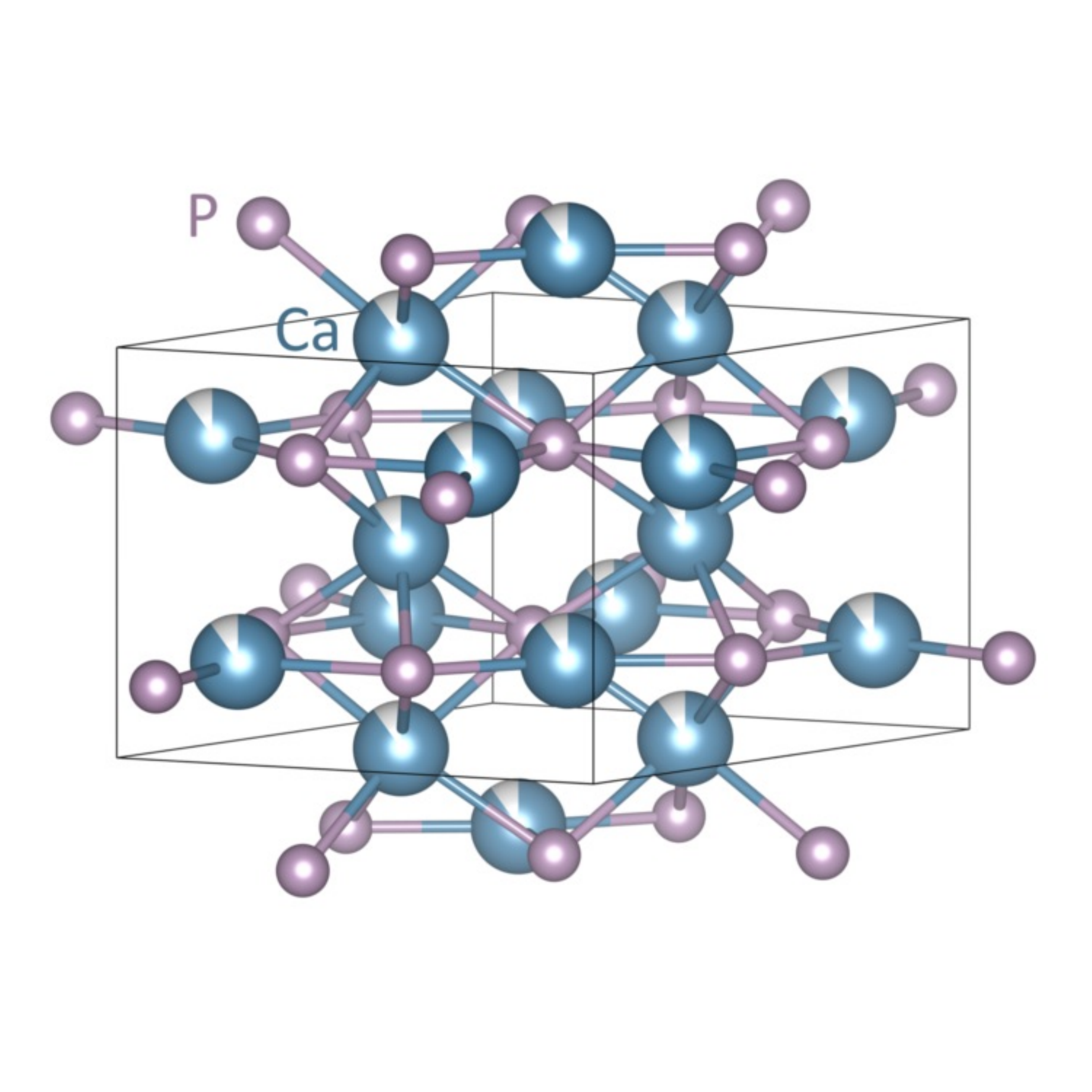} &
\includegraphics[width=1.6in]{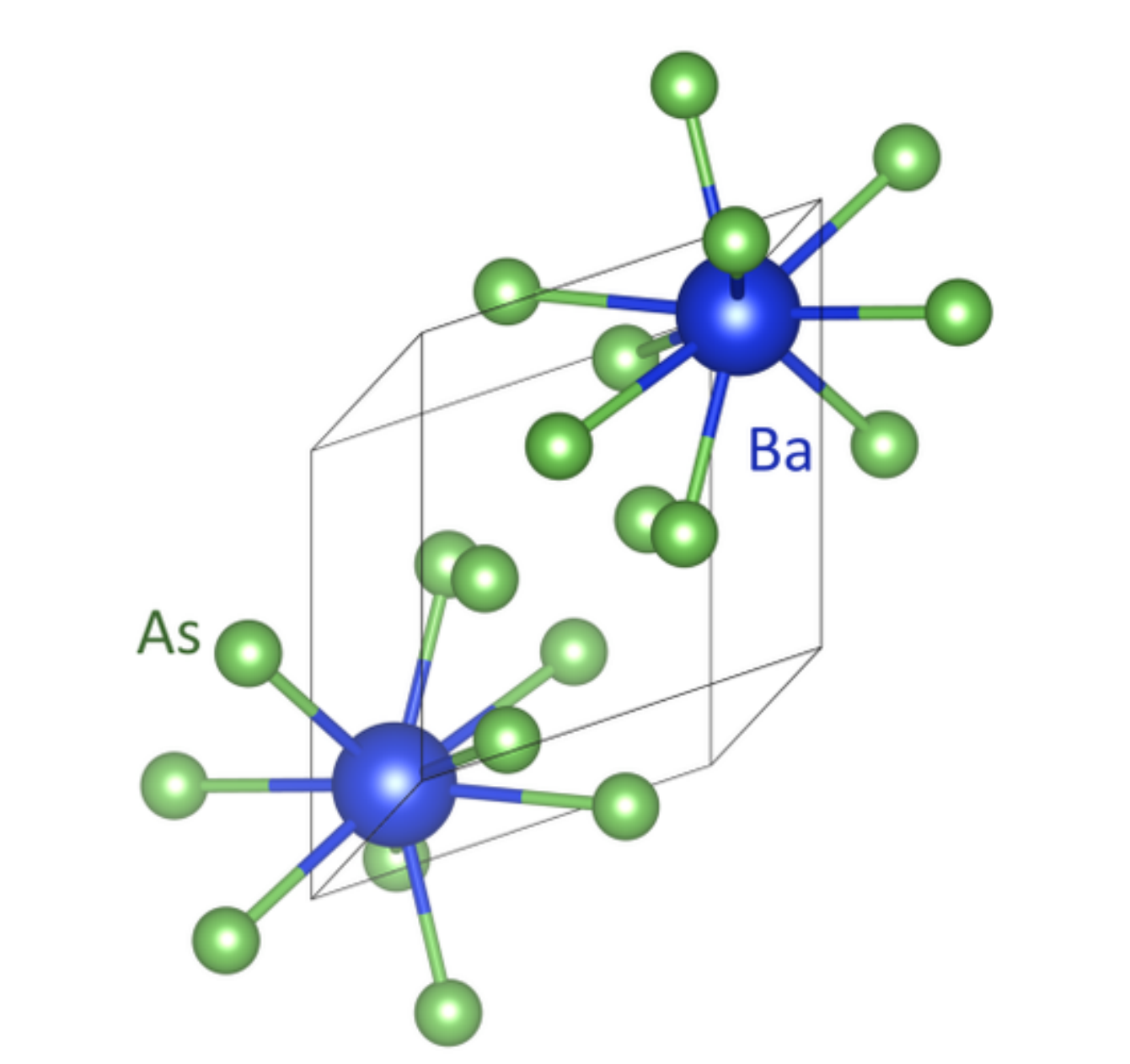} &
\includegraphics[width=1.6in]{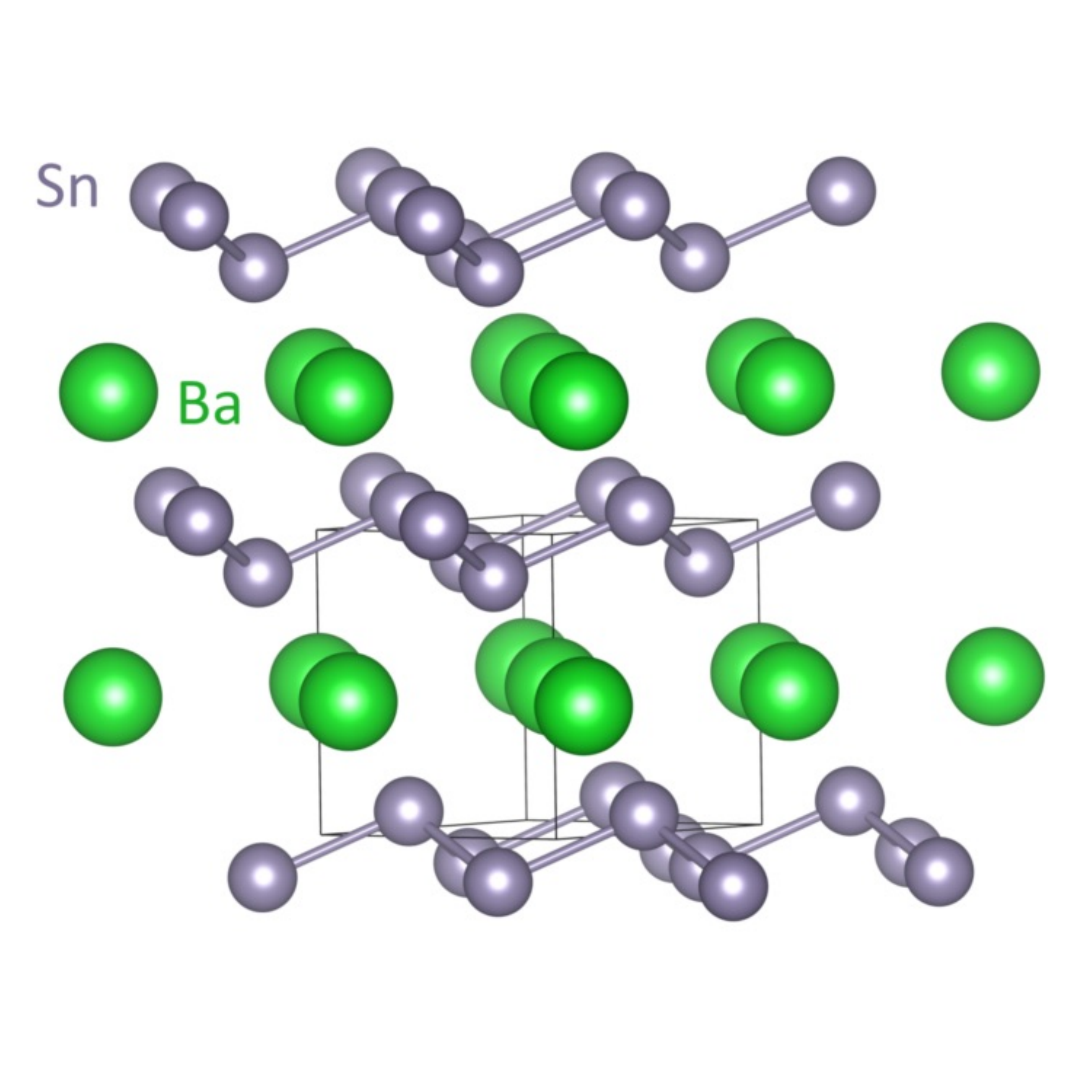} &
\includegraphics[width=1.6in]{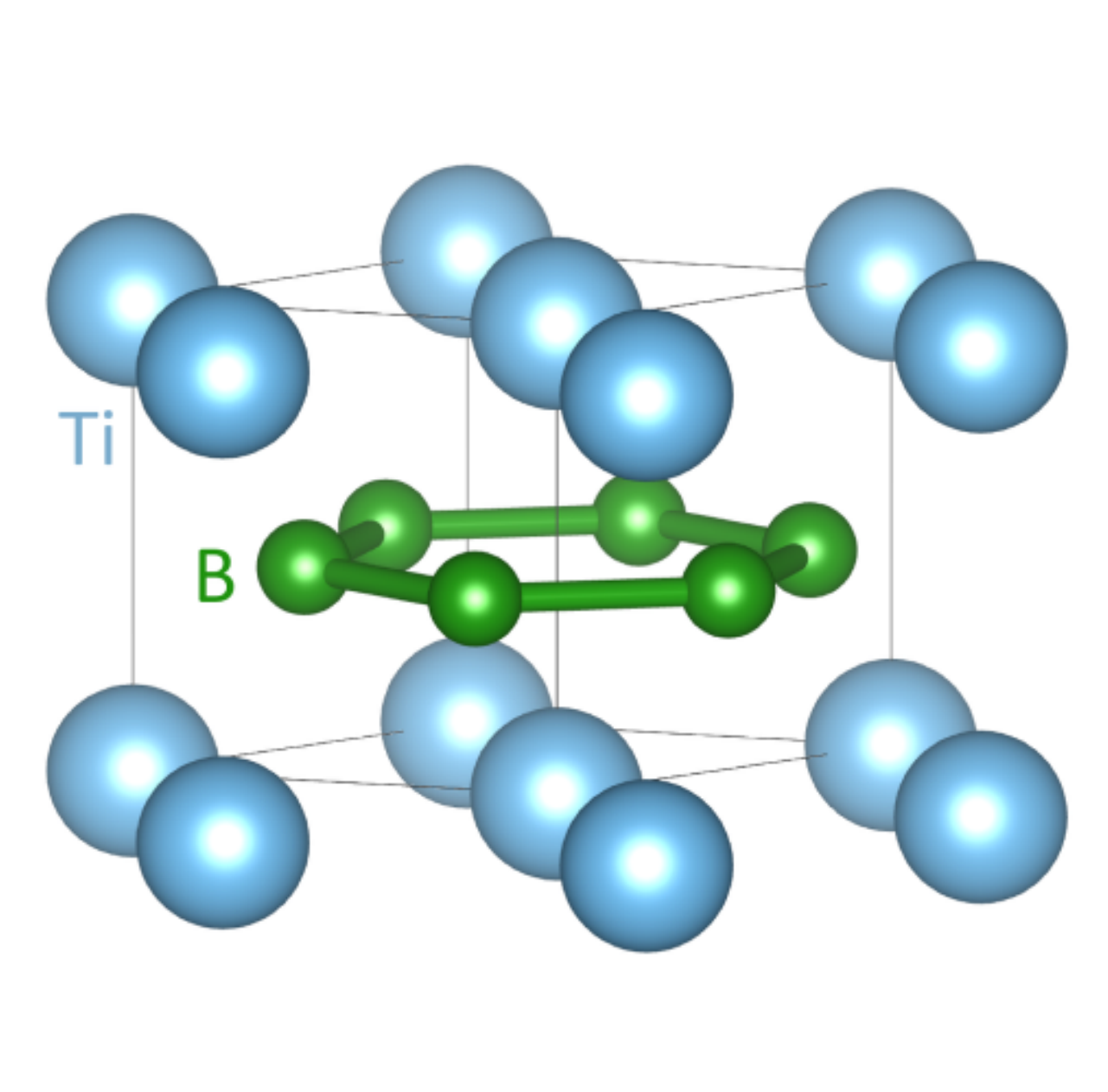} \\
\hline

\raisebox{2\totalheight}{\parbox[c|]{2cm}{\raggedright Band structure without SOC}} &
\includegraphics[width=1.55in]{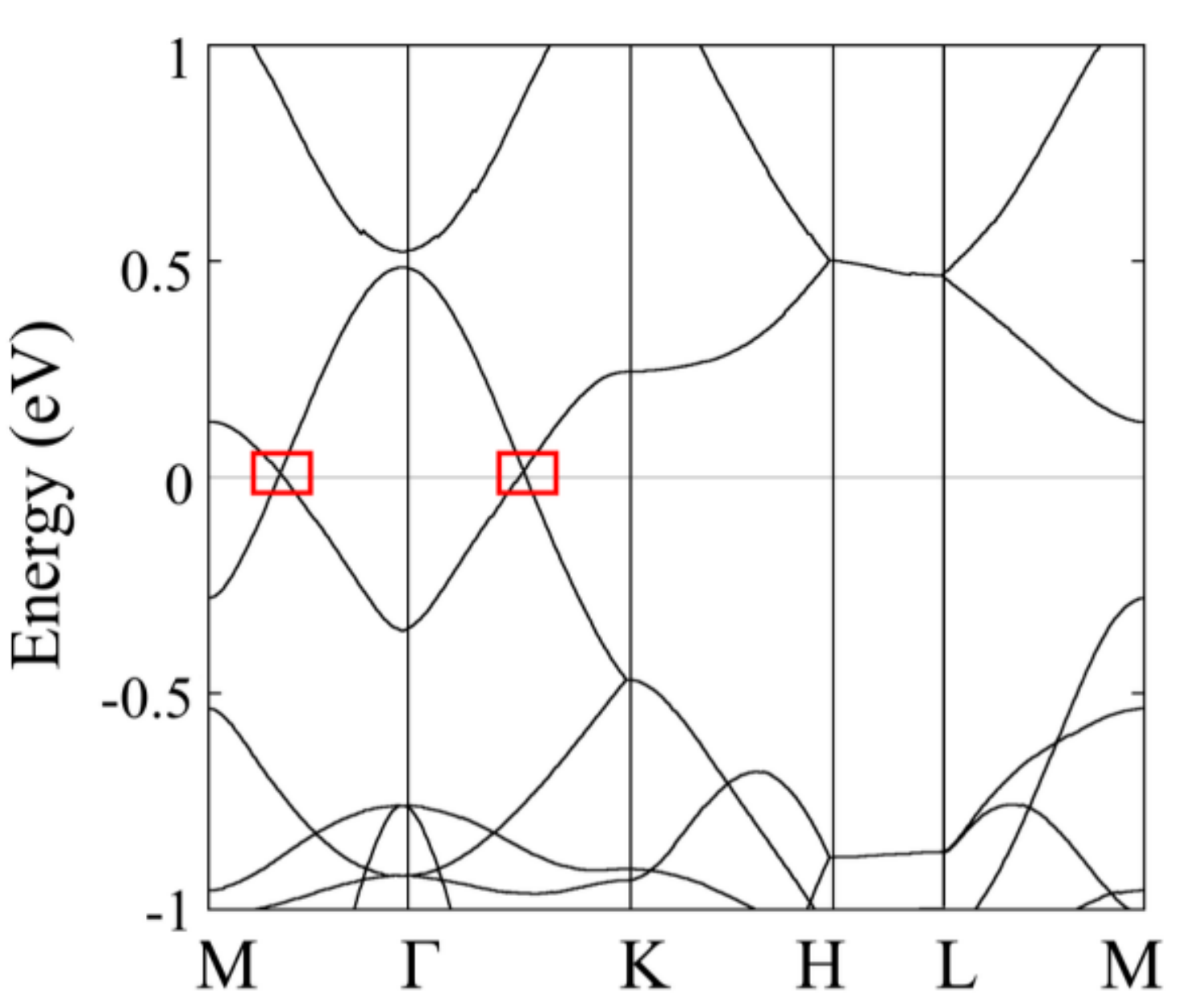} &
\includegraphics[width=1.55in]{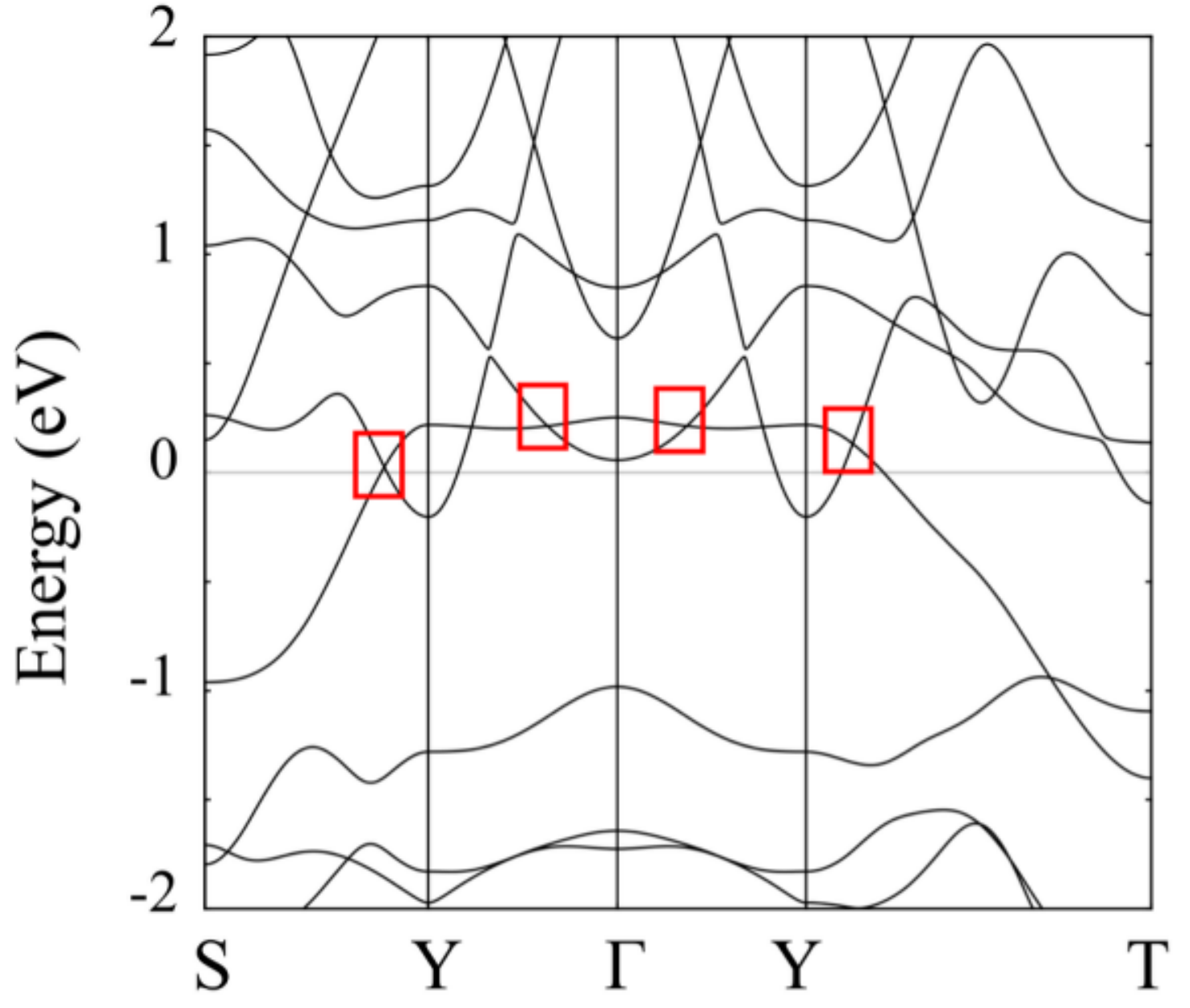} &
\includegraphics[width=1.5in]{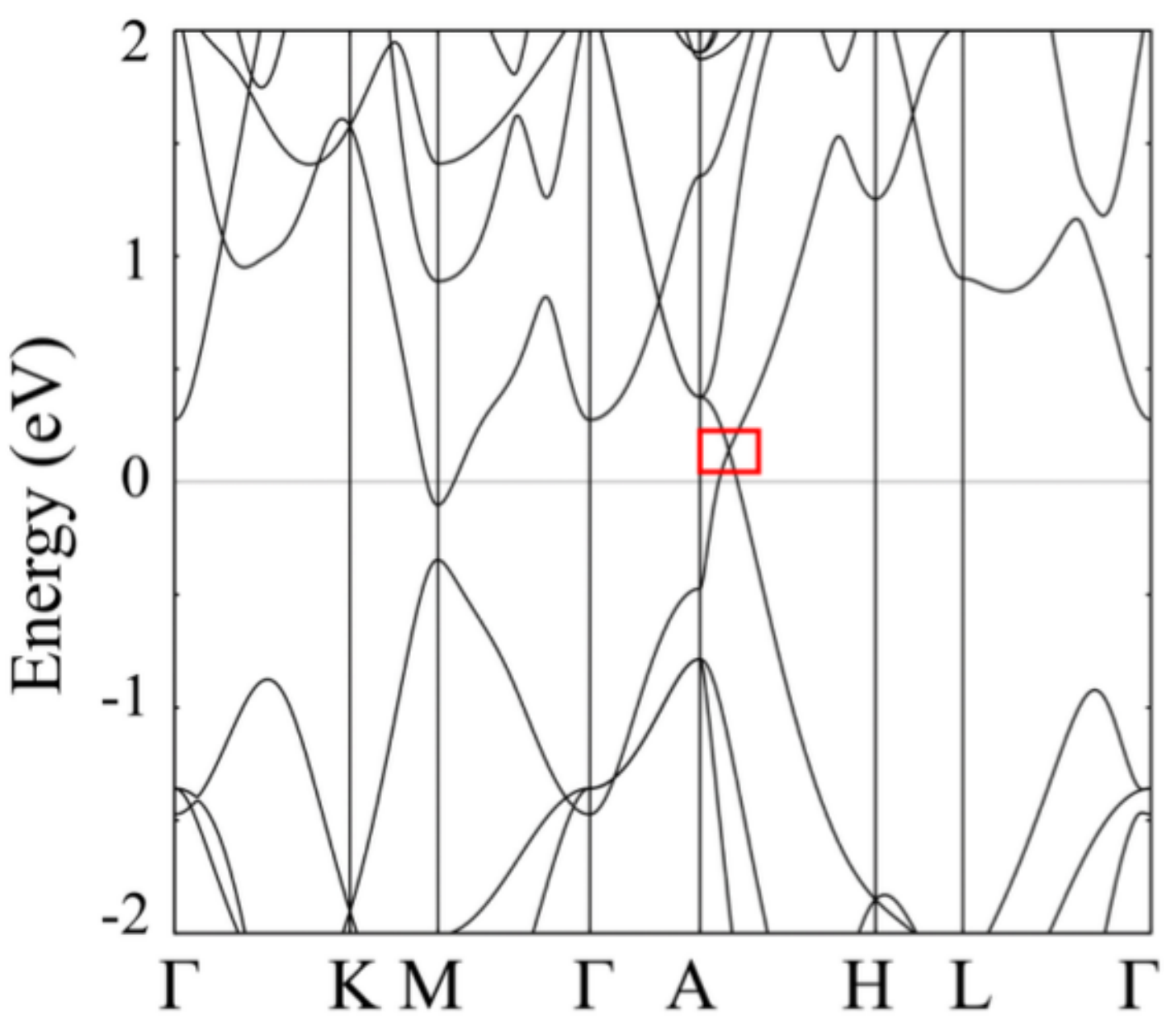} &
\includegraphics[width=1.58In]{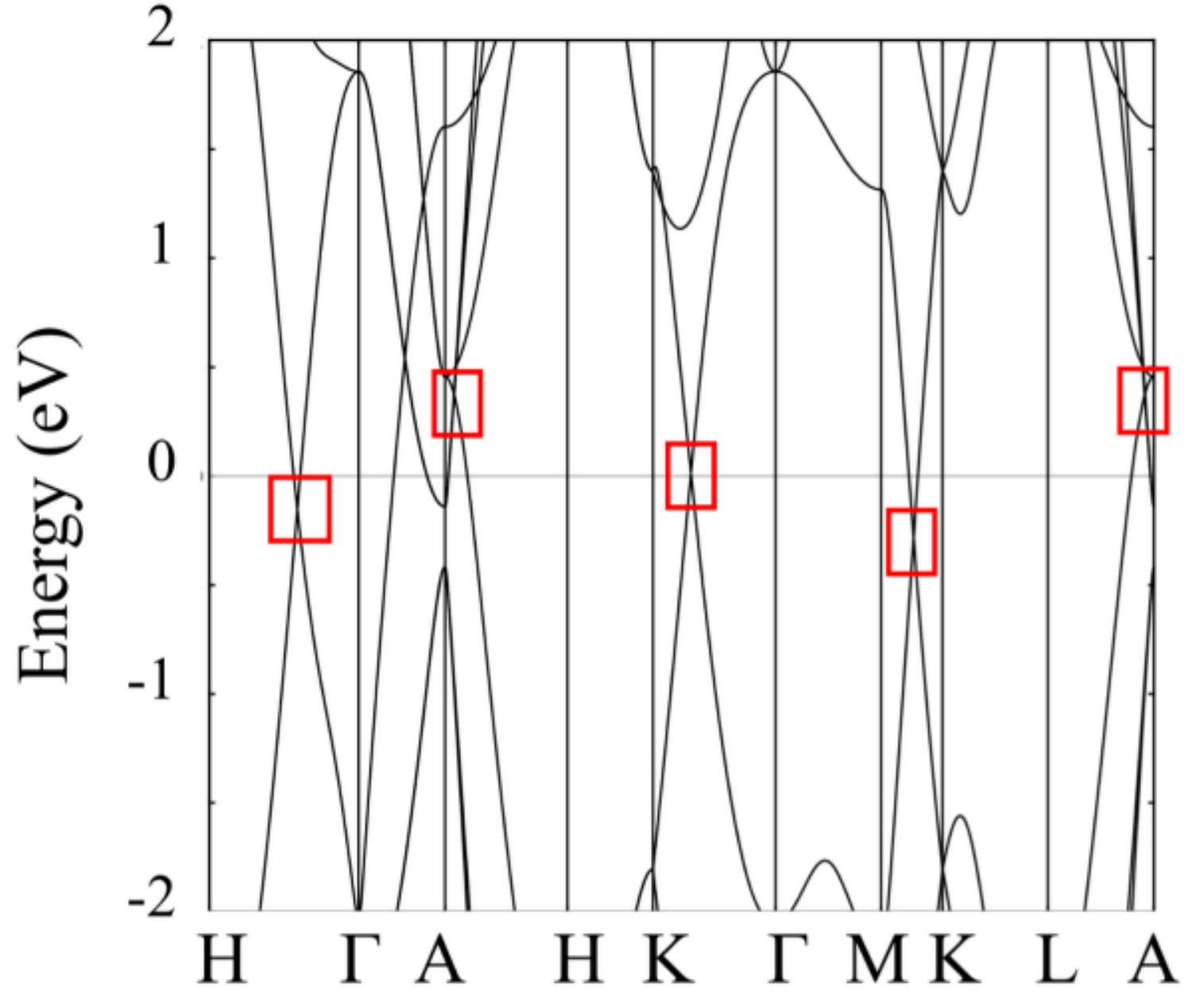} \\
\hline

\raisebox{2.5\totalheight}{\parbox[c|]{2cm}{\raggedright Band structure with SOC}} &
\includegraphics[width=1.55in]{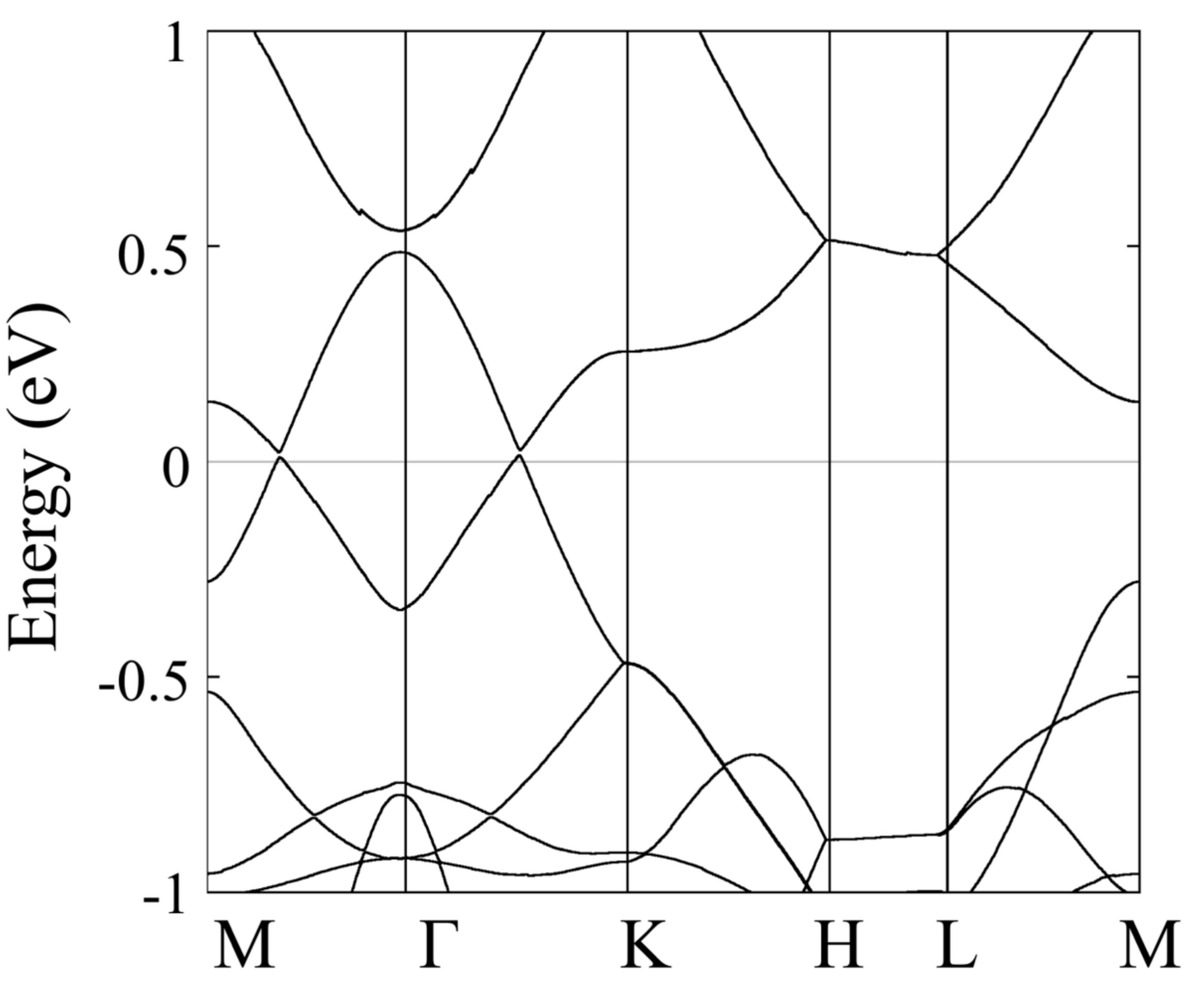} &
\includegraphics[width=1.5in]{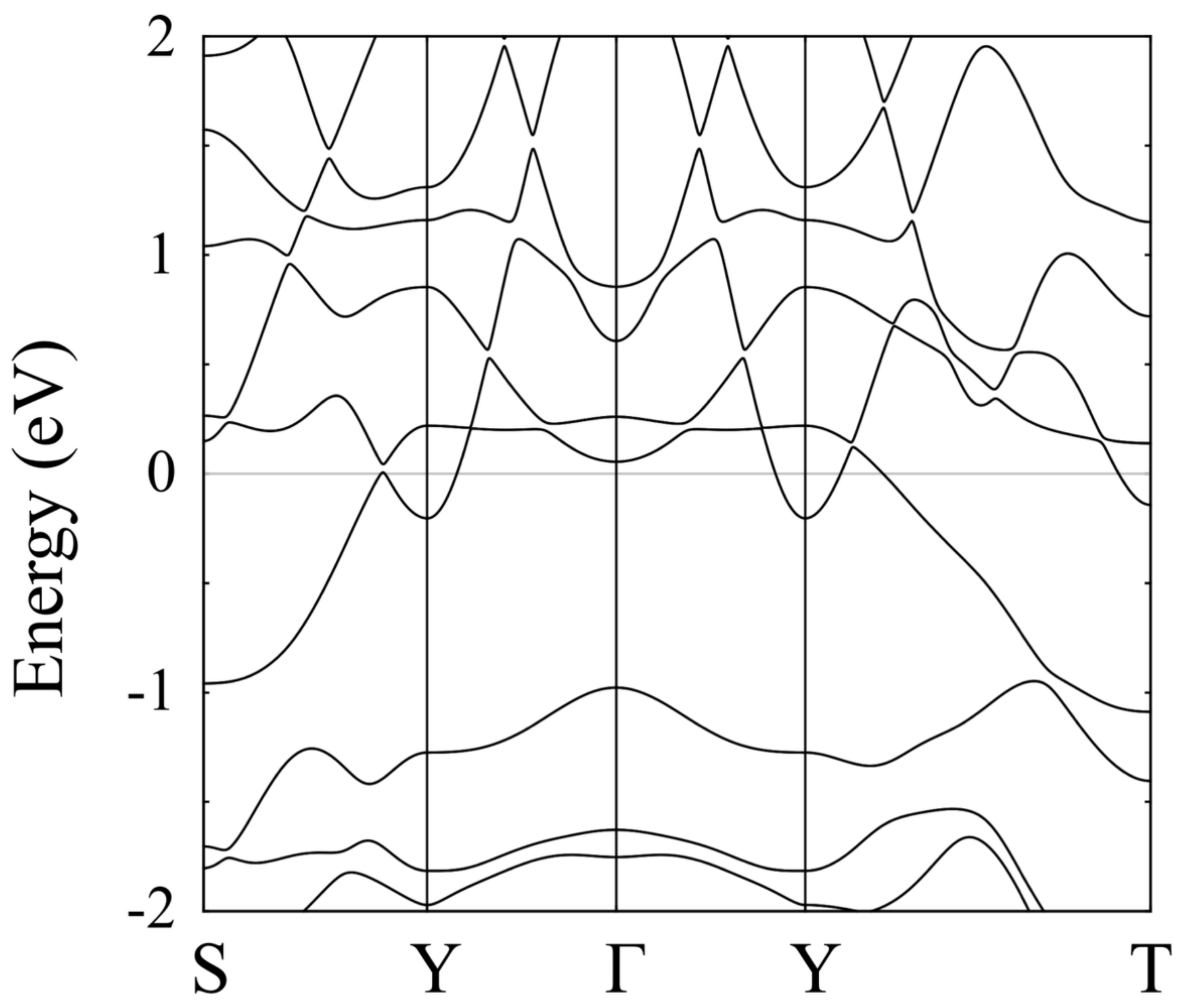} &
\includegraphics[width=1.5in]{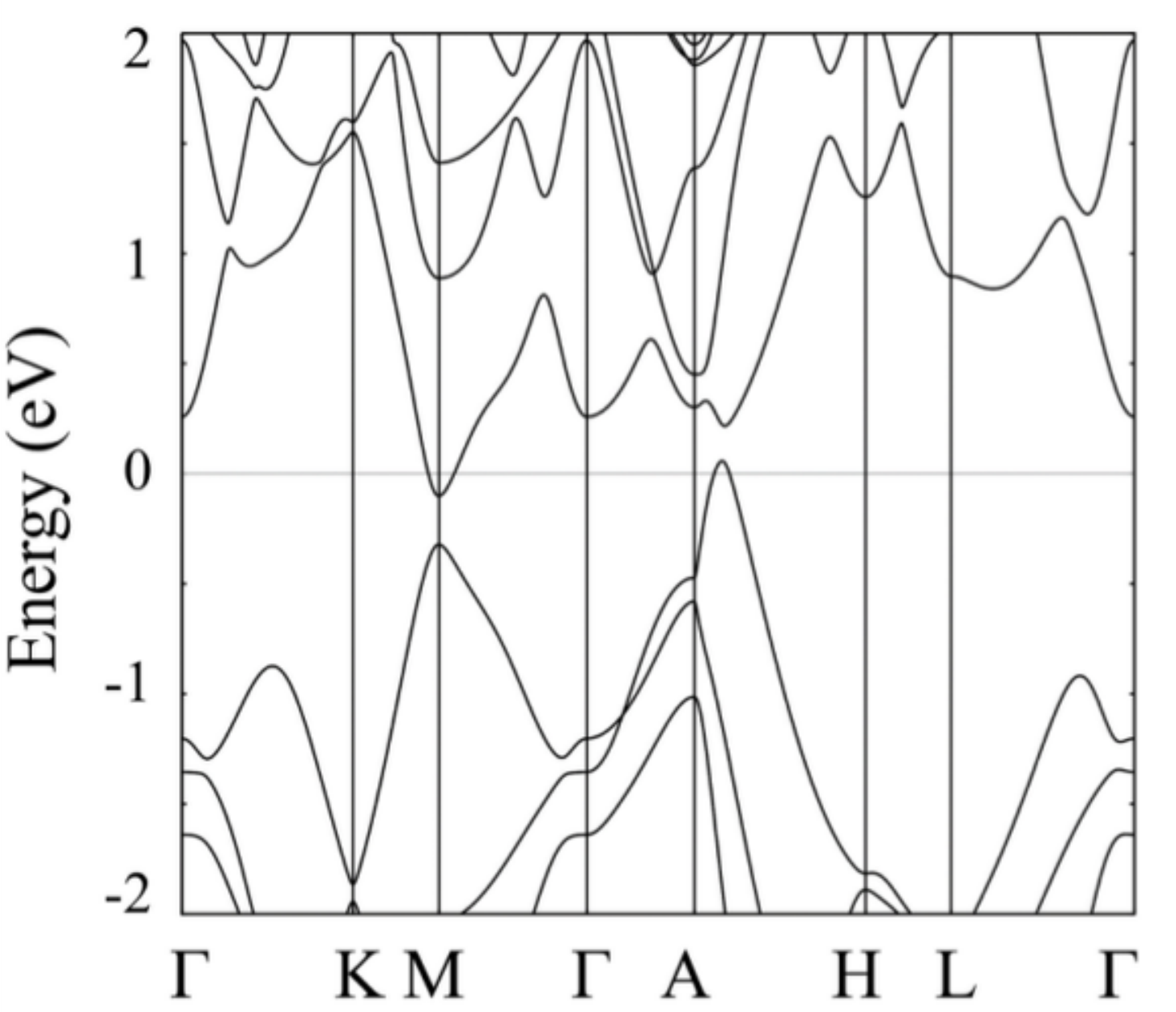} &
\includegraphics[width=1.58in]{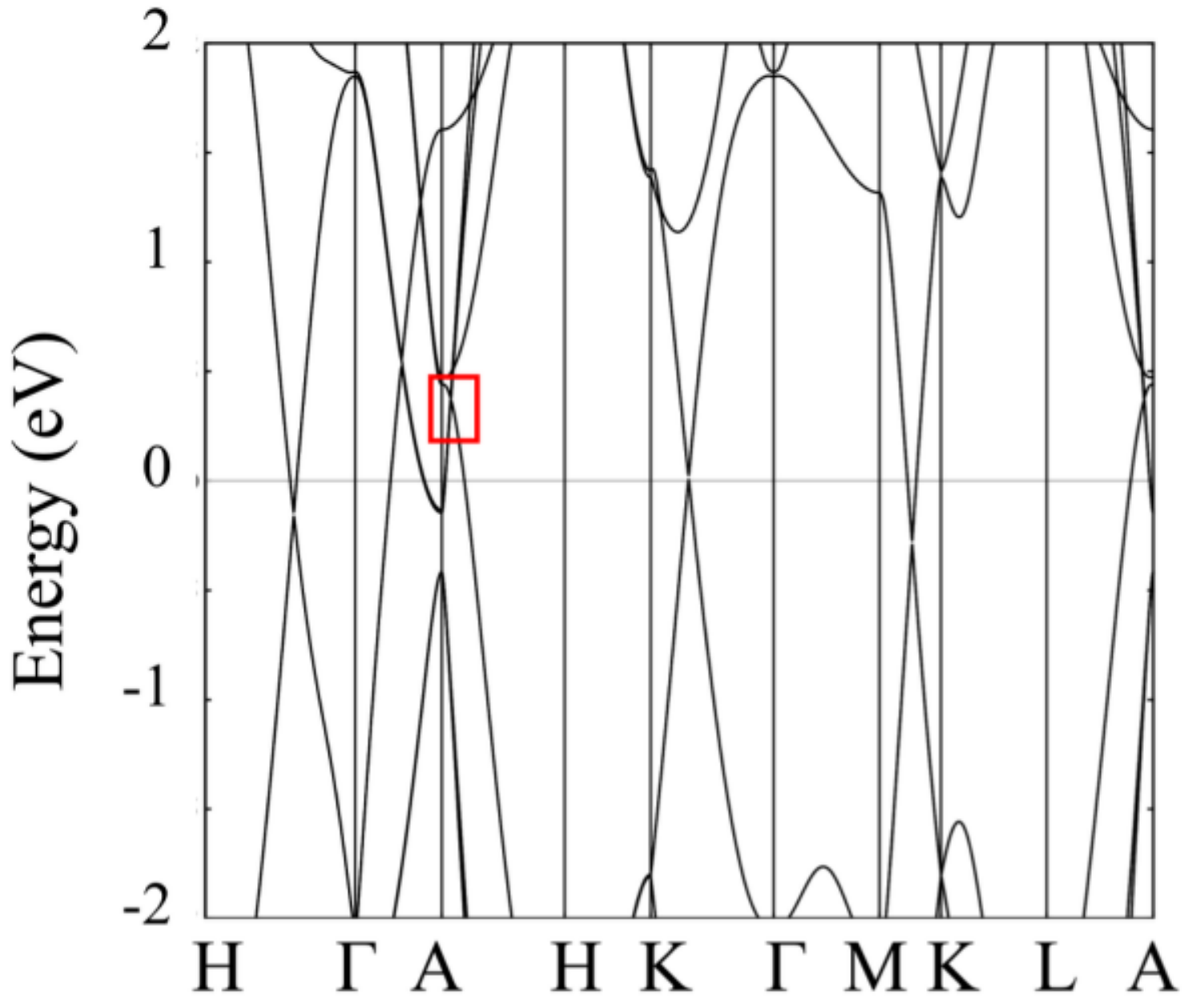} \\
\hline

\raisebox{1.5\totalheight}{\parbox[c|]{2cm}{\raggedright Nodal line distribution without SOC}} &
\includegraphics[width=1.1in]{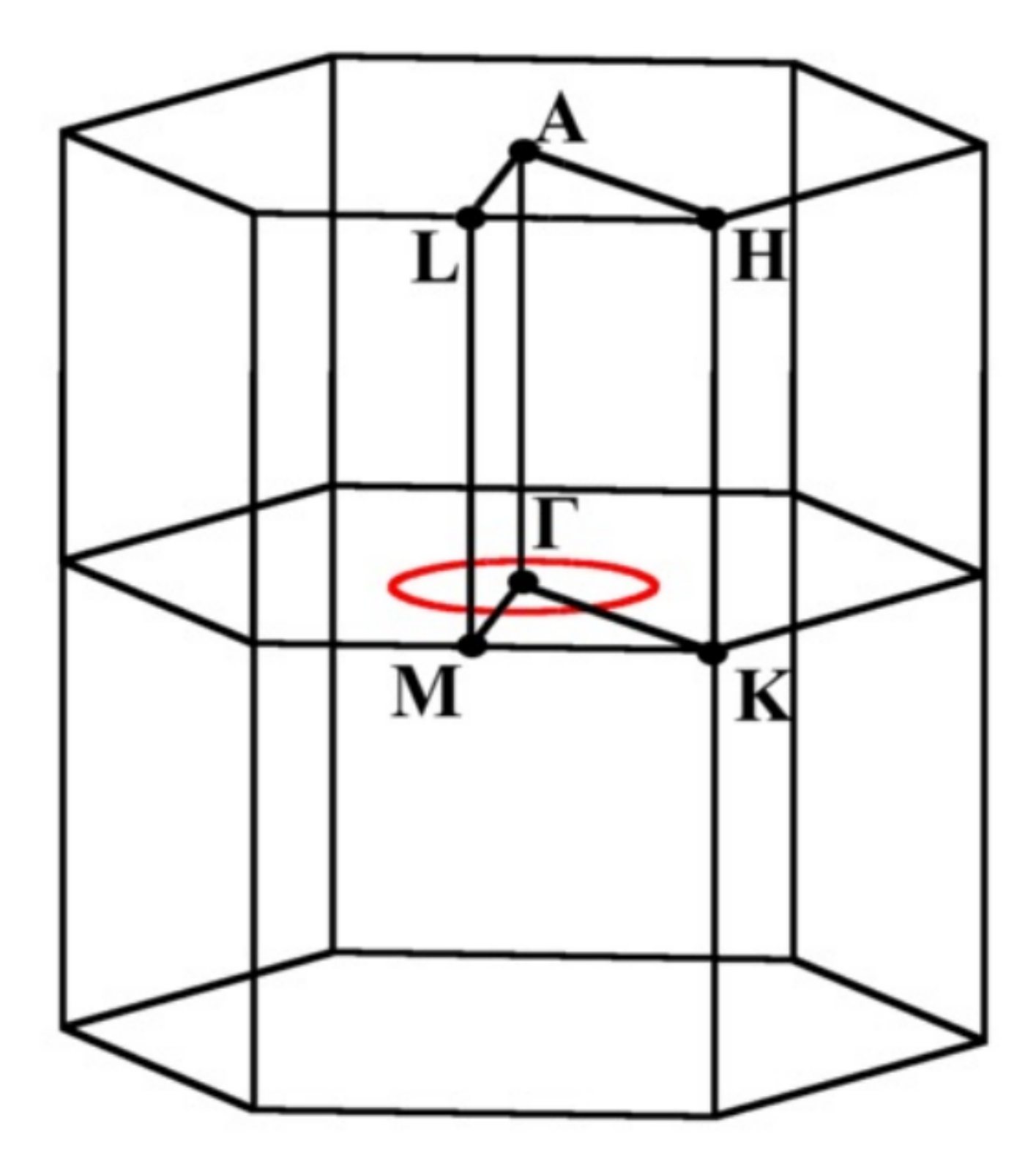} &
\includegraphics[width=1.3in]{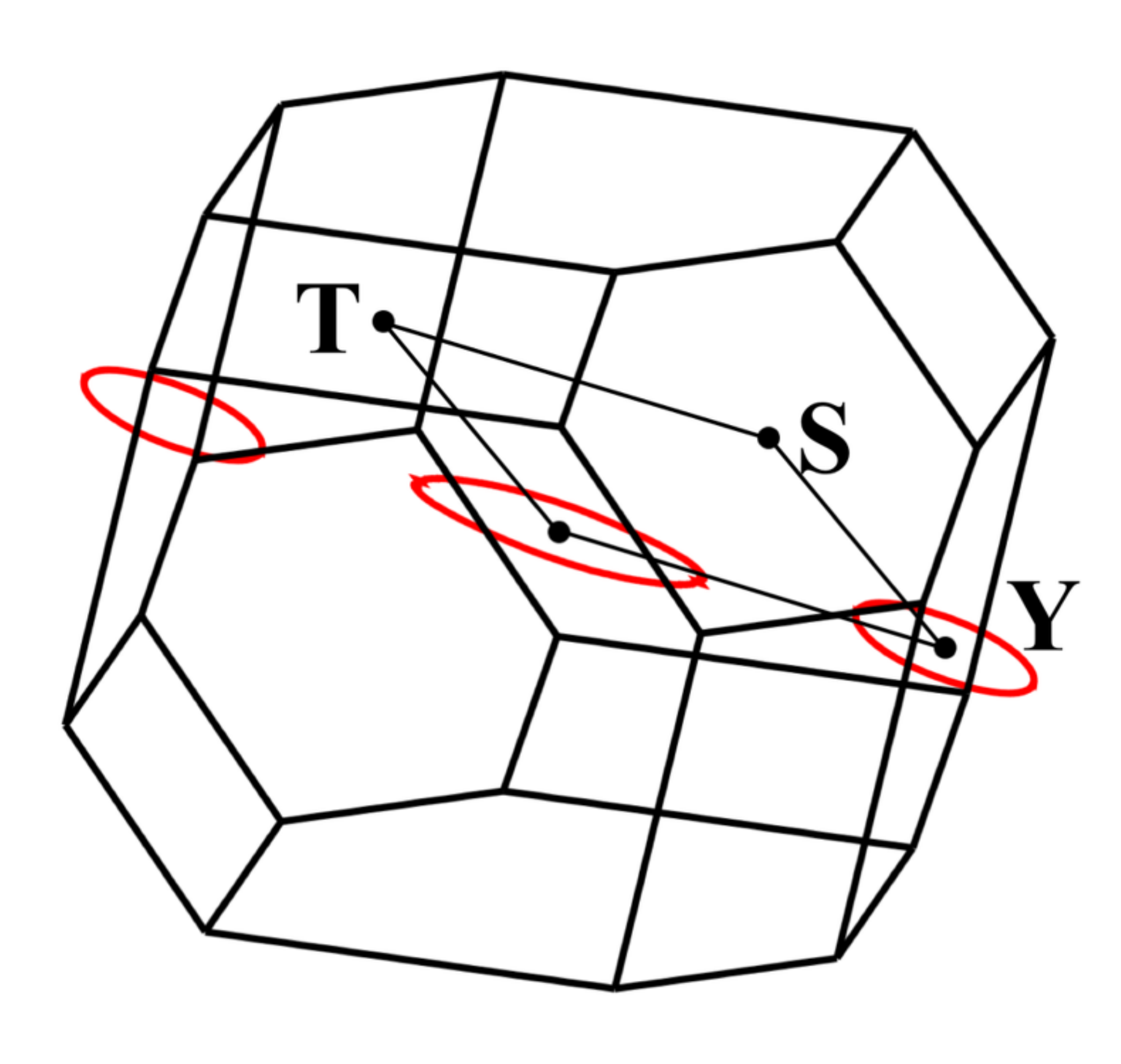}  &
\includegraphics[width=1.35in]{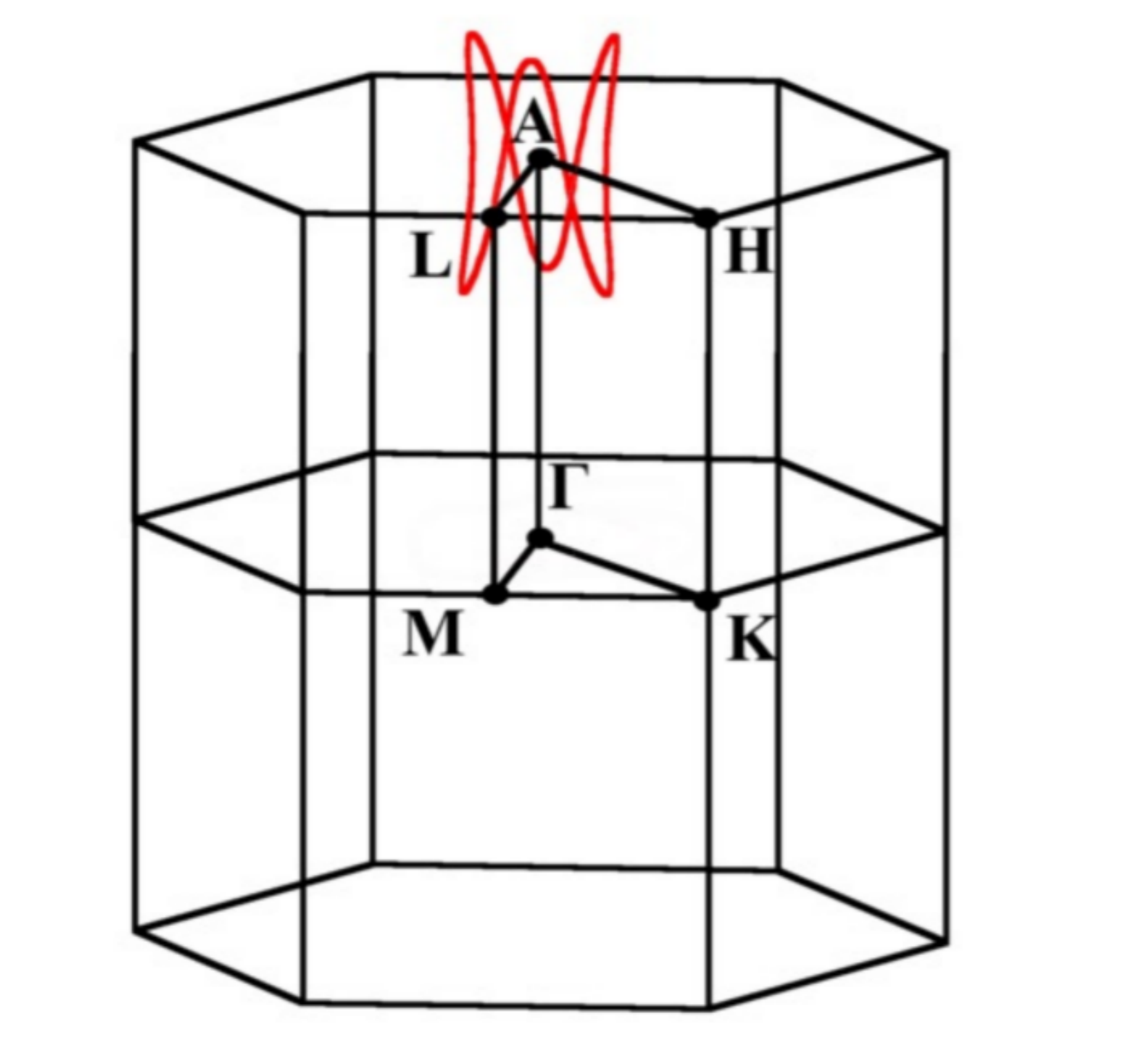}  &
\includegraphics[width=1.5in]{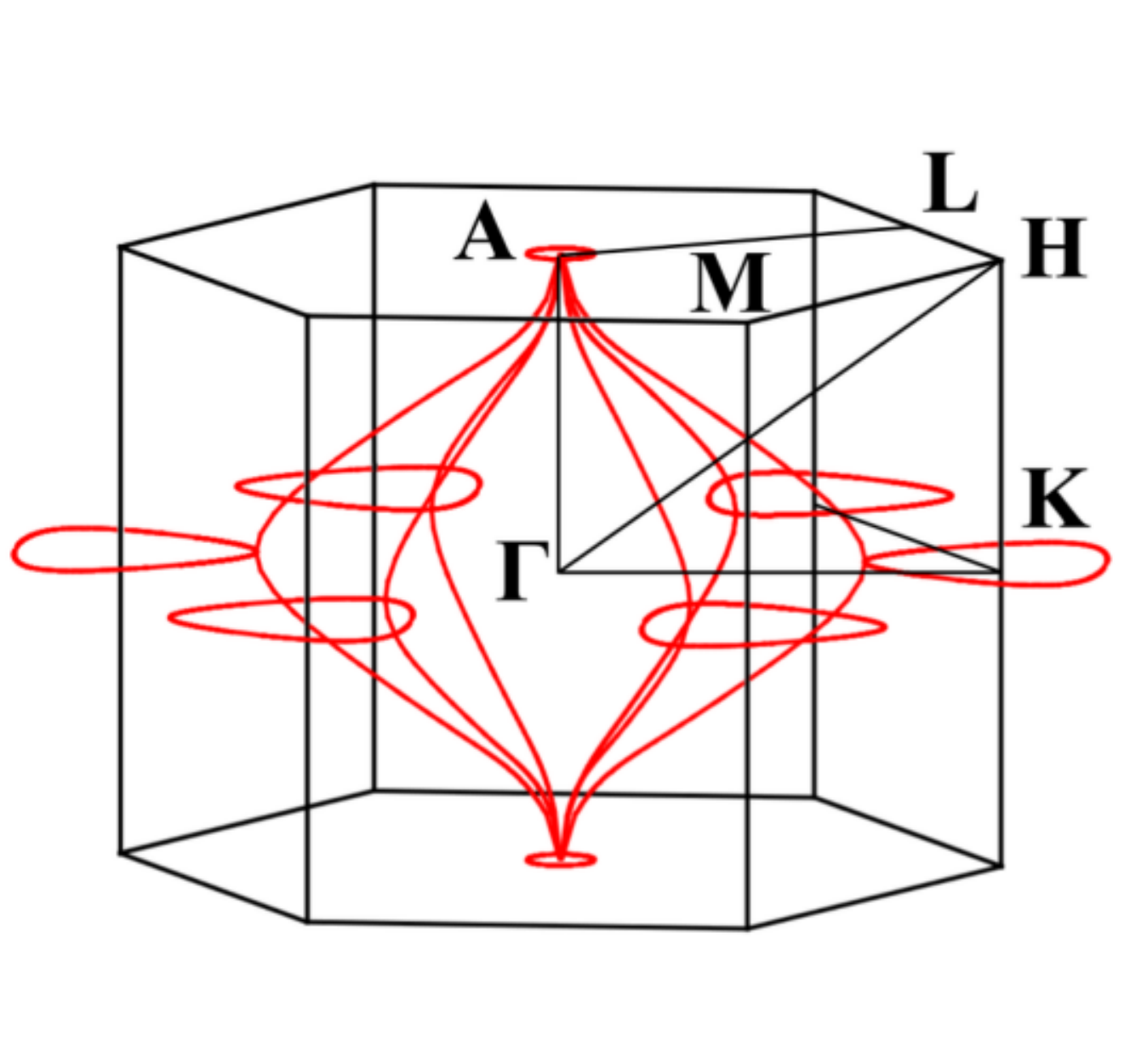} \\
\hline

\raisebox{0.1\totalheight}{\parbox[c|]{2cm}{\raggedright References}} &
{[89], [94]} &
{[95]} &
{[96], [97]} &
{[98-100]} \\
\hline

\bottomrule

\end{tabular}
\label{tab:gt}
\end{adjustbox}

\end{table*}

\begin{table*}[htbp]
\caption{Time-reversal and inversion symmetry protected nodal line materials (continued)}
\centering

\begin{adjustbox}{width=0.9\textwidth}

\newcommand*{\TitleParbox}[1]{\parbox[c]{1.75cm}{\raggedright #1}}%

\makeatletter
\newcommand{\thickhline}{%
    \noalign {\ifnum 0=`}\fi \hrule height 1pt
    \futurelet \reserved@a \@xhline
}
\newcolumntype{"}{@{\hskip\tabcolsep\vrule width 1pt\hskip\tabcolsep}}
\makeatother

\newcolumntype{?}{!{\vrule width 1pt}}

\hspace*{-1cm}
\begin{tabular}{c?c c c  c} 
 
 \toprule

\textbf{Materials} & \textbf{MTC} & \textbf{Hyperhoneycomb lattice}  &  \textbf{Cu$_2$Si} & \textbf{Hg$_3$As$_2$} \\
\thickhline

\midrule

\raisebox{3\totalheight}{\parbox[c|]{2cm}{\raggedright Crystal structure \cite{mullen2015line}}} &
\includegraphics[width=1.4in]{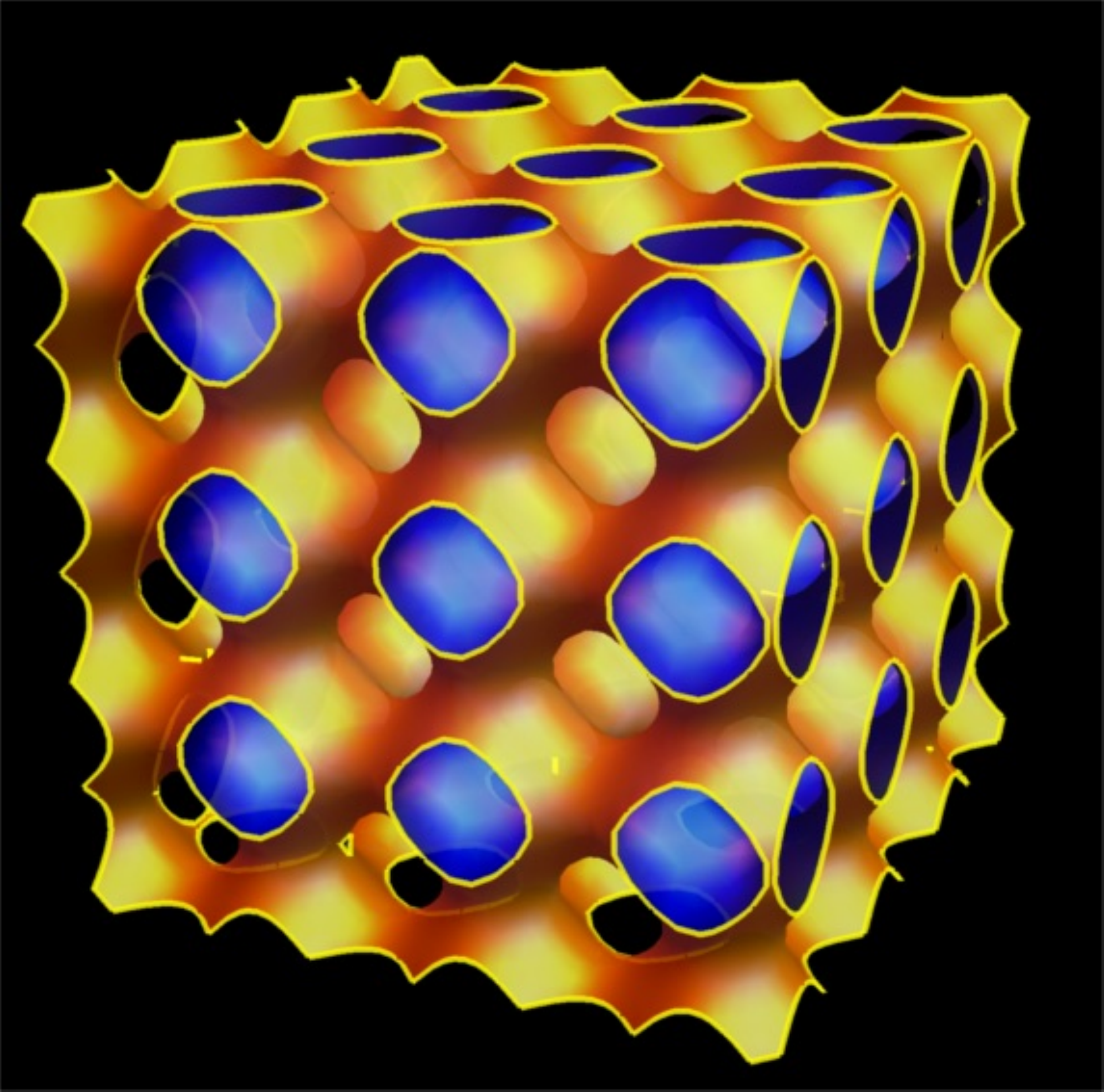} &
\includegraphics[width=1.5in]{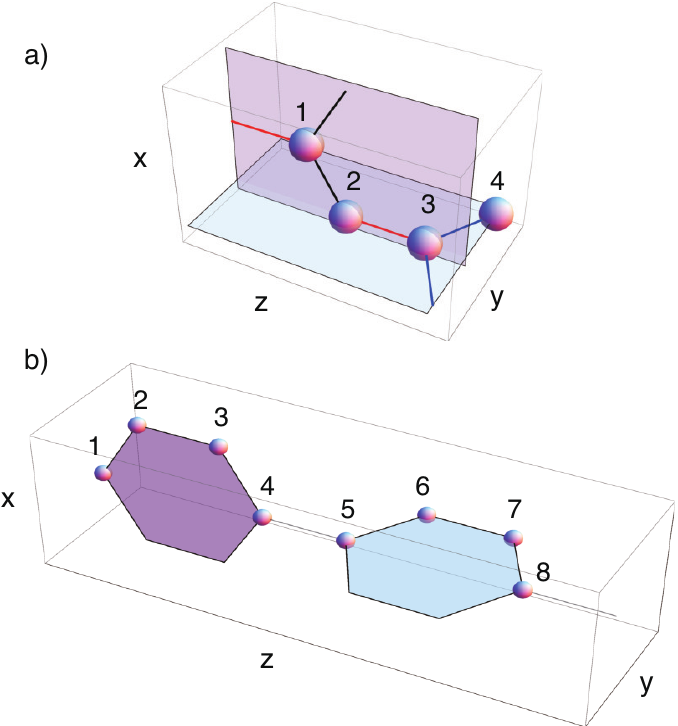} &
\includegraphics[width=1.5in]{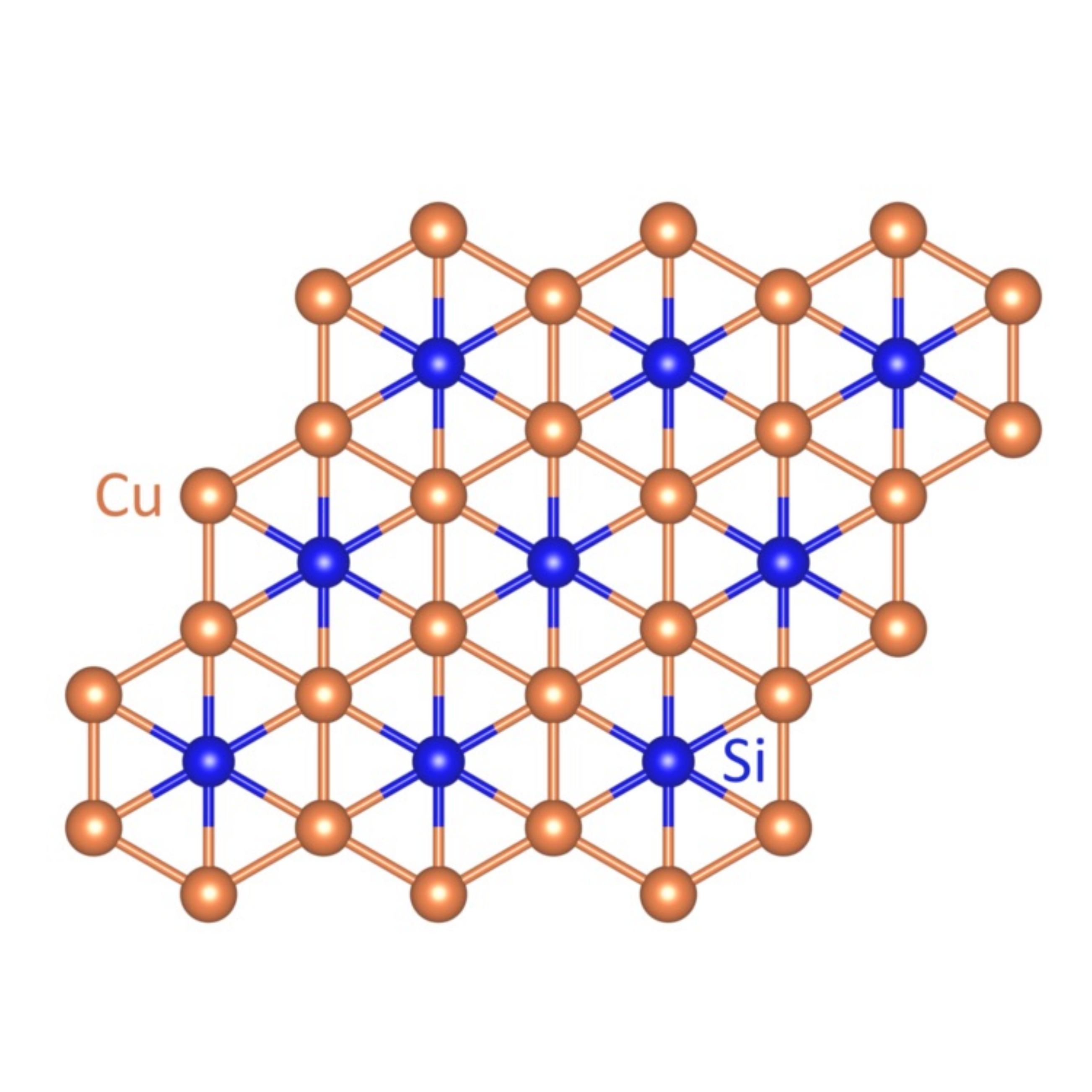} &
\includegraphics[width=1.5in]{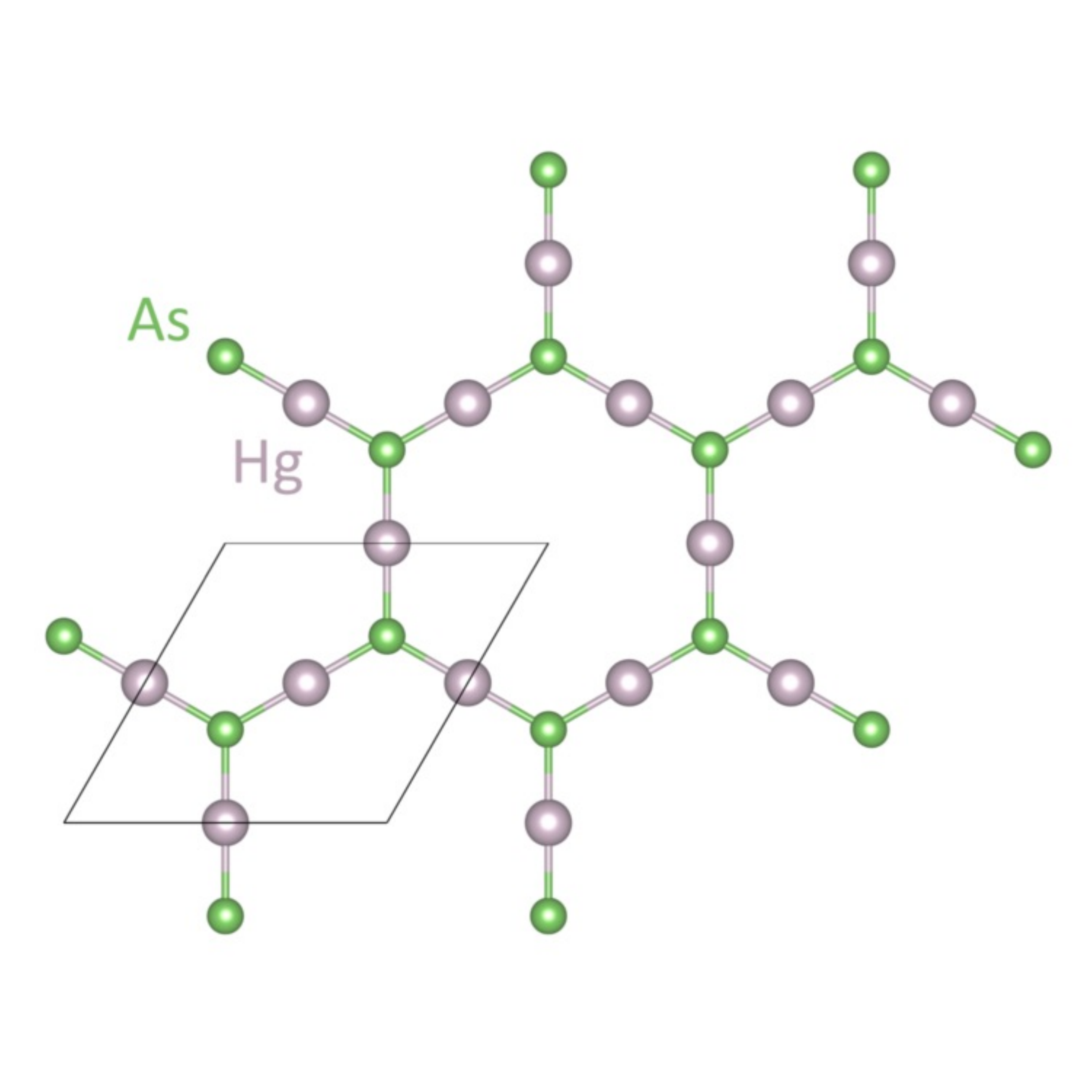} \\
\hline

\raisebox{2\totalheight}{\parbox[c|]{2cm}{\raggedright Band structure without SOC\cite{weng2015topological, mullen2015line}}} &
\includegraphics[width=1.5in]{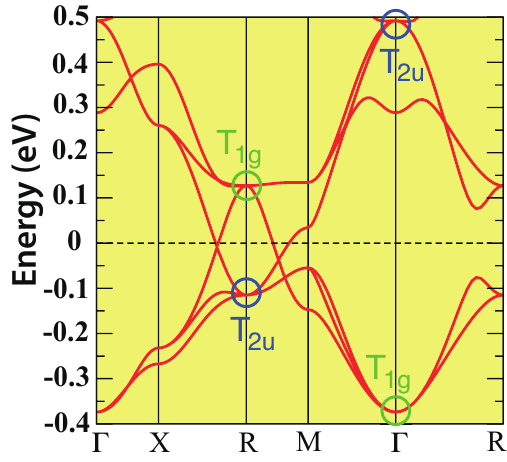} &
\includegraphics[width=1.5in]{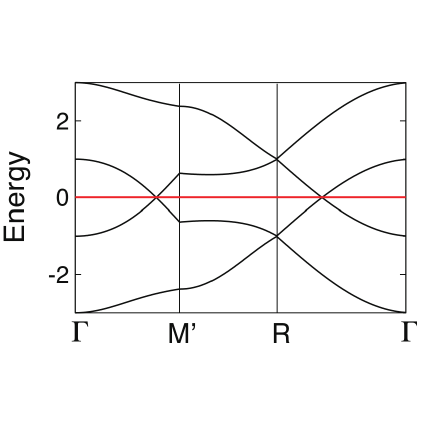} &
\includegraphics[width=1.5in]{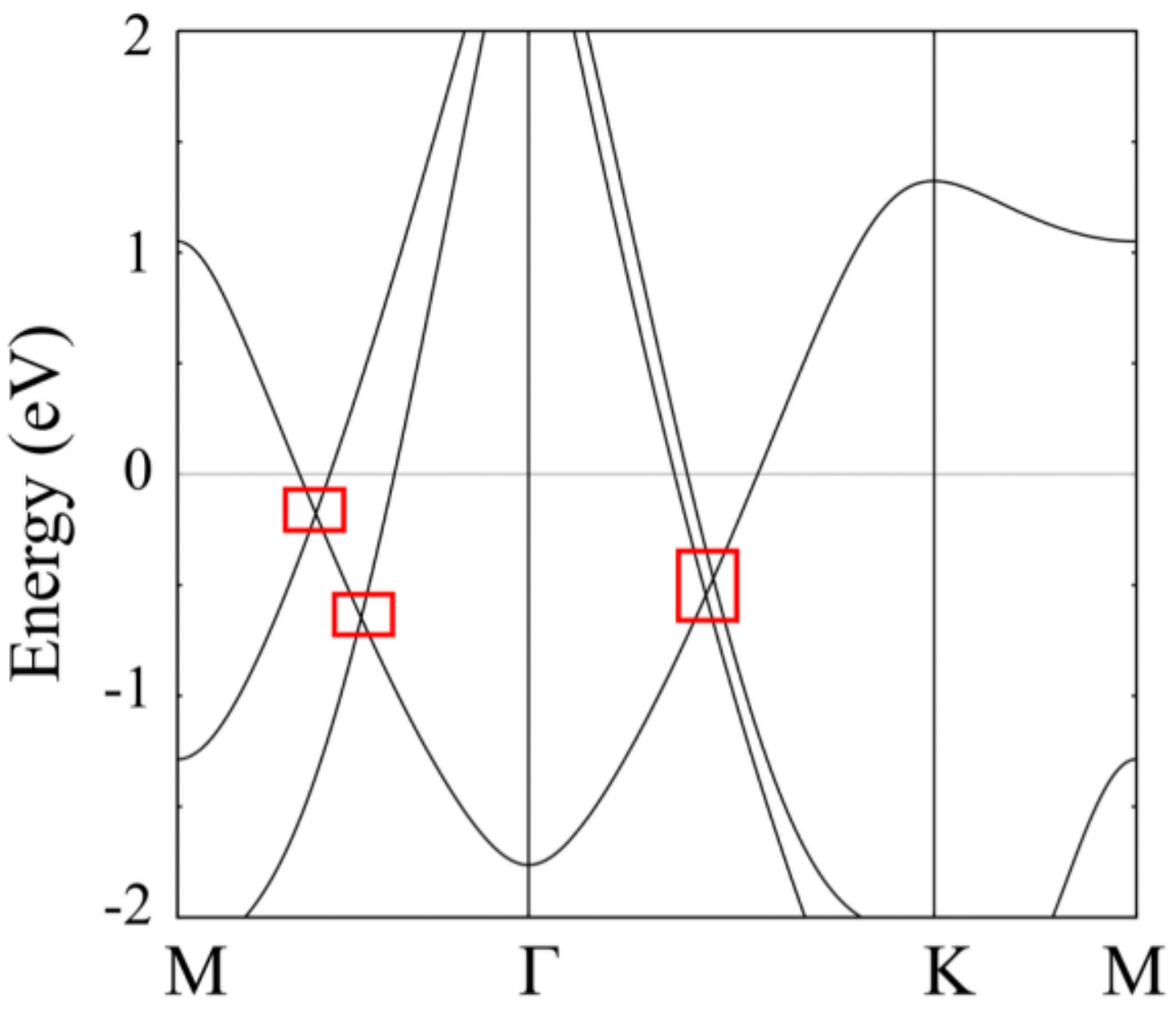} &
\includegraphics[width=1.5in]{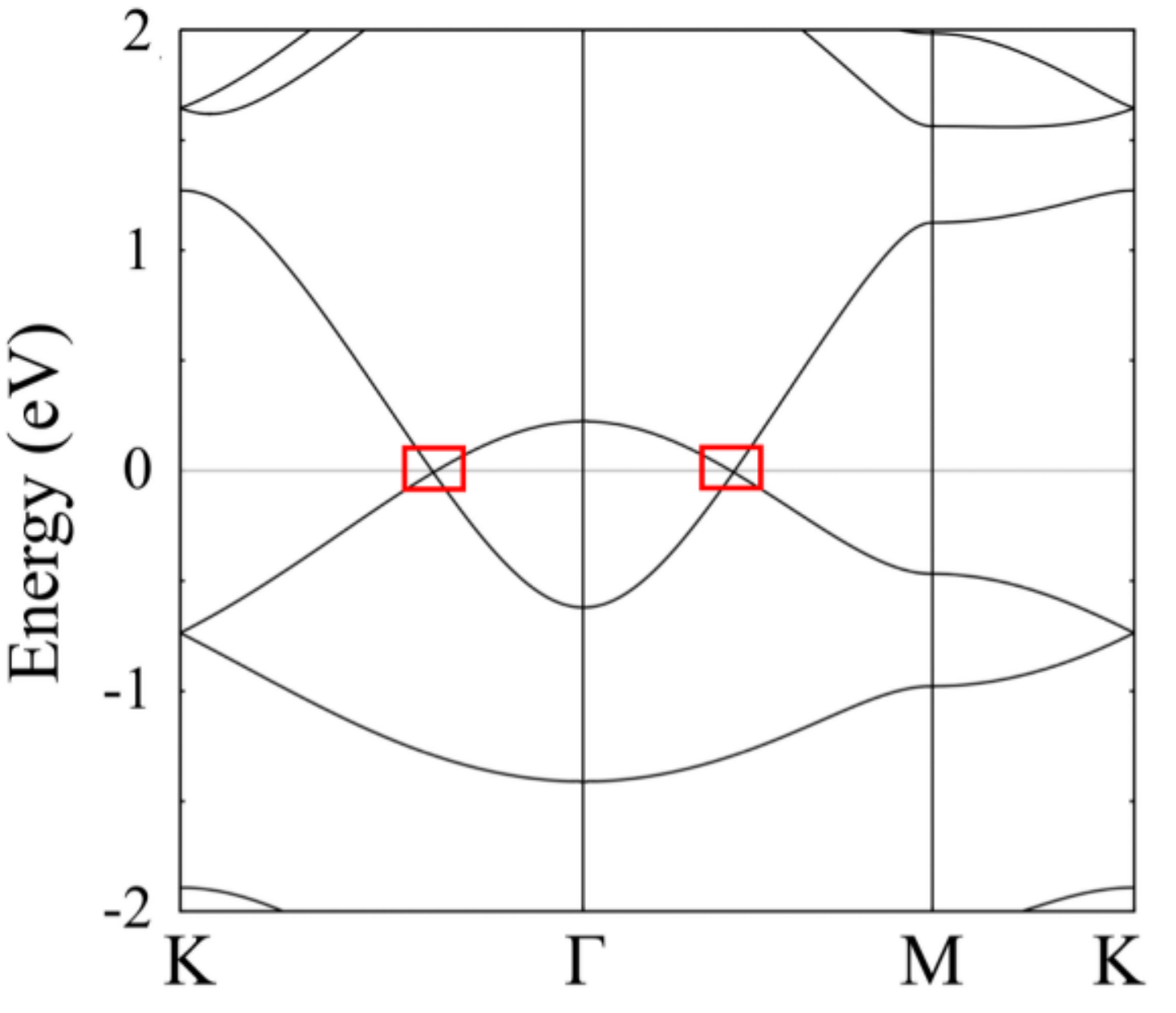} \\

\hline

\raisebox{2.5\totalheight}{\parbox[c|]{2cm}{\raggedright Band structure with SOC}} &
\raisebox{4\totalheight}{N/A} &
\raisebox{4\totalheight}{N/A} &
\includegraphics[width=1.5in]{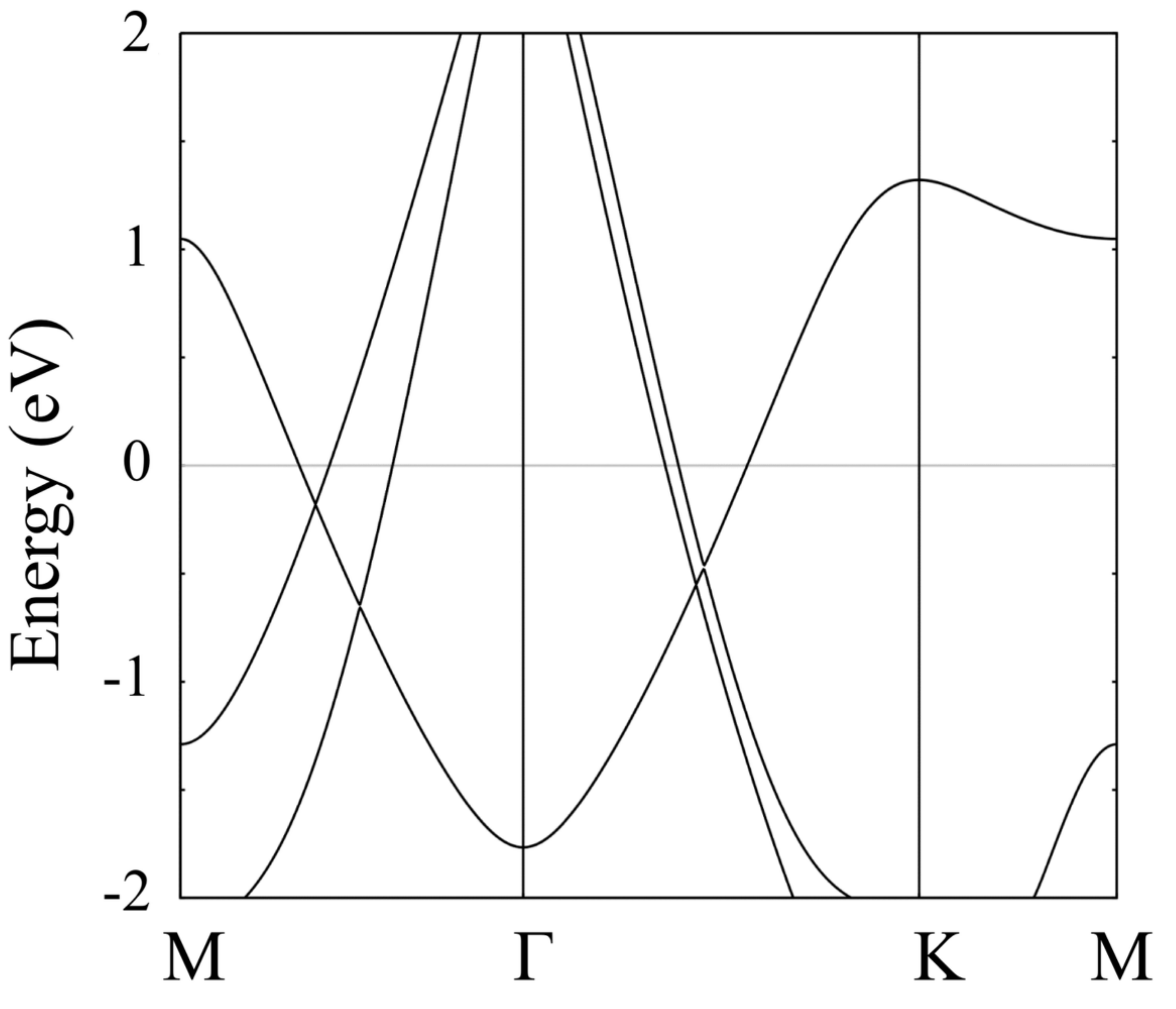} &
\includegraphics[width=1.5in]{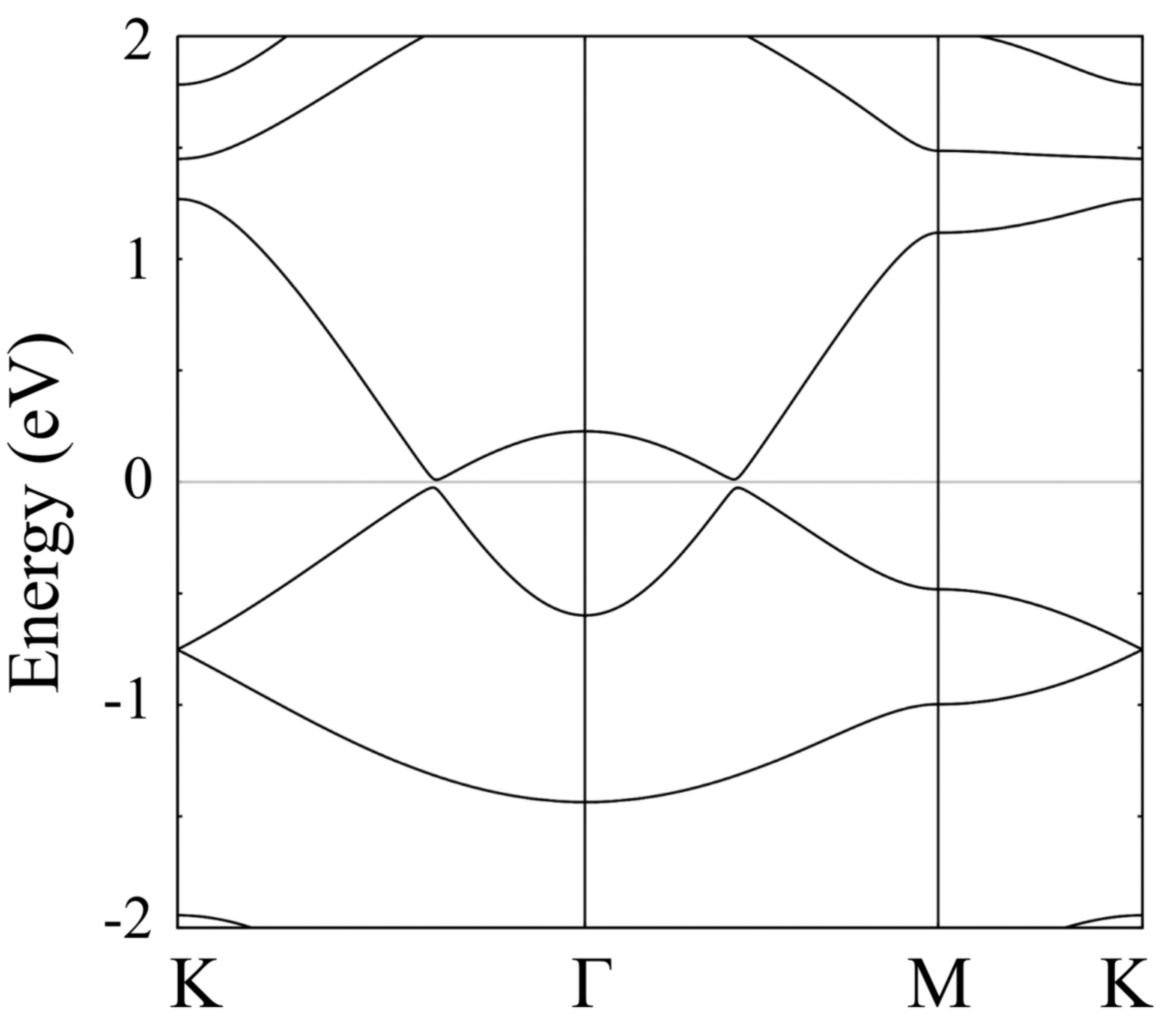} \\
\hline

\raisebox{1.2\totalheight}{\parbox[c|]{2cm}{\raggedright Nodal line distribution without SOC\cite{weng2015topological, mullen2015line}}} &
\includegraphics[width=1.3in]{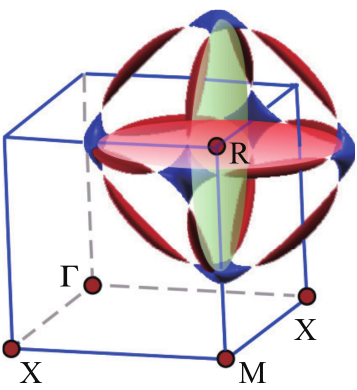} &
\includegraphics[width=1.5in]{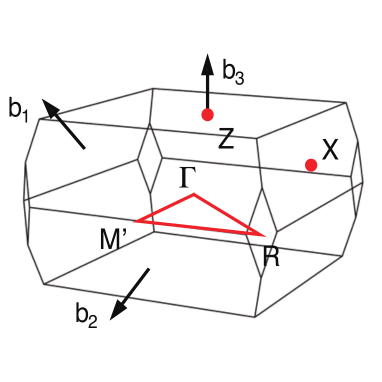} &
\includegraphics[width=1.3in]{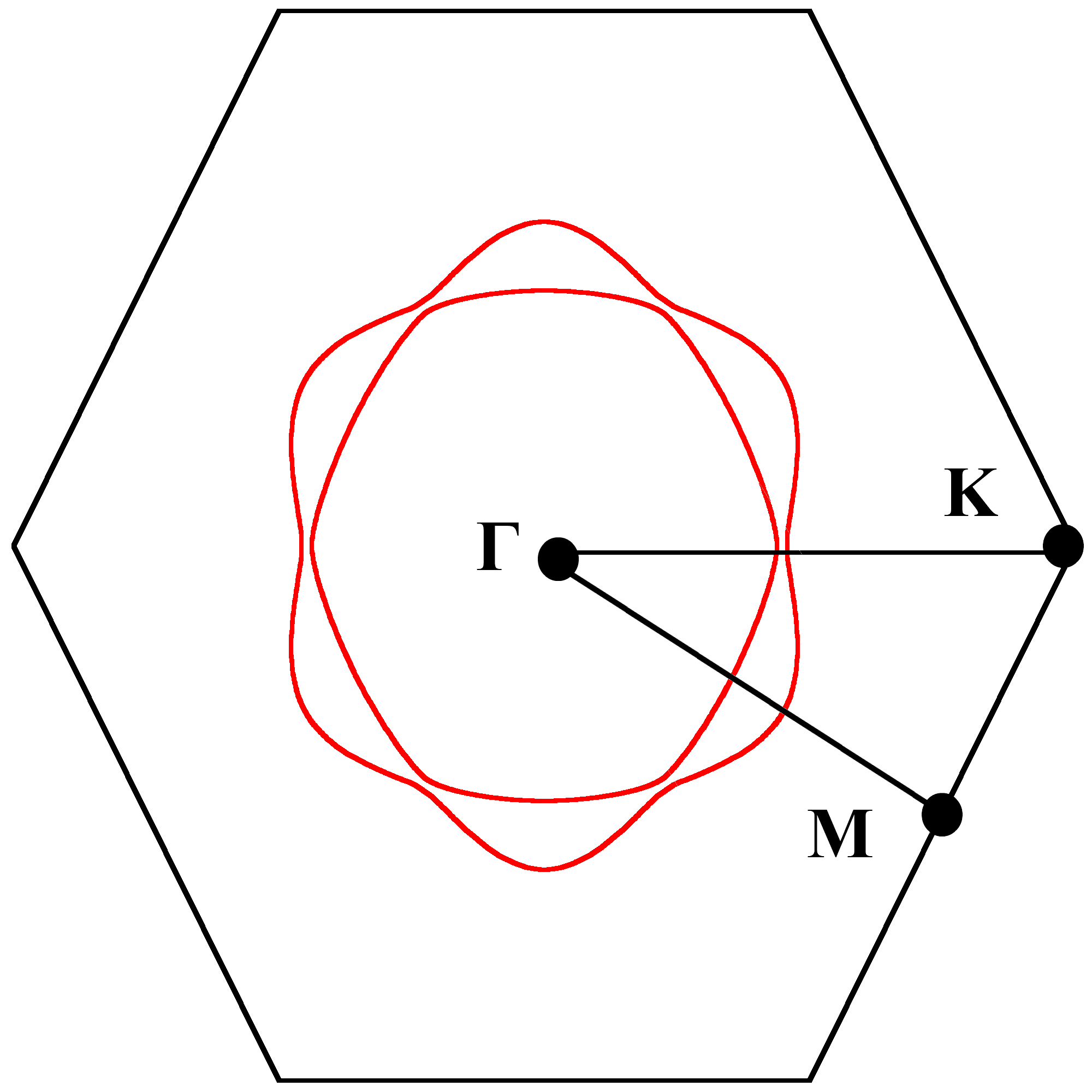} &
\includegraphics[width=1.4in]{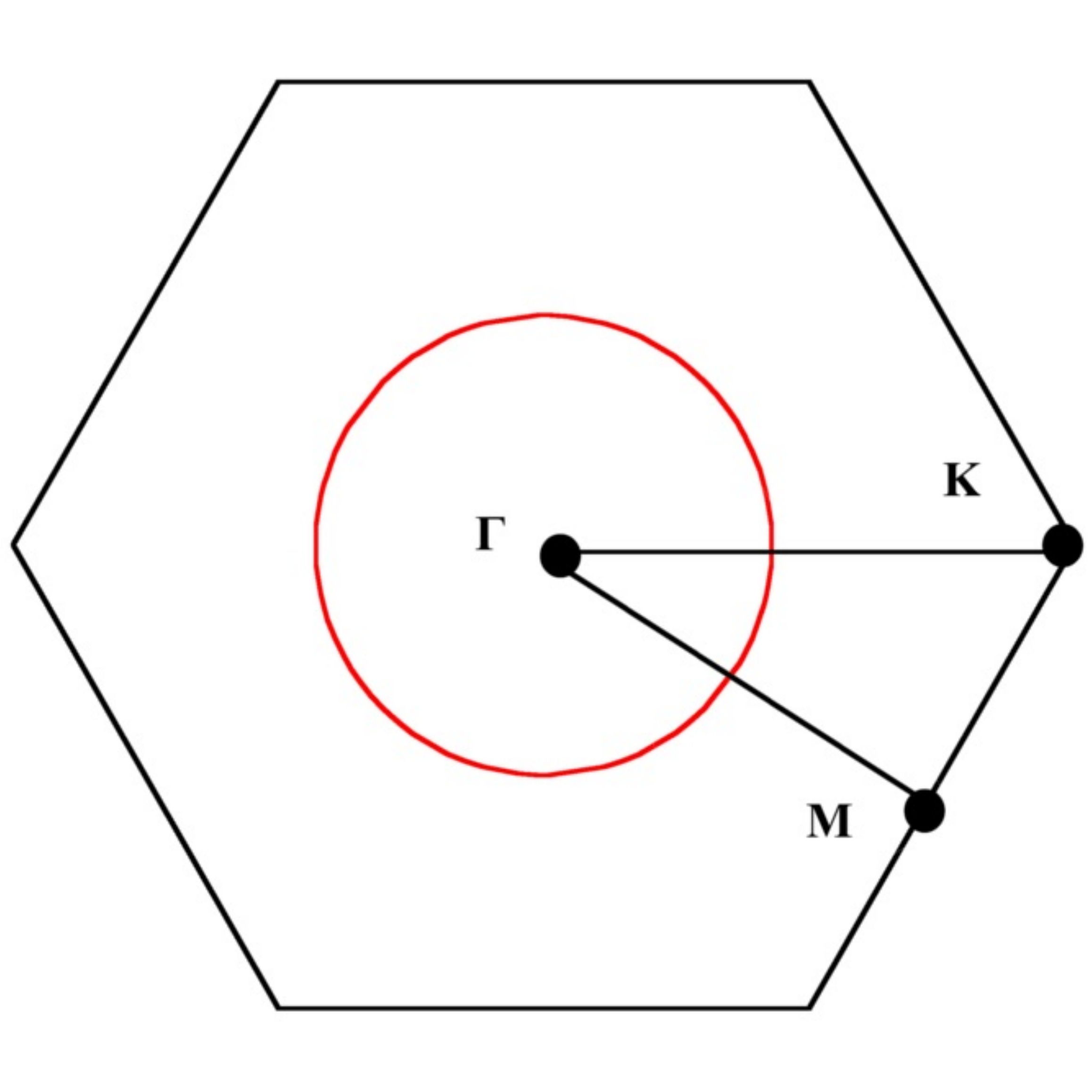} \\
\hline

\raisebox{0.1\totalheight}{\parbox[c|]{2cm}{\raggedright References}} &
{[54]} &
{[101], [111], [112]} &
{[102]} &
{[103]} \\
\hline

\bottomrule

\end{tabular}
\label{tab:gt}
\end{adjustbox}

\end{table*}

\textbf{Cu$_3$N$X$}\cite{kim2015dirac, yu2015topological}: Cu$_3$N crystallizes in the cubic anti-RuO$_3$ structure in space group $Pm\overline{3}m$ No. 221. At the center of the unit cell there is a void, which can host intercalate atoms such as Ni, Cu, Pd, An, Ag, and Cd. By doping it with extrinsic nonmagnetic atoms, a Dirac nodal ring occurs near the E$_F$ without SOC. As an example, Cu$_3$NPd is an antiperovskite material (shown in Table I) in which the conduction and valence bands are inverted at the R point. Without SOC, the bands are six-fold degenerate around R, forming three nodal rings perpendicular to each other. When SOC is included, C$_4$ symmetry along the R-M line protects the Dirac point on the line, making Cu$_3$NPd, in reality, a 3D DSM with three pairs of Dirac points. Since the heavy Pd in Cu$_3$NPd brings a large SOC, the effect of SOC can be diminished by replacing Pd with lighter elements such as Ag and Ni.\par

\textbf{CaTe}\cite{du2016cate}: CaTe is a CsCl-type structured alkaline-earth chalcogenide also in space group $Pm\overline{3}m$ No. 221. Without SOC, CaTe is a topological nodal ring semimetal: three nodal rings are perpendicular to each other at M point with the drumhead surface states embedded within each ring (Table I). When SOC is considered, each nodal ring evolves into two Dirac points along the M-R line, which are protected by the C$_4$ rotation symmetry, making CaTe a 3D DSM.\par

\textbf{La$X$}\cite{zeng2015topological, nayak2017multiple, tafti2016temperature, tafti2016resistivity, kumar2016observation, zeng2016compensated}: The family of lanthanum monopnictides La$X$ has the rock salt structure, where $X$ could be N, P, As, Sb, or Bi. It is in space group $Fm$-$3m$ No. 225 (shown in Table I). When SOC is ignored in LaN, band crossings form three intersecting nodal rings centered at the X point, which are protected by inversion symmetry, time-reversal symmetry, as well as C$_4$ rotation symmetry. When SOC is considered, spin rotation symmetry is broken, resulting in a lift of degeneracy everywhere except two Dirac points. Therefore, SOC turns the nodal rings into three sets of Dirac points at three X points. However, the other compounds in the pnictogen group are TIs with SOC because of larger lattice constant and enhanced SOC\cite{zeng2015topological}. \par

Although the DFT calculations show the compounds in the La$X$ family other than LaN to be TIs with SOC, some recent transport measurements indicate that they behave like semimetals\cite{tafti2016temperature, tafti2016resistivity, kumar2016observation}. In particular, both LaSb and LaBi show extreme MR, which are possibly originated from a combination of electron-hole compensation and a special orbital texture on the electron pocket\cite{tafti2016temperature, tafti2016resistivity}. Their anisotropic characteristic of MR is the result of ellipsoidal electron pockets centered at the X point \cite{kumar2016observation}. However, ARPES measurements show that LaSb is topological trivial without any surface state observed \cite{zeng2016compensated}, and LaBi with an usual surface state which has a linear conduction band and a parabolic valence band\cite{wu2016asymmetric}. \par 

\textbf{Ca$_3$P$_2$}\cite{xie2015new, chan20163}: Ca$_3$P$_2$ has the hexagonal Mn$_5$Si$_3$ type crystal structure with space group $P$$6$$_3$/$mcm$ No. 193 (shown in Table II). Without SOC, Ca$_3$P$_2$ shows a four-fold degenerate nodal ring. Band crossings are protected by k$_z$=0 mirror plane and a combination of inversion and time-reversal symmetry. The SOC is very small for the light elements Ca and P and when SOC is included, spin-rotation symmetry is conserved resulting in a very small band gap opening in the crossing of Ca$_3$P$_2$.\par

\textbf{CaP$_3$ family} \cite{xu2017topological}: The crystal structure of CaP$_3$ families (CaP$_3$, CaAs$_3$, SrP$_3$, SrAs$_3$, and BaAs$_3$) can be seen as a series of 2D puckered polyanionic layers stacking on top of each other: within each layer, 14 P atoms form a 2D circle, with 2 Ca atoms sitting inside the circle; in between the layers, different layers stack along the $b$ axis. Among the CaP$_3$ family of materials, CaP$_3$ and CaAs$_3$ belong to space group $P\overline{1}$; SrP$_3$, SrAs$_3$, and BaAs$_3$ belong to space group $C2/m$. The family has both inversion and time-reversal symmetry. Without SOC, the conduction and valence bands with opposite eigenvalues are inverted around Y point near the E$_F$, giving rise to the nodal lines for SrP$_3$, SrAs$_3$, and BaAs$_3$ at the $\Gamma$-Y-S plane, and the nodal lines for CaP$_3$ and CaAs$_3$ slightly off the plane. Taking into account of SOC, all the crossings are gapped along the nodal line, making them TIs. Among all materials within the family, SrP$_3$ gives the smallest gap (6.11 meV along S-Y and 1.76 meV along Y-$\Gamma$). Table II shows an example of BaAs$_3$. The SOC induced gap in BaAs$_3$ is 38.97 meV along S-Y direction and 6.22 meV along Y-$\Gamma$ direction. \par

\textbf{Alkaline-earth stannides, germanides, and silicides} \cite{huang2016topological, nayak2017multiple}: The alkaline-earth $A$$X$$_2$ family of compounds have similar crystal and electronic structures. BaSn$_2$ (Table II), for example, crystalizes in the trigonal structure. It is in space group $P\overline{3}m1$ No. 164. In BaSn$_2$, Sn atoms form a honeycomb lattice and the Ba layer intercalates between two adjacent Sn layers. When SOC is ignored, these materials are characterized by a snakelike closed nodal ring winding around the A point, extending above and below the k$_z$ = $\pi$ plane. The nodal ring is protected by a combination of inversion and time-reversal symmetry. The special shape of the nodal line is determined by additional symmetries, such as the C$_2$ operation along A-H and the C$_3$ rotation symmetry along K$_z$. BaSn$_2$ has only one nodal ring within the BZ; most other nodal ring systems typically have multiple circles or ellipses. When SOC is included, the nodal line is gapped and the system becomes a TI. \par

\textbf{AlB$_2$-type diborides} \cite{feng2017topological, kumar2012electronic, zhang2017coexistence}: Both TiB$_2$ and ZrB$_2$ crystallize in centrosymmetric layered hexagonal structure with space group P6/mmm No. 191 (Table II). It consists graphene-like boron layers and alternating hexagonal titanium or zirconium layers. Both compounds have been successfully synthesized in laboratory, and have brought much interests owing to the unique combination of properties such as high melting point, high bonding strength, high thermal and electrical conductivity \cite{waskowska2011thermoelastic, okamoto2010anisotropic}. \par

Some electronic properties of the AlB$_2$ diborides are embedded in their band structures, such as a nodal net structure including triple point, nexus, and nodal link\cite{feng2017topological, kumar2012electronic}. Take TiB$_2$ as an example, without SOC, six band crossings emerge around the E$_F$, along the H-$\Gamma$, $\Gamma$-A, A-H, K-$\Gamma$, M-K, and L-A directions. These band crossings give rise to four different kinds of nodal line structures, including 1) a nodal ring in the M$_x$$_y$ plane centered around K point \cite{zhang2017coexistence}; 2) a nodal ring in the in the M$_x$$_y$ plane centered around A point; 3) nodal lines in three vertical mirror planes $\sigma _{v1}$, $\sigma _{v2}$, and $\sigma _{v3}$; 4) nodal line along $\Gamma$-A starting from a triple point. All the four classes of nodal lines together compose a complex nodal net structure. In the presence of SOC, all the crossings except the one along $\Gamma$-A are gapped out, with the gap size ranging from 18 meV to 26 meV. SOC also generates a new Dirac point along $\Gamma$-A. \par

\textbf{Mackay-Terrones crystal}\cite{weng2015topological}: A Mackay-Terrones crystal (MTC) is a 3D network formed by tessellation of 4, 6, or 8 member carbon rings on top of a primitive Schwarz minimal surface. Table III shows a 6-membered example with space group $Pm\overline{3}m$. Without SOC, the coexistence of spacial inversion and time-reversal symmetry generate and protect three orthogonal nodal rings. The nodal rings are in the k$_x$ = $\pi$/a, k$_y$ = $\pi$/a, and k$_z$ = $\pi$/a planes around the R point. Nodal line distributions in Table II show the electron pockets contribution (in blue) and hole pocket contribution (in red) to the nodal lines. Including SOC, a gap opens up leading to a 3D TI. However, similar to graphene, the computed SOC is very small (around 0.13 meV at 1.5 K). \par

\textbf{3D Honeycomb lattice}\cite{mullen2015line, ezawa2016loop, lee2014topological}: A 3D honeycomb lattice is a 3D analog to a 2D honeycomb lattice. Two typical 3D honeycomb lattices include the hyperhoneycomb and stripy-honeycomb lattice, which can be realized in $\beta$-Li$_2$IrO$_3$ and $\gamma$-Li$_2$IrO$_3$, respectively\cite{takayama2015hyperhoneycomb, biffin2014unconventional, modic2014new}. Both of these lattices have been predicted to host nodal loops without SOC. For example, in the hyperhoneycomb lattice, shown in Table III, all atoms are connected by three coplanar bonds spaced by 120 degrees. The planar trigonal connectivity of sites and its sublattice symmetry give rise to Dirac nodes. A tight-binding model reveals the Fermi surface to be a highly anisotropic torus shape, and is expected to show quantized Hall conductivity when applying magnetic field along the torus direction \cite{mullen2015line}. Various antiferromagnetic (AFM) orders are reported in the hyperhoneycomb lattice\cite{lee2014heisenberg, kimchi2014three, lee2014order}, some of which can destroy the nodal loop\cite{ezawa2016loop}. Including SOC in these systems, regardless of magnetism, evolves them into strong TIs\cite{mullen2015line, ezawa2016loop}. \par

\textbf{Cu$_2$Si}\cite{feng2016discovery}: Cu$_2$Si is composed of a honeycomb Cu lattice and a triangular Si lattice, in which Si and Cu atoms are coplanar in monolayer (Table III). It is in space group $P$63/$mmc$ No. 194. Without SOC, the Dirac nodal rings in Cu$_2$Si are shown as two concentric rings centered at $\Gamma$ point. Introducing buckling in the Cu lattice, shifting Si atoms out of plane, or artificially increasing the strength of SOC gaps out the nodal ring, confirming that mirror symmetry protects the Dirac nodal ring in the absence of SOC. However, because SOC is intrinsically small in Cu$_2$Si, its effect is also small (gap openings $\leq$15 meV). Unlike many other proposed materials which are hard to synthesize, especially in thin film form, the experimental synthesis of monolayer Cu$_2$Si on Cu(111) surface by chemical vapor deposition has been realized decades ago, and ARPES has been measured confirming the band crossings at both sides of $\Gamma$ \cite{feng2016discovery}. \par

\textbf{Honeycomb-kagome(HK) lattice}\cite{lu2016two}: HK lattice is a composite lattice composed by interpenetrating honeycomb and kagome sublattices. It can be realized in $A$$_3$$B$$_2$ compounds where $A$ is a group-IIB cation and $B$ is a group-VA anion. $A$$_3$$B$$_2$ compounds have been predicted to host a 2D nodal ring. For example, in Hg$_3$As$_2$ (Table III) with space group $P$63/$mmc$ No. 194, a band inversion happens between the Hg s-orbital and the As p$_z$-orbital without SOC. With respect to $xy$ mirror plane, the eigenstates of two bands have same parity but opposite eigenvalues, therefore resulting in a band inversion. In the presence of SOC, the nodal ring turns into a 2D TI, which has been confirmed by the existence of nontrivial helical edge states\cite{lu2016two}. \par


\subsection{Mirror reflection symmetry protected nodal line materials}

\begin{table*}[htbp]
\caption{Mirror reflection symmetry protected nodal line materials}
\centering

\begin{adjustbox}{width=0.9\textwidth}

\newcommand*{\TitleParbox}[1]{\parbox[c]{1.75cm}{\raggedright #1}}%

\makeatletter
\newcommand{\thickhline}{%
    \noalign {\ifnum 0=`}\fi \hrule height 1pt
    \futurelet \reserved@a \@xhline
}
\newcolumntype{"}{@{\hskip\tabcolsep\vrule width 1pt\hskip\tabcolsep}}
\makeatother

\newcolumntype{?}{!{\vrule width 1pt}}

\hspace*{-1cm}
\begin{tabular}{c?cc|cc} 
 
 \toprule

\textbf{Materials} & \multicolumn{2}{c|}{\textbf{PbTaSe$_2$}} &  \multicolumn{2}{c}{\textbf{TaAs}} \\
\thickhline

\midrule

\raisebox{3\totalheight}{\parbox[c|]{2cm}{\raggedright Crystal structure}} &

\multicolumn{2}{c|}{\includegraphics[width=1.5in]{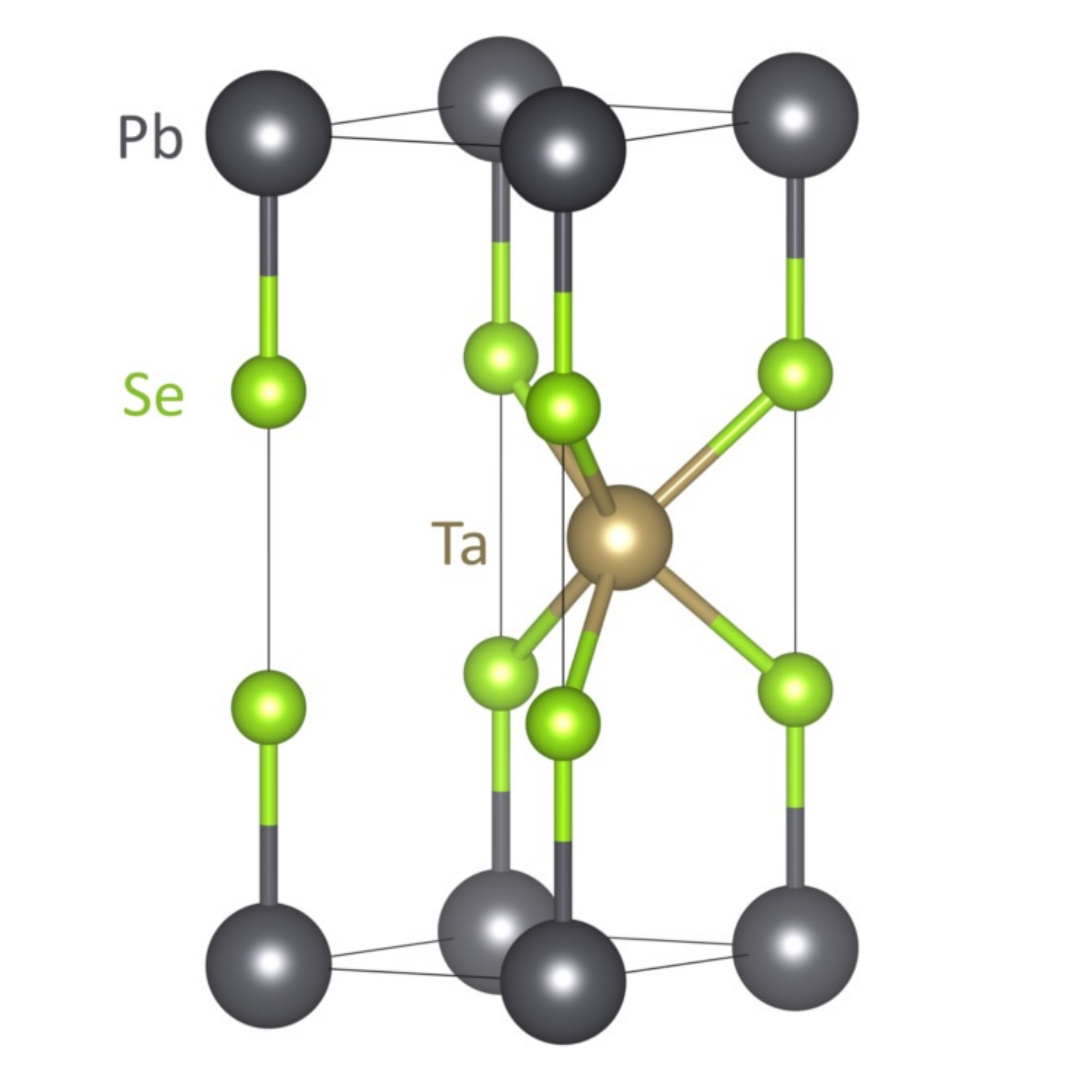}} &
\multicolumn{2}{c}{\includegraphics[width=1.6in]{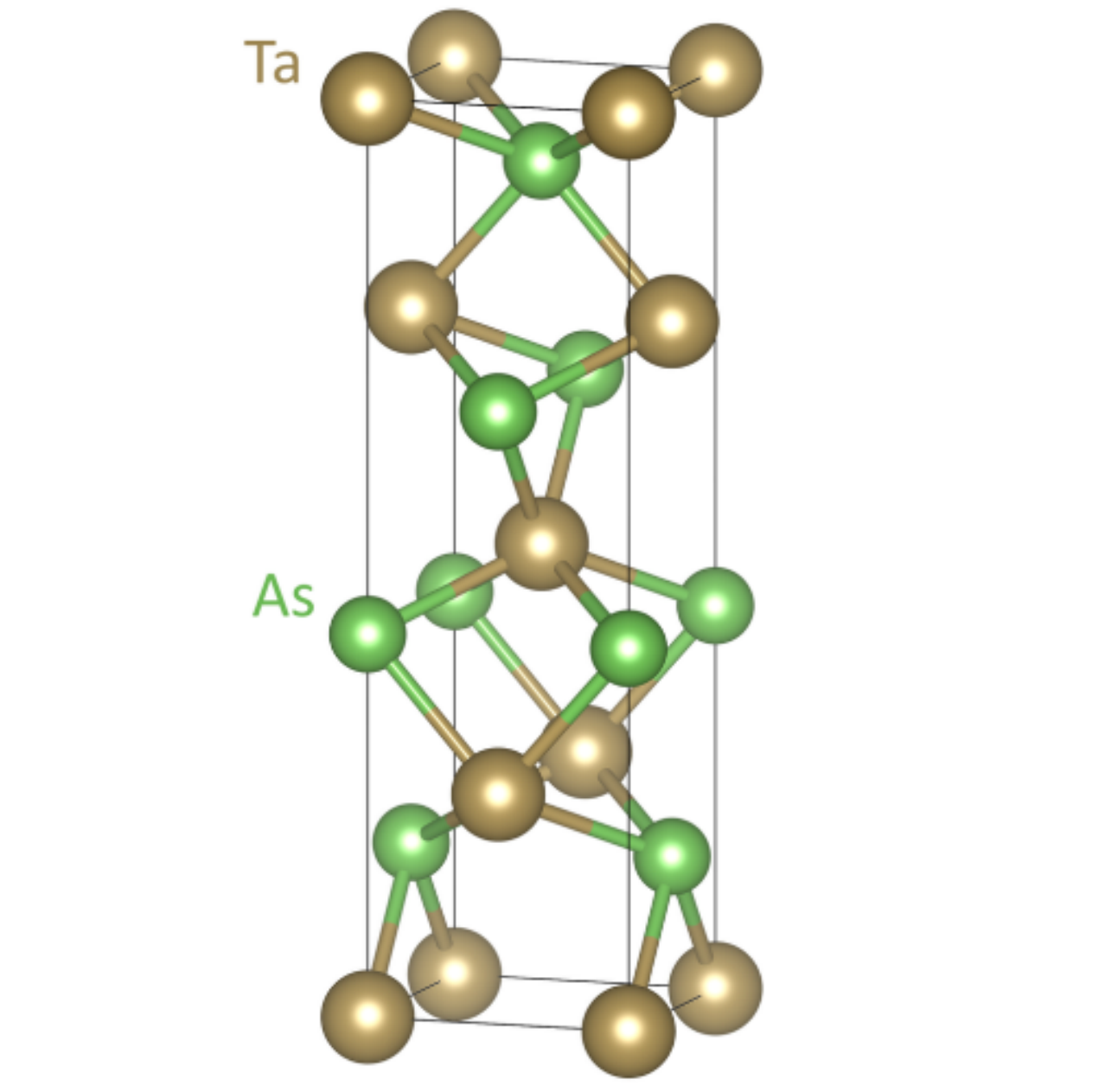}}\\
\hline

\raisebox{1.5\totalheight}{\parbox[c|]{2cm}{\raggedright Band structure without and with SOC}} &
\includegraphics[width=1.5in]{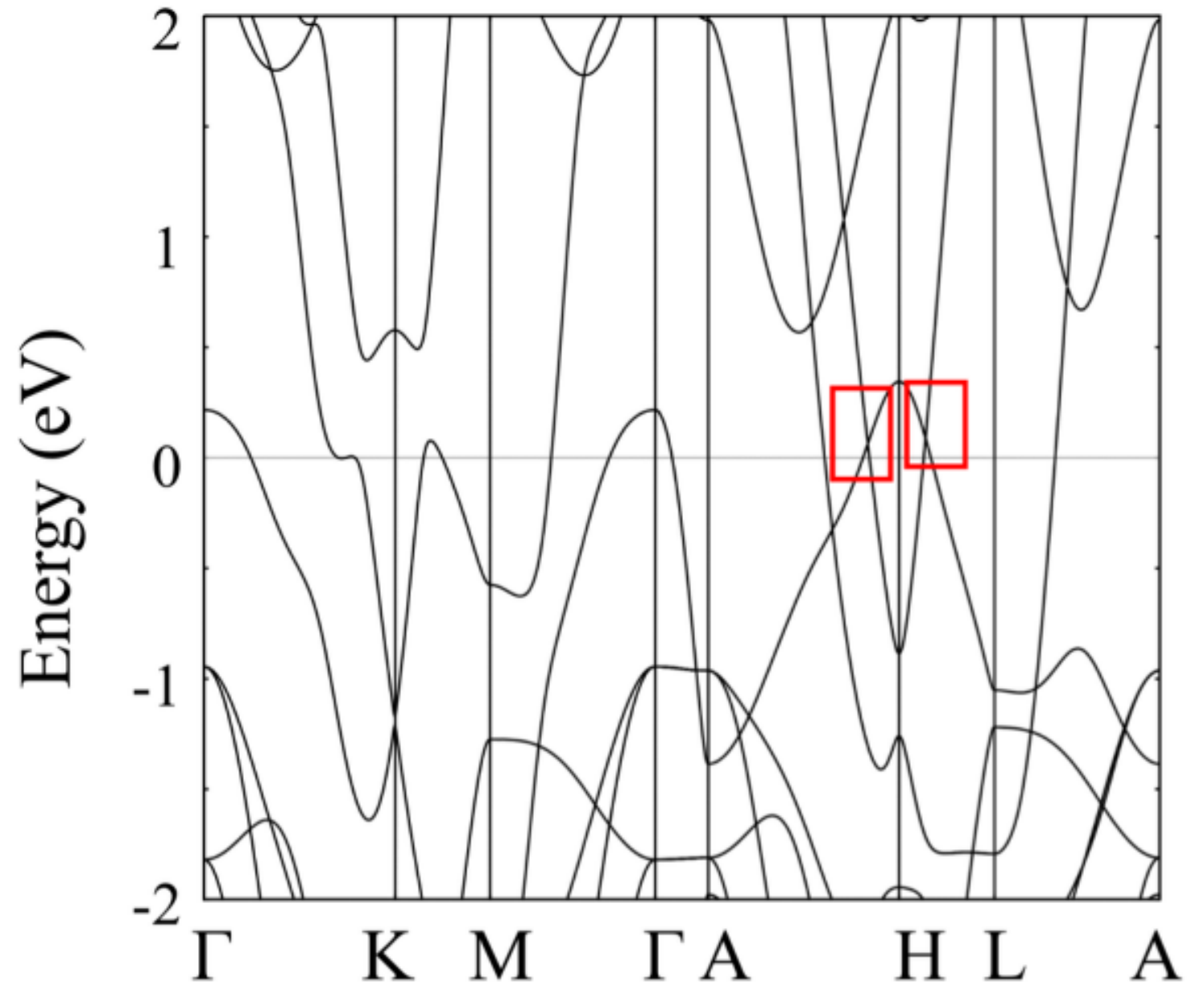} &
\includegraphics[width=1.5in]{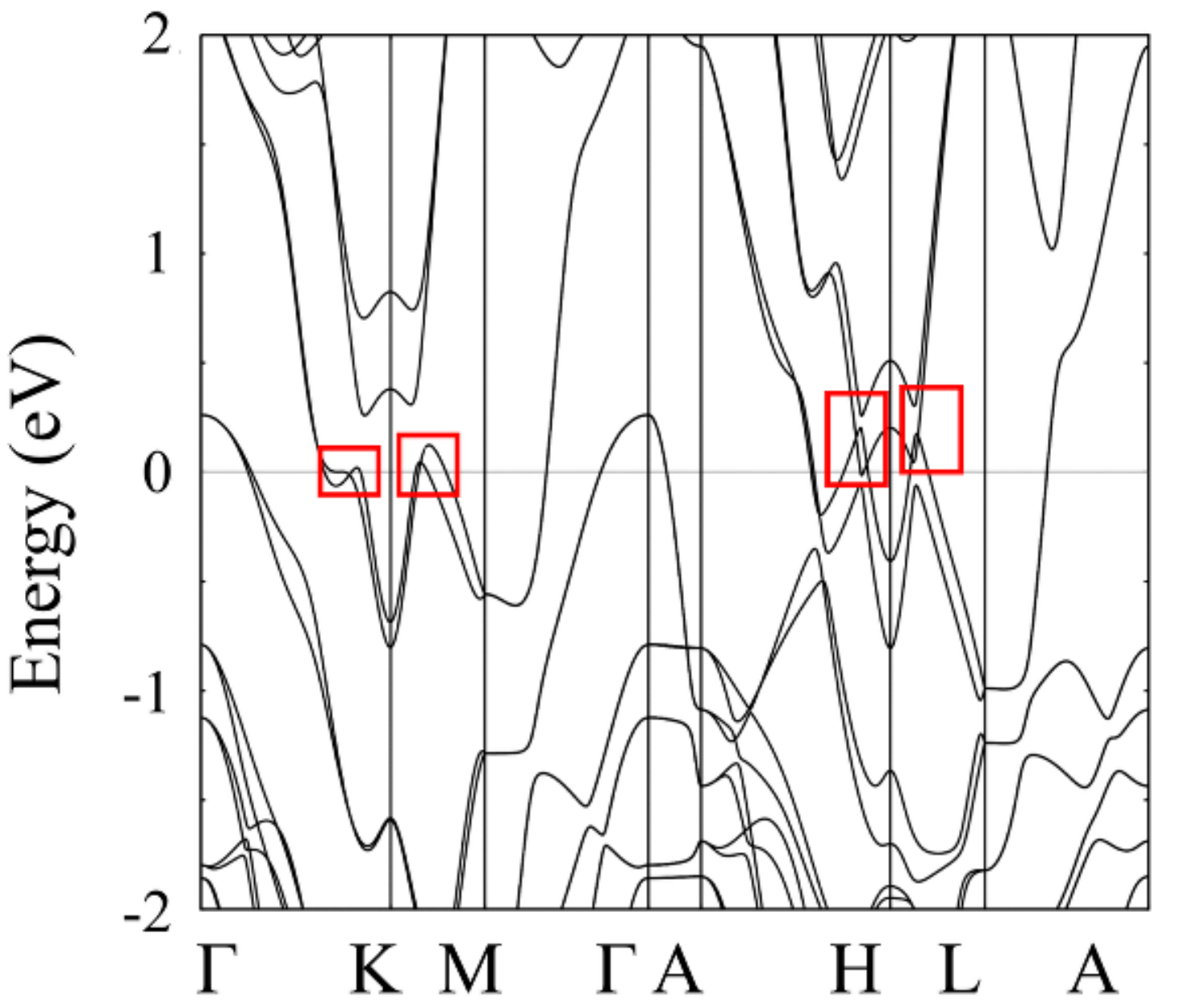}&
\includegraphics[width=1.5in]{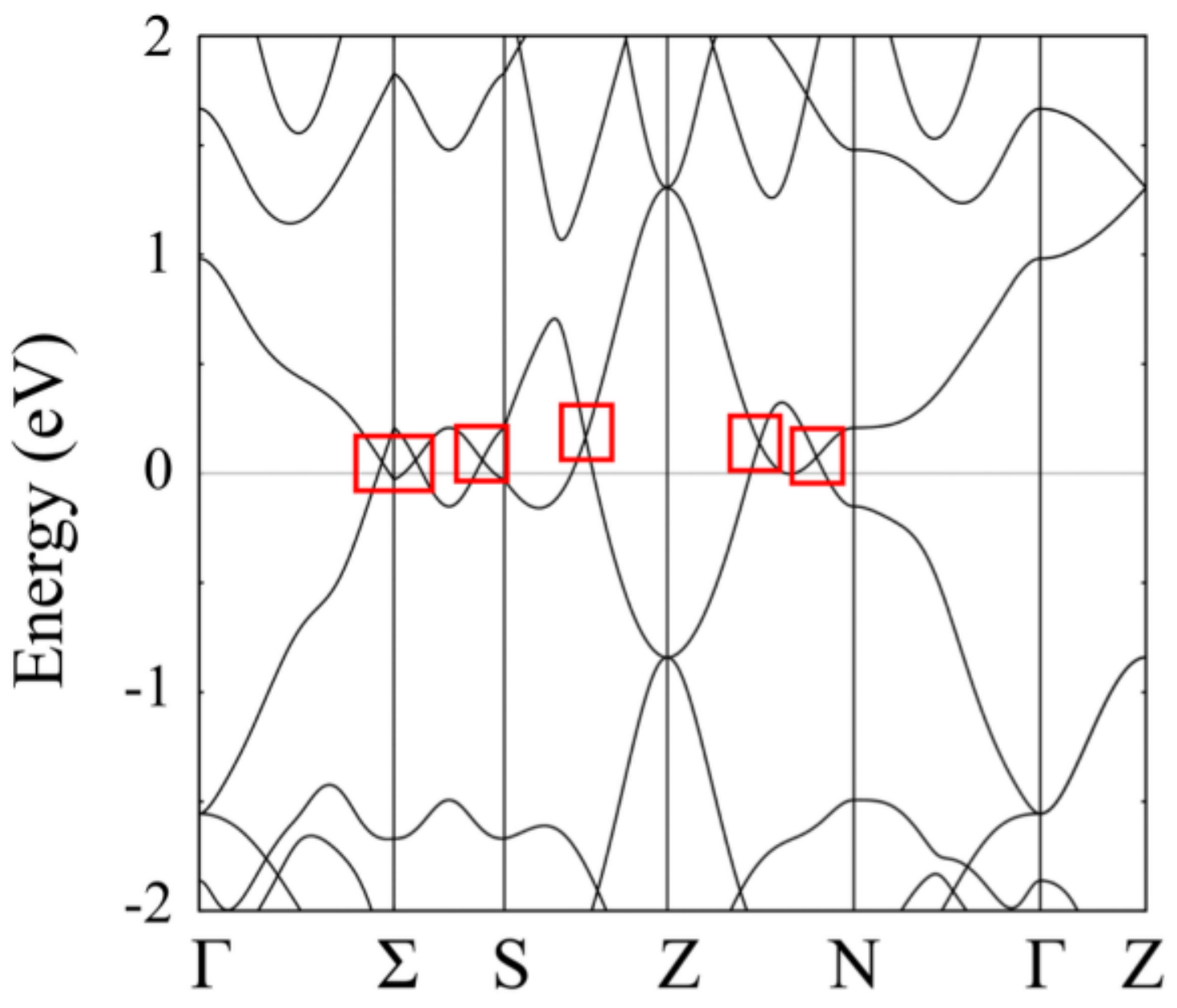} &
\includegraphics[width=1.5in]{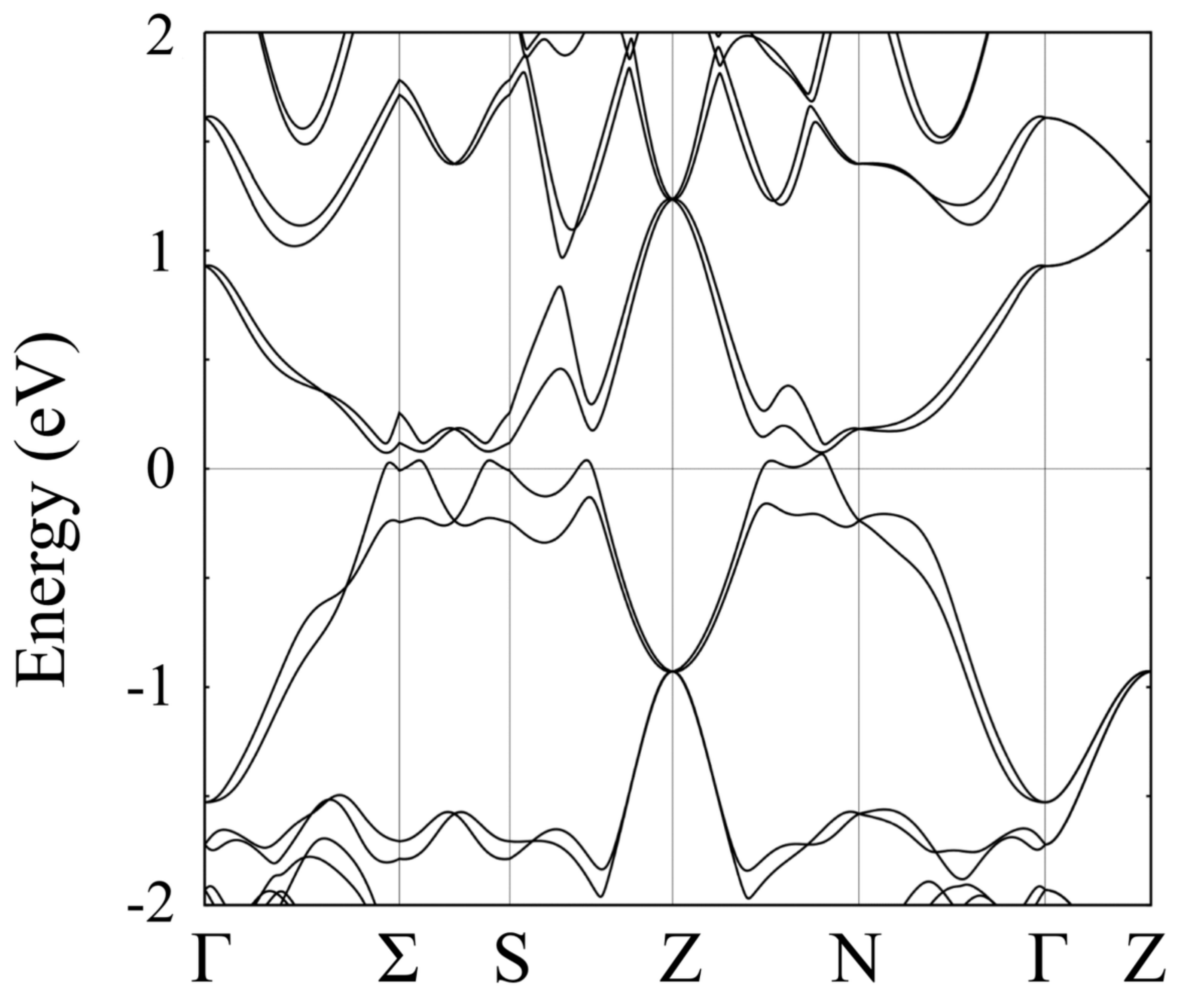}\\
\hline

\raisebox{1\totalheight}{\parbox[c|]{2cm}{\raggedright Nodal line distribution without and with SOC}} &
\includegraphics[width=1.5in]{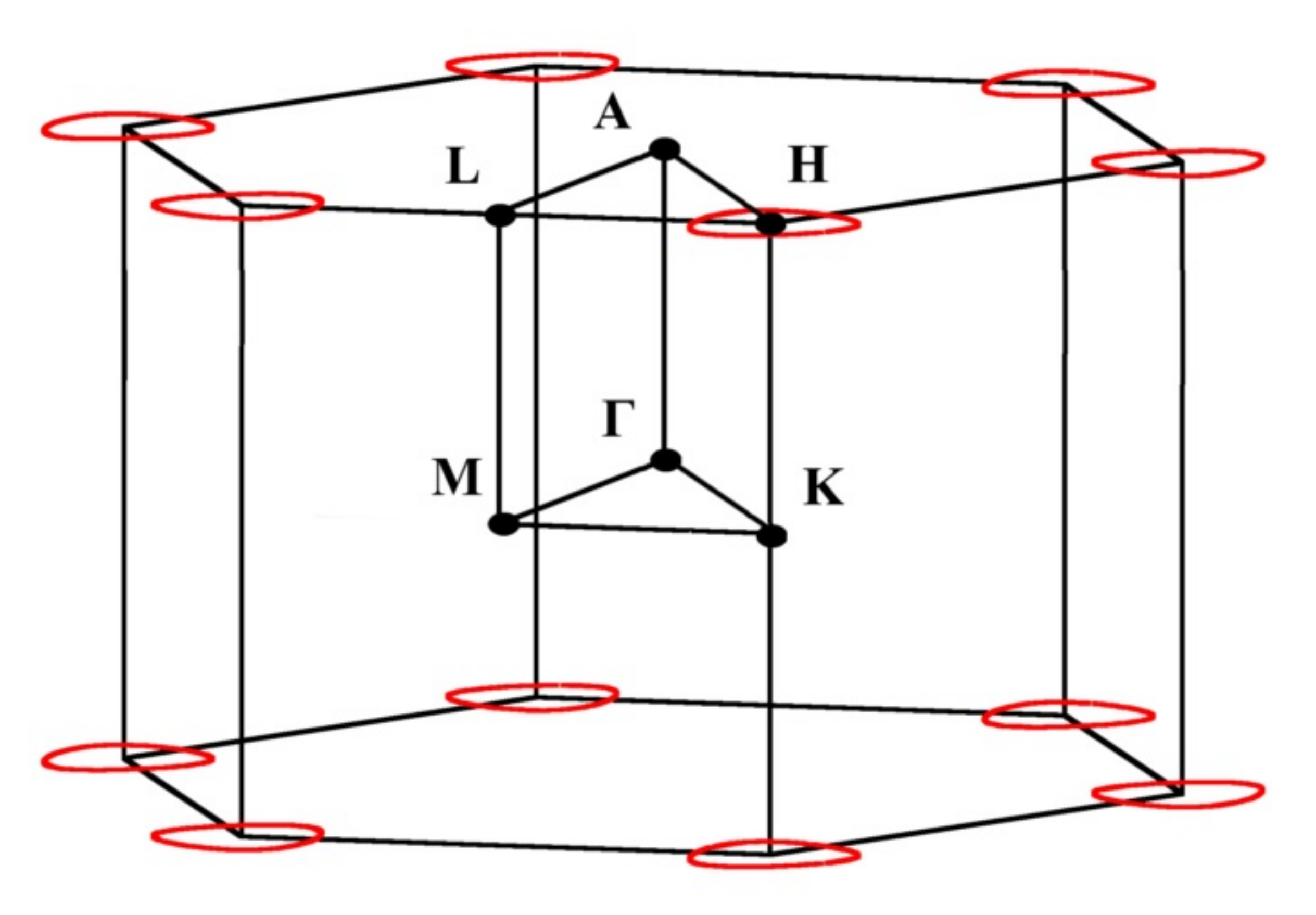} &
\includegraphics[width=1.5in]{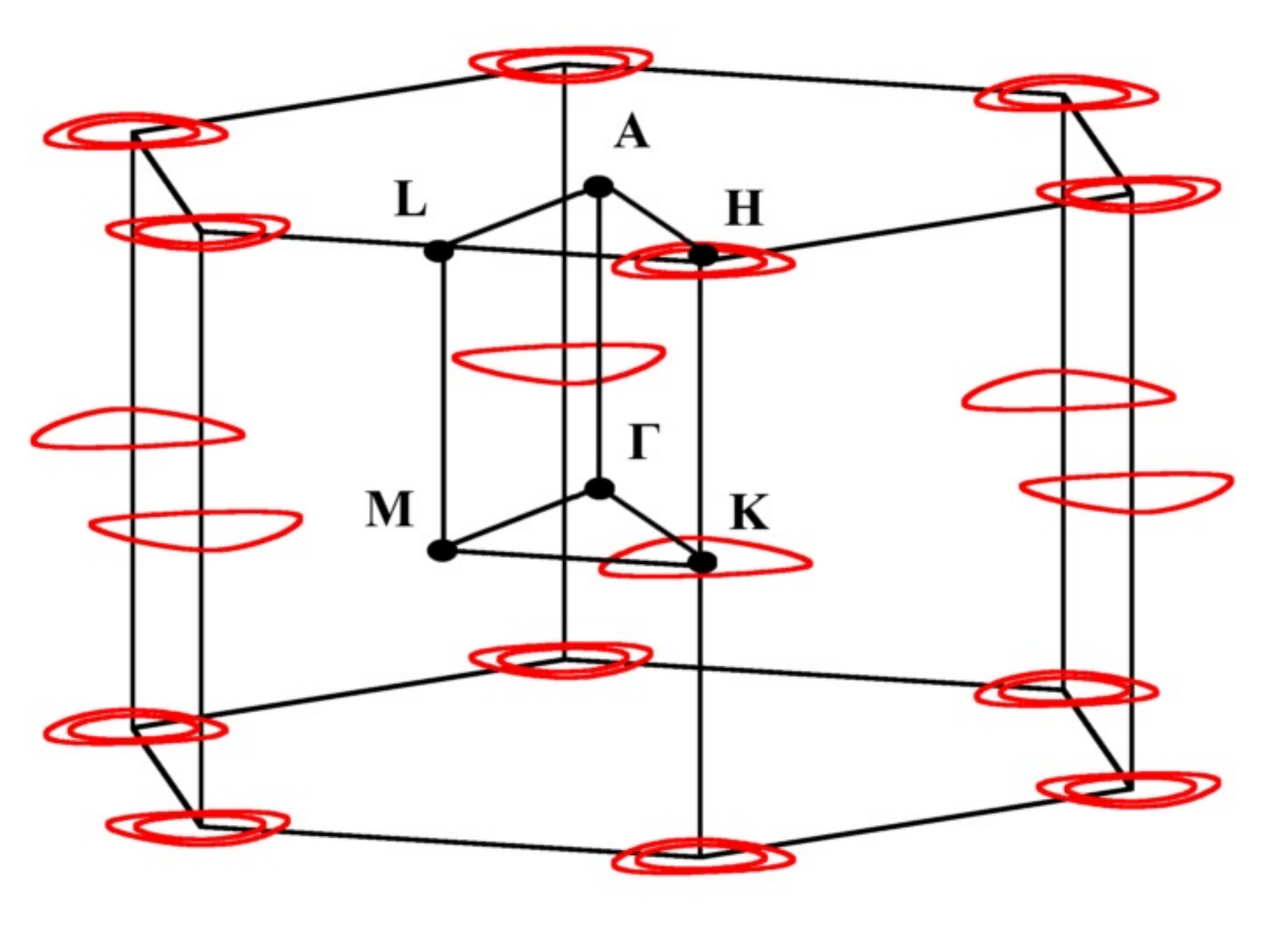} &
\includegraphics[width=1.5in]{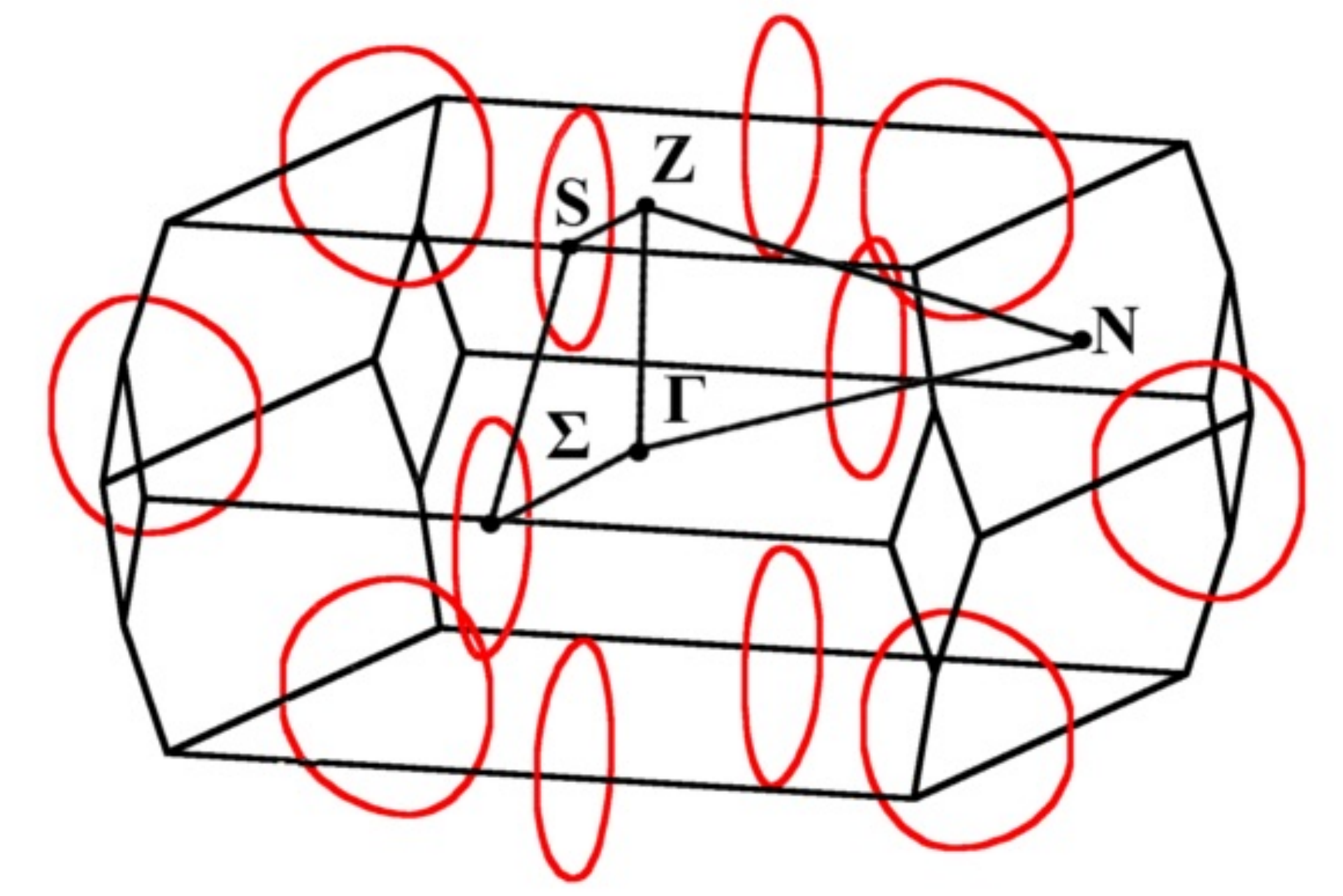} &
\includegraphics[width=1.5in]{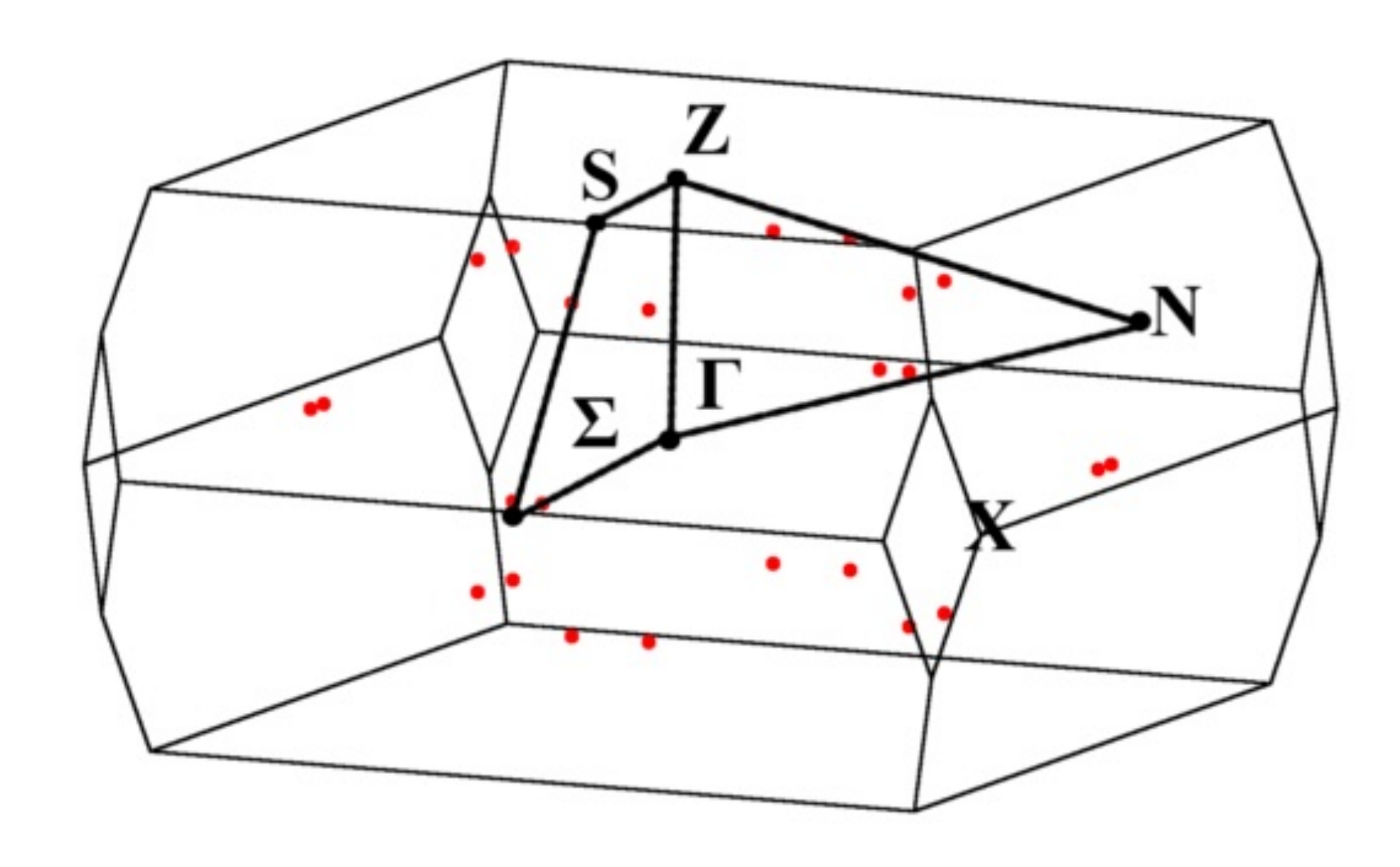}\\
\hline

\raisebox{9\totalheight}{\parbox[c|]{2cm}{\raggedright ARPES \cite{bian2016topological, yang2015weyl}}} &

\multicolumn{2}{c|}{\includegraphics[width=2.6in]{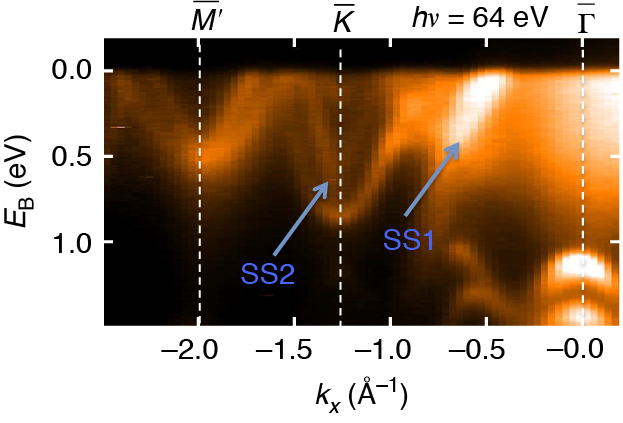}} &
\multicolumn{2}{c}{\includegraphics[width=2.5in]{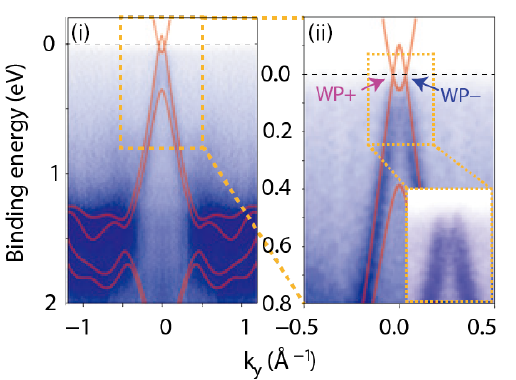}} \\
\hline

\raisebox{0.1\totalheight}{\parbox[c|]{2cm}{\raggedright References}} &
\multicolumn{2}{c|}{[73], [79-82], [119-121]} &
\multicolumn{2}{c}{[21], [30], [35-38], [40], [48], [52], [122-129]} \\
\hline

\bottomrule

\end{tabular}
\label{tab:gt}
\end{adjustbox}

\end{table*}

When a material has a mirror reflection symmetry which commutes with its Hamiltonian, without SOC, nodal lines can be generated by the mirror operation. With SOC, depending on the strength of SOC as well as the atomic orbitals making up the electronic states, the spin degeneracy can be lifted, evolving a DNLS into a WNLS (PbTaSe$_2$), a WSM (TaAs and HfC), or a TI (CaAg$X$($X$ = P, As)). These three classes are discussed below in this section. \par

Table IV to V show the crystal structures, electronic band structures without and with SOC, nodal line (Weyl points) distributions without and with SOC within a BZ, as well as the available ARPES measurements confirming the presence of nodal lines (Weyl points). Without SOC, the Dirac points corresponding to the nodal lines are coded with red square boxes in the band structure. With SOC, the red square boxes indicate the mirror symmetry protected crossings and SOC created accidental crossings. The nodal line distribution without SOC, as well as the nodal line/ Weyl points distributions with SOC are also shown in the tables. \par

\begin{table*}[htbp]
\caption{Mirror reflection symmetry protected nodal line materials}
\centering

\begin{adjustbox}{width=0.9\textwidth}

\newcommand*{\TitleParbox}[1]{\parbox[c]{1.75cm}{\raggedright #1}}%

\makeatletter
\newcommand{\thickhline}{%
    \noalign {\ifnum 0=`}\fi \hrule height 1pt
    \futurelet \reserved@a \@xhline
}
\newcolumntype{"}{@{\hskip\tabcolsep\vrule width 1pt\hskip\tabcolsep}}
\makeatother

\newcolumntype{?}{!{\vrule width 1pt}}

\hspace*{-1cm}
\begin{tabular}{c?cc|cc} 
 
 \toprule

\textbf{Materials} &  \multicolumn{2}{c|}{\textbf{HfC}} &  \multicolumn{2}{c}{\textbf{CaAgAs}} \\
\thickhline

\midrule

\raisebox{2.8\totalheight}{\parbox[c|]{2cm}{\raggedright Crystal structure}} &

\multicolumn{2}{c|}{\includegraphics[width=1.4in]{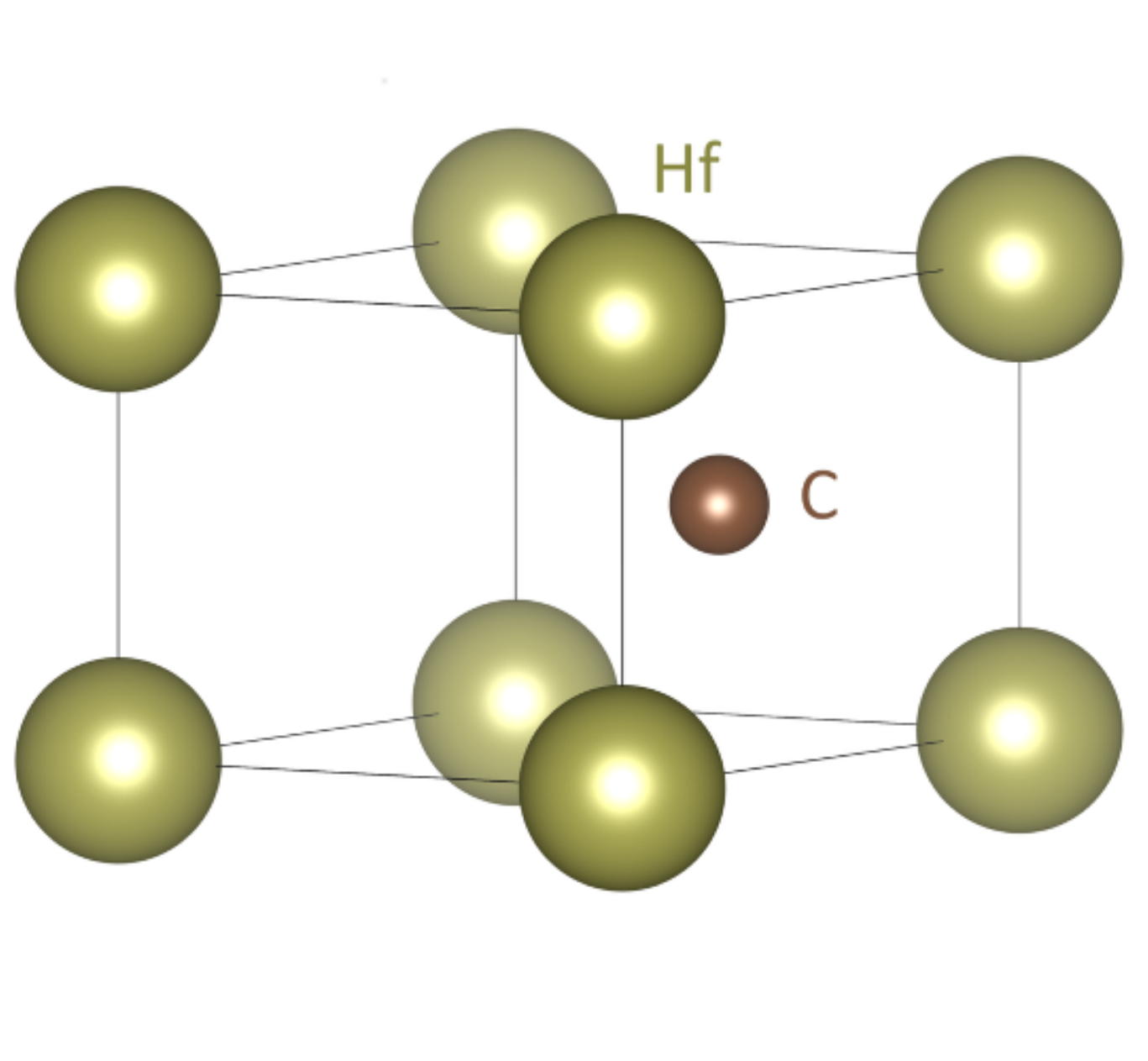}} &
\multicolumn{2}{c}{\includegraphics[width=1.5in]{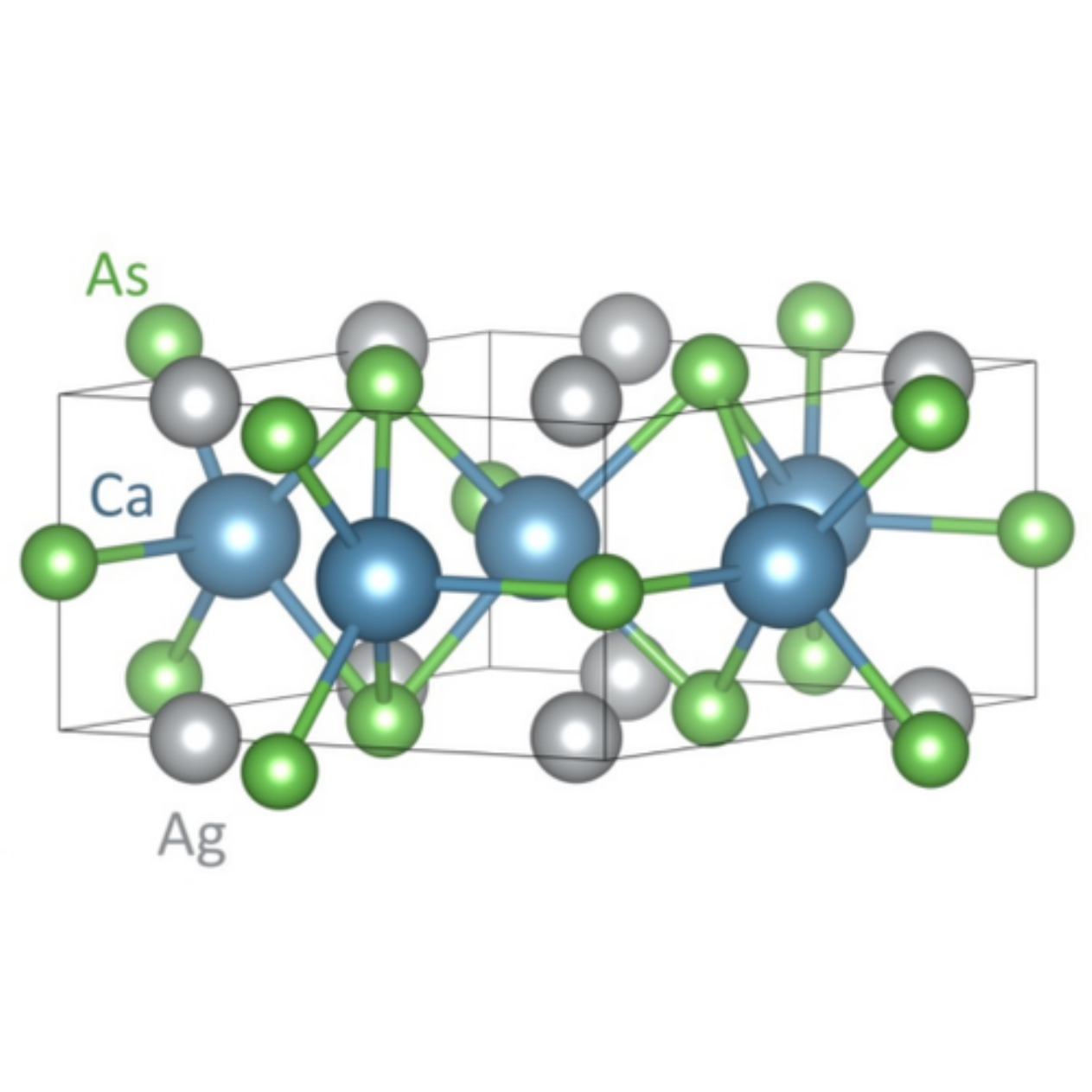}} \\
\hline

\raisebox{1.5\totalheight}{\parbox[c|]{2cm}{\raggedright Band structure without and with SOC}} &
\includegraphics[width=1.6in]{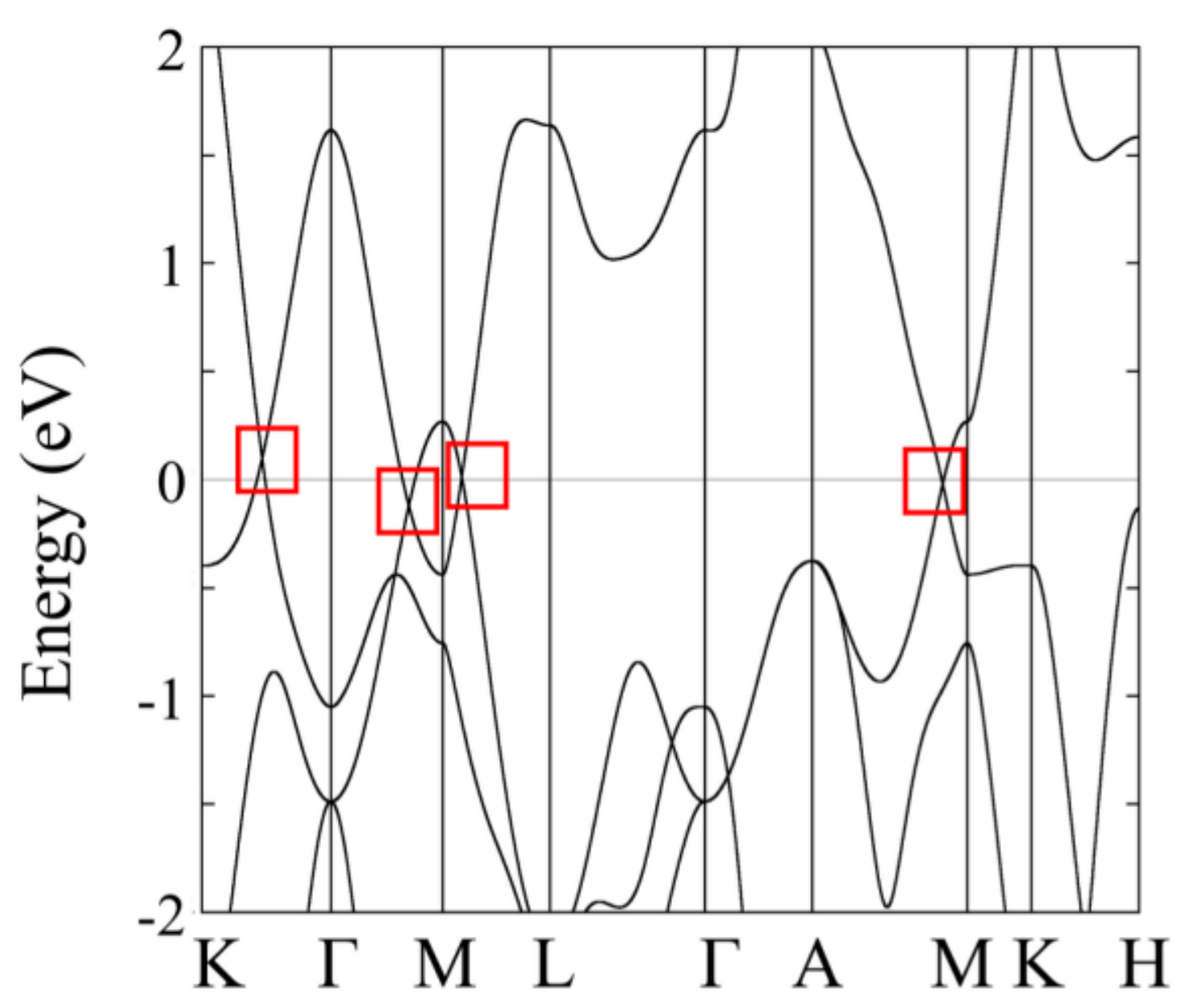} &
\includegraphics[width=1.6in]{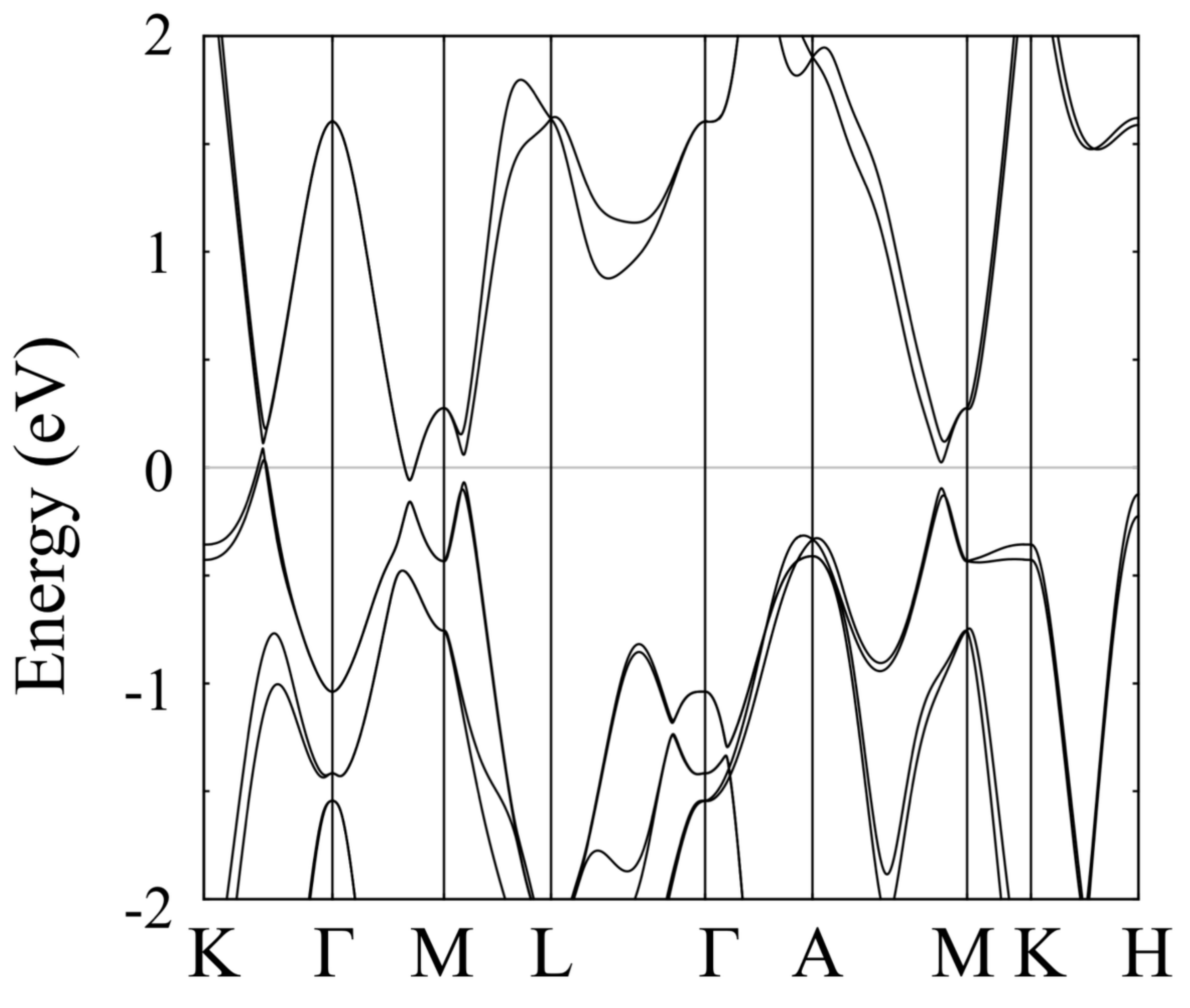} &
\includegraphics[width=1.55in]{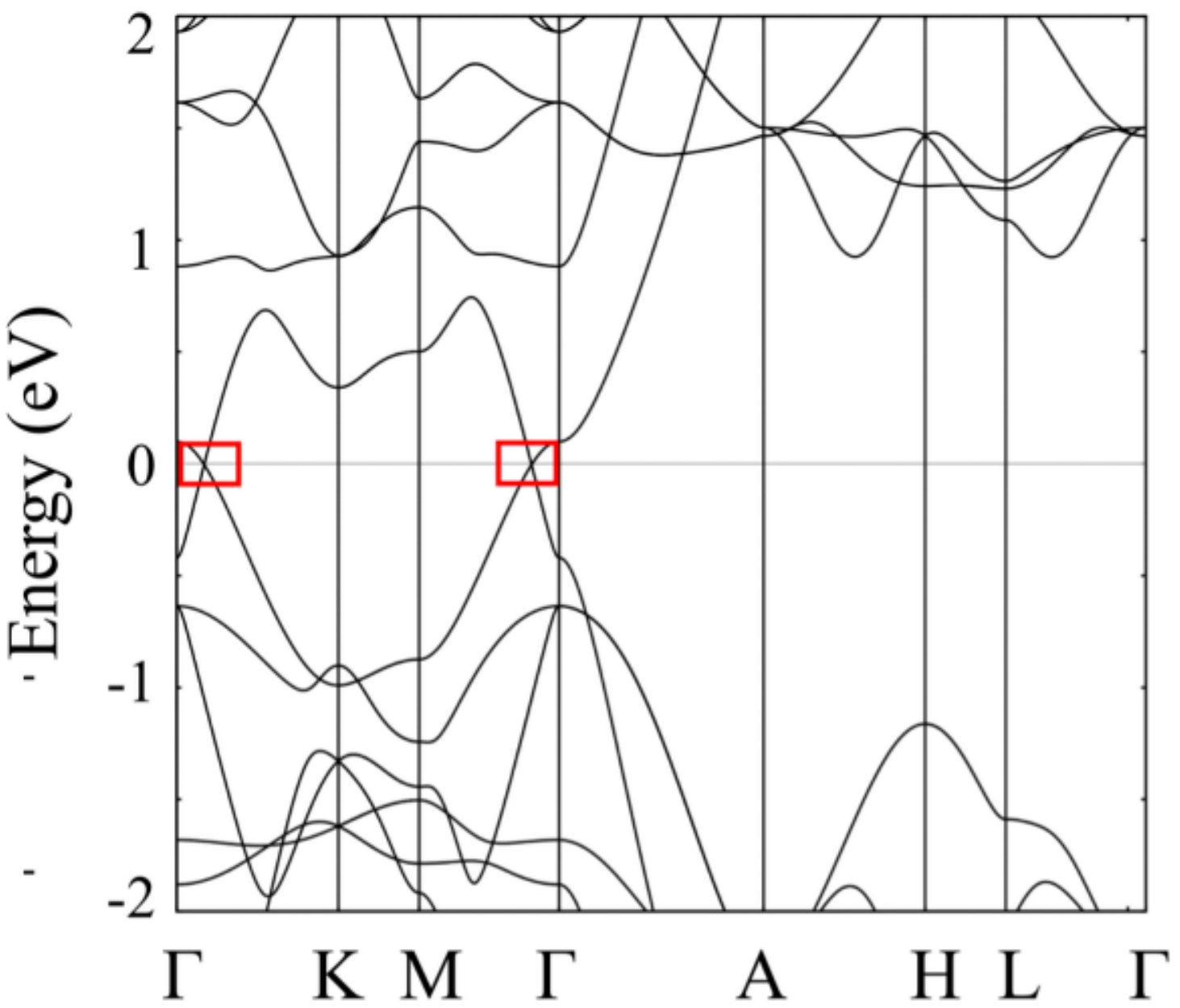} &
\includegraphics[width=1.55in]{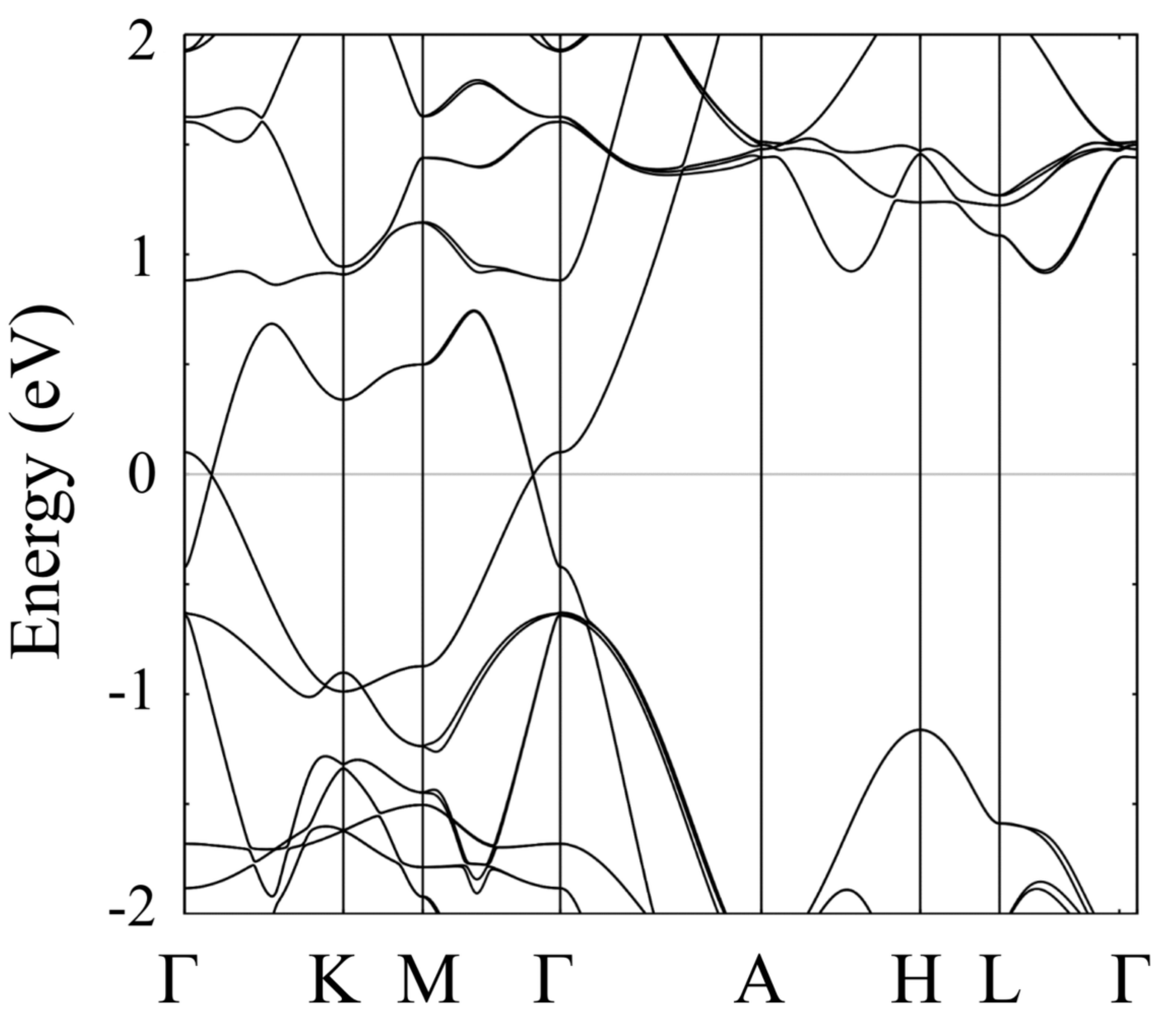} \\
\hline

\raisebox{1.2\totalheight}{\parbox[c|]{2cm}{\raggedright Nodal line distribution without and with SOC}} &
\includegraphics[width=1.3in]{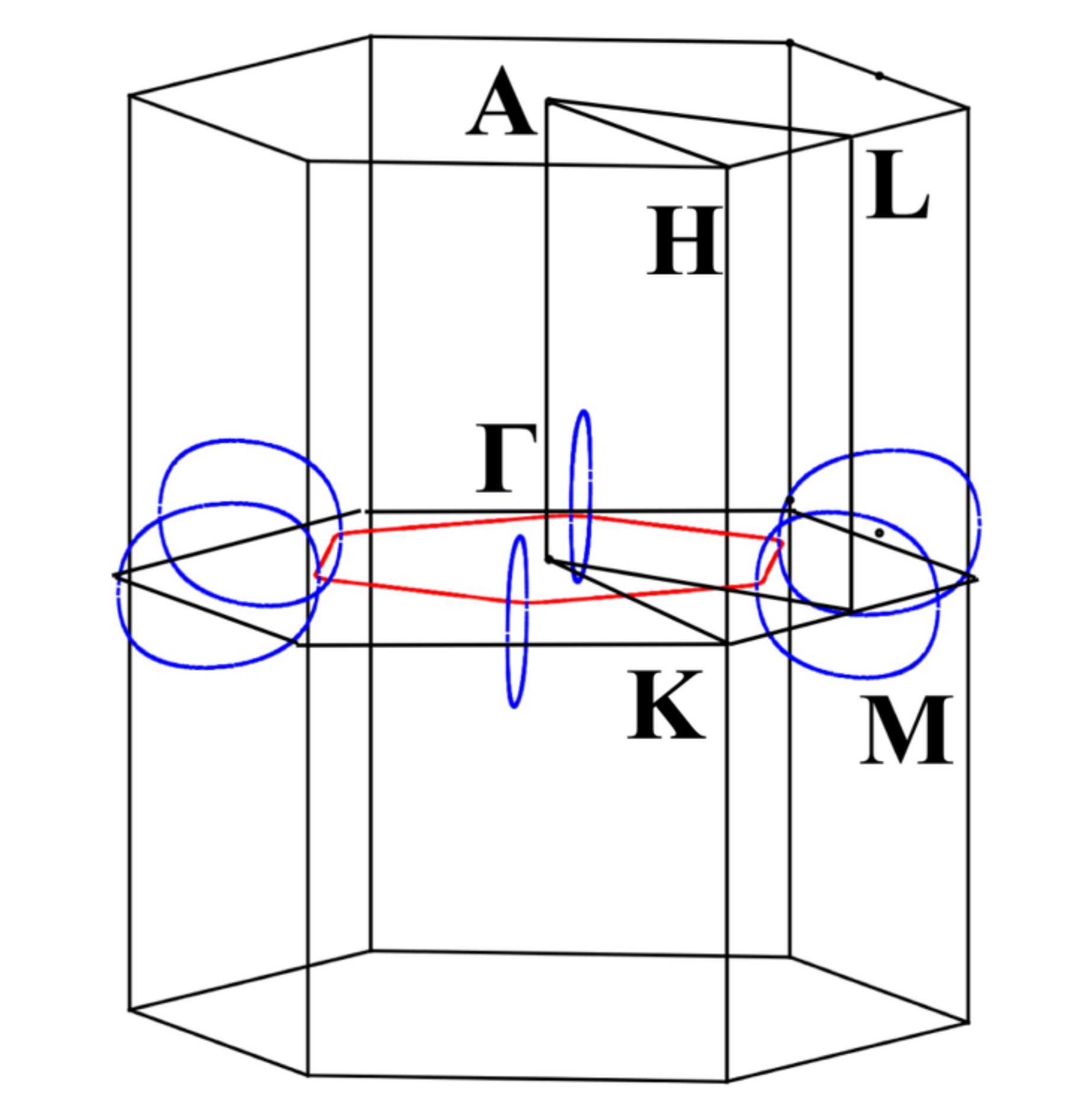} &
\includegraphics[width=1.28in]{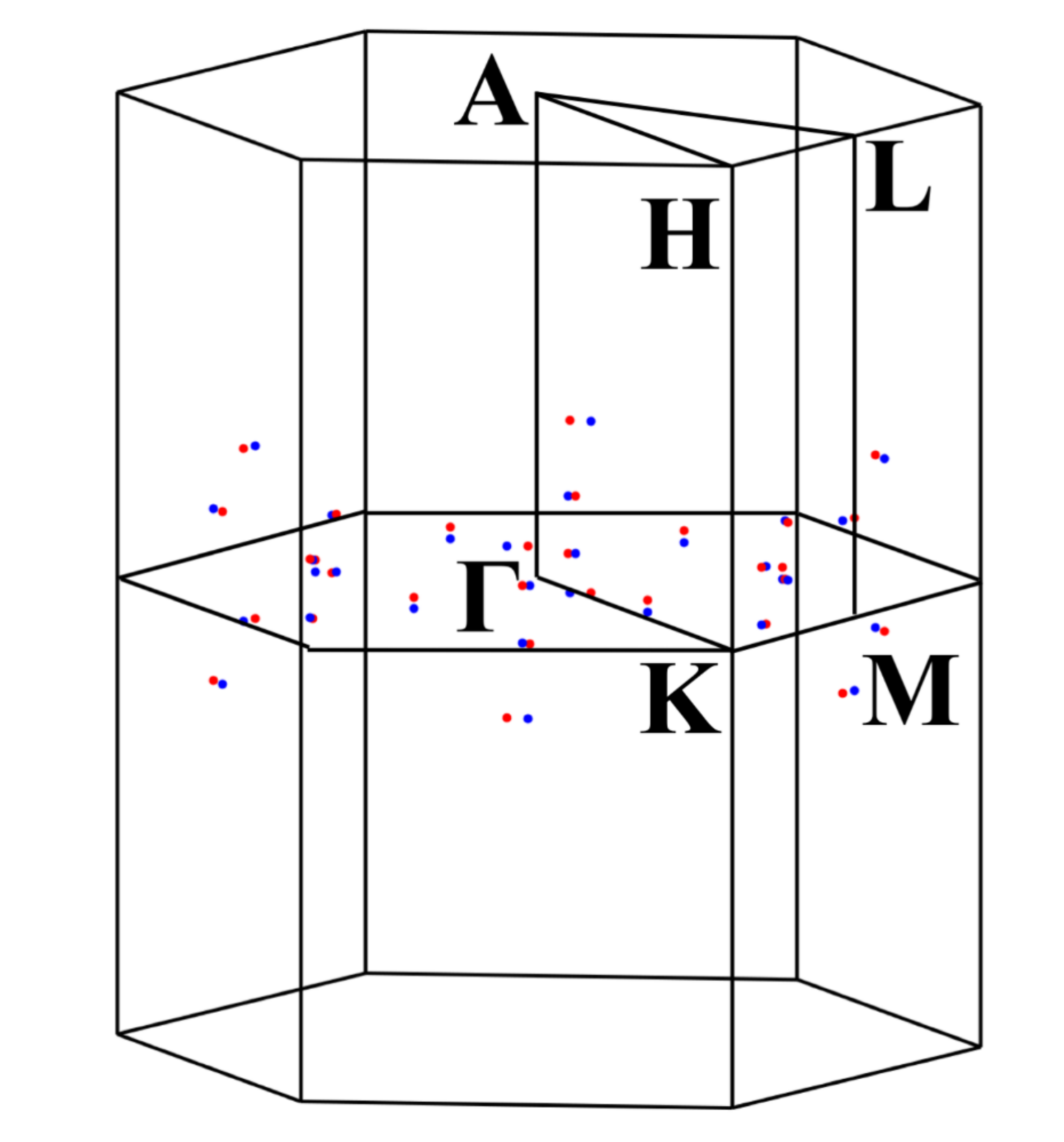} &
\includegraphics[width=1.5in]{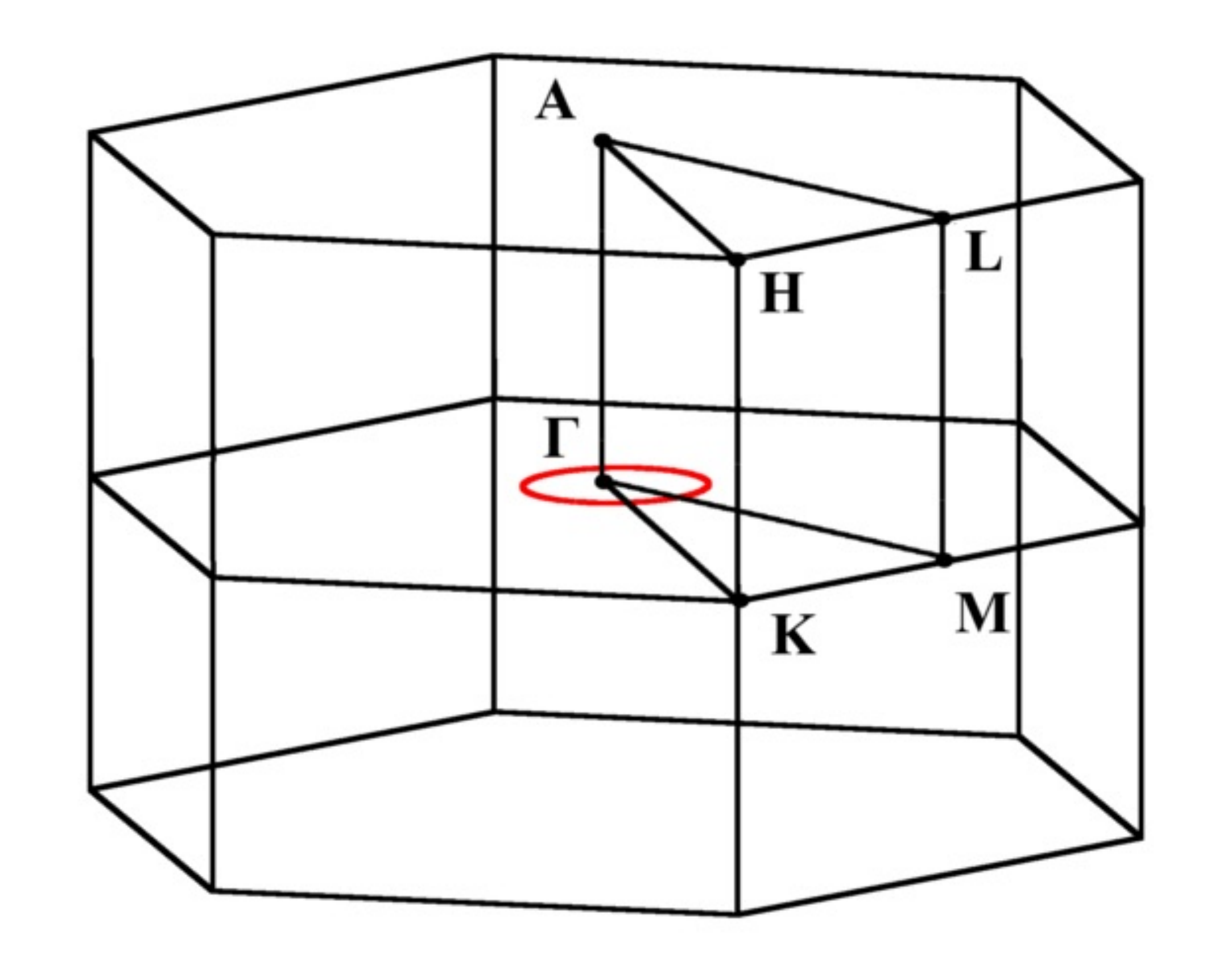} &
\raisebox{7\totalheight}{TI with SOC} \\
\hline

\raisebox{0.1\totalheight}{\parbox[c|]{2cm}{\raggedright References}} &
\multicolumn{2}{c|}{[85]} &
\multicolumn{2}{c}{[130]} \\
\hline

\bottomrule

\end{tabular}
\label{tab:gt}
\end{adjustbox}

\end{table*}

\textbf{Pb(Tl)TaSe$_2$} \cite{bian2016topological, ali2014noncentrosymmetric, guan2016superconducting, chang2016topological, bian2016drumhead, pang2016nodeless, zhang2016superconducting, sankar2017anisotropic}: PbTaSe$_2$, shown in Table IV, is a layered, non-centrosymmetric compound in space group $P$-$6m2$ No.187\cite{ali2014noncentrosymmetric}. Specific heat, electrical resistivity, and magnetic susceptibility, London penetration depth measurements have shown it to be a anisotropic type-II BCS superconductor \cite{ali2014noncentrosymmetric, pang2016nodeless, zhang2016superconducting, sankar2017anisotropic}. The crystal structure of PbTaSe$_2$ can be thought of as alternating stacks of TaSe$_2$ and hexagonal Pb layers, with Pb atoms directly sitting above the Se atoms. It contains a mirror plane going through the Ta atomic plane that reflects k$_z$ to -k$_z$, which plays the essential role in protecting the topological nodal line. \par

Without SOC, the conduction and valence bands of both Pb and Se belong to different representations with opposite eigenvalues, forming spinless nodal rings around H and H' in the plane of k$_z$=$\pi$. When SOC is included, due to the spin degeneracy being lifted because of the lack of inversion symmetry, each band splits into two bands with opposite spin polarization and mirror reflection eigenvalues. Consequently, the spinless Dirac nodal ring turns into four spinful Weyl nodal rings. Of these four spinful nodal rings, only two are protected by mirror reflection, resulting in two nodal rings around H protected by the reflection symmetry with respect to Ta atom plane. In addition, SOC also gives rise to another accidental crossing generating a nodal ring at the K point on the k$_z$=0 plane that does not exist without SOC. The ARPES measurements (Table IV) along M-K-$\Gamma$ show the nodal ring around K point, with the surface states (SS) indicated by the blue arrows\cite{bian2016topological, chang2016topological}. This represents a rare case where SOC creates a Weyl nodal ring by generating `new' crossings rather than by just splitting Dirac nodal rings apart. It also serves as a good example that systems with 2D massive Dirac fermions can be interfaced with a transition metal dichalcogenide to create thermodynamically stable 3D massive Dirac fermions. \par

Aside from the mirror-reflection protected and the accidental band crossing generated nodal lines, spin momentum locked spin texture is another interesting feature of PbTaSe$_2$, as it is one of the most prominent characteristics of topological surface states\cite{guan2016superconducting, chang2016topological}. Two superconducting topological surface state (TSS) have been predicted by DFT calculations with opposite helical spin polarization existing within the bulk gap at $\Gamma$, and have been confirmed by quasiparticle scattering interference results\cite{guan2016superconducting}. The fully gapped superconducting order of the TSSs in the presence of helical spin polarization make PbTaSe$_2$, and related compounds, promising candidates to be topological superconductors. \par

\textbf{TaAs family}\cite{weng2015weyl, yang2015weyl, xu2015discovery, lv2015experimental, yang2015weyl, liu2016evolution, xu2015discovery1, belopolski2016criteria, inoue2016quasiparticle, batabyal2016visualizing, shekhar2015extremely, ghimire2015magnetotransport, wang2016helicity, zhang2016signatures, luo2015electron, huang2015observation, zhang2016signatures, arnold2016negative, wang2016helicity, moll2016magnetic, zhang2016signatures, arnold2016negative, shekhar2015extremely, wang2016helicity}: TaAs is in a body-centered tetragonal structure with non-symmorphic space group $I$4$_1$$md$ No. 109, which lacks inversion symmetry (Table IV). Its structure can be thought of as planes of face sharing trigonal prisms stacked on top of each other but rotated by 90 degrees every layer. It contains two mirror planes, M$_x$ and M$_y$, and two glide planes, M$_x$$_y$ and M$_-$$_x$$_y$. \par

Without SOC, band crossings within the ZN$\Gamma$ plane are protected by M$_y$ because of the opposite mirror eigenvalues, therefore creating nodal rings within the plane. It is important to note here that although TaAs also has glide mirrors, the non-symmorphic mirror operations do not generate the crossings in this system. Including SOC, the nodal rings within the ZN$\Gamma$ plane are fully gapped, resulting in two pairs of Weyl points located 2 meV above the E$_F$ and four pairs of Weyl points located about 21 meV below the E$_F$. All Weyl points are located slightly off the mirror planes. \par 

After the theoretical prediction of this Weyl semimetal family (TaAs, TaP, NbAs, and NbP), a series of photoemission, tunneling spectroscopy, and transport measurements have been carried out to probe the characteristic of Weyl fermions. These include the observation of Fermi arcs by ARPES \cite{xu2015discovery, lv2015experimental, yang2015weyl, liu2016evolution, xu2015discovery1, belopolski2016criteria} and by scanning tunneling microscope (STM) \cite{inoue2016quasiparticle, batabyal2016visualizing}. For example, Table IV shows an an ARPES measurement of a broadband dispersion (i) and a zoomed-in plot (ii) across the two Weyl points, denoted by WP+ and WP-.  

The linear band dispersion and non-trivial Berry phase are characterized by chiral magnetotransport properties, such as ultra-high mobility, large transverse MR \cite{shekhar2015extremely, ghimire2015magnetotransport, wang2016helicity, zhang2016signatures, luo2015electron}, and negative longitudinal MR \cite{huang2015observation, zhang2016signatures, arnold2016negative, wang2016helicity, moll2016magnetic}, and the studies of chiral magnetotransport properties \cite{huang2015observation, zhang2016signatures, arnold2016negative, shekhar2015extremely, wang2016helicity}. \par

\textbf{HfC}\cite{yu2017nodal}:  The WC-type HfC is in hexagonal space group $P\overline{6}m2$ No. 187 (Shown in Table V). It has time-reversal symmetry and two mirror planes, including M$_x$$_y$ and M$_x$$_z$, but lacks inversion symmetry. Protected by mirror symmetry M$_x$$_y$ and M$_x$$_z$, two types of nodal rings emerge in the absence of SOC, one lying in the M$_x$$_y$ plane around the $\Gamma$ point, and the other one lying in the M$_x$$_z$ plane around the M point. With SOC, the crossings along $\Gamma$-M, M-L, and M-A are fully gapped, split the nodal chain into 30 pairs of Weyl points off the mirror planes. \par

\textbf{CaAg$X$($X$=P, As)}\cite{yamakage2015line}: CaAg$X$ crystallizes in the ZrNiAl-type structure with space group $P$-$62m$ No. 189 (Table V). The Ag$X$$_4$ is in kagome-triangular lattice, which is formed by edge and corner sharing tetrahedra with intercalated Ca atoms. It has mirror reflection symmetry (M$_x$$_y$) but lacks inversion symmetry. Without SOC, the bulk DNL is protected by M$_x$$_y$, creating a nodal ring around the $\Gamma$ point in the BZ. Turning on SOC, the degeneracy in dihedral point group symmetry is lifted, transforming the system into a TI. For CaAgP, however, the effect of SOC is expected to be less than 1 meV\cite{yamakage2015line}.  \par


\subsection{Non-symmorphic symmetry protected Dirac nodal line materials}

Non-symmorphic symmetry is a global symmetry that is robust against SOC. As discussed earlier, non-symmorphic symmetry creates band folding, resulting in otherwise singly degenerate points to cross at certain doubly degenerate points at the BZ boundaries. \par

Table VI to VII show the crystal structures, electronic band structures without and with SOC, nodal line distributions without and with SOC within a BZ, as well as the available ARPES measurements confirming the presence of non-symmorphic symmetry protected nodal lines and gapped unprotected nodal lines. In the electronic band structure figures, the Dirac points corresponding to the symmorphic symmetry generated nodal lines are coded with red square boxes, and those protected by non-symmorphic symmetry are coded with green squares (ovals). In the nodal line distribution figures, the nodal lines generated by symmorphic symmetry are shown in red lines, and those generated and protected by non-symmorphic symmetry are shown with green lines. Only non-symmorphic symmetry generated nodal lines survive in the presence of SOC. \par

\textbf{$M$SiS} \cite{young2015dirac, schoop2016dirac, neupane2016observation, chen2017dirac, takane2016dirac, ali2016butterfly, wang2016evidence, singha2016titanic, hu2016evidence1, kumar2016unusual, topp2016non, lv2016extremely}: Zr(Hf)SiS is a tetragonal PbFCl-type compound with space group $P4/nmm$ No. 129. It has a two-dimensional layered structure consisting of Zr(Hf) and two neighboring S layers sandwiched between Si layers. The square nets of Si atoms are located on a glide plane (Table VI). \par

The band structures without and with SOC clearly show that the DNL generated by non-symmorphic symmetry of the square lattice explained earlier is protected from gap opening even in the presence of SOC \cite{young2015dirac}. The bands responsible for the crossings linearly disperse as high as 2 eV above and below the E$_F$, which is a much larger range compared with the other known Dirac materials. For ZrSiS, without SOC, several Dirac-like crossings can be observed along $\Gamma$-X, $\Gamma$-M, Z-R and Z-A, and are protected by C$_2v$. The nodal loop generated by these crossings is shown in red in Table VI. In addition, there are two other Dirac crossings at the X and R points, which are located 0.7 eV and 0.5 eV below the E$_F$ as well at M and A, located at 2.3 eV below the E$_F$. These crossings, shown by the solid green and light blue lines in nodal line distribution without SOC in table VI, are located within the glide mirror planes, protected by the non-symmorphic symmetry. With SOC, the C$_2v$ point group allows only one irreducible representation, and therefore gaps out the $\Gamma$-X, $\Gamma$-M, Z-R and Z-A crossings in both ZrSiS and HfSiS. However, the non-symmorphic protected crossings at M, X, A, and R are robust against SOC, although these are buried below the E$_F$. \par

\begin{table*}[htbp]
\caption{Non-symmorphic symmetry protected nodal line materials}
\centering

\begin{adjustbox}{width=0.9\textwidth}

\newcommand*{\TitleParbox}[1]{\parbox[c]{1.75cm}{\raggedright #1}}%

\makeatletter
\newcommand{\thickhline}{%
    \noalign {\ifnum 0=`}\fi \hrule height 1pt
    \futurelet \reserved@a \@xhline
}
\newcolumntype{"}{@{\hskip\tabcolsep\vrule width 1pt\hskip\tabcolsep}}
\makeatother

\newcolumntype{?}{!{\vrule width 1pt}}

\hspace*{-1cm}
\begin{tabular}{c ? cc|cc} 
 
 \toprule

\textbf{Materials} & \multicolumn{2}{c|}{\textbf{ZrSiS}} &  \multicolumn{2}{c}{\textbf{HfSiS}} \\
\thickhline

\midrule

\raisebox{2\totalheight}{\parbox[c|]{2cm}{\raggedright Crystal structure}} &

\multicolumn{2}{c|}{\includegraphics[width=1.1in]{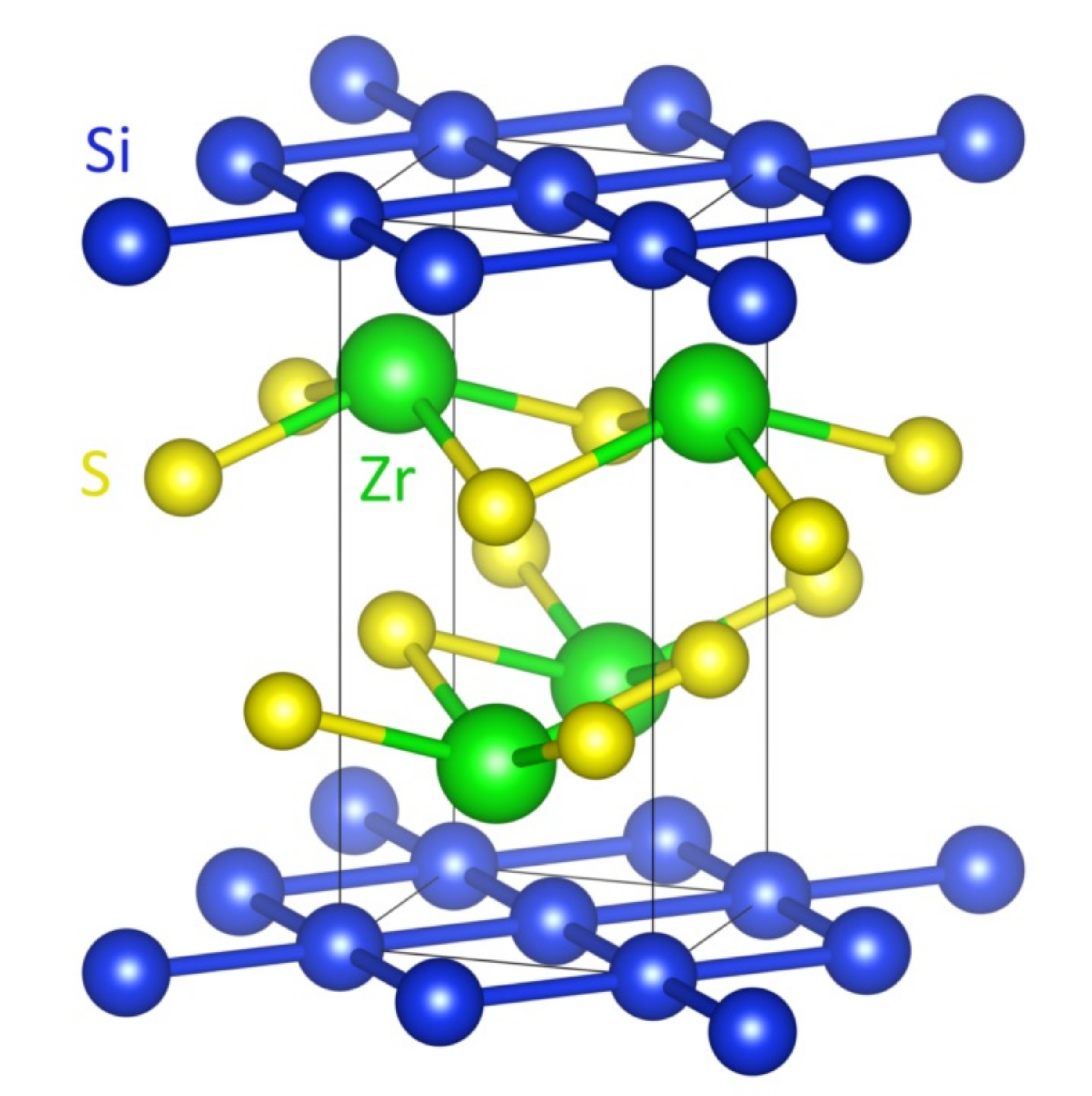}} &
\multicolumn{2}{c}{\includegraphics[width=1.1in]{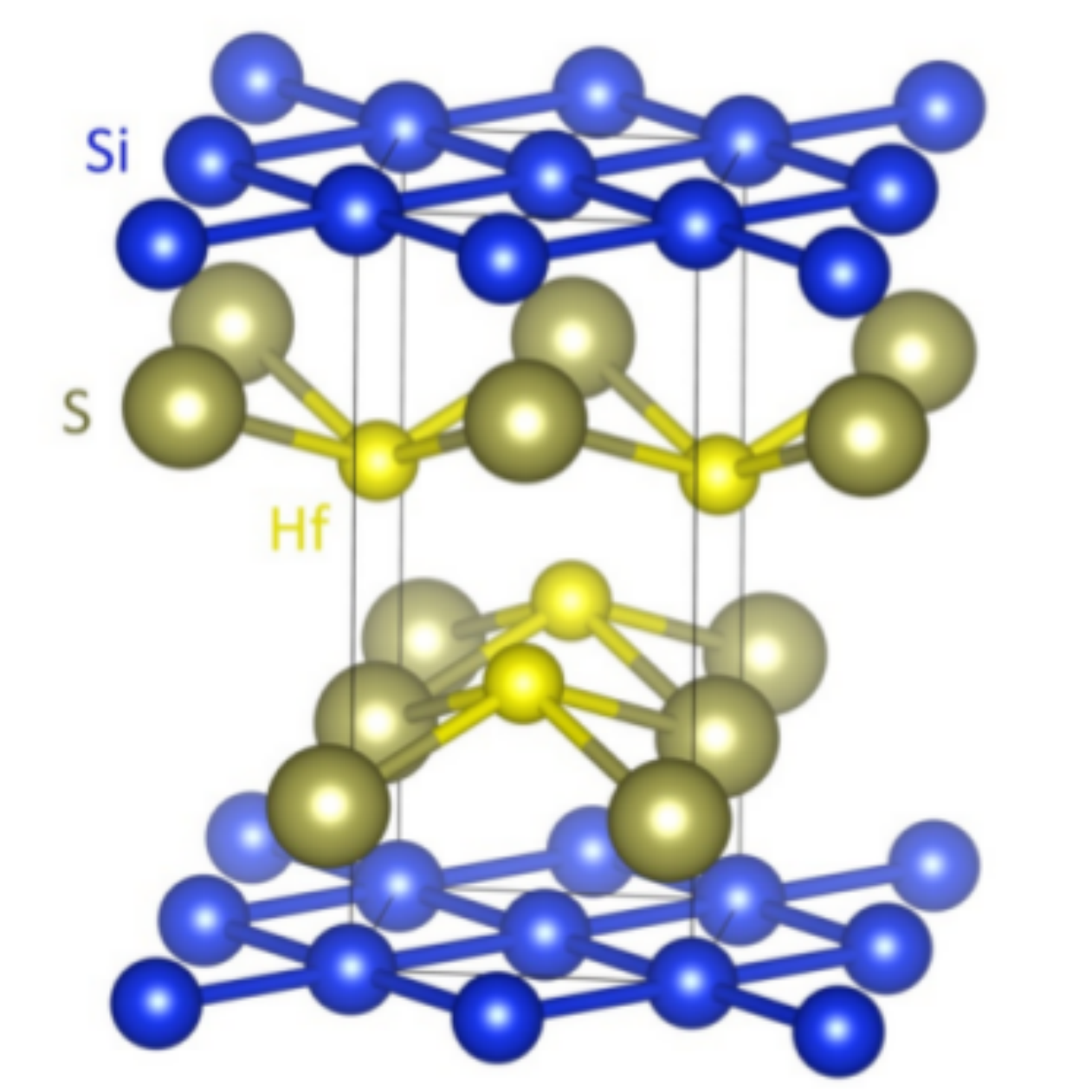}}\\
\hline

\raisebox{1.5\totalheight}{\parbox[c|]{2cm}{\raggedright Band structure without and with SOC}} &
\includegraphics[width=1.5in]{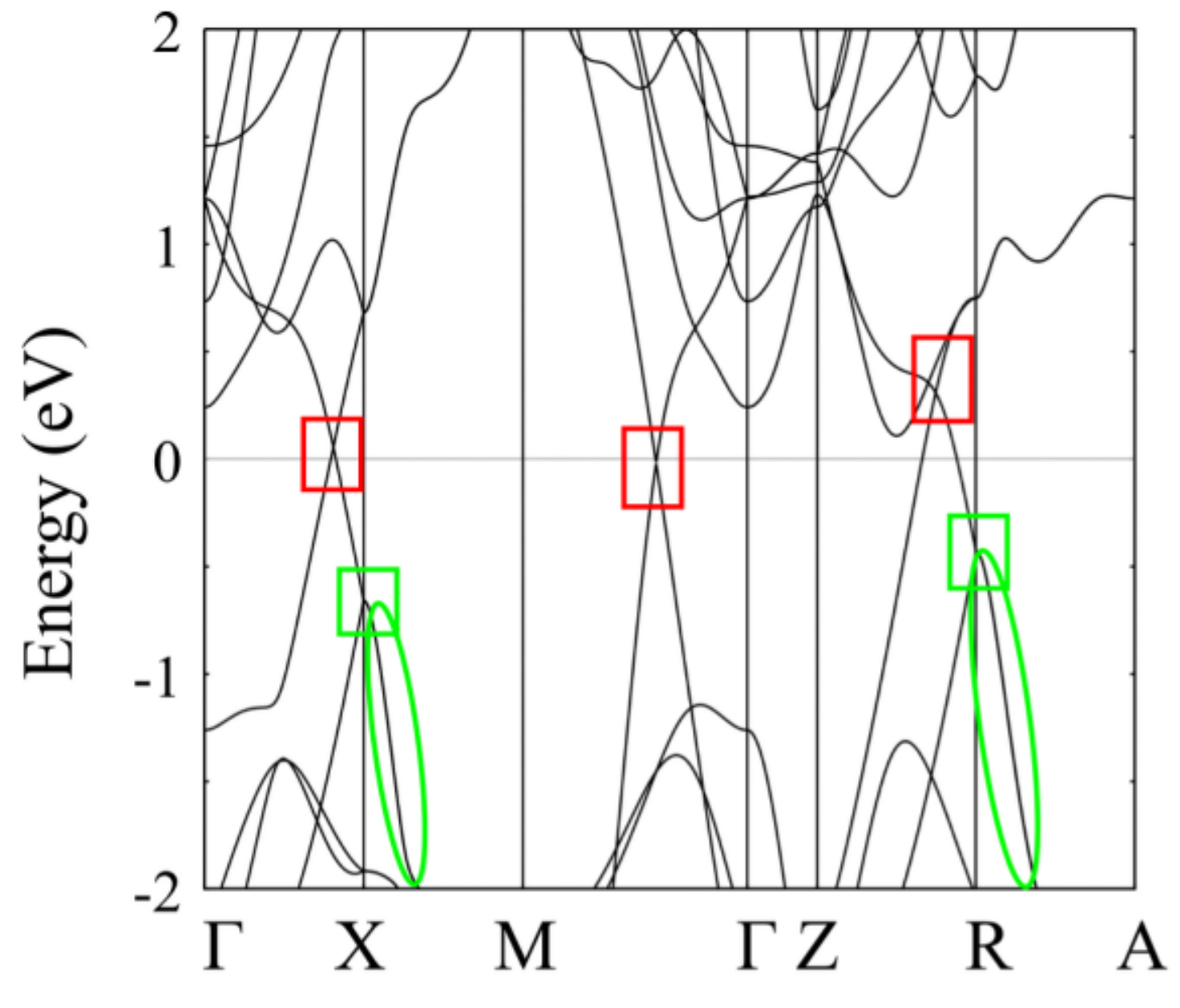} &
\includegraphics[width=1.5in]{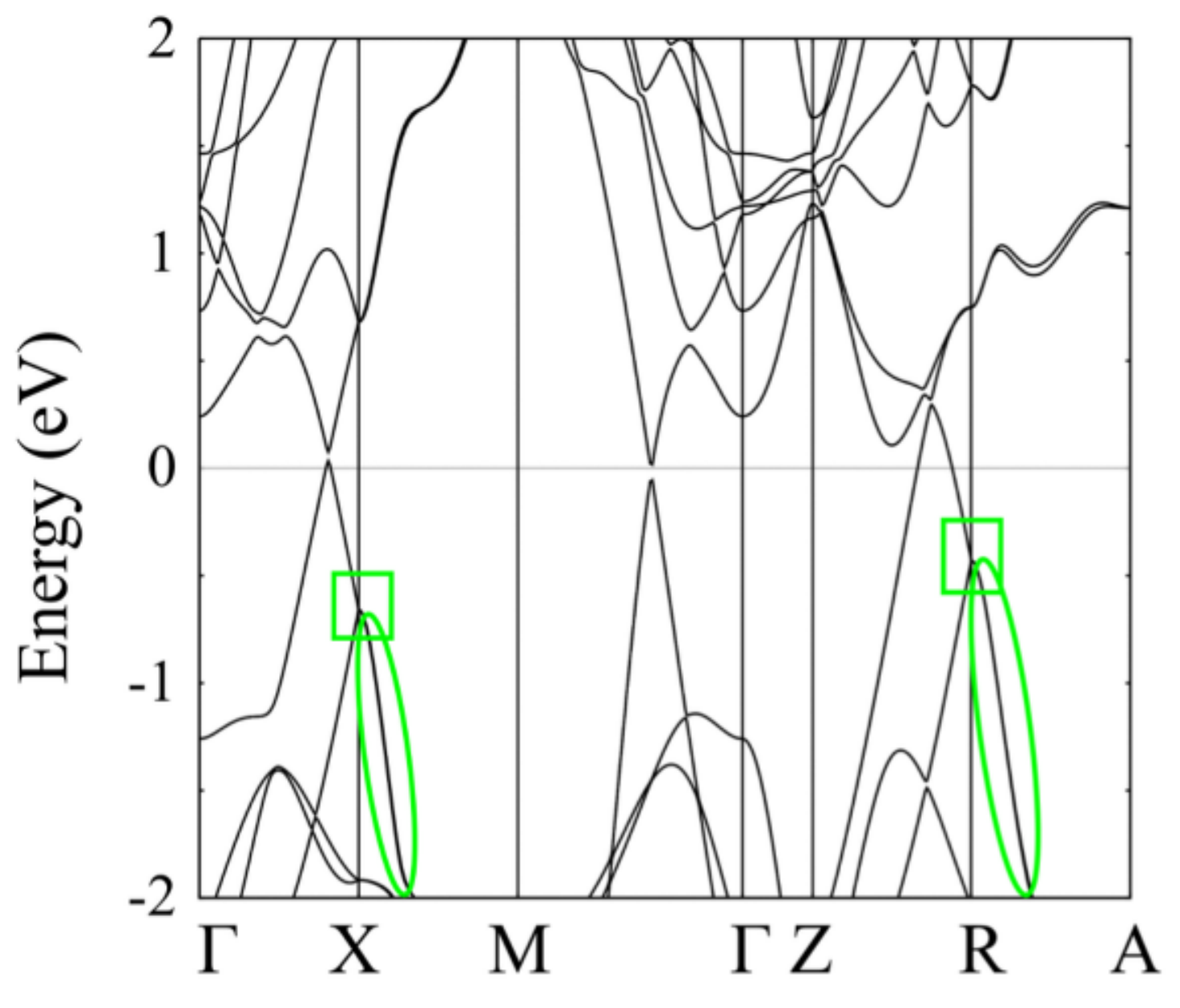}&
\includegraphics[width=1.5in]{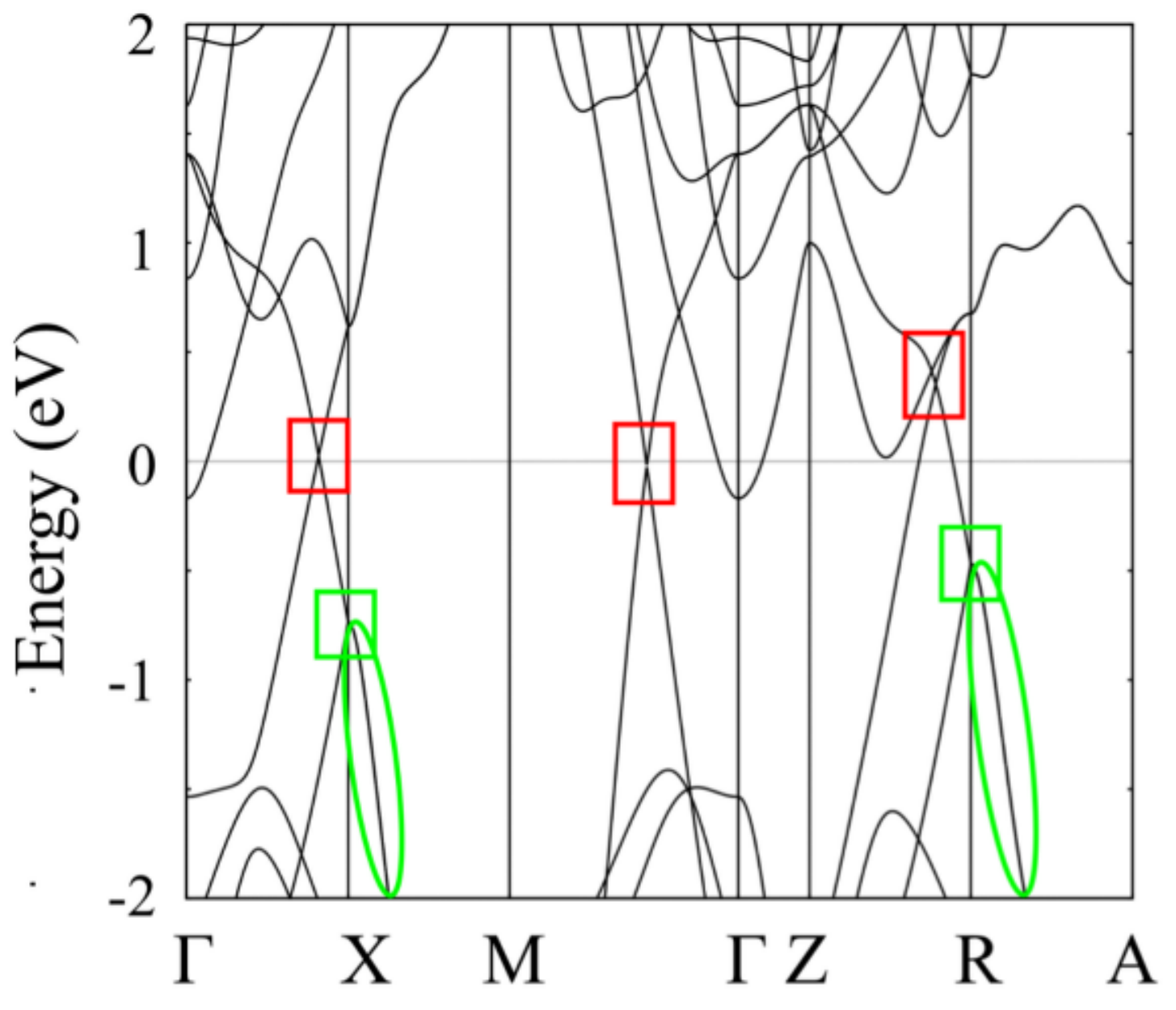} &
\includegraphics[width=1.5in]{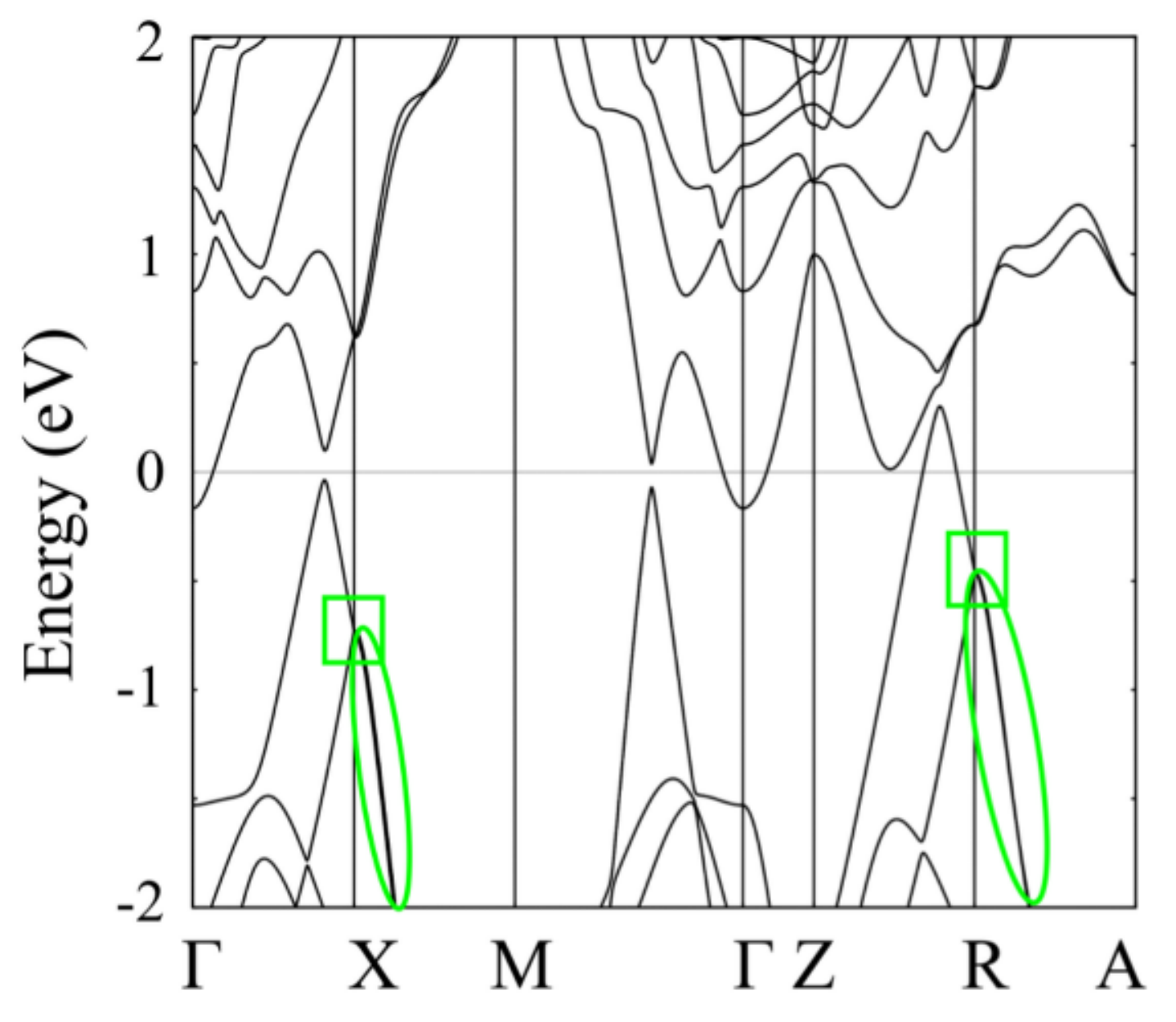} \\
\hline

\raisebox{1\totalheight}{\parbox[c|]{2cm}{\raggedright Nodal line distribution without and with SOC}} &
\includegraphics[width=1in]{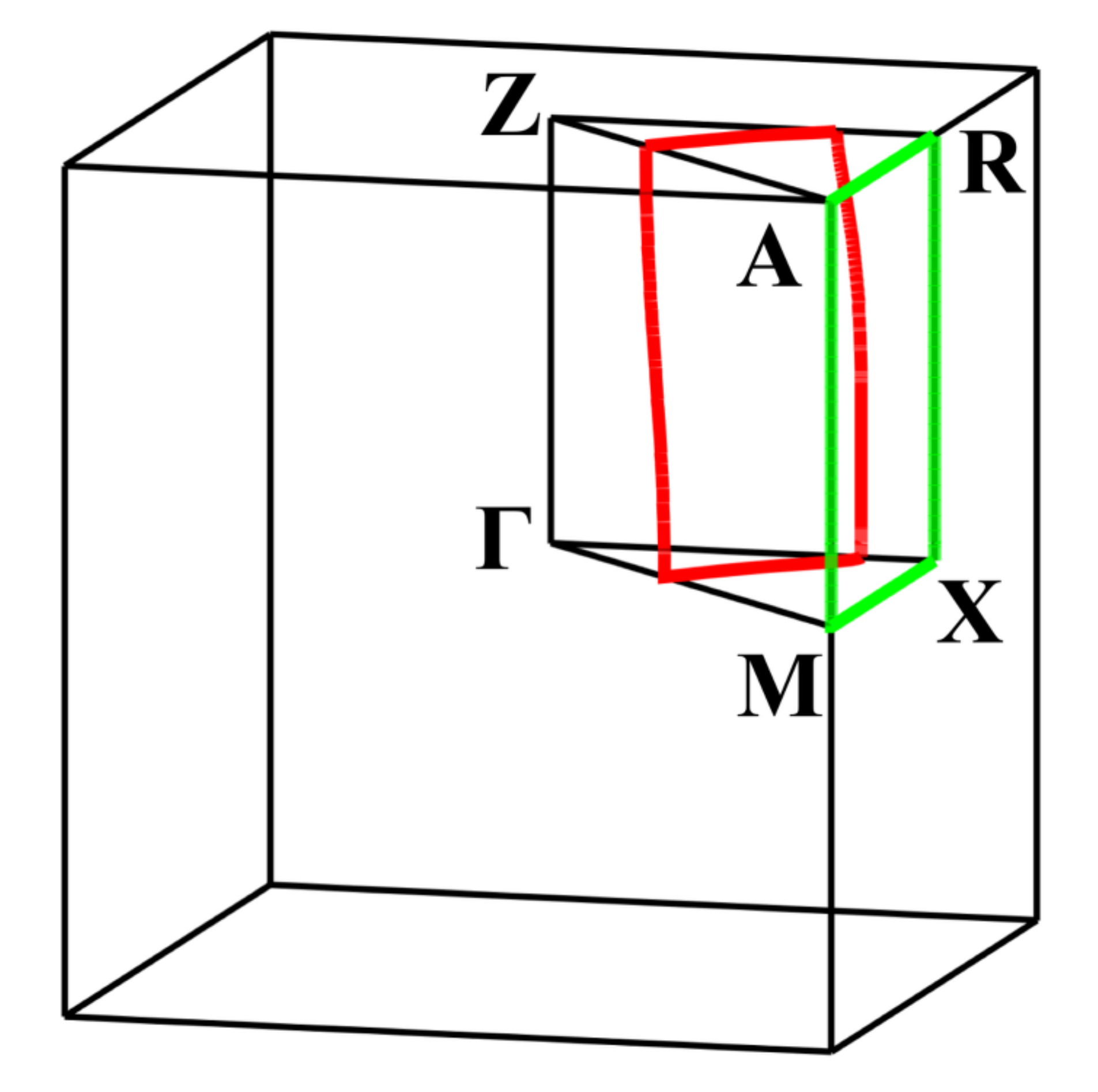} &
\includegraphics[width=1in]{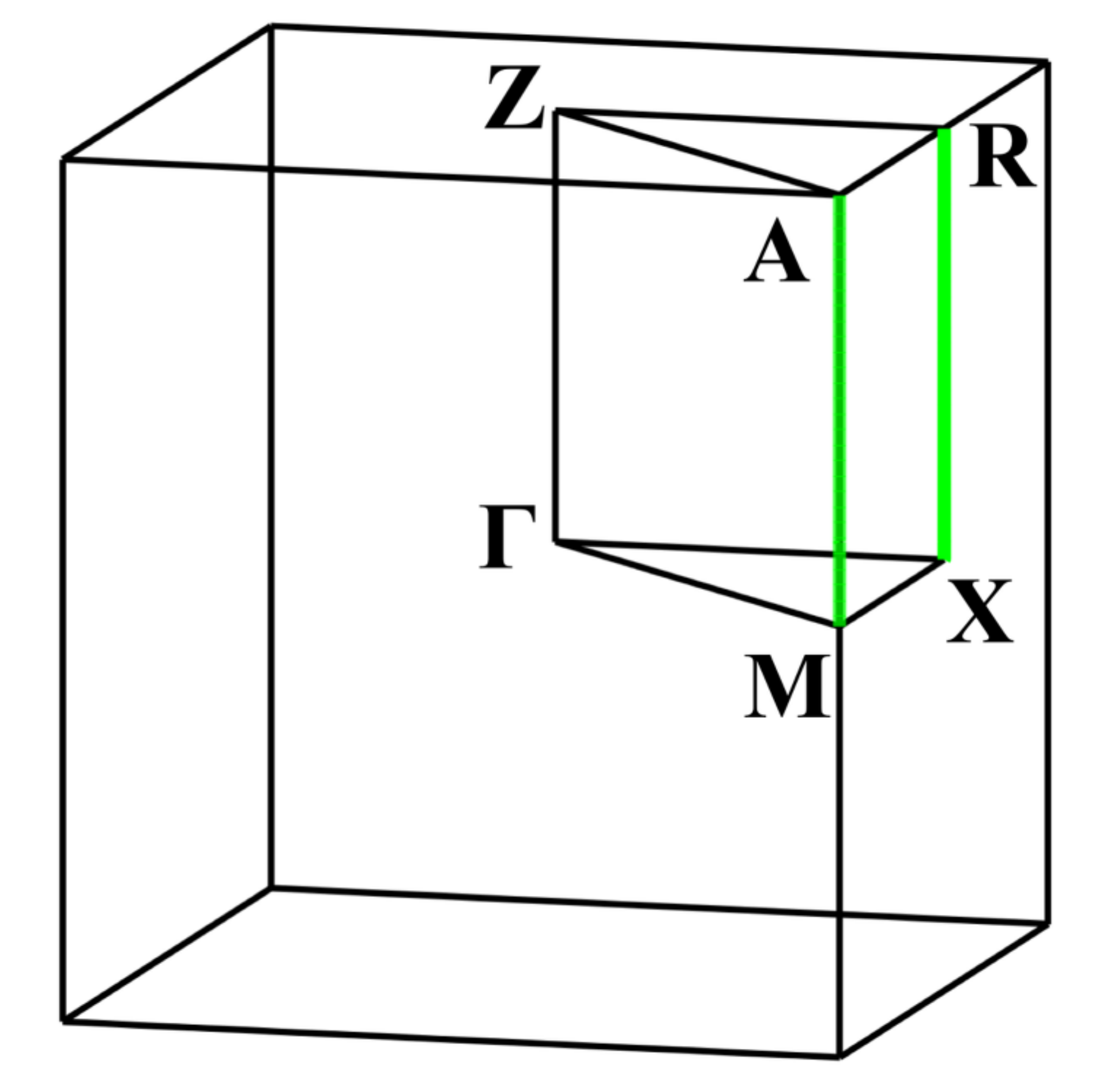} &
\includegraphics[width=1in]{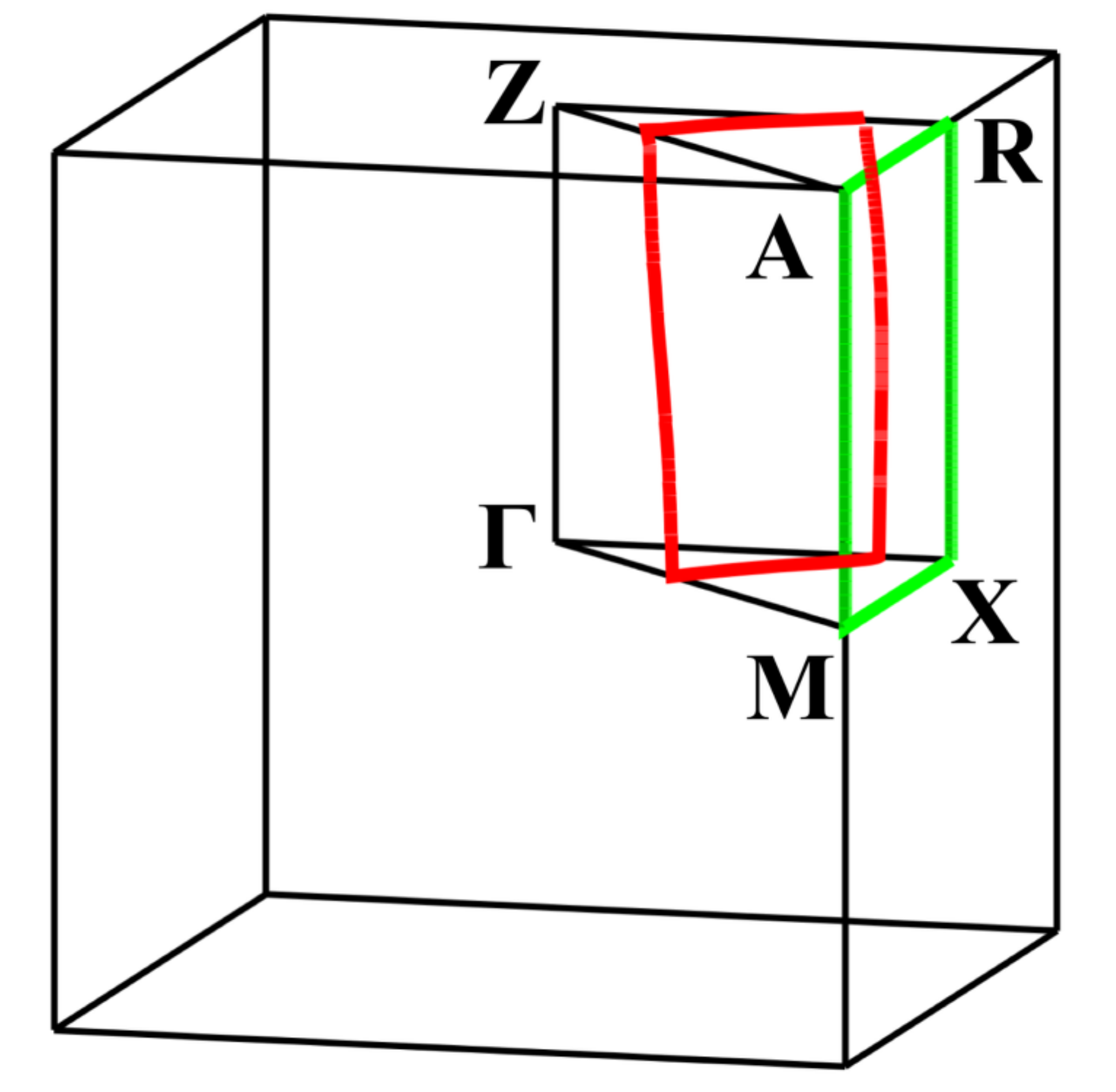} &
\includegraphics[width=1in]{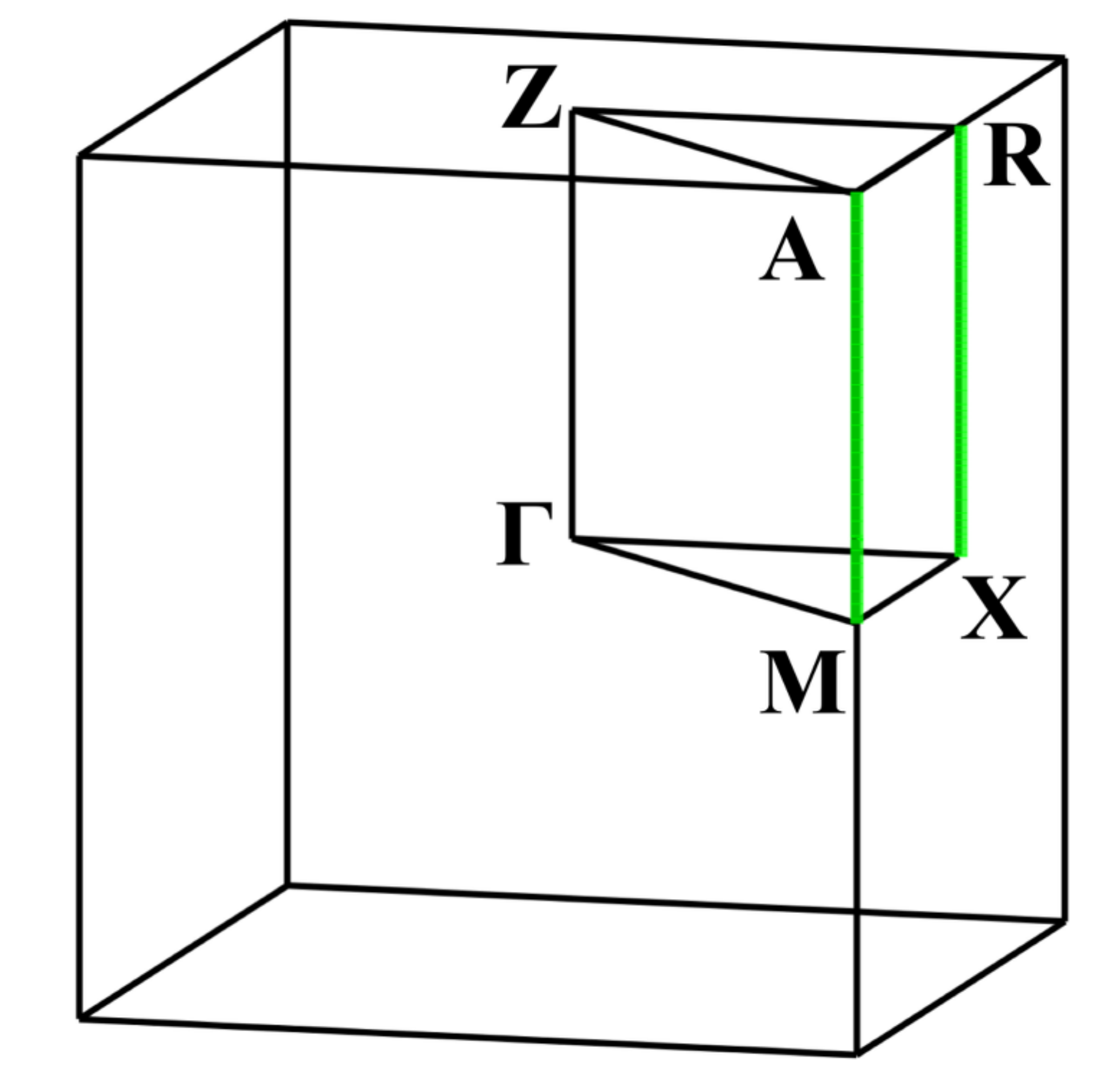} \\
\hline

\raisebox{\totalheight}{\parbox[c|]{2cm}{\raggedright ARPES\cite{chen2017dirac}}} &

\multicolumn{2}{c|}{\includegraphics[width=3in]{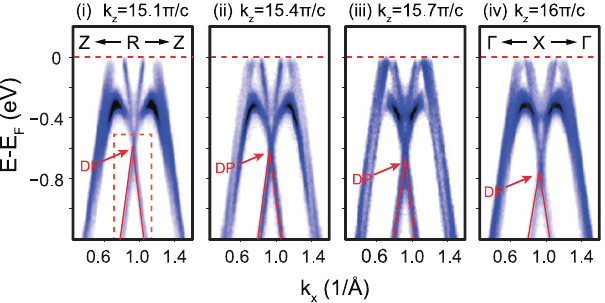}} &
\multicolumn{2}{c}{\includegraphics[width=3in]{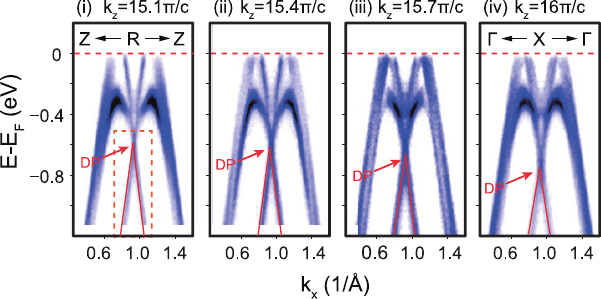}} \\

{} &
\multicolumn{2}{c|}{\includegraphics[width=1.2in]{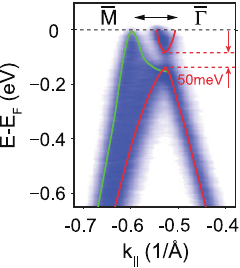}} &
\multicolumn{2}{c}{\includegraphics[width=1.2in]{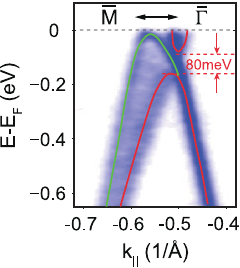}} \\
\hline

\raisebox{0.1\totalheight}{\parbox[c|]{2cm}{\raggedright References}} &
\multicolumn{4}{c}{[74], [77], [78], [87], [131-138]} \\
\hline

\bottomrule

\end{tabular}
\label{tab:gt}
\end{adjustbox}

\end{table*}

\begin{table*}[htbp]
\caption{Non-symmorphic symmetry protected nodal line materials (Continued)}
\centering

\begin{adjustbox}{width=0.9\textwidth}

\newcommand*{\TitleParbox}[1]{\parbox[c]{1.75cm}{\raggedright #1}}%

\makeatletter
\newcommand{\thickhline}{%
    \noalign {\ifnum 0=`}\fi \hrule height 1pt
    \futurelet \reserved@a \@xhline
}
\newcolumntype{"}{@{\hskip\tabcolsep\vrule width 1pt\hskip\tabcolsep}}
\makeatother

\newcolumntype{?}{!{\vrule width 1pt}}

\hspace*{-1cm}
\begin{tabular}{c ? cc|cc} 
 
 \toprule

\textbf{Materials} & \multicolumn{2}{c|}{\textbf{SrIrO$_3$}} &  \multicolumn{2}{c}{\textbf{IrO$_2$}} \\
\thickhline

\midrule

\raisebox{2\totalheight}{\parbox[c|]{2cm}{\raggedright Crystal structure}} &

\multicolumn{2}{c|}{\includegraphics[width=1.3in]{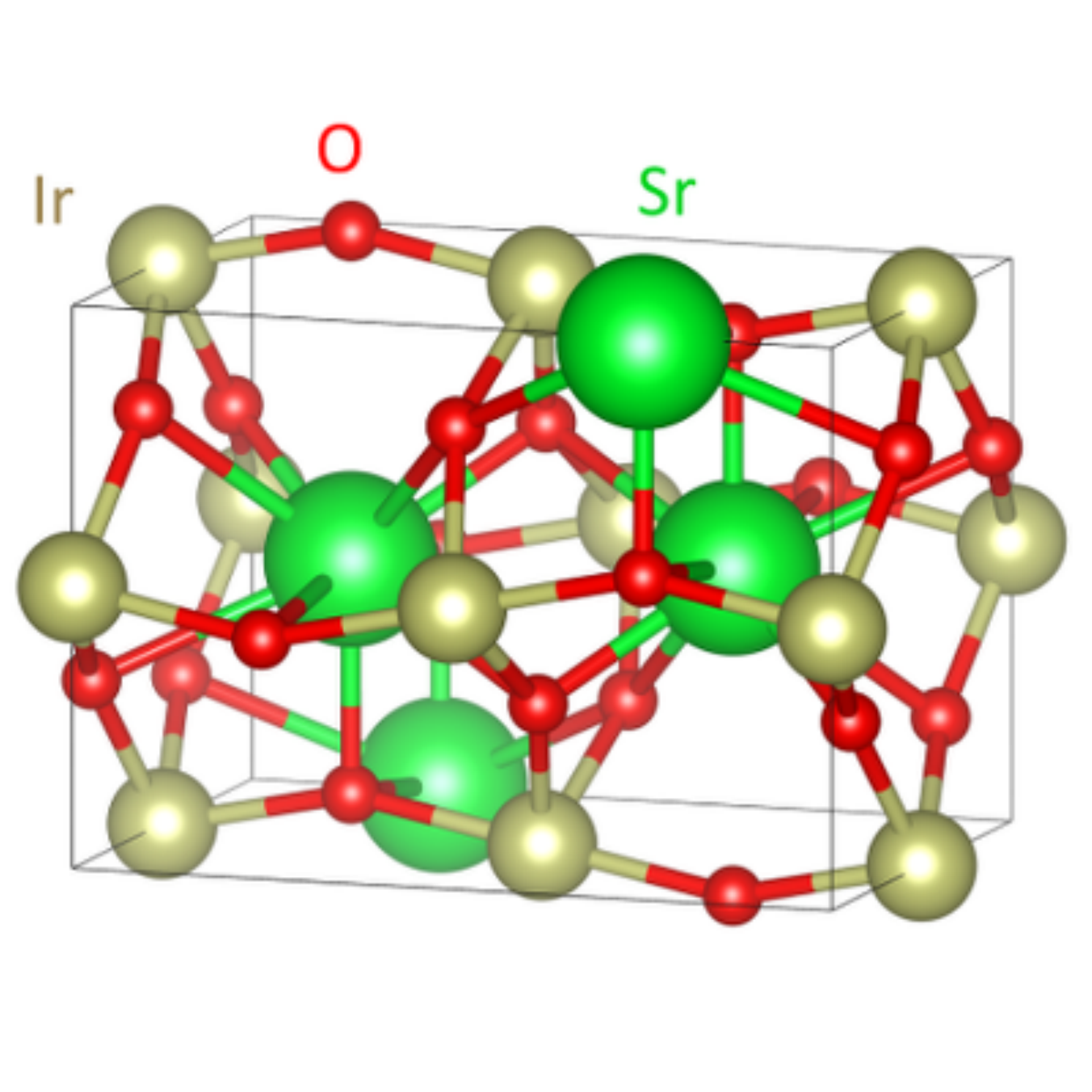}} &
\multicolumn{2}{c}{\includegraphics[width=1.3in]{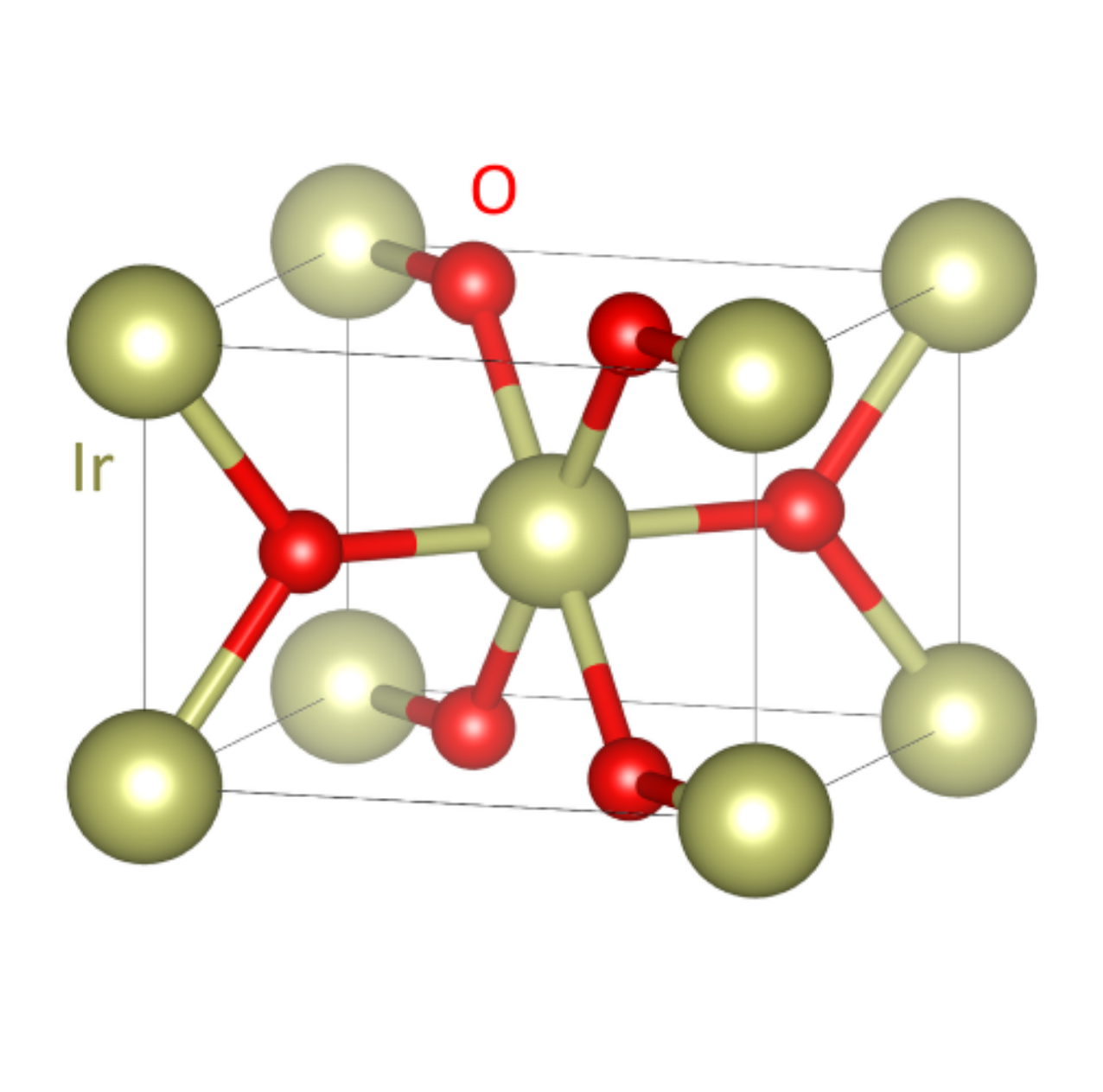}} \\
\hline

\raisebox{1.5\totalheight}{\parbox[c|]{2cm}{\raggedright Band structure without and with SOC}} &
\includegraphics[width=1.5in]{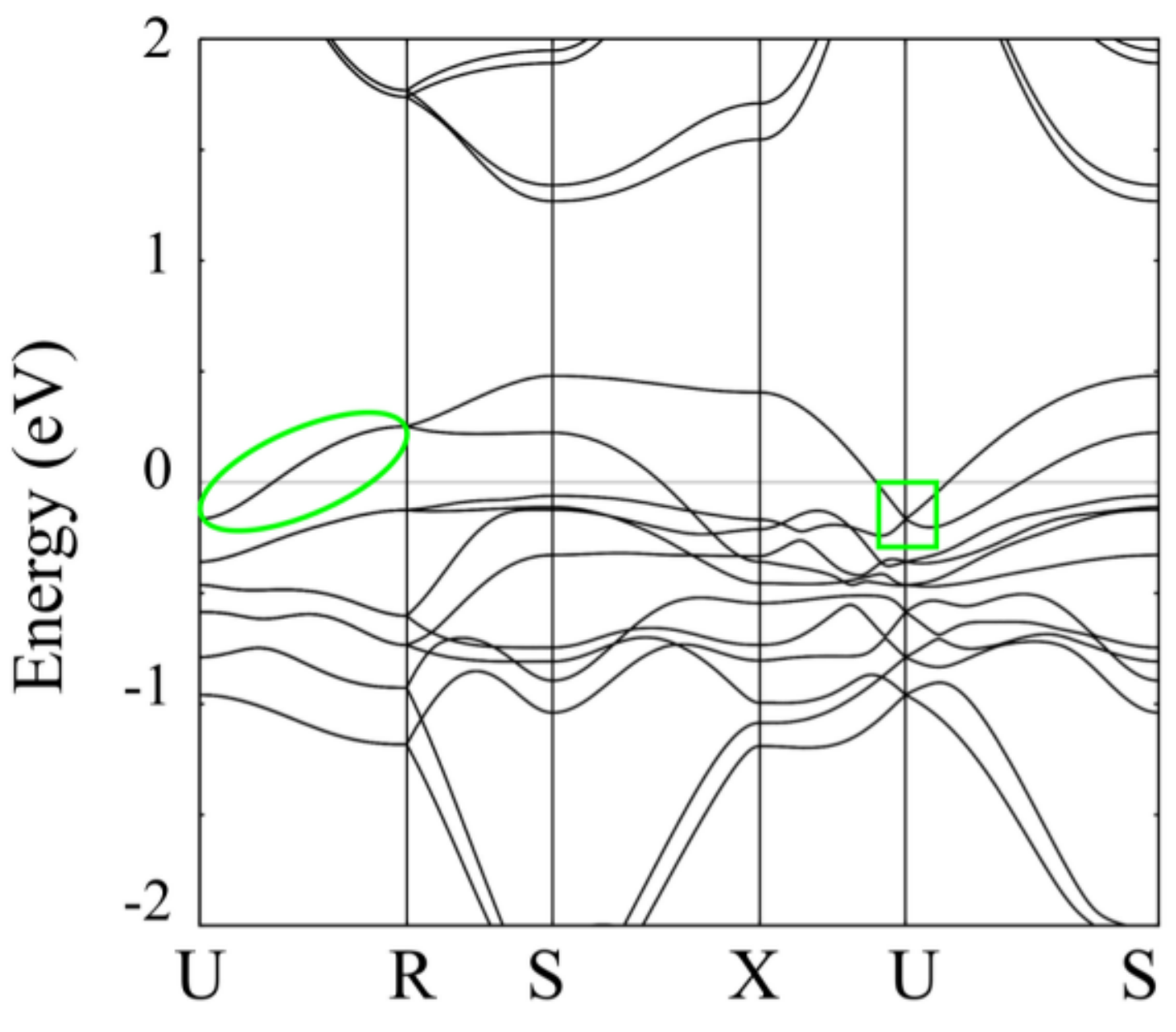} &ta
\includegraphics[width=1.5in]{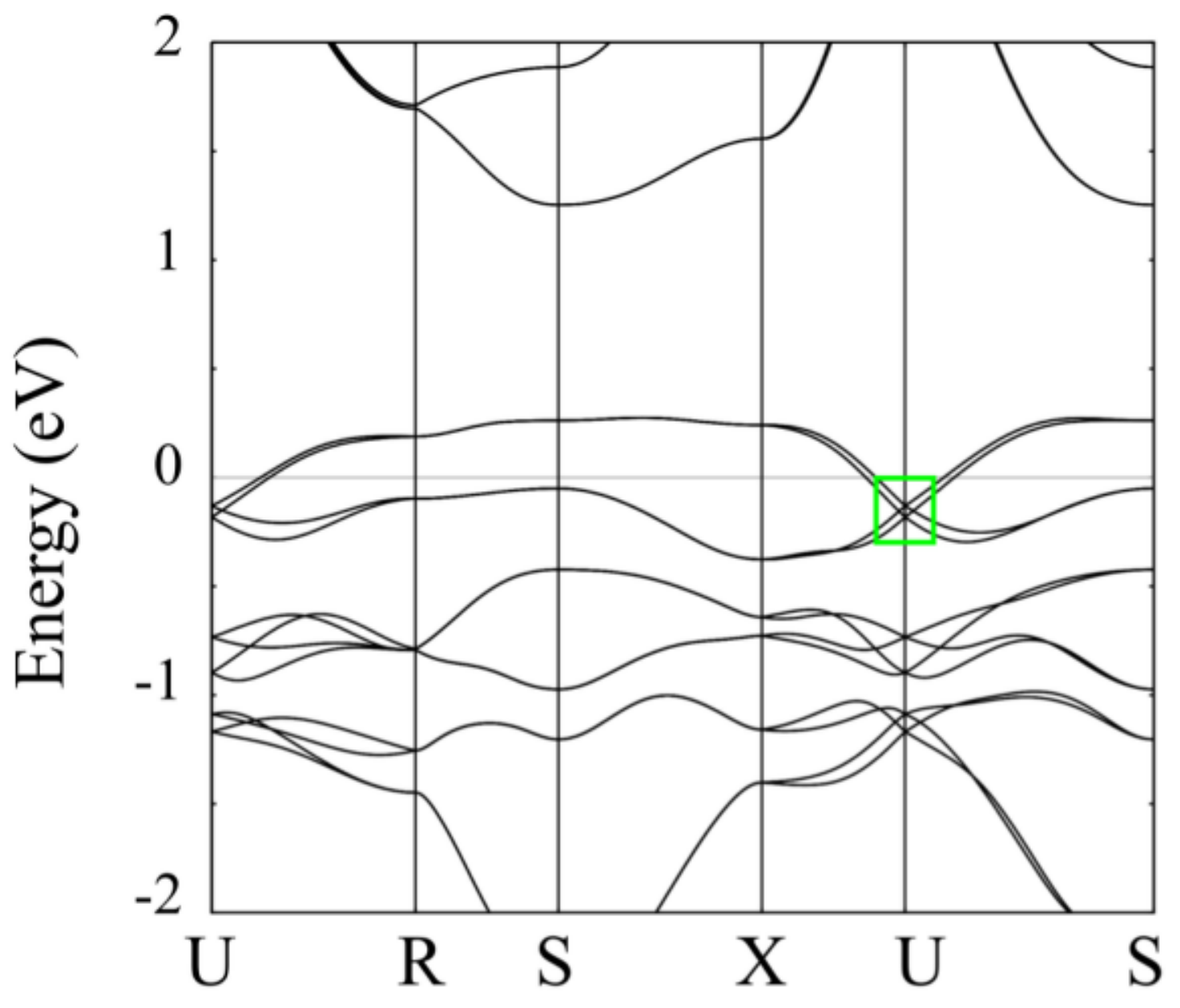}  &
\includegraphics[width=1.5in]{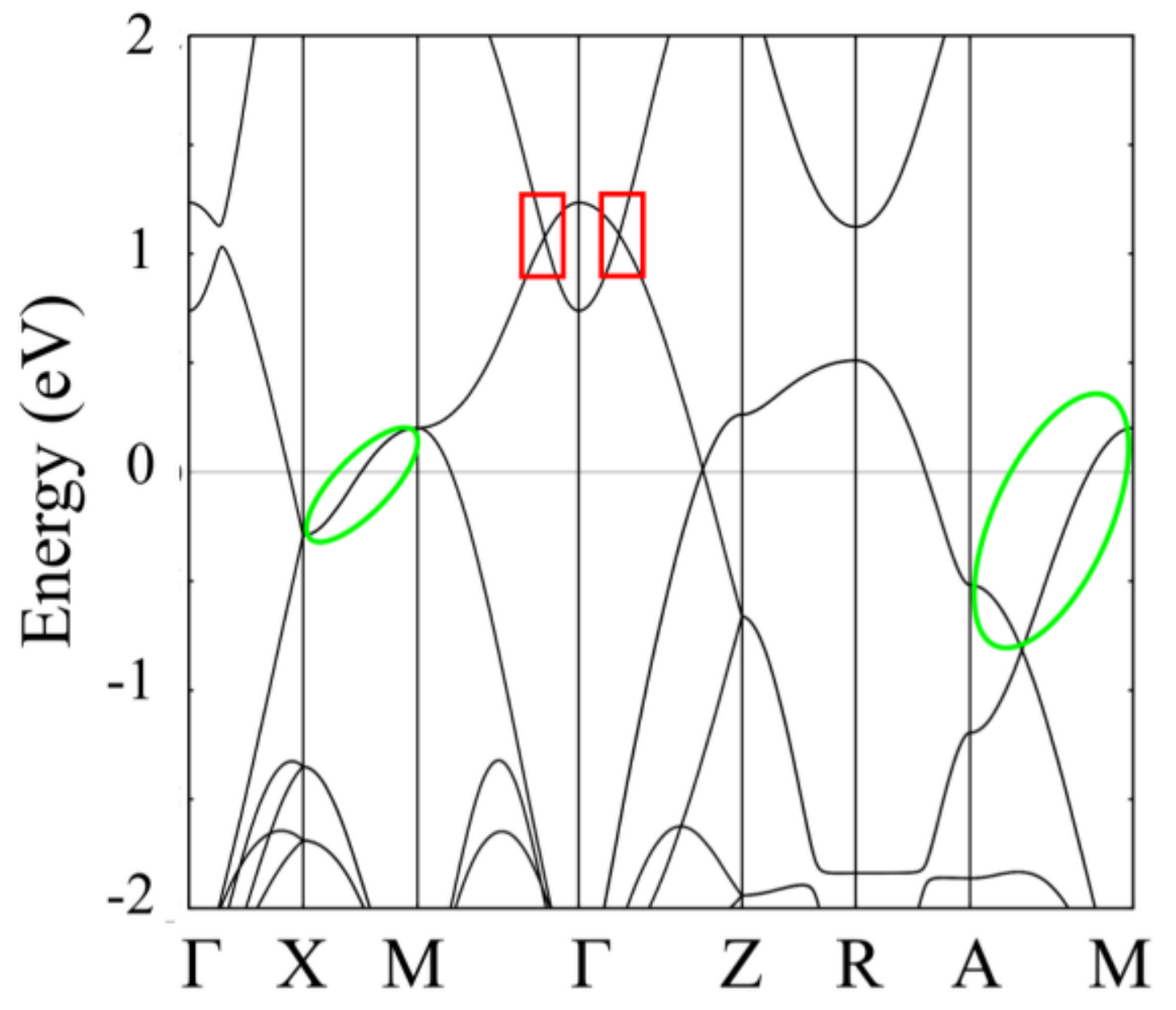} &
\includegraphics[width=1.5in]{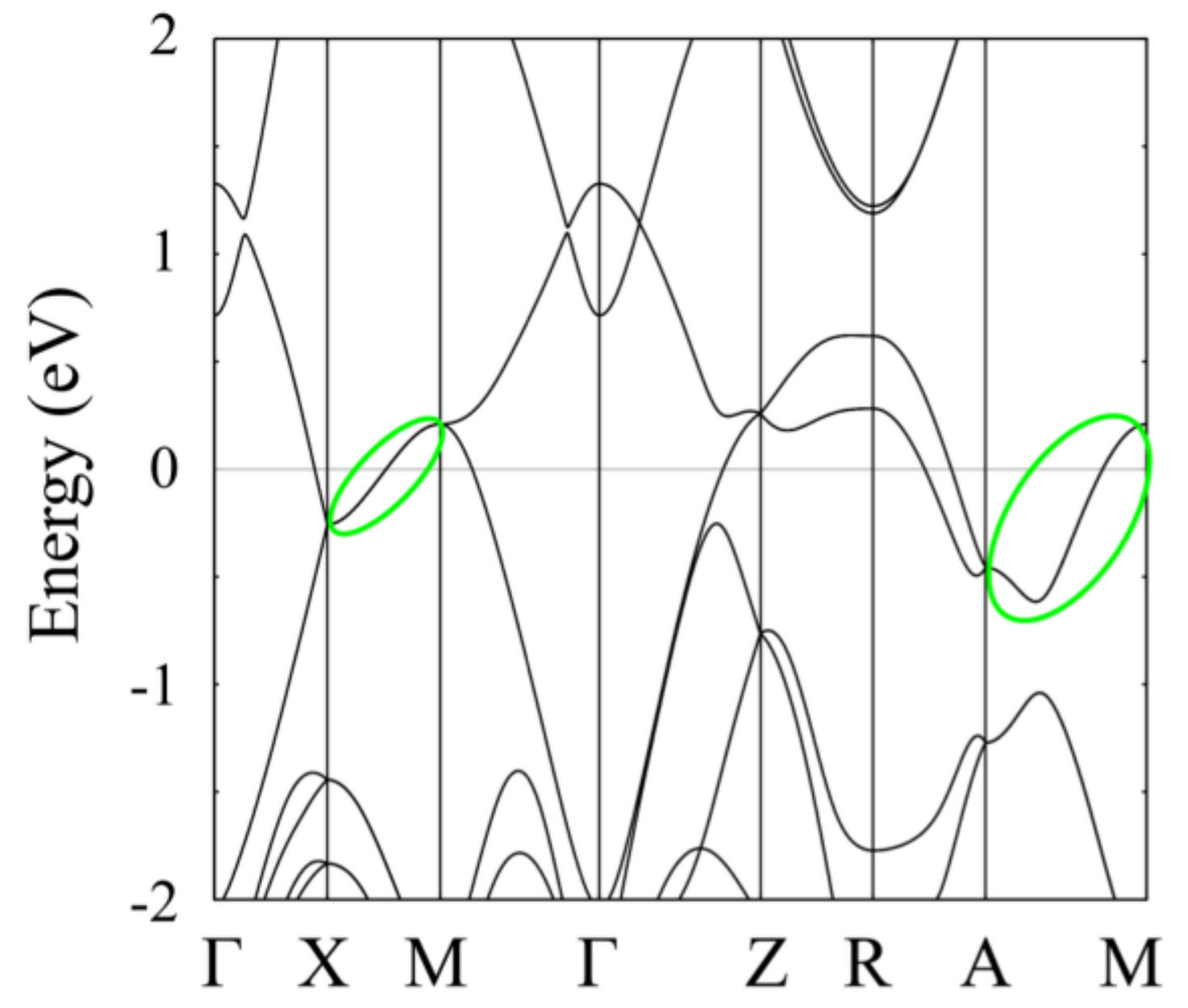} \\
\hline

\raisebox{1\totalheight}{\parbox[c|]{2cm}{\raggedright Nodal line distribution without and with SOC}} &
\includegraphics[width=1.3in]{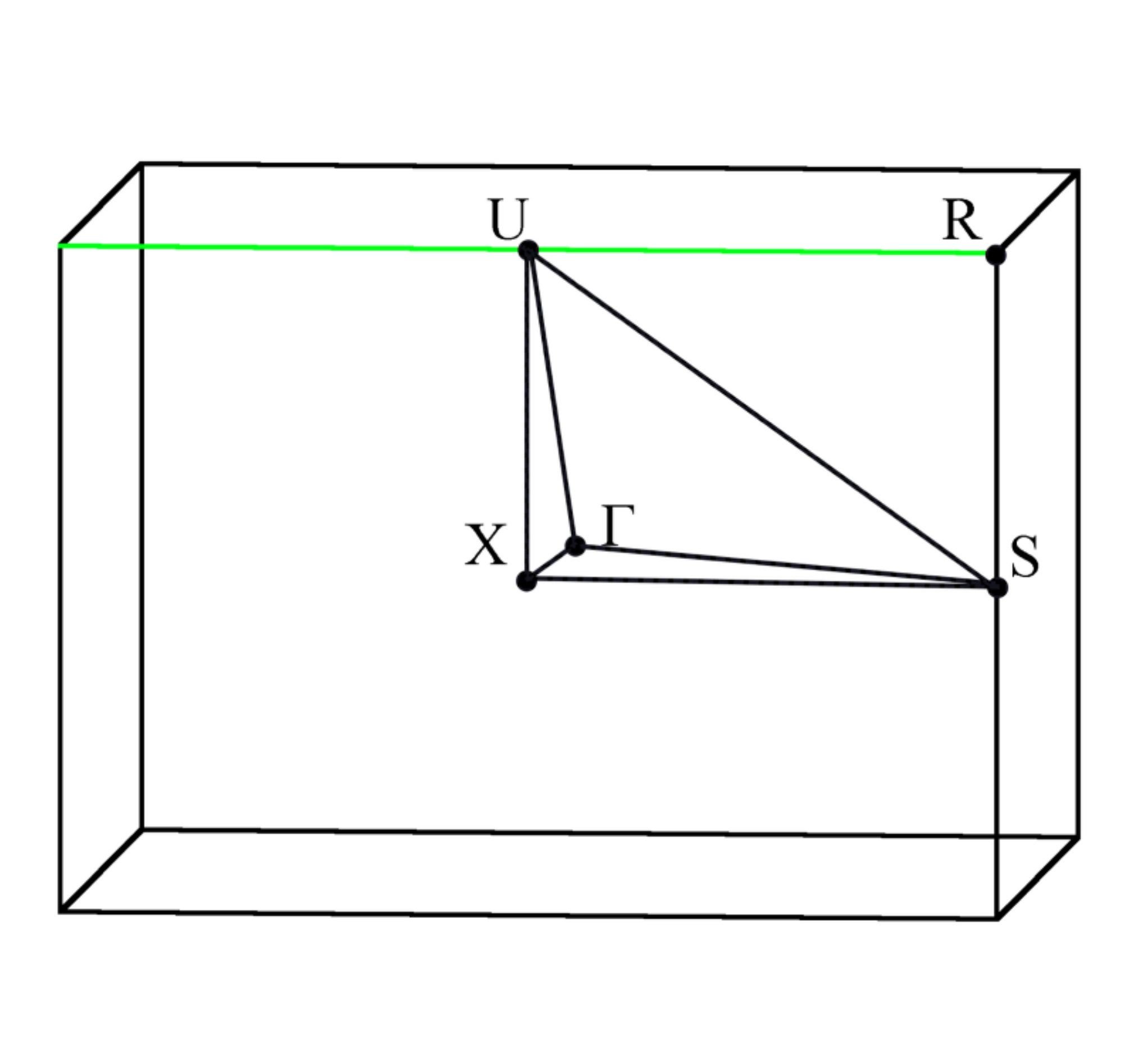} &
\includegraphics[width=1.3in]{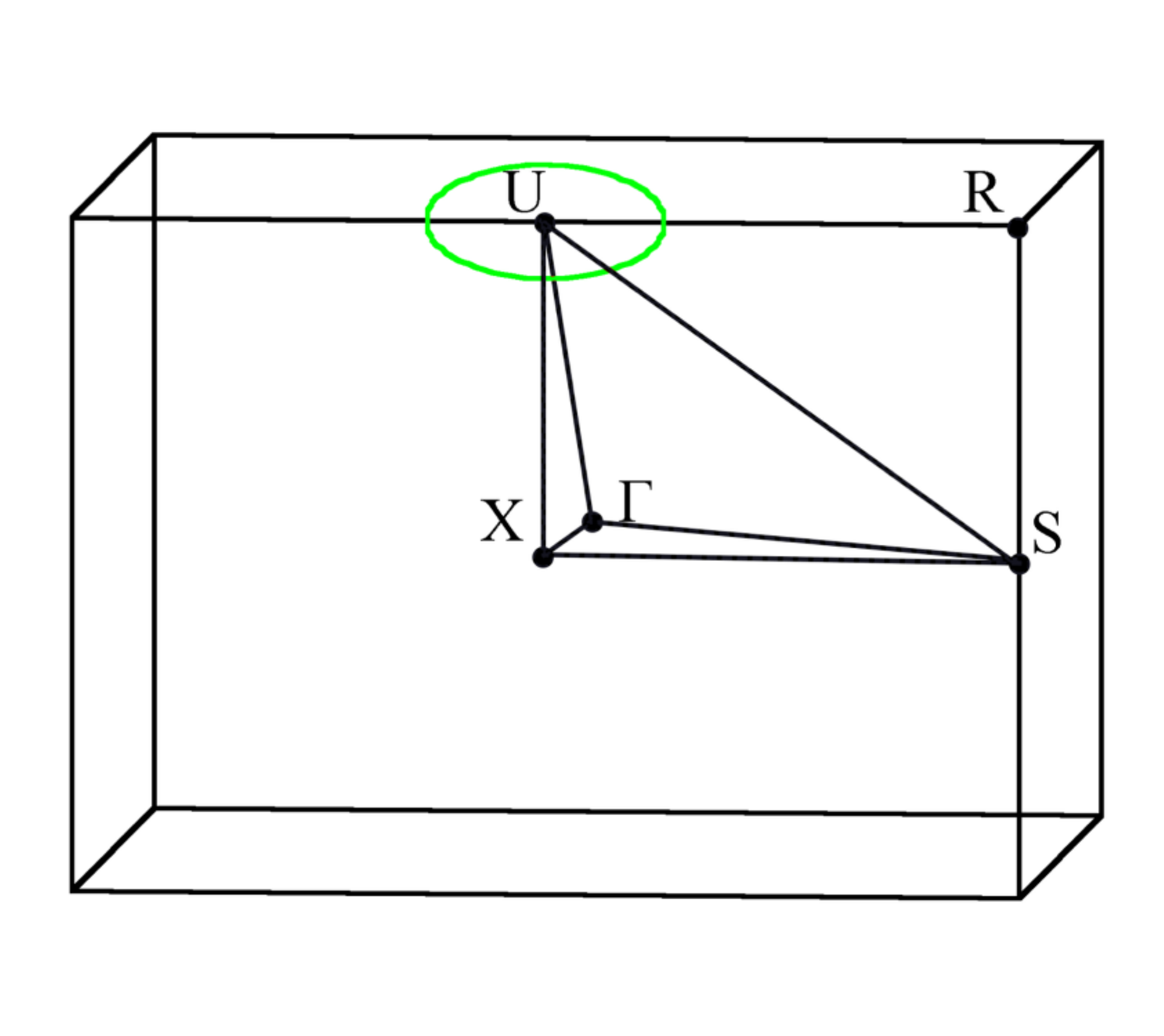} &
\includegraphics[width=1.2in]{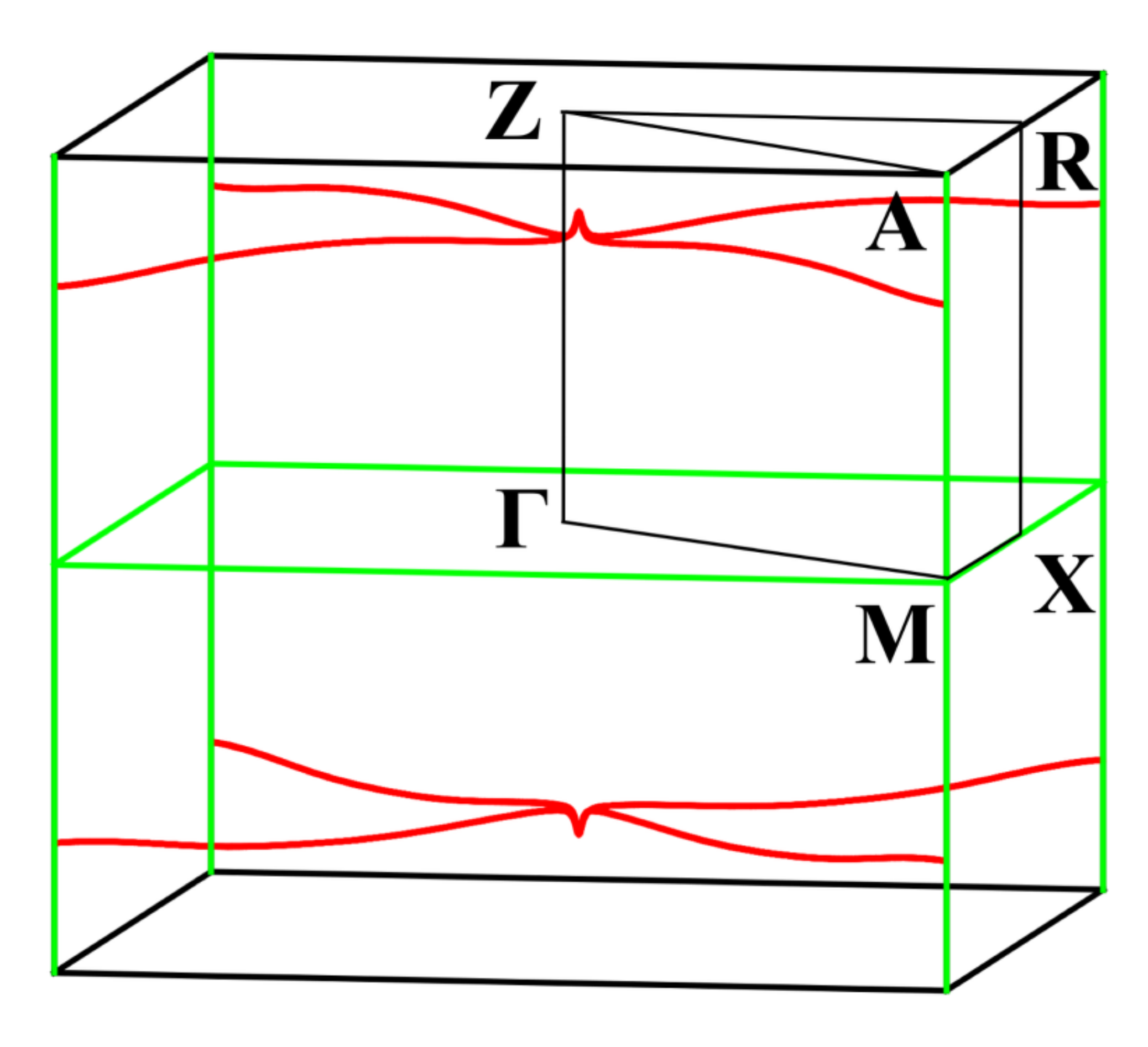} &
\includegraphics[width=1.2in]{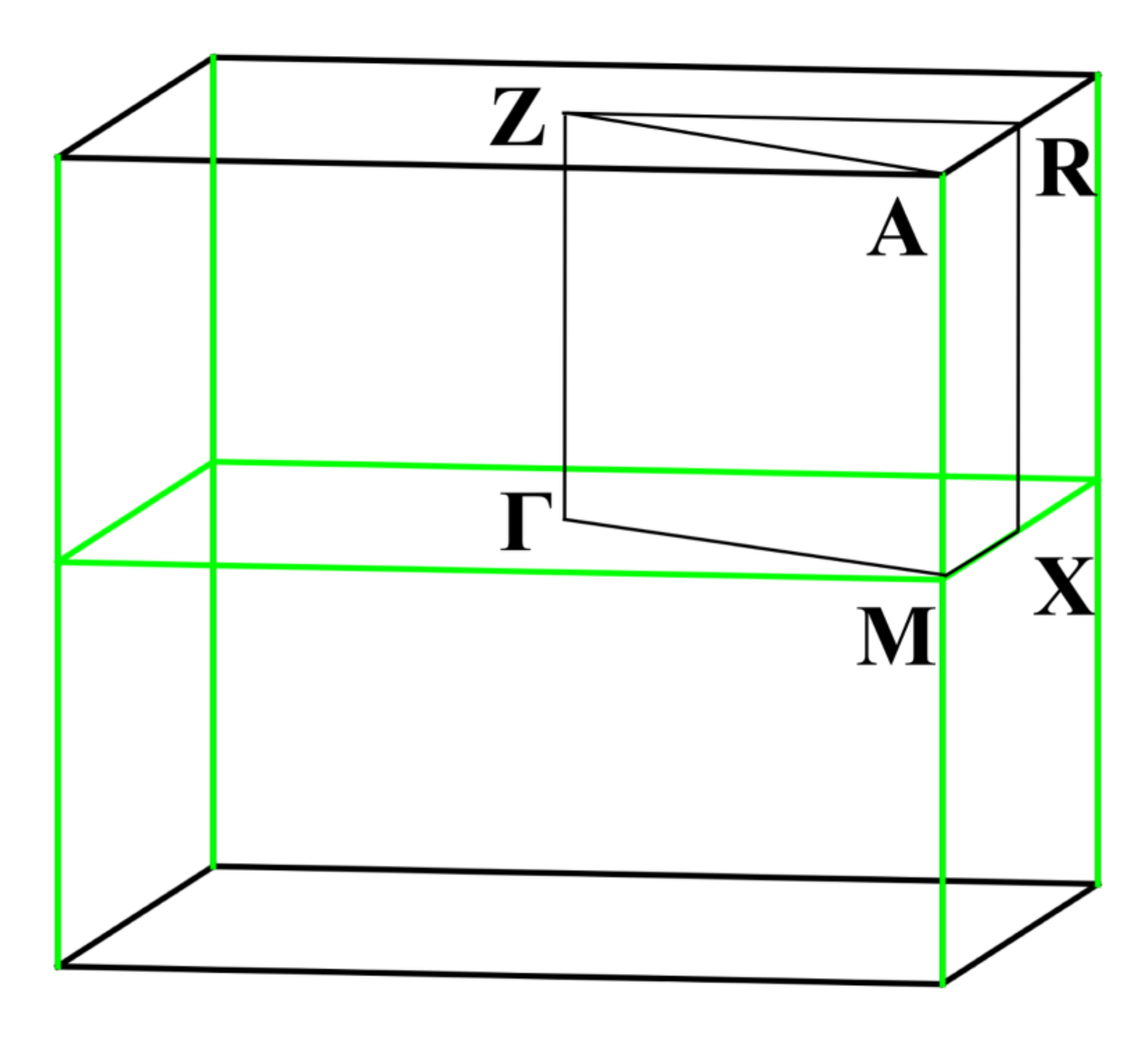} \\
\hline

\raisebox{6\totalheight}{\parbox[c|]{2cm}{\raggedright ARPES \cite{nie2015interplay}}} &

\multicolumn{2}{c|}{\includegraphics[width=2.5in]{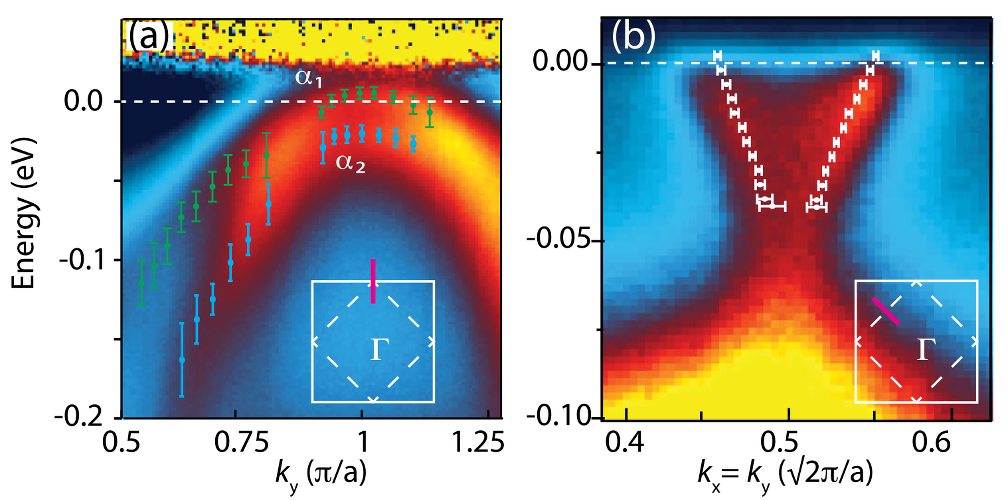}} &
\multicolumn{2}{c}{N/A} \\
\hline

\raisebox{0.1\totalheight}{\parbox[c|]{2cm}{\raggedright References}} &
\multicolumn{2}{c|}{[66], [67], [76], [142-150]} &
\multicolumn{2}{c}{[75], [152], [153]} \\
\hline

\bottomrule

\end{tabular}
\label{tab:gt}
\end{adjustbox}

\end{table*}

The non-symmorphic symmetry protected and the unprotected symmorphic DNL have been verified experimentally by ARPES in the $M$SiS compound family \cite{schoop2016dirac, neupane2016observation, chen2017dirac, takane2016dirac}. In ZrSiS, the DNL protected by non-symmorphic symmetry can be clearly seen at X-R 0.5 eV below the E$_F$. The Dirac points (DP) responsible for the non-symmorphic nodal lines are shown with red arrow in ARPES data in Table VI \cite{schoop2016dirac, neupane2016observation, chen2017dirac}. Similarly, the non-symmorphic symmetry generated DNL was also been measured in HfSiS \cite{chen2017dirac, takane2016dirac}. However, the C$_2v$ protected DNL along $\Gamma$-X and $\Gamma$-M are gapped out by SOC in both materials. Consistent with the electronic calculation, the SOC splitting is larger in HfSiS (80 meV) compared with ZrSiS (15 meV). In addition, larger Rashba splitting has also been seen at the time-reversal invariant X point in HfSiS \cite{takane2016dirac}, likely due to the heavier atomic mass of Hf compared to that of Zr. \par

Extraordinary magneto-electronic transport phenomenon have been found in this material system\cite{ali2016butterfly, wang2016evidence, singha2016titanic, hu2016evidence1, kumar2016unusual, lv2016extremely}. The MR is large and positive for both materials, reaching 1.8 $\times$ 10$^5$ percent at 9T and 2K for ZrSiS without saturation \cite{wang2016evidence, singha2016titanic, hu2016evidence1, lv2016extremely}. The MR are both highly anisotropic and both compounds show a `butterfly'-shaped angular MR, originating from the 2D and 3D Dirac pockets comprising Fermi surface \cite{ali2016butterfly, kumar2016unusual, lv2016extremely}. In addition, ZrSiS also shows evidence of a topological phase transition with changing magnetic field angle\cite{ali2016butterfly}.  \par

ZrSiS and HfSiS are two well-studied examples in the larger ternary material family, $WHM$, ($W$=Zr, Hf, or La; $H$=Si, Ge, Sn, or Sb; $M$=O, S, Se, and Te) \cite{lou2016emergence, hu2016evidence, xu2015two}, which share the same crystal structure and have topological non-trivial phases. For example, ZrSiTe has the non-symmorphic crossings at X and R lie very close to the E$_F$, making it a good candidate system to study non-symmorphic line nodes\cite{topp2016non}. \par

\textbf{Orthorhombic perovskite iridates}\cite{chen2015topological, zeb2012interplay, carter2012semimetal, fang2016topological, fang2015topological, liu2016strain, liu2016direct, nie2015interplay, fujioka2017correlated, zhang2014sensitively, gruenewald2014compressive, biswas2014metal}: $A$IrO$_3$, where $A$ is an alkaline-earth metal, is a class of orthorhombic perovskite iridates in space group $Pbnm$ No. 62. A typical unit cell of $A$IrO$_3$ has four Ir atoms with different oxygen octahedral environments. It contains three different mirror planes: b-glide, n-glide, and M$_z$. \par

Taking SrIrO$_3$ as an example (shown in Figure VII), without SOC, singly degenerate DNL are presence along U-R. In the presence of strong SOC from Ir, a Dirac double nodal ring emerges around the U point close to the E$_F$, protected by n-glide non-symmorphic symmetry. The double nodal ring exhibits a non-trivial topology that leads to localized surface zero mode protected by chiral (a combination of time-reversal and particle-hole symmetry\cite{schnyder2008classification}) and mirror reflection symmetry, and a pair of counter-propagating helical modes localized in a dislocation line, making it a topological crystalline metal (TCM)\cite{chen2015topological}. A direct observation of nodal lines in this correlated oxide system is still under investigation\cite{liu2016direct}. When the n-glide symmetry is broken, for example, by epitaxy-induced symmetry breaking, the double-nodal line breaks into two Dirac points \cite{liu2016strain, fang2016topological}. When considering magnetism, the Dirac nodal ring evolves into a Weyl nodal ring, Weyl points, or a TI, depending on the direction of the field\cite{chen2015topological}. \par

Unusual magnetotransport properties are expected to arise from such topological transitions. However, the MR of the epitaxially-grown thin films are shown to be small \cite{zhang2014sensitively, gruenewald2014compressive, biswas2014metal}. One study on strain-free bulk SrIrO$_3$ shows the quadratic transverse and linear longitudinal MR to be 100 percent and 50 percent at 14 T and 2 K, which are 2 to 3 orders of magnitude higher than the MR on strained thin films\cite{fujioka2017correlated}. Furthermore, a change in optical conductivity spectra induced by interband transition of Dirac cones has been observed. Such filed-induced topological transition of Dirac nodes could be a result of enhanced paramagnetic susceptibility\cite{fujioka2017correlated}. \par

\textbf{Rutile oxides} \cite{sun2017dirac, lin2004low, ryden1972high}: $R$O$_2$, where $R$ could be Ir, Os, and Ru, crystalizes in rutile-type lattice structures in space group P4$_2$/mnm No. 136. It has one $R$ atom sitting at each corner of a unit cell, as well as one $R$ atom at the center. Each $R$ atom is surrounded by six O atoms that form a distorted octahedron. Rutile oxides have both inversion and time-reversal symmetry. They also have mirror reflection symmetry M$_y$$_z$ and M$_x$$_y$, four-fold rotation symmetry C$_4$$_z$, as well as non-symmorphic symmetry n$_x$ and n$_4$$_z$. Two types of DNLs exist in this system, which will be discussed below. \par 

The first type of DNLs exist only in the absence of SOC. Without SOC, six and eight fold crossing points appear along $\Gamma$-Z and M-A directions, which are denoted as hexatruple point (HP) and octuple point (OP). The HP and OP points are connected by DNLs inside the BZ, forming a network as shown in blue in Table VII. These nodal line networks exist in the mirror reflection planes (110) and ($\bar{1}$10). They are protected by mirror reflection symmetry without SOC, and can be transformed to each other by either inversion or time-reversal symmetry. With SOC, both HPs and OPs got gapped out, as well as the DNLs connecting them. \par

The second type of DNLs are stable against SOC, because of the non-symmorphic symmetry. As discussed earlier, the combination of inversion and time-reversal symmetry give rise to doubly degenerate crossings considering spins. In addition, the existence of non-symmorphic symmetry generates even higher degeneracies at some specific $k$ point in the BZ. For example, in IrO$_2$ (Table VII), without SOC, k$_x$ = $\pi$ and k$_y$ = $\pi$ appear to be Dirac nodal planes. With SOC, the combination of inversion, time-reversal symmetry, mirror reflection symmetry as well as non-symmorphic symmetry guarantees four degenerate orthogonal eigenstates along both the X-M and M-A direction. DNLs protected by non-symmorphic symmetry in IrO$_2$ are at the E$_F$, which are possibly the reason for the high conductivity and large MR in the system\cite{lin2004low, ryden1972high}. 


\section{KCu$_2$EuTe$_4$: A new mirror symmetry protected WNL}

\begin{figure*}[t!]
 \begin{center}
 \includegraphics[width=1\textwidth]{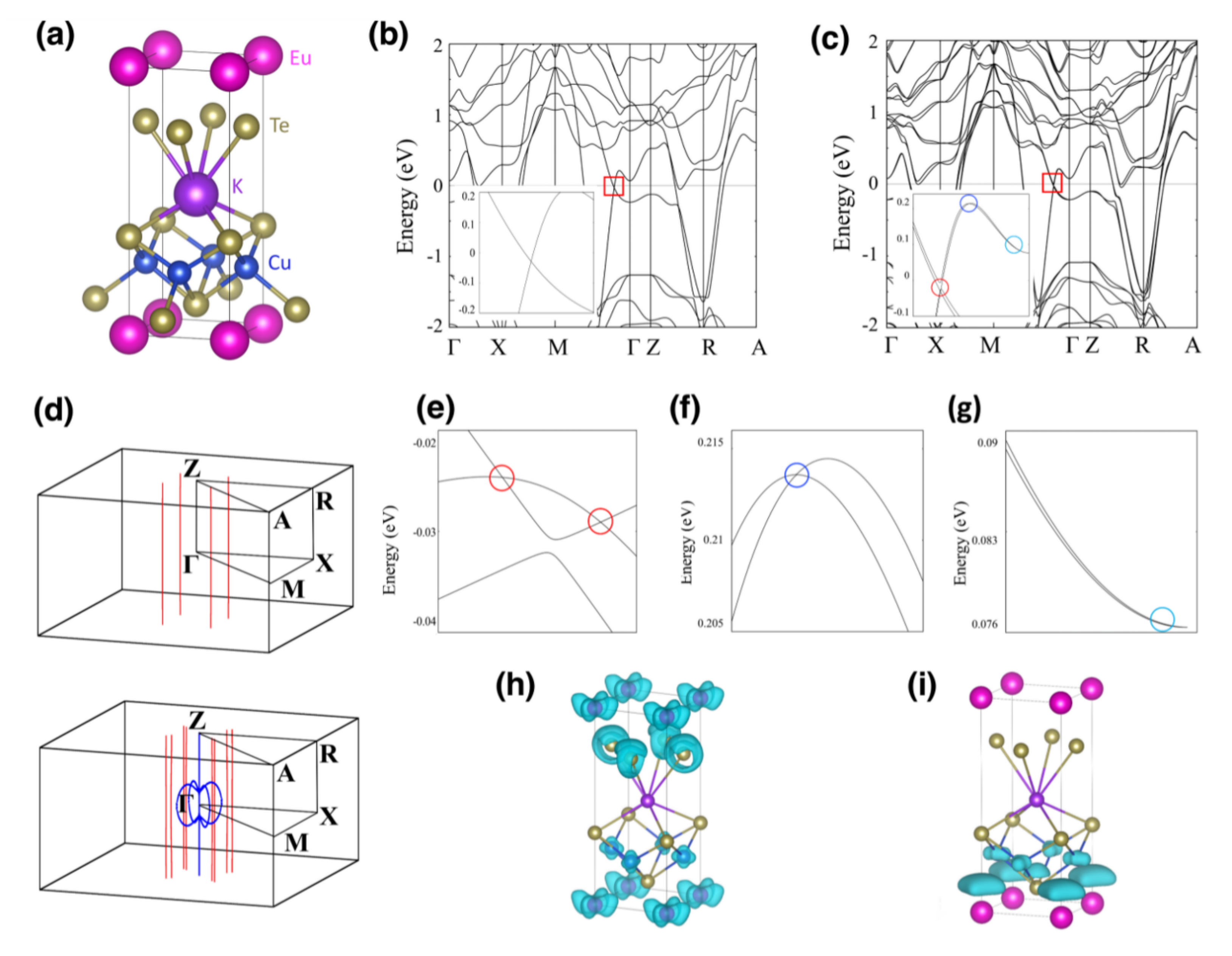}
 \end{center}
 
      \captionsetup{justification   = raggedright,
              singlelinecheck = false}
              
 \caption{Crystal structure, band structure, and nodal line distribution of KCu$_2$EuTe$_4$. (a) Crystal structure of KCu$_2$EuTe$_4$. (b,c) Calculated bulk band structure of KCu$_2$EuTe$_4$ in paramagnetic phase without SOC and with SOC. (d) Nodal line distribution without (upper) and with SOC (lower). The red lines represent DNLs that are created and protected by mirror symmetry; the blue lines show the `flower-shaped' WNLs that are generated by accidental band crossings upon considering SOC. (e-g) Zoom-in band structure along M-$\Gamma$ with the inclusion of SOC. The crossings shown in (e) is responsible for the four Weyl nodal lines, in (f) corresponds to accidental `flower-shaped' nodal rings, and in (g) is responsible for the accidental nodal line along C$_4$ axis. (h) Partial charge distribution of the states corresponding to the WNLs in (e). (i) Partial charge distribution of the states corresponding to the WNLs in (f) and (g). }
\end{figure*}

KCu$_2$EuTe$_4$ is one of the large family of Te square net compounds and its crystal structure is shown in Figure 5a. Crystallizing in the tetragonal space group $P4mm$ No. 99, it can be built up in the following way: layers of corner sharing EuTe$_8$ square anti-prisms stack in edge-sharing manners with tetrahedral CuTe$_4$ layers below and square anti-prismatic KTe$_8$ layers above. The result is a non-centrosymmetric structure with several square nets and a particularly dense Te-square net shared by the KTe and EuTe layers. This lack of inversion, unlike the ZrSiS family of compounds, allows SOC to lift the spin degeneracies. KCu$_2$EuTe$_4$ has two sets of mirror planes, M$_x$ and M$_x$$_y$, as well as C$_4$ rotation symmetry, which are responsible for creating the nodal line/rings in this compound explained below.\par

The overview of the band structures of KCu$_2$EuTe$_4$ in the paramagnetic state without and with SOC are shown in Figure 5b and 5c. As expected for square net compounds, without SOC, highly linearly dispersive bands are observed to cross the E$_F$ and almost exclusively create the Fermi surface. Among them, one set of conduction and valence bands with opposite parity invert nearly at the E$_F$, resulting in four two-fold spinless DNL protected by M$_x$$_y$ mirror symmetry. The distribution of this set of nodal lines in the first BZ is shown in red in the upper panel of Figure 5d. The energy dispersion of these nodal lines are almost flat. \par 

Upon introducing SOC, each band splits into two singly degenerate bands, giving rise to more band crossings in the M$_x$$_y$ mirror plane. However, as noted earlier, only if the bands have opposite eigenvalues ($\pm i$) under mirror operation are the band crossings maintained. In KCu$_2$EuTe$_4$, the spinless DNL splits into two spinful WNLs as shown in red in the lower panel of Figure 5d. The band projections of these states show that the WNLs are attributed to both the Eu and Te sublattices, shown in Figure 5h. Similar to the inversion plus SOC generated new crossing in PbTaSe$_2$, SOC in KCu$_2$EuTe$_4$ also generates a new accidental band crossing and a set of `flower-shaped' Weyl nodal rings in the M$_x$$_y$ plane, appearing above the E$_F$ shown in blue in the lower panel of Figure 5d. Band projections show that mainly Eu sublattices contribute to these states (Figure 5i). A zoom-in view of the band between M and G with SOC are shown in Figures 5e to 5g. Among them, the band structure shown in Figure 5e corresponds to the DNL split WNLs, Figure 5f corresponds to the crossing responsible for the `flower shaped' nodal ring, and Figure 5g corresponds to inversion plus SOC newly generated nodal lines along the C$_4$ rotation axis. \par

\begin{table*}[]
\centering
\caption{Experimentally confirmed and theoretically predicted topological nodal line semimetals}
\label{my-label}
\begin{tabular}{|c|c|c|c|l|}
\hline
\textbf{Materials}                   & \textbf{Symmetry}     & \textbf{\begin{tabular}[c]{@{}c@{}}Topological type without SOC\\ (+ SU(2))\end{tabular}} & \textbf{Topological type with SOC} & \multicolumn{1}{c|}{\textbf{Gap Size}}                                                     \\ \hline
$M$SiS ($M$ = Zr, Hf)                    & I + TR + mirror glide & \multirow{18}{*}{DNLS}                                                                    & \multirow{3}{*}{DNLS}              & \multirow{10}{*}{}                                                                         \\ \cline{1-2}
$A$IrO$_3$                                & I + TR + mirror glide &                                                                                           &                                    &                                                                                            \\ \cline{1-2}
IrO$_2$                                 & I + TR + mirror glide &                                                                                           &                                    &                                                                                            \\ \cline{1-2} \cline{4-4}
Pb(Tl)TaSe$_2$                          & Mirror                &                                                                                           & \multirow{2}{*}{WNLS}              &                                                                                            \\ \cline{1-2}
KCu$_2$EuTe$_4$                            & Mirror                &                                                                                           &                                    &                                                                                            \\ \cline{1-2} \cline{4-4}
Cu$_3$N$X$ ($X$ = Ni, Cu, Pd, An, Ag, Cd)   & I + TR + C$_4$           &                                                                                           & \multirow{3}{*}{DSM}               &                                                                                            \\ \cline{1-2}
CaTe                                 & I + TR + C$_4$           &                                                                                           &                                    &                                                                                            \\ \cline{1-2}
La$X$ ($X$ = N, P, As, Sb, Bi)           & I + TR + C$_4$           &                                                                                           &                                    &                                                                                            \\ \cline{1-2} \cline{4-4}
TaAs                                 & Mirror                &                                                                                           & \multirow{2}{*}{WSM}               &                                                                                            \\ \cline{1-2}
HfC                                  & Mirror                &                                                                                           &                                    &                                                                                            \\ \cline{1-2} \cline{4-5} 
MTC                                  & I + TR                &                                                                                           & \multirow{8}{*}{TI}                & \multicolumn{1}{c|}{0.13 meV at 1.5 K}                                                     \\ \cline{1-2} \cline{5-5} 
$A$$X$$_2$ ($A$ = Ca, Sr; Ba; $X$ = Si, Ge, Sn) & I + TR                &                                                                                           &                                    & \multicolumn{1}{c|}{$<$ 160 meV}                                                               \\ \cline{1-2} \cline{5-5} 
Hyperhoneycomb lattice               & I + TR                &                                                                                           &                                    & \multicolumn{1}{c|}{N/A}                                                                   \\ \cline{1-2} \cline{5-5} 
Ca$_3$P family                          & I + TR                &                                                                                           &                                    & \multicolumn{1}{c|}{\begin{tabular}[c]{@{}c@{}}From 1.76 meV \\ to 47.14 meV\end{tabular}} \\ \cline{1-2} \cline{5-5} 
Ca$_3$P$_2$                                & I + TR + Mirror       &                                                                                           &                                    & \multicolumn{1}{c|}{9 meV}                                                                 \\ \cline{1-2} \cline{5-5} 
Cu$_2$Si                                & I + TR + Mirror       &                                                                                           &                                    & \multicolumn{1}{c|}{$<$15 meV}                                                                 \\ \cline{1-2} \cline{5-5} 
Hg$_3$As$_2$                               & I + TR + Mirror       &                                                                                           &                                    & \multicolumn{1}{c|}{34 meV}                                                                \\ \cline{1-2} \cline{5-5} 
CaAg$X$ ($X$ = P, As)                    & TR + Mirror           &                                                                                           &                                    & \multicolumn{1}{c|}{P = 1 meV; As = 75 meV}                                                 \\ \hline
\end{tabular}
\end{table*}

It is also worth mentioning that aside from the nodal lines and rings discussed above, more crossings both above and below the E$_F$, within about 1 eV, are present, implying more interesting nodal lines/rings to be explored in this family of compounds. Tuning the E$_F$ and band structure via chemical control, similar to the ZrSiS family, should be possible. To the best of our knowledge, KCu$_2$EuTe$_4$ is the first predicted WNLS which have nodal lines almost exactly at the E$_F$, making it a good candidate to study the exotic physics related to WNLS. Transport and magnetic measurements by Kanatzidis et al indicate KCu$_2$EuTe$_4$ to be a paramagnetic metal or semimetal down to 2 K\cite{patschke1999cu}. Their electron diffraction experiments revealed large amounts of twinning with possible superstructure peaks indicating a small structural distortion. However this has not been fully solved and is planned for future work. In addition, since some distorted square nets can be topologically equivalent to perfect square nets similar to how hyper-honeycomb lattices are topologically equivalent to perfect honeycombs, a rigorous study of the robustness of the various crossings to likely distortion modes is currently under preparation\cite{mullen2015line}.\par

\section{Summary and Experimental Outlook} 

Topological nodal line semimetals, as precursors of many other TSMs or TIs, have drawn great interest in the field of condensed matter physics in the last couple of years. The combination of inversion and time-reversal symmetry often generates nodal lines/rings, however their protection against SOC requires one or more spatial symmetries, in particular, mirror reflection or non-symmorphic symmetry. Rotation symmetry can make DNLSs into DSMs or TIs. Mirror reflection symmetry can protect the degeneracy on the plane of the mirror, and non-symmorphic symmetry forces band degeneracies at high symmetry points. Mirror reflections can generate new crossings rather than just split four-fold crossings when combined with a lack of inversion and large SOC. Table V summarizes the experimentally confirmed and theoretically predicted topological nodal line semimetals, their topological category without and with SOC, their gap size (if applicable) as well as their corresponding crystalline symmetries. \par

With several materials now predicted to house the exotic electronic structures of NLSs, future studies will focus on experimental investigation of the properties stemming from those electronic structures. Firstly, there is a large opportunity to characterize the NLSs using probes other than photoemisssion. As topological surface states is a characteristic signature of the nontrivial topology, direct probing of drumhead surface states is the next experimental step. The dispersion of the drumhead surface states is smaller than that of typical bulk valence and conduction bands, making them good candidates for studying correlation effects at surfaces\cite{liu2017correlation}. It is proposed that in Dirac nodal ring systems, with coexistance of inversion and time-reversal symmetries as well as negligible SOC, and in the small Hubbard interaction region, surface ferromagnetism can be obtained and characterized by the surface mode divergence of the spin susceptibility\cite{liu2017correlation}. Increasing the Hubbard interaction should drive the system into a surface charge-ordered phase through a continuous quantum phase transition \cite{liu2017correlation}. In addition, the bulk DNLs can also be investigated via transport measurements; Novel Landau level structures in nodal ring semimetals are predicted as a function of the strength and direction of applied magnetic fields\cite{rhim2015landau, mullen2015line}. Almost non-dispersive Landau levels, as a function of momentum, have been predicted when the magnetic field is applied in the plane of the ring. Near the center of the ring, almost flat Dirac zero modes are expected\cite{rhim2015landau}.\par

Aside from just the fundamental physics of topological nodal line semimetals, more applied investigations of these materials have also been proposed, for example, making future spintronic devices. One route to generating spin currents (necessary in spintronics) is to use the spin Hall effect (SHE), where electric current generates a transverse spin current. The intrinsic spin Hall conductivity (SHC) can be calculated by integrating the spin Berry curvature over the BZ. Since anti-crossings in the electronic structure induced by SOC can lead to a large Berry curvature, to maximize SHC, one needs to maximize the number of band anti-crossings (gapped by SOC) at E$_F$, among other parameters. A DNL consists of an infinite number of Dirac points along the line of degeneracy, and, unless protected by spatial symmetries as described earlier, are gapped due to SOC. Therefore it is natural to think that gapped DNL materials have large intrinsic SHEs. It has recently been found the DNL in metallic rutile oxides IrO$_2$, OsO$_2$, and RuO$_2$ contribute to the large SHC in these materials\cite{sun2017dirac}. \par

The interaction of nodal line materials with light is another avenue for future work. It has been studied in DSMs and WSMs that circularly polarized light can couple with electrons and break the time-reversal symmetry of the material through the Floquet effect\cite{ebihara2016chiral, chan2016chiral, taguchi2016photovoltaic}. This light-induced interaction gives rise to a photovoltaic anomalous Hall effect as well as large photocurrents\cite{chan2017photocurrents}. It has been theorized that such an interaction can also happen in NLSs, that circularly polarized light could break a nodal line into Weyl points accompanied by a signature of anomalous Hall conductivity tunable by the incident light \cite{taguchi2016photovoltaic1, yan2016tunable, ezawa2017photoinduced}. The anomalous Hall conductivity depends largely on the location of E$_F$: when the degeneracy is at the E$_F$, the induced Hall conductivity depends on the radius of the nodal ring and not on intensity of the light. However, when the E$_F$ is far away from the line node, the Hall conductivity does vary with the light intensity56. This dependence of Hall conductivity on E$_F$ could have potential applications in phototransistors based on nodal line materials  \cite{taguchi2016photovoltaic1}. \par

The recent advances in the theoretical understanding of NLS and specifically the underlying symmetry causes of DNLs and WNLs have led to the discovery of several good candidate materials. Now experimental investigation of the properties stemming from this novel physics is expected to develop and drive the next wave of topological physics studies, both in a fundamental as well as applied manner. \par

\vspace{5mm} 

\textbf{Acknowledgments}:
We acknowledge the support of the Alexander von Humboldt Foundation, their Sofia Kovalevskaja Award, the German Federal Ministry of Education and Research as well as the Minerva Stiftung and the Max Plank Group. Also, we acknowledge support by the Ruth and Herman Albert Scholars Program for New Scientists in Weizmann Institute of Science, Israel as well as the German-Israeli Foundation for Scientific Research and Development. \par

\textbf{Author Contributions}:
S.Y. Yang led the research effort and wrote the majority of the manuscript. H. Yang carried out the majority of the electronic structure calculations and identified the nodal lines of KCu$_2$EuTe$_4$. E. Derunova assisted in calculations and discussions. S.S.P Parkin, B. Yan, and M. N. Ali directed the research effort. All authors contributed to writing the manuscript. M. N. Ali is the principal investigator. \par

\textbf{Disclosure Statement}:
No potential conflict of interest was reported by the authors. \par

%

\begin{thebibliography}{100}


\makeatletter
\providecommand \@ifxundefined [1]{%
 \@ifx{#1\undefined}
}%
\providecommand \@ifnum [1]{%
 \ifnum #1\expandafter \@firstoftwo
 \else \expandafter \@secondoftwo
 \fi
}%
\providecommand \@ifx [1]{%
 \ifx #1\expandafter \@firstoftwo
 \else \expandafter \@secondoftwo
 \fi
}%
\providecommand \natexlab [1]{#1}%
\providecommand \enquote  [1]{``#1''}%
\providecommand \bibnamefont  [1]{#1}%
\providecommand \bibfnamefont [1]{#1}%
\providecommand \citenamefont [1]{#1}%
\providecommand \href@noop [0]{\@secondoftwo}%
\providecommand \href [0]{\begingroup \@sanitize@url \@href}%
\providecommand \@href[1]{\@@startlink{#1}\@@href}%
\providecommand \@@href[1]{\endgroup#1\@@endlink}%
\providecommand \@sanitize@url [0]{\catcode `\\12\catcode `\$12\catcode
  `\&12\catcode `\#12\catcode `\^12\catcode `\_12\catcode `\%12\relax}%
\providecommand \@@startlink[1]{}%
\providecommand \@@endlink[0]{}%
\providecommand \url  [0]{\begingroup\@sanitize@url \@url }%
\providecommand \@url [1]{\endgroup\@href {#1}{\urlprefix }}%
\providecommand \urlprefix  [0]{URL }%
\providecommand \Eprint [0]{\href }%
\providecommand \doibase [0]{http://dx.doi.org/}%
\providecommand \selectlanguage [0]{\@gobble}%
\providecommand \bibinfo  [0]{\@secondoftwo}%
\providecommand \bibfield  [0]{\@secondoftwo}%
\providecommand \translation [1]{[#1]}%
\providecommand \BibitemOpen [0]{}%
\providecommand \bibitemStop [0]{}%
\providecommand \bibitemNoStop [0]{.\EOS\space}%
\providecommand \EOS [0]{\spacefactor3000\relax}%
\providecommand \BibitemShut  [1]{\csname bibitem#1\endcsname}%
\let\auto@bib@innerbib\@empty
\bibitem [{\citenamefont {Hasan}\ and\ \citenamefont
  {Kane}(2010)}]{hasan2010colloquium}%
  \BibitemOpen
  \bibfield  {author} {\bibinfo {author} {\bibfnamefont {M.~Z.}\ \bibnamefont
  {Hasan}}\ and\ \bibinfo {author} {\bibfnamefont {C.~L.}\ \bibnamefont
  {Kane}},\ }\href@noop {} {\bibfield  {journal} {\bibinfo  {journal} {Reviews
  of Modern Physics}\ }\textbf {\bibinfo {volume} {82}},\ \bibinfo {pages}
  {3045} (\bibinfo {year} {2010})}\BibitemShut {NoStop}%
\bibitem [{\citenamefont {Qi}\ and\ \citenamefont
  {Zhang}(2011)}]{qi2011topological}%
  \BibitemOpen
  \bibfield  {author} {\bibinfo {author} {\bibfnamefont {X.-L.}\ \bibnamefont
  {Qi}}\ and\ \bibinfo {author} {\bibfnamefont {S.-C.}\ \bibnamefont {Zhang}},\
  }\href@noop {} {\bibfield  {journal} {\bibinfo  {journal} {Reviews of Modern
  Physics}\ }\textbf {\bibinfo {volume} {83}},\ \bibinfo {pages} {1057}
  (\bibinfo {year} {2011})}\BibitemShut {NoStop}%
\bibitem [{\citenamefont {Armitage}\ \emph {et~al.}(2017)\citenamefont
  {Armitage}, \citenamefont {Mele},\ and\ \citenamefont
  {Vishwanath}}]{armitage2017weyl}%
  \BibitemOpen
  \bibfield  {author} {\bibinfo {author} {\bibfnamefont {N.}~\bibnamefont
  {Armitage}}, \bibinfo {author} {\bibfnamefont {E.}~\bibnamefont {Mele}}, \
  and\ \bibinfo {author} {\bibfnamefont {A.}~\bibnamefont {Vishwanath}},\
  }\href@noop {} {\bibfield  {journal} {\bibinfo  {journal} {arXiv preprint
  arXiv:1705.01111}\ } (\bibinfo {year} {2017})}\BibitemShut {NoStop}%
\bibitem [{\citenamefont {Wang}\ \emph {et~al.}(2012)\citenamefont {Wang},
  \citenamefont {Sun}, \citenamefont {Chen}, \citenamefont {Franchini},
  \citenamefont {Xu}, \citenamefont {Weng}, \citenamefont {Dai},\ and\
  \citenamefont {Fang}}]{wang2012dirac}%
  \BibitemOpen
  \bibfield  {author} {\bibinfo {author} {\bibfnamefont {Z.}~\bibnamefont
  {Wang}}, \bibinfo {author} {\bibfnamefont {Y.}~\bibnamefont {Sun}}, \bibinfo
  {author} {\bibfnamefont {X.-Q.}\ \bibnamefont {Chen}}, \bibinfo {author}
  {\bibfnamefont {C.}~\bibnamefont {Franchini}}, \bibinfo {author}
  {\bibfnamefont {G.}~\bibnamefont {Xu}}, \bibinfo {author} {\bibfnamefont
  {H.}~\bibnamefont {Weng}}, \bibinfo {author} {\bibfnamefont {X.}~\bibnamefont
  {Dai}}, \ and\ \bibinfo {author} {\bibfnamefont {Z.}~\bibnamefont {Fang}},\
  }\href@noop {} {\bibfield  {journal} {\bibinfo  {journal} {Physical Review
  B}\ }\textbf {\bibinfo {volume} {85}},\ \bibinfo {pages} {195320} (\bibinfo
  {year} {2012})}\BibitemShut {NoStop}%
\bibitem [{\citenamefont {Liu}\ \emph {et~al.}(2014{\natexlab{a}})\citenamefont
  {Liu}, \citenamefont {Zhou}, \citenamefont {Zhang}, \citenamefont {Wang},
  \citenamefont {Weng}, \citenamefont {Prabhakaran}, \citenamefont {Mo},
  \citenamefont {Shen}, \citenamefont {Fang}, \citenamefont {Dai} \emph
  {et~al.}}]{liu2014discovery}%
  \BibitemOpen
  \bibfield  {author} {\bibinfo {author} {\bibfnamefont {Z.}~\bibnamefont
  {Liu}}, \bibinfo {author} {\bibfnamefont {B.}~\bibnamefont {Zhou}}, \bibinfo
  {author} {\bibfnamefont {Y.}~\bibnamefont {Zhang}}, \bibinfo {author}
  {\bibfnamefont {Z.}~\bibnamefont {Wang}}, \bibinfo {author} {\bibfnamefont
  {H.}~\bibnamefont {Weng}}, \bibinfo {author} {\bibfnamefont {D.}~\bibnamefont
  {Prabhakaran}}, \bibinfo {author} {\bibfnamefont {S.-K.}\ \bibnamefont {Mo}},
  \bibinfo {author} {\bibfnamefont {Z.}~\bibnamefont {Shen}}, \bibinfo {author}
  {\bibfnamefont {Z.}~\bibnamefont {Fang}}, \bibinfo {author} {\bibfnamefont
  {X.}~\bibnamefont {Dai}},  \emph {et~al.},\ }\href@noop {} {\bibfield
  {journal} {\bibinfo  {journal} {Science}\ }\textbf {\bibinfo {volume}
  {343}},\ \bibinfo {pages} {864} (\bibinfo {year}
  {2014}{\natexlab{a}})}\BibitemShut {NoStop}%
\bibitem [{\citenamefont {Neupane}\ \emph {et~al.}(2014)\citenamefont
  {Neupane}, \citenamefont {Xu}, \citenamefont {Sankar}, \citenamefont
  {Alidoust}, \citenamefont {Bian}, \citenamefont {Liu}, \citenamefont
  {Belopolski}, \citenamefont {Chang}, \citenamefont {Jeng}, \citenamefont
  {Lin} \emph {et~al.}}]{neupane2014observation}%
  \BibitemOpen
  \bibfield  {author} {\bibinfo {author} {\bibfnamefont {M.}~\bibnamefont
  {Neupane}}, \bibinfo {author} {\bibfnamefont {S.}~\bibnamefont {Xu}},
  \bibinfo {author} {\bibfnamefont {R.}~\bibnamefont {Sankar}}, \bibinfo
  {author} {\bibfnamefont {N.}~\bibnamefont {Alidoust}}, \bibinfo {author}
  {\bibfnamefont {G.}~\bibnamefont {Bian}}, \bibinfo {author} {\bibfnamefont
  {C.}~\bibnamefont {Liu}}, \bibinfo {author} {\bibfnamefont {I.}~\bibnamefont
  {Belopolski}}, \bibinfo {author} {\bibfnamefont {T.}~\bibnamefont {Chang}},
  \bibinfo {author} {\bibfnamefont {H.}~\bibnamefont {Jeng}}, \bibinfo {author}
  {\bibfnamefont {H.}~\bibnamefont {Lin}},  \emph {et~al.},\ }\href@noop {}
  {\bibfield  {journal} {\bibinfo  {journal} {Nature communications}\ }\textbf
  {\bibinfo {volume} {5}},\ \bibinfo {pages} {3786} (\bibinfo {year}
  {2014})}\BibitemShut {NoStop}%
\bibitem [{\citenamefont {Liu}\ \emph {et~al.}(2014{\natexlab{b}})\citenamefont
  {Liu}, \citenamefont {Jiang}, \citenamefont {Zhou}, \citenamefont {Wang},
  \citenamefont {Zhang}, \citenamefont {Weng}, \citenamefont {Prabhakaran},
  \citenamefont {Mo}, \citenamefont {Peng}, \citenamefont {Dudin} \emph
  {et~al.}}]{liu2014stable}%
  \BibitemOpen
  \bibfield  {author} {\bibinfo {author} {\bibfnamefont {Z.}~\bibnamefont
  {Liu}}, \bibinfo {author} {\bibfnamefont {J.}~\bibnamefont {Jiang}}, \bibinfo
  {author} {\bibfnamefont {B.}~\bibnamefont {Zhou}}, \bibinfo {author}
  {\bibfnamefont {Z.}~\bibnamefont {Wang}}, \bibinfo {author} {\bibfnamefont
  {Y.}~\bibnamefont {Zhang}}, \bibinfo {author} {\bibfnamefont
  {H.}~\bibnamefont {Weng}}, \bibinfo {author} {\bibfnamefont {D.}~\bibnamefont
  {Prabhakaran}}, \bibinfo {author} {\bibfnamefont {S.}~\bibnamefont {Mo}},
  \bibinfo {author} {\bibfnamefont {H.}~\bibnamefont {Peng}}, \bibinfo {author}
  {\bibfnamefont {P.}~\bibnamefont {Dudin}},  \emph {et~al.},\ }\href@noop {}
  {\bibfield  {journal} {\bibinfo  {journal} {Nature materials}\ }\textbf
  {\bibinfo {volume} {13}},\ \bibinfo {pages} {677} (\bibinfo {year}
  {2014}{\natexlab{b}})}\BibitemShut {NoStop}%
\bibitem [{\citenamefont {Borisenko}\ \emph {et~al.}(2014)\citenamefont
  {Borisenko}, \citenamefont {Gibson}, \citenamefont {Evtushinsky},
  \citenamefont {Zabolotnyy}, \citenamefont {B{\"u}chner},\ and\ \citenamefont
  {Cava}}]{borisenko2014experimental}%
  \BibitemOpen
  \bibfield  {author} {\bibinfo {author} {\bibfnamefont {S.}~\bibnamefont
  {Borisenko}}, \bibinfo {author} {\bibfnamefont {Q.}~\bibnamefont {Gibson}},
  \bibinfo {author} {\bibfnamefont {D.}~\bibnamefont {Evtushinsky}}, \bibinfo
  {author} {\bibfnamefont {V.}~\bibnamefont {Zabolotnyy}}, \bibinfo {author}
  {\bibfnamefont {B.}~\bibnamefont {B{\"u}chner}}, \ and\ \bibinfo {author}
  {\bibfnamefont {R.~J.}\ \bibnamefont {Cava}},\ }\href@noop {} {\bibfield
  {journal} {\bibinfo  {journal} {Physical review letters}\ }\textbf {\bibinfo
  {volume} {113}},\ \bibinfo {pages} {027603} (\bibinfo {year}
  {2014})}\BibitemShut {NoStop}%
\bibitem [{\citenamefont {Wang}\ \emph {et~al.}(2013)\citenamefont {Wang},
  \citenamefont {Weng}, \citenamefont {Wu}, \citenamefont {Dai},\ and\
  \citenamefont {Fang}}]{wang2013three}%
  \BibitemOpen
  \bibfield  {author} {\bibinfo {author} {\bibfnamefont {Z.}~\bibnamefont
  {Wang}}, \bibinfo {author} {\bibfnamefont {H.}~\bibnamefont {Weng}}, \bibinfo
  {author} {\bibfnamefont {Q.}~\bibnamefont {Wu}}, \bibinfo {author}
  {\bibfnamefont {X.}~\bibnamefont {Dai}}, \ and\ \bibinfo {author}
  {\bibfnamefont {Z.}~\bibnamefont {Fang}},\ }\href@noop {} {\bibfield
  {journal} {\bibinfo  {journal} {Physical Review B}\ }\textbf {\bibinfo
  {volume} {88}},\ \bibinfo {pages} {125427} (\bibinfo {year}
  {2013})}\BibitemShut {NoStop}%
\bibitem [{\citenamefont {Ali}\ \emph {et~al.}(2014{\natexlab{a}})\citenamefont
  {Ali}, \citenamefont {Gibson}, \citenamefont {Jeon}, \citenamefont {Zhou},
  \citenamefont {Yazdani},\ and\ \citenamefont {Cava}}]{ali2014crystal}%
  \BibitemOpen
  \bibfield  {author} {\bibinfo {author} {\bibfnamefont {M.~N.}\ \bibnamefont
  {Ali}}, \bibinfo {author} {\bibfnamefont {Q.}~\bibnamefont {Gibson}},
  \bibinfo {author} {\bibfnamefont {S.}~\bibnamefont {Jeon}}, \bibinfo {author}
  {\bibfnamefont {B.~B.}\ \bibnamefont {Zhou}}, \bibinfo {author}
  {\bibfnamefont {A.}~\bibnamefont {Yazdani}}, \ and\ \bibinfo {author}
  {\bibfnamefont {R.}~\bibnamefont {Cava}},\ }\href@noop {} {\bibfield
  {journal} {\bibinfo  {journal} {Inorganic chemistry}\ }\textbf {\bibinfo
  {volume} {53}},\ \bibinfo {pages} {4062} (\bibinfo {year}
  {2014}{\natexlab{a}})}\BibitemShut {NoStop}%
\bibitem [{\citenamefont {Jeon}\ \emph {et~al.}(2014)\citenamefont {Jeon},
  \citenamefont {Zhou}, \citenamefont {Gyenis}, \citenamefont {Feldman},
  \citenamefont {Kimchi}, \citenamefont {Potter}, \citenamefont {Gibson},
  \citenamefont {Cava}, \citenamefont {Vishwanath},\ and\ \citenamefont
  {Yazdani}}]{jeon2014landau}%
  \BibitemOpen
  \bibfield  {author} {\bibinfo {author} {\bibfnamefont {S.}~\bibnamefont
  {Jeon}}, \bibinfo {author} {\bibfnamefont {B.~B.}\ \bibnamefont {Zhou}},
  \bibinfo {author} {\bibfnamefont {A.}~\bibnamefont {Gyenis}}, \bibinfo
  {author} {\bibfnamefont {B.~E.}\ \bibnamefont {Feldman}}, \bibinfo {author}
  {\bibfnamefont {I.}~\bibnamefont {Kimchi}}, \bibinfo {author} {\bibfnamefont
  {A.~C.}\ \bibnamefont {Potter}}, \bibinfo {author} {\bibfnamefont {Q.~D.}\
  \bibnamefont {Gibson}}, \bibinfo {author} {\bibfnamefont {R.~J.}\
  \bibnamefont {Cava}}, \bibinfo {author} {\bibfnamefont {A.}~\bibnamefont
  {Vishwanath}}, \ and\ \bibinfo {author} {\bibfnamefont {A.}~\bibnamefont
  {Yazdani}},\ }\href@noop {} {\bibfield  {journal} {\bibinfo  {journal}
  {Nature materials}\ }\textbf {\bibinfo {volume} {13}},\ \bibinfo {pages}
  {851} (\bibinfo {year} {2014})}\BibitemShut {NoStop}%
\bibitem [{\citenamefont {Feng}\ \emph {et~al.}(2015)\citenamefont {Feng},
  \citenamefont {Pang}, \citenamefont {Wu}, \citenamefont {Wang}, \citenamefont
  {Weng}, \citenamefont {Li}, \citenamefont {Dai}, \citenamefont {Fang},
  \citenamefont {Shi},\ and\ \citenamefont {Lu}}]{feng2015large}%
  \BibitemOpen
  \bibfield  {author} {\bibinfo {author} {\bibfnamefont {J.}~\bibnamefont
  {Feng}}, \bibinfo {author} {\bibfnamefont {Y.}~\bibnamefont {Pang}}, \bibinfo
  {author} {\bibfnamefont {D.}~\bibnamefont {Wu}}, \bibinfo {author}
  {\bibfnamefont {Z.}~\bibnamefont {Wang}}, \bibinfo {author} {\bibfnamefont
  {H.}~\bibnamefont {Weng}}, \bibinfo {author} {\bibfnamefont {J.}~\bibnamefont
  {Li}}, \bibinfo {author} {\bibfnamefont {X.}~\bibnamefont {Dai}}, \bibinfo
  {author} {\bibfnamefont {Z.}~\bibnamefont {Fang}}, \bibinfo {author}
  {\bibfnamefont {Y.}~\bibnamefont {Shi}}, \ and\ \bibinfo {author}
  {\bibfnamefont {L.}~\bibnamefont {Lu}},\ }\href@noop {} {\bibfield  {journal}
  {\bibinfo  {journal} {Physical Review B}\ }\textbf {\bibinfo {volume} {92}},\
  \bibinfo {pages} {081306} (\bibinfo {year} {2015})}\BibitemShut {NoStop}%
\bibitem [{\citenamefont {Liang}\ \emph {et~al.}(2015)\citenamefont {Liang},
  \citenamefont {Gibson}, \citenamefont {Ali}, \citenamefont {Liu},
  \citenamefont {Cava},\ and\ \citenamefont {Ong}}]{liang2015ultrahigh}%
  \BibitemOpen
  \bibfield  {author} {\bibinfo {author} {\bibfnamefont {T.}~\bibnamefont
  {Liang}}, \bibinfo {author} {\bibfnamefont {Q.}~\bibnamefont {Gibson}},
  \bibinfo {author} {\bibfnamefont {M.~N.}\ \bibnamefont {Ali}}, \bibinfo
  {author} {\bibfnamefont {M.}~\bibnamefont {Liu}}, \bibinfo {author}
  {\bibfnamefont {R.}~\bibnamefont {Cava}}, \ and\ \bibinfo {author}
  {\bibfnamefont {N.}~\bibnamefont {Ong}},\ }\href@noop {} {\bibfield
  {journal} {\bibinfo  {journal} {Nature materials}\ }\textbf {\bibinfo
  {volume} {14}},\ \bibinfo {pages} {280} (\bibinfo {year} {2015})}\BibitemShut
  {NoStop}%
\bibitem [{\citenamefont {Wang}\ \emph
  {et~al.}(2016{\natexlab{a}})\citenamefont {Wang}, \citenamefont {Li},
  \citenamefont {Yu},\ and\ \citenamefont {Liao}}]{wang2016aharonov}%
  \BibitemOpen
  \bibfield  {author} {\bibinfo {author} {\bibfnamefont {L.-X.}\ \bibnamefont
  {Wang}}, \bibinfo {author} {\bibfnamefont {C.-Z.}\ \bibnamefont {Li}},
  \bibinfo {author} {\bibfnamefont {D.-P.}\ \bibnamefont {Yu}}, \ and\ \bibinfo
  {author} {\bibfnamefont {Z.-M.}\ \bibnamefont {Liao}},\ }\href@noop {}
  {\bibfield  {journal} {\bibinfo  {journal} {Nature communications}\ }\textbf
  {\bibinfo {volume} {7}} (\bibinfo {year} {2016}{\natexlab{a}})}\BibitemShut
  {NoStop}%
\bibitem [{\citenamefont {Nielsen}\ and\ \citenamefont
  {Ninomiya}(1981{\natexlab{a}})}]{nielsen1981absence}%
  \BibitemOpen
  \bibfield  {author} {\bibinfo {author} {\bibfnamefont {H.~B.}\ \bibnamefont
  {Nielsen}}\ and\ \bibinfo {author} {\bibfnamefont {M.}~\bibnamefont
  {Ninomiya}},\ }\href@noop {} {\bibfield  {journal} {\bibinfo  {journal}
  {Nuclear Physics B}\ }\textbf {\bibinfo {volume} {185}},\ \bibinfo {pages}
  {20} (\bibinfo {year} {1981}{\natexlab{a}})}\BibitemShut {NoStop}%
\bibitem [{\citenamefont {Nielsen}\ and\ \citenamefont
  {Ninomiya}(1981{\natexlab{b}})}]{nielsen1981absence1}%
  \BibitemOpen
  \bibfield  {author} {\bibinfo {author} {\bibfnamefont {H.~B.}\ \bibnamefont
  {Nielsen}}\ and\ \bibinfo {author} {\bibfnamefont {M.}~\bibnamefont
  {Ninomiya}},\ }\href@noop {} {\bibfield  {journal} {\bibinfo  {journal}
  {Nuclear Physics B}\ }\textbf {\bibinfo {volume} {193}},\ \bibinfo {pages}
  {173} (\bibinfo {year} {1981}{\natexlab{b}})}\BibitemShut {NoStop}%
\bibitem [{\citenamefont {Fang}\ \emph {et~al.}(2003)\citenamefont {Fang},
  \citenamefont {Nagaosa}, \citenamefont {Takahashi}, \citenamefont {Asamitsu},
  \citenamefont {Mathieu}, \citenamefont {Ogasawara}, \citenamefont {Yamada},
  \citenamefont {Kawasaki}, \citenamefont {Tokura},\ and\ \citenamefont
  {Terakura}}]{fang2003anomalous}%
  \BibitemOpen
  \bibfield  {author} {\bibinfo {author} {\bibfnamefont {Z.}~\bibnamefont
  {Fang}}, \bibinfo {author} {\bibfnamefont {N.}~\bibnamefont {Nagaosa}},
  \bibinfo {author} {\bibfnamefont {K.~S.}\ \bibnamefont {Takahashi}}, \bibinfo
  {author} {\bibfnamefont {A.}~\bibnamefont {Asamitsu}}, \bibinfo {author}
  {\bibfnamefont {R.}~\bibnamefont {Mathieu}}, \bibinfo {author} {\bibfnamefont
  {T.}~\bibnamefont {Ogasawara}}, \bibinfo {author} {\bibfnamefont
  {H.}~\bibnamefont {Yamada}}, \bibinfo {author} {\bibfnamefont
  {M.}~\bibnamefont {Kawasaki}}, \bibinfo {author} {\bibfnamefont
  {Y.}~\bibnamefont {Tokura}}, \ and\ \bibinfo {author} {\bibfnamefont
  {K.}~\bibnamefont {Terakura}},\ }\href@noop {} {\bibfield  {journal}
  {\bibinfo  {journal} {Science}\ }\textbf {\bibinfo {volume} {302}},\ \bibinfo
  {pages} {92} (\bibinfo {year} {2003})}\BibitemShut {NoStop}%
\bibitem [{\citenamefont {Wan}\ \emph {et~al.}(2011)\citenamefont {Wan},
  \citenamefont {Turner}, \citenamefont {Vishwanath},\ and\ \citenamefont
  {Savrasov}}]{wan2011topological}%
  \BibitemOpen
  \bibfield  {author} {\bibinfo {author} {\bibfnamefont {X.}~\bibnamefont
  {Wan}}, \bibinfo {author} {\bibfnamefont {A.~M.}\ \bibnamefont {Turner}},
  \bibinfo {author} {\bibfnamefont {A.}~\bibnamefont {Vishwanath}}, \ and\
  \bibinfo {author} {\bibfnamefont {S.~Y.}\ \bibnamefont {Savrasov}},\
  }\href@noop {} {\bibfield  {journal} {\bibinfo  {journal} {Physical Review
  B}\ }\textbf {\bibinfo {volume} {83}},\ \bibinfo {pages} {205101} (\bibinfo
  {year} {2011})}\BibitemShut {NoStop}%
\bibitem [{\citenamefont {Balents}(2011)}]{balents2011viewpoint}%
  \BibitemOpen
  \bibfield  {author} {\bibinfo {author} {\bibfnamefont {L.}~\bibnamefont
  {Balents}},\ }\href@noop {} {\bibfield  {journal} {\bibinfo  {journal}
  {Physics}\ }\textbf {\bibinfo {volume} {4}},\ \bibinfo {pages} {36} (\bibinfo
  {year} {2011})}\BibitemShut {NoStop}%
\bibitem [{\citenamefont {Xu}\ \emph {et~al.}(2011)\citenamefont {Xu},
  \citenamefont {Weng}, \citenamefont {Wang}, \citenamefont {Dai},\ and\
  \citenamefont {Fang}}]{xu2011chern}%
  \BibitemOpen
  \bibfield  {author} {\bibinfo {author} {\bibfnamefont {G.}~\bibnamefont
  {Xu}}, \bibinfo {author} {\bibfnamefont {H.}~\bibnamefont {Weng}}, \bibinfo
  {author} {\bibfnamefont {Z.}~\bibnamefont {Wang}}, \bibinfo {author}
  {\bibfnamefont {X.}~\bibnamefont {Dai}}, \ and\ \bibinfo {author}
  {\bibfnamefont {Z.}~\bibnamefont {Fang}},\ }\href@noop {} {\bibfield
  {journal} {\bibinfo  {journal} {Physical review letters}\ }\textbf {\bibinfo
  {volume} {107}},\ \bibinfo {pages} {186806} (\bibinfo {year}
  {2011})}\BibitemShut {NoStop}%
\bibitem [{\citenamefont {Weng}\ \emph
  {et~al.}(2015{\natexlab{a}})\citenamefont {Weng}, \citenamefont {Fang},
  \citenamefont {Fang}, \citenamefont {Bernevig},\ and\ \citenamefont
  {Dai}}]{weng2015weyl}%
  \BibitemOpen
  \bibfield  {author} {\bibinfo {author} {\bibfnamefont {H.}~\bibnamefont
  {Weng}}, \bibinfo {author} {\bibfnamefont {C.}~\bibnamefont {Fang}}, \bibinfo
  {author} {\bibfnamefont {Z.}~\bibnamefont {Fang}}, \bibinfo {author}
  {\bibfnamefont {B.~A.}\ \bibnamefont {Bernevig}}, \ and\ \bibinfo {author}
  {\bibfnamefont {X.}~\bibnamefont {Dai}},\ }\href@noop {} {\bibfield
  {journal} {\bibinfo  {journal} {Physical Review X}\ }\textbf {\bibinfo
  {volume} {5}},\ \bibinfo {pages} {011029} (\bibinfo {year}
  {2015}{\natexlab{a}})}\BibitemShut {NoStop}%
\bibitem [{\citenamefont {Huang}\ \emph
  {et~al.}(2015{\natexlab{a}})\citenamefont {Huang}, \citenamefont {Xu},
  \citenamefont {Belopolski}, \citenamefont {Lee}, \citenamefont {Chang},
  \citenamefont {Wang}, \citenamefont {Alidoust}, \citenamefont {Bian},
  \citenamefont {Neupane}, \citenamefont {Zhang} \emph
  {et~al.}}]{huang2015weyl}%
  \BibitemOpen
  \bibfield  {author} {\bibinfo {author} {\bibfnamefont {S.-M.}\ \bibnamefont
  {Huang}}, \bibinfo {author} {\bibfnamefont {S.-Y.}\ \bibnamefont {Xu}},
  \bibinfo {author} {\bibfnamefont {I.}~\bibnamefont {Belopolski}}, \bibinfo
  {author} {\bibfnamefont {C.-C.}\ \bibnamefont {Lee}}, \bibinfo {author}
  {\bibfnamefont {G.}~\bibnamefont {Chang}}, \bibinfo {author} {\bibfnamefont
  {B.}~\bibnamefont {Wang}}, \bibinfo {author} {\bibfnamefont {N.}~\bibnamefont
  {Alidoust}}, \bibinfo {author} {\bibfnamefont {G.}~\bibnamefont {Bian}},
  \bibinfo {author} {\bibfnamefont {M.}~\bibnamefont {Neupane}}, \bibinfo
  {author} {\bibfnamefont {C.}~\bibnamefont {Zhang}},  \emph {et~al.},\
  }\href@noop {} {\bibfield  {journal} {\bibinfo  {journal} {Nature
  communications}\ }\textbf {\bibinfo {volume} {6}},\ \bibinfo {pages} {7373}
  (\bibinfo {year} {2015}{\natexlab{a}})}\BibitemShut {NoStop}%
\bibitem [{\citenamefont {Nielsen}\ and\ \citenamefont
  {Ninomiya}(1983)}]{nielsen1983adler}%
  \BibitemOpen
  \bibfield  {author} {\bibinfo {author} {\bibfnamefont {H.~B.}\ \bibnamefont
  {Nielsen}}\ and\ \bibinfo {author} {\bibfnamefont {M.}~\bibnamefont
  {Ninomiya}},\ }\href@noop {} {\bibfield  {journal} {\bibinfo  {journal}
  {Physics Letters B}\ }\textbf {\bibinfo {volume} {130}},\ \bibinfo {pages}
  {389} (\bibinfo {year} {1983})}\BibitemShut {NoStop}%
\bibitem [{\citenamefont {Hosur}\ \emph {et~al.}(2012)\citenamefont {Hosur},
  \citenamefont {Parameswaran},\ and\ \citenamefont
  {Vishwanath}}]{hosur2012charge}%
  \BibitemOpen
  \bibfield  {author} {\bibinfo {author} {\bibfnamefont {P.}~\bibnamefont
  {Hosur}}, \bibinfo {author} {\bibfnamefont {S.}~\bibnamefont {Parameswaran}},
  \ and\ \bibinfo {author} {\bibfnamefont {A.}~\bibnamefont {Vishwanath}},\
  }\href@noop {} {\bibfield  {journal} {\bibinfo  {journal} {Physical review
  letters}\ }\textbf {\bibinfo {volume} {108}},\ \bibinfo {pages} {046602}
  (\bibinfo {year} {2012})}\BibitemShut {NoStop}%
\bibitem [{\citenamefont {Son}\ and\ \citenamefont
  {Spivak}(2013)}]{son2013chiral}%
  \BibitemOpen
  \bibfield  {author} {\bibinfo {author} {\bibfnamefont {D.}~\bibnamefont
  {Son}}\ and\ \bibinfo {author} {\bibfnamefont {B.}~\bibnamefont {Spivak}},\
  }\href@noop {} {\bibfield  {journal} {\bibinfo  {journal} {Physical Review
  B}\ }\textbf {\bibinfo {volume} {88}},\ \bibinfo {pages} {104412} (\bibinfo
  {year} {2013})}\BibitemShut {NoStop}%
\bibitem [{\citenamefont {Kim}\ \emph {et~al.}(2013)\citenamefont {Kim},
  \citenamefont {Kim}, \citenamefont {Wang}, \citenamefont {Sasaki},
  \citenamefont {Satoh}, \citenamefont {Ohnishi}, \citenamefont {Kitaura},
  \citenamefont {Yang},\ and\ \citenamefont {Li}}]{kim2013dirac}%
  \BibitemOpen
  \bibfield  {author} {\bibinfo {author} {\bibfnamefont {H.-J.}\ \bibnamefont
  {Kim}}, \bibinfo {author} {\bibfnamefont {K.-S.}\ \bibnamefont {Kim}},
  \bibinfo {author} {\bibfnamefont {J.-F.}\ \bibnamefont {Wang}}, \bibinfo
  {author} {\bibfnamefont {M.}~\bibnamefont {Sasaki}}, \bibinfo {author}
  {\bibfnamefont {N.}~\bibnamefont {Satoh}}, \bibinfo {author} {\bibfnamefont
  {A.}~\bibnamefont {Ohnishi}}, \bibinfo {author} {\bibfnamefont
  {M.}~\bibnamefont {Kitaura}}, \bibinfo {author} {\bibfnamefont
  {M.}~\bibnamefont {Yang}}, \ and\ \bibinfo {author} {\bibfnamefont
  {L.}~\bibnamefont {Li}},\ }\href@noop {} {\bibfield  {journal} {\bibinfo
  {journal} {Physical review letters}\ }\textbf {\bibinfo {volume} {111}},\
  \bibinfo {pages} {246603} (\bibinfo {year} {2013})}\BibitemShut {NoStop}%
\bibitem [{\citenamefont {Parameswaran}\ \emph {et~al.}(2014)\citenamefont
  {Parameswaran}, \citenamefont {Grover}, \citenamefont {Abanin}, \citenamefont
  {Pesin},\ and\ \citenamefont {Vishwanath}}]{parameswaran2014probing}%
  \BibitemOpen
  \bibfield  {author} {\bibinfo {author} {\bibfnamefont {S.}~\bibnamefont
  {Parameswaran}}, \bibinfo {author} {\bibfnamefont {T.}~\bibnamefont
  {Grover}}, \bibinfo {author} {\bibfnamefont {D.}~\bibnamefont {Abanin}},
  \bibinfo {author} {\bibfnamefont {D.}~\bibnamefont {Pesin}}, \ and\ \bibinfo
  {author} {\bibfnamefont {A.}~\bibnamefont {Vishwanath}},\ }\href@noop {}
  {\bibfield  {journal} {\bibinfo  {journal} {Physical Review X}\ }\textbf
  {\bibinfo {volume} {4}},\ \bibinfo {pages} {031035} (\bibinfo {year}
  {2014})}\BibitemShut {NoStop}%
\bibitem [{\citenamefont {Xiong}\ \emph {et~al.}(2015)\citenamefont {Xiong},
  \citenamefont {Kushwaha}, \citenamefont {Liang}, \citenamefont {Krizan},
  \citenamefont {Hirschberger}, \citenamefont {Wang}, \citenamefont {Cava},\
  and\ \citenamefont {Ong}}]{xiong2015evidence}%
  \BibitemOpen
  \bibfield  {author} {\bibinfo {author} {\bibfnamefont {J.}~\bibnamefont
  {Xiong}}, \bibinfo {author} {\bibfnamefont {S.~K.}\ \bibnamefont {Kushwaha}},
  \bibinfo {author} {\bibfnamefont {T.}~\bibnamefont {Liang}}, \bibinfo
  {author} {\bibfnamefont {J.~W.}\ \bibnamefont {Krizan}}, \bibinfo {author}
  {\bibfnamefont {M.}~\bibnamefont {Hirschberger}}, \bibinfo {author}
  {\bibfnamefont {W.}~\bibnamefont {Wang}}, \bibinfo {author} {\bibfnamefont
  {R.}~\bibnamefont {Cava}}, \ and\ \bibinfo {author} {\bibfnamefont
  {N.}~\bibnamefont {Ong}},\ }\href@noop {} {\bibfield  {journal} {\bibinfo
  {journal} {Science}\ }\textbf {\bibinfo {volume} {350}},\ \bibinfo {pages}
  {413} (\bibinfo {year} {2015})}\BibitemShut {NoStop}%
\bibitem [{\citenamefont {Li}\ \emph {et~al.}(2015)\citenamefont {Li},
  \citenamefont {Wang}, \citenamefont {Liu}, \citenamefont {Wang},
  \citenamefont {Liao},\ and\ \citenamefont {Yu}}]{li2015giant}%
  \BibitemOpen
  \bibfield  {author} {\bibinfo {author} {\bibfnamefont {C.-Z.}\ \bibnamefont
  {Li}}, \bibinfo {author} {\bibfnamefont {L.-X.}\ \bibnamefont {Wang}},
  \bibinfo {author} {\bibfnamefont {H.}~\bibnamefont {Liu}}, \bibinfo {author}
  {\bibfnamefont {J.}~\bibnamefont {Wang}}, \bibinfo {author} {\bibfnamefont
  {Z.-M.}\ \bibnamefont {Liao}}, \ and\ \bibinfo {author} {\bibfnamefont
  {D.-P.}\ \bibnamefont {Yu}},\ }\href@noop {} {\bibfield  {journal} {\bibinfo
  {journal} {Nature communications}\ }\textbf {\bibinfo {volume} {6}},\
  \bibinfo {pages} {10137} (\bibinfo {year} {2015})}\BibitemShut {NoStop}%
\bibitem [{\citenamefont {Arnold}\ \emph {et~al.}(2016)\citenamefont {Arnold},
  \citenamefont {Shekhar}, \citenamefont {Wu}, \citenamefont {Sun},
  \citenamefont {Dos~Reis}, \citenamefont {Kumar}, \citenamefont {Naumann},
  \citenamefont {Ajeesh}, \citenamefont {Schmidt}, \citenamefont {Grushin}
  \emph {et~al.}}]{arnold2016negative}%
  \BibitemOpen
  \bibfield  {author} {\bibinfo {author} {\bibfnamefont {F.}~\bibnamefont
  {Arnold}}, \bibinfo {author} {\bibfnamefont {C.}~\bibnamefont {Shekhar}},
  \bibinfo {author} {\bibfnamefont {S.-C.}\ \bibnamefont {Wu}}, \bibinfo
  {author} {\bibfnamefont {Y.}~\bibnamefont {Sun}}, \bibinfo {author}
  {\bibfnamefont {R.~D.}\ \bibnamefont {Dos~Reis}}, \bibinfo {author}
  {\bibfnamefont {N.}~\bibnamefont {Kumar}}, \bibinfo {author} {\bibfnamefont
  {M.}~\bibnamefont {Naumann}}, \bibinfo {author} {\bibfnamefont {M.~O.}\
  \bibnamefont {Ajeesh}}, \bibinfo {author} {\bibfnamefont {M.}~\bibnamefont
  {Schmidt}}, \bibinfo {author} {\bibfnamefont {A.~G.}\ \bibnamefont
  {Grushin}},  \emph {et~al.},\ }\href@noop {} {\bibfield  {journal} {\bibinfo
  {journal} {Nature communications}\ }\textbf {\bibinfo {volume} {7}},\
  \bibinfo {pages} {11615} (\bibinfo {year} {2016})}\BibitemShut {NoStop}%
\bibitem [{\citenamefont {Zhang}\ \emph
  {et~al.}(2017{\natexlab{a}})\citenamefont {Zhang}, \citenamefont {Zhang},
  \citenamefont {Wang}, \citenamefont {Liu}, \citenamefont {Chen},
  \citenamefont {Lu}, \citenamefont {Liang}, \citenamefont {Cao}, \citenamefont
  {Yuan}, \citenamefont {Tang} \emph {et~al.}}]{zhang2017room}%
  \BibitemOpen
  \bibfield  {author} {\bibinfo {author} {\bibfnamefont {C.}~\bibnamefont
  {Zhang}}, \bibinfo {author} {\bibfnamefont {E.}~\bibnamefont {Zhang}},
  \bibinfo {author} {\bibfnamefont {W.}~\bibnamefont {Wang}}, \bibinfo {author}
  {\bibfnamefont {Y.}~\bibnamefont {Liu}}, \bibinfo {author} {\bibfnamefont
  {Z.-G.}\ \bibnamefont {Chen}}, \bibinfo {author} {\bibfnamefont
  {S.}~\bibnamefont {Lu}}, \bibinfo {author} {\bibfnamefont {S.}~\bibnamefont
  {Liang}}, \bibinfo {author} {\bibfnamefont {J.}~\bibnamefont {Cao}}, \bibinfo
  {author} {\bibfnamefont {X.}~\bibnamefont {Yuan}}, \bibinfo {author}
  {\bibfnamefont {L.}~\bibnamefont {Tang}},  \emph {et~al.},\ }\href@noop {}
  {\bibfield  {journal} {\bibinfo  {journal} {Nature communications}\ }\textbf
  {\bibinfo {volume} {8}},\ \bibinfo {pages} {13741} (\bibinfo {year}
  {2017}{\natexlab{a}})}\BibitemShut {NoStop}%
\bibitem [{\citenamefont {Hirschberger}\ \emph {et~al.}(2016)\citenamefont
  {Hirschberger}, \citenamefont {Kushwaha}, \citenamefont {Wang}, \citenamefont
  {Gibson}, \citenamefont {Liang}, \citenamefont {Belvin}, \citenamefont
  {Bernevig}, \citenamefont {Cava}, \citenamefont {Ong} \emph
  {et~al.}}]{hirschberger2016chiral}%
  \BibitemOpen
  \bibfield  {author} {\bibinfo {author} {\bibfnamefont {M.}~\bibnamefont
  {Hirschberger}}, \bibinfo {author} {\bibfnamefont {S.}~\bibnamefont
  {Kushwaha}}, \bibinfo {author} {\bibfnamefont {Z.}~\bibnamefont {Wang}},
  \bibinfo {author} {\bibfnamefont {Q.}~\bibnamefont {Gibson}}, \bibinfo
  {author} {\bibfnamefont {S.}~\bibnamefont {Liang}}, \bibinfo {author}
  {\bibfnamefont {C.}~\bibnamefont {Belvin}}, \bibinfo {author} {\bibfnamefont
  {B.}~\bibnamefont {Bernevig}}, \bibinfo {author} {\bibfnamefont
  {R.}~\bibnamefont {Cava}}, \bibinfo {author} {\bibfnamefont {N.}~\bibnamefont
  {Ong}},  \emph {et~al.},\ }\href@noop {} {\bibfield  {journal} {\bibinfo
  {journal} {Nature Materials}\ }\textbf {\bibinfo {volume} {15}},\ \bibinfo
  {pages} {1161} (\bibinfo {year} {2016})}\BibitemShut {NoStop}%
\bibitem [{\citenamefont {Hosur}\ and\ \citenamefont
  {Qi}(2013)}]{hosur2013recent}%
  \BibitemOpen
  \bibfield  {author} {\bibinfo {author} {\bibfnamefont {P.}~\bibnamefont
  {Hosur}}\ and\ \bibinfo {author} {\bibfnamefont {X.}~\bibnamefont {Qi}},\
  }\href@noop {} {\bibfield  {journal} {\bibinfo  {journal} {Comptes Rendus
  Physique}\ }\textbf {\bibinfo {volume} {14}},\ \bibinfo {pages} {857}
  (\bibinfo {year} {2013})}\BibitemShut {NoStop}%
\bibitem [{\citenamefont {Volovik}(2015)}]{volovik2015standard}%
  \BibitemOpen
  \bibfield  {author} {\bibinfo {author} {\bibfnamefont {G.}~\bibnamefont
  {Volovik}},\ }\href@noop {} {\bibfield  {journal} {\bibinfo  {journal}
  {Physica Scripta}\ }\textbf {\bibinfo {volume} {2015}},\ \bibinfo {pages}
  {014014} (\bibinfo {year} {2015})}\BibitemShut {NoStop}%
\bibitem [{\citenamefont {Xu}\ \emph {et~al.}(2015{\natexlab{a}})\citenamefont
  {Xu}, \citenamefont {Belopolski}, \citenamefont {Alidoust}, \citenamefont
  {Neupane}, \citenamefont {Bian}, \citenamefont {Zhang}, \citenamefont
  {Sankar}, \citenamefont {Chang}, \citenamefont {Yuan}, \citenamefont {Lee}
  \emph {et~al.}}]{xu2015discovery}%
  \BibitemOpen
  \bibfield  {author} {\bibinfo {author} {\bibfnamefont {S.-Y.}\ \bibnamefont
  {Xu}}, \bibinfo {author} {\bibfnamefont {I.}~\bibnamefont {Belopolski}},
  \bibinfo {author} {\bibfnamefont {N.}~\bibnamefont {Alidoust}}, \bibinfo
  {author} {\bibfnamefont {M.}~\bibnamefont {Neupane}}, \bibinfo {author}
  {\bibfnamefont {G.}~\bibnamefont {Bian}}, \bibinfo {author} {\bibfnamefont
  {C.}~\bibnamefont {Zhang}}, \bibinfo {author} {\bibfnamefont
  {R.}~\bibnamefont {Sankar}}, \bibinfo {author} {\bibfnamefont
  {G.}~\bibnamefont {Chang}}, \bibinfo {author} {\bibfnamefont
  {Z.}~\bibnamefont {Yuan}}, \bibinfo {author} {\bibfnamefont {C.-C.}\
  \bibnamefont {Lee}},  \emph {et~al.},\ }\href@noop {} {\bibfield  {journal}
  {\bibinfo  {journal} {Science}\ }\textbf {\bibinfo {volume} {349}},\ \bibinfo
  {pages} {613} (\bibinfo {year} {2015}{\natexlab{a}})}\BibitemShut {NoStop}%
\bibitem [{\citenamefont {Lv}\ \emph {et~al.}(2015)\citenamefont {Lv},
  \citenamefont {Weng}, \citenamefont {Fu}, \citenamefont {Wang}, \citenamefont
  {Miao}, \citenamefont {Ma}, \citenamefont {Richard}, \citenamefont {Huang},
  \citenamefont {Zhao}, \citenamefont {Chen} \emph
  {et~al.}}]{lv2015experimental}%
  \BibitemOpen
  \bibfield  {author} {\bibinfo {author} {\bibfnamefont {B.}~\bibnamefont
  {Lv}}, \bibinfo {author} {\bibfnamefont {H.}~\bibnamefont {Weng}}, \bibinfo
  {author} {\bibfnamefont {B.}~\bibnamefont {Fu}}, \bibinfo {author}
  {\bibfnamefont {X.}~\bibnamefont {Wang}}, \bibinfo {author} {\bibfnamefont
  {H.}~\bibnamefont {Miao}}, \bibinfo {author} {\bibfnamefont {J.}~\bibnamefont
  {Ma}}, \bibinfo {author} {\bibfnamefont {P.}~\bibnamefont {Richard}},
  \bibinfo {author} {\bibfnamefont {X.}~\bibnamefont {Huang}}, \bibinfo
  {author} {\bibfnamefont {L.}~\bibnamefont {Zhao}}, \bibinfo {author}
  {\bibfnamefont {G.}~\bibnamefont {Chen}},  \emph {et~al.},\ }\href@noop {}
  {\bibfield  {journal} {\bibinfo  {journal} {Physical Review X}\ }\textbf
  {\bibinfo {volume} {5}},\ \bibinfo {pages} {031013} (\bibinfo {year}
  {2015})}\BibitemShut {NoStop}%
\bibitem [{\citenamefont {Yang}\ \emph
  {et~al.}(2015{\natexlab{a}})\citenamefont {Yang}, \citenamefont {Liu},
  \citenamefont {Sun}, \citenamefont {Peng}, \citenamefont {Yang},
  \citenamefont {Zhang}, \citenamefont {Zhou}, \citenamefont {Zhang},
  \citenamefont {Guo}, \citenamefont {Rahn} \emph {et~al.}}]{yang2015weyl}%
  \BibitemOpen
  \bibfield  {author} {\bibinfo {author} {\bibfnamefont {L.}~\bibnamefont
  {Yang}}, \bibinfo {author} {\bibfnamefont {Z.}~\bibnamefont {Liu}}, \bibinfo
  {author} {\bibfnamefont {Y.}~\bibnamefont {Sun}}, \bibinfo {author}
  {\bibfnamefont {H.}~\bibnamefont {Peng}}, \bibinfo {author} {\bibfnamefont
  {H.}~\bibnamefont {Yang}}, \bibinfo {author} {\bibfnamefont {T.}~\bibnamefont
  {Zhang}}, \bibinfo {author} {\bibfnamefont {B.}~\bibnamefont {Zhou}},
  \bibinfo {author} {\bibfnamefont {Y.}~\bibnamefont {Zhang}}, \bibinfo
  {author} {\bibfnamefont {Y.}~\bibnamefont {Guo}}, \bibinfo {author}
  {\bibfnamefont {M.}~\bibnamefont {Rahn}},  \emph {et~al.},\ }\href@noop {}
  {\bibfield  {journal} {\bibinfo  {journal} {Nature physics}\ }\textbf
  {\bibinfo {volume} {11}},\ \bibinfo {pages} {728} (\bibinfo {year}
  {2015}{\natexlab{a}})}\BibitemShut {NoStop}%
\bibitem [{\citenamefont {Xu}\ \emph {et~al.}(2015{\natexlab{b}})\citenamefont
  {Xu}, \citenamefont {Alidoust}, \citenamefont {Belopolski}, \citenamefont
  {Yuan}, \citenamefont {Bian}, \citenamefont {Chang}, \citenamefont {Zheng},
  \citenamefont {Strocov}, \citenamefont {Sanchez}, \citenamefont {Chang} \emph
  {et~al.}}]{xu2015discovery1}%
  \BibitemOpen
  \bibfield  {author} {\bibinfo {author} {\bibfnamefont {S.-Y.}\ \bibnamefont
  {Xu}}, \bibinfo {author} {\bibfnamefont {N.}~\bibnamefont {Alidoust}},
  \bibinfo {author} {\bibfnamefont {I.}~\bibnamefont {Belopolski}}, \bibinfo
  {author} {\bibfnamefont {Z.}~\bibnamefont {Yuan}}, \bibinfo {author}
  {\bibfnamefont {G.}~\bibnamefont {Bian}}, \bibinfo {author} {\bibfnamefont
  {T.-R.}\ \bibnamefont {Chang}}, \bibinfo {author} {\bibfnamefont
  {H.}~\bibnamefont {Zheng}}, \bibinfo {author} {\bibfnamefont {V.~N.}\
  \bibnamefont {Strocov}}, \bibinfo {author} {\bibfnamefont {D.~S.}\
  \bibnamefont {Sanchez}}, \bibinfo {author} {\bibfnamefont {G.}~\bibnamefont
  {Chang}},  \emph {et~al.},\ }\href@noop {} {\bibfield  {journal} {\bibinfo
  {journal} {Nature Physics}\ }\textbf {\bibinfo {volume} {11}},\ \bibinfo
  {pages} {748} (\bibinfo {year} {2015}{\natexlab{b}})}\BibitemShut {NoStop}%
\bibitem [{\citenamefont {Souma}\ \emph {et~al.}(2016)\citenamefont {Souma},
  \citenamefont {Wang}, \citenamefont {Kotaka}, \citenamefont {Sato},
  \citenamefont {Nakayama}, \citenamefont {Tanaka}, \citenamefont {Kimizuka},
  \citenamefont {Takahashi}, \citenamefont {Yamauchi}, \citenamefont {Oguchi}
  \emph {et~al.}}]{souma2016direct}%
  \BibitemOpen
  \bibfield  {author} {\bibinfo {author} {\bibfnamefont {S.}~\bibnamefont
  {Souma}}, \bibinfo {author} {\bibfnamefont {Z.}~\bibnamefont {Wang}},
  \bibinfo {author} {\bibfnamefont {H.}~\bibnamefont {Kotaka}}, \bibinfo
  {author} {\bibfnamefont {T.}~\bibnamefont {Sato}}, \bibinfo {author}
  {\bibfnamefont {K.}~\bibnamefont {Nakayama}}, \bibinfo {author}
  {\bibfnamefont {Y.}~\bibnamefont {Tanaka}}, \bibinfo {author} {\bibfnamefont
  {H.}~\bibnamefont {Kimizuka}}, \bibinfo {author} {\bibfnamefont
  {T.}~\bibnamefont {Takahashi}}, \bibinfo {author} {\bibfnamefont
  {K.}~\bibnamefont {Yamauchi}}, \bibinfo {author} {\bibfnamefont
  {T.}~\bibnamefont {Oguchi}},  \emph {et~al.},\ }\href@noop {} {\bibfield
  {journal} {\bibinfo  {journal} {Physical Review B}\ }\textbf {\bibinfo
  {volume} {93}},\ \bibinfo {pages} {161112} (\bibinfo {year}
  {2016})}\BibitemShut {NoStop}%
\bibitem [{\citenamefont {Liu}\ \emph {et~al.}(2016{\natexlab{a}})\citenamefont
  {Liu}, \citenamefont {Yang}, \citenamefont {Sun}, \citenamefont {Zhang},
  \citenamefont {Peng}, \citenamefont {Yang}, \citenamefont {Chen},
  \citenamefont {Zhang}, \citenamefont {Guo}, \citenamefont {Prabhakaran} \emph
  {et~al.}}]{liu2016evolution}%
  \BibitemOpen
  \bibfield  {author} {\bibinfo {author} {\bibfnamefont {Z.}~\bibnamefont
  {Liu}}, \bibinfo {author} {\bibfnamefont {L.}~\bibnamefont {Yang}}, \bibinfo
  {author} {\bibfnamefont {Y.}~\bibnamefont {Sun}}, \bibinfo {author}
  {\bibfnamefont {T.}~\bibnamefont {Zhang}}, \bibinfo {author} {\bibfnamefont
  {H.}~\bibnamefont {Peng}}, \bibinfo {author} {\bibfnamefont {H.}~\bibnamefont
  {Yang}}, \bibinfo {author} {\bibfnamefont {C.}~\bibnamefont {Chen}}, \bibinfo
  {author} {\bibfnamefont {Y.}~\bibnamefont {Zhang}}, \bibinfo {author}
  {\bibfnamefont {Y.}~\bibnamefont {Guo}}, \bibinfo {author} {\bibfnamefont
  {D.}~\bibnamefont {Prabhakaran}},  \emph {et~al.},\ }\href@noop {} {\bibfield
   {journal} {\bibinfo  {journal} {Nature materials}\ }\textbf {\bibinfo
  {volume} {15}},\ \bibinfo {pages} {27} (\bibinfo {year}
  {2016}{\natexlab{a}})}\BibitemShut {NoStop}%
\bibitem [{\citenamefont {Lu}\ \emph {et~al.}(2015)\citenamefont {Lu},
  \citenamefont {Wang}, \citenamefont {Ye}, \citenamefont {Ran}, \citenamefont
  {Fu}, \citenamefont {Joannopoulos},\ and\ \citenamefont
  {Solja{\v{c}}i{\'c}}}]{lu2015experimental}%
  \BibitemOpen
  \bibfield  {author} {\bibinfo {author} {\bibfnamefont {L.}~\bibnamefont
  {Lu}}, \bibinfo {author} {\bibfnamefont {Z.}~\bibnamefont {Wang}}, \bibinfo
  {author} {\bibfnamefont {D.}~\bibnamefont {Ye}}, \bibinfo {author}
  {\bibfnamefont {L.}~\bibnamefont {Ran}}, \bibinfo {author} {\bibfnamefont
  {L.}~\bibnamefont {Fu}}, \bibinfo {author} {\bibfnamefont {J.~D.}\
  \bibnamefont {Joannopoulos}}, \ and\ \bibinfo {author} {\bibfnamefont
  {M.}~\bibnamefont {Solja{\v{c}}i{\'c}}},\ }\href@noop {} {\bibfield
  {journal} {\bibinfo  {journal} {Science}\ }\textbf {\bibinfo {volume}
  {349}},\ \bibinfo {pages} {622} (\bibinfo {year} {2015})}\BibitemShut
  {NoStop}%
\bibitem [{\citenamefont {Soluyanov}\ \emph {et~al.}(2015)\citenamefont
  {Soluyanov}, \citenamefont {Gresch}, \citenamefont {Wang}, \citenamefont
  {Wu}, \citenamefont {Troyer}, \citenamefont {Dai},\ and\ \citenamefont
  {Bernevig}}]{soluyanov2015type}%
  \BibitemOpen
  \bibfield  {author} {\bibinfo {author} {\bibfnamefont {A.~A.}\ \bibnamefont
  {Soluyanov}}, \bibinfo {author} {\bibfnamefont {D.}~\bibnamefont {Gresch}},
  \bibinfo {author} {\bibfnamefont {Z.}~\bibnamefont {Wang}}, \bibinfo {author}
  {\bibfnamefont {Q.}~\bibnamefont {Wu}}, \bibinfo {author} {\bibfnamefont
  {M.}~\bibnamefont {Troyer}}, \bibinfo {author} {\bibfnamefont
  {X.}~\bibnamefont {Dai}}, \ and\ \bibinfo {author} {\bibfnamefont {B.~A.}\
  \bibnamefont {Bernevig}},\ }\href@noop {} {\bibfield  {journal} {\bibinfo
  {journal} {Nature}\ }\textbf {\bibinfo {volume} {527}},\ \bibinfo {pages}
  {495} (\bibinfo {year} {2015})}\BibitemShut {NoStop}%
\bibitem [{\citenamefont {Sun}\ \emph {et~al.}(2015)\citenamefont {Sun},
  \citenamefont {Wu}, \citenamefont {Ali}, \citenamefont {Felser},\ and\
  \citenamefont {Yan}}]{sun2015prediction}%
  \BibitemOpen
  \bibfield  {author} {\bibinfo {author} {\bibfnamefont {Y.}~\bibnamefont
  {Sun}}, \bibinfo {author} {\bibfnamefont {S.-C.}\ \bibnamefont {Wu}},
  \bibinfo {author} {\bibfnamefont {M.~N.}\ \bibnamefont {Ali}}, \bibinfo
  {author} {\bibfnamefont {C.}~\bibnamefont {Felser}}, \ and\ \bibinfo {author}
  {\bibfnamefont {B.}~\bibnamefont {Yan}},\ }\href@noop {} {\bibfield
  {journal} {\bibinfo  {journal} {Physical Review B}\ }\textbf {\bibinfo
  {volume} {92}},\ \bibinfo {pages} {161107} (\bibinfo {year}
  {2015})}\BibitemShut {NoStop}%
\bibitem [{\citenamefont {Huang}\ \emph
  {et~al.}(2016{\natexlab{a}})\citenamefont {Huang}, \citenamefont {McCormick},
  \citenamefont {Ochi}, \citenamefont {Zhao}, \citenamefont {Suzuki},
  \citenamefont {Arita}, \citenamefont {Wu}, \citenamefont {Mou}, \citenamefont
  {Cao}, \citenamefont {Yan} \emph {et~al.}}]{huang2016spectroscopic}%
  \BibitemOpen
  \bibfield  {author} {\bibinfo {author} {\bibfnamefont {L.}~\bibnamefont
  {Huang}}, \bibinfo {author} {\bibfnamefont {T.~M.}\ \bibnamefont
  {McCormick}}, \bibinfo {author} {\bibfnamefont {M.}~\bibnamefont {Ochi}},
  \bibinfo {author} {\bibfnamefont {Z.}~\bibnamefont {Zhao}}, \bibinfo {author}
  {\bibfnamefont {M.-T.}\ \bibnamefont {Suzuki}}, \bibinfo {author}
  {\bibfnamefont {R.}~\bibnamefont {Arita}}, \bibinfo {author} {\bibfnamefont
  {Y.}~\bibnamefont {Wu}}, \bibinfo {author} {\bibfnamefont {D.}~\bibnamefont
  {Mou}}, \bibinfo {author} {\bibfnamefont {H.}~\bibnamefont {Cao}}, \bibinfo
  {author} {\bibfnamefont {J.}~\bibnamefont {Yan}},  \emph {et~al.},\
  }\href@noop {} {\bibfield  {journal} {\bibinfo  {journal} {Nature materials}\
  }\textbf {\bibinfo {volume} {15}},\ \bibinfo {pages} {1155} (\bibinfo {year}
  {2016}{\natexlab{a}})}\BibitemShut {NoStop}%
\bibitem [{\citenamefont {Deng}\ \emph {et~al.}(2016)\citenamefont {Deng},
  \citenamefont {Wan}, \citenamefont {Deng}, \citenamefont {Zhang},
  \citenamefont {Ding}, \citenamefont {Wang}, \citenamefont {Yan},
  \citenamefont {Huang}, \citenamefont {Zhang}, \citenamefont {Xu} \emph
  {et~al.}}]{deng2016experimental}%
  \BibitemOpen
  \bibfield  {author} {\bibinfo {author} {\bibfnamefont {K.}~\bibnamefont
  {Deng}}, \bibinfo {author} {\bibfnamefont {G.}~\bibnamefont {Wan}}, \bibinfo
  {author} {\bibfnamefont {P.}~\bibnamefont {Deng}}, \bibinfo {author}
  {\bibfnamefont {K.}~\bibnamefont {Zhang}}, \bibinfo {author} {\bibfnamefont
  {S.}~\bibnamefont {Ding}}, \bibinfo {author} {\bibfnamefont {E.}~\bibnamefont
  {Wang}}, \bibinfo {author} {\bibfnamefont {M.}~\bibnamefont {Yan}}, \bibinfo
  {author} {\bibfnamefont {H.}~\bibnamefont {Huang}}, \bibinfo {author}
  {\bibfnamefont {H.}~\bibnamefont {Zhang}}, \bibinfo {author} {\bibfnamefont
  {Z.}~\bibnamefont {Xu}},  \emph {et~al.},\ }\href@noop {} {\bibfield
  {journal} {\bibinfo  {journal} {Nature Physics}\ }\textbf {\bibinfo {volume}
  {12}},\ \bibinfo {pages} {1105} (\bibinfo {year} {2016})}\BibitemShut
  {NoStop}%
\bibitem [{\citenamefont {Jiang}\ \emph {et~al.}(2017)\citenamefont {Jiang},
  \citenamefont {Liu}, \citenamefont {Sun}, \citenamefont {Yang}, \citenamefont
  {Rajamathi}, \citenamefont {Qi}, \citenamefont {Yang}, \citenamefont {Chen},
  \citenamefont {Peng}, \citenamefont {Hwang} \emph
  {et~al.}}]{jiang2017signature}%
  \BibitemOpen
  \bibfield  {author} {\bibinfo {author} {\bibfnamefont {J.}~\bibnamefont
  {Jiang}}, \bibinfo {author} {\bibfnamefont {Z.}~\bibnamefont {Liu}}, \bibinfo
  {author} {\bibfnamefont {Y.}~\bibnamefont {Sun}}, \bibinfo {author}
  {\bibfnamefont {H.}~\bibnamefont {Yang}}, \bibinfo {author} {\bibfnamefont
  {C.}~\bibnamefont {Rajamathi}}, \bibinfo {author} {\bibfnamefont
  {Y.}~\bibnamefont {Qi}}, \bibinfo {author} {\bibfnamefont {L.}~\bibnamefont
  {Yang}}, \bibinfo {author} {\bibfnamefont {C.}~\bibnamefont {Chen}}, \bibinfo
  {author} {\bibfnamefont {H.}~\bibnamefont {Peng}}, \bibinfo {author}
  {\bibfnamefont {C.}~\bibnamefont {Hwang}},  \emph {et~al.},\ }\href@noop {}
  {\bibfield  {journal} {\bibinfo  {journal} {Nature communications}\ }\textbf
  {\bibinfo {volume} {8}},\ \bibinfo {pages} {13973} (\bibinfo {year}
  {2017})}\BibitemShut {NoStop}%
\bibitem [{\citenamefont {Tamai}\ \emph {et~al.}(2016)\citenamefont {Tamai},
  \citenamefont {Wu}, \citenamefont {Cucchi}, \citenamefont {Bruno},
  \citenamefont {Ricc{\`o}}, \citenamefont {Kim}, \citenamefont {Hoesch},
  \citenamefont {Barreteau}, \citenamefont {Giannini}, \citenamefont {Besnard}
  \emph {et~al.}}]{tamai2016fermi}%
  \BibitemOpen
  \bibfield  {author} {\bibinfo {author} {\bibfnamefont {A.}~\bibnamefont
  {Tamai}}, \bibinfo {author} {\bibfnamefont {Q.}~\bibnamefont {Wu}}, \bibinfo
  {author} {\bibfnamefont {I.}~\bibnamefont {Cucchi}}, \bibinfo {author}
  {\bibfnamefont {F.~Y.}\ \bibnamefont {Bruno}}, \bibinfo {author}
  {\bibfnamefont {S.}~\bibnamefont {Ricc{\`o}}}, \bibinfo {author}
  {\bibfnamefont {T.}~\bibnamefont {Kim}}, \bibinfo {author} {\bibfnamefont
  {M.}~\bibnamefont {Hoesch}}, \bibinfo {author} {\bibfnamefont
  {C.}~\bibnamefont {Barreteau}}, \bibinfo {author} {\bibfnamefont
  {E.}~\bibnamefont {Giannini}}, \bibinfo {author} {\bibfnamefont
  {C.}~\bibnamefont {Besnard}},  \emph {et~al.},\ }\href@noop {} {\bibfield
  {journal} {\bibinfo  {journal} {Physical Review X}\ }\textbf {\bibinfo
  {volume} {6}},\ \bibinfo {pages} {031021} (\bibinfo {year}
  {2016})}\BibitemShut {NoStop}%
\bibitem [{\citenamefont {Huang}\ \emph
  {et~al.}(2015{\natexlab{b}})\citenamefont {Huang}, \citenamefont {Zhao},
  \citenamefont {Long}, \citenamefont {Wang}, \citenamefont {Chen},
  \citenamefont {Yang}, \citenamefont {Liang}, \citenamefont {Xue},
  \citenamefont {Weng}, \citenamefont {Fang} \emph
  {et~al.}}]{huang2015observation}%
  \BibitemOpen
  \bibfield  {author} {\bibinfo {author} {\bibfnamefont {X.}~\bibnamefont
  {Huang}}, \bibinfo {author} {\bibfnamefont {L.}~\bibnamefont {Zhao}},
  \bibinfo {author} {\bibfnamefont {Y.}~\bibnamefont {Long}}, \bibinfo {author}
  {\bibfnamefont {P.}~\bibnamefont {Wang}}, \bibinfo {author} {\bibfnamefont
  {D.}~\bibnamefont {Chen}}, \bibinfo {author} {\bibfnamefont {Z.}~\bibnamefont
  {Yang}}, \bibinfo {author} {\bibfnamefont {H.}~\bibnamefont {Liang}},
  \bibinfo {author} {\bibfnamefont {M.}~\bibnamefont {Xue}}, \bibinfo {author}
  {\bibfnamefont {H.}~\bibnamefont {Weng}}, \bibinfo {author} {\bibfnamefont
  {Z.}~\bibnamefont {Fang}},  \emph {et~al.},\ }\href@noop {} {\bibfield
  {journal} {\bibinfo  {journal} {Physical Review X}\ }\textbf {\bibinfo
  {volume} {5}},\ \bibinfo {pages} {031023} (\bibinfo {year}
  {2015}{\natexlab{b}})}\BibitemShut {NoStop}%
\bibitem [{\citenamefont {Du}\ \emph {et~al.}(2016{\natexlab{a}})\citenamefont
  {Du}, \citenamefont {Wang}, \citenamefont {Chen}, \citenamefont {Mao},
  \citenamefont {Khan}, \citenamefont {Xu}, \citenamefont {Zhou}, \citenamefont
  {Zhang}, \citenamefont {Yang}, \citenamefont {Chen} \emph
  {et~al.}}]{du2016large}%
  \BibitemOpen
  \bibfield  {author} {\bibinfo {author} {\bibfnamefont {J.}~\bibnamefont
  {Du}}, \bibinfo {author} {\bibfnamefont {H.}~\bibnamefont {Wang}}, \bibinfo
  {author} {\bibfnamefont {Q.}~\bibnamefont {Chen}}, \bibinfo {author}
  {\bibfnamefont {Q.}~\bibnamefont {Mao}}, \bibinfo {author} {\bibfnamefont
  {R.}~\bibnamefont {Khan}}, \bibinfo {author} {\bibfnamefont {B.}~\bibnamefont
  {Xu}}, \bibinfo {author} {\bibfnamefont {Y.}~\bibnamefont {Zhou}}, \bibinfo
  {author} {\bibfnamefont {Y.}~\bibnamefont {Zhang}}, \bibinfo {author}
  {\bibfnamefont {J.}~\bibnamefont {Yang}}, \bibinfo {author} {\bibfnamefont
  {B.}~\bibnamefont {Chen}},  \emph {et~al.},\ }\href@noop {} {\bibfield
  {journal} {\bibinfo  {journal} {SCIENCE CHINA Physics, Mechanics \&
  Astronomy}\ }\textbf {\bibinfo {volume} {59}},\ \bibinfo {pages} {1}
  (\bibinfo {year} {2016}{\natexlab{a}})}\BibitemShut {NoStop}%
\bibitem [{\citenamefont {Yang}\ \emph
  {et~al.}(2015{\natexlab{b}})\citenamefont {Yang}, \citenamefont {Liu},
  \citenamefont {Wang}, \citenamefont {Zheng},\ and\ \citenamefont
  {Xu}}]{yang2015chiral}%
  \BibitemOpen
  \bibfield  {author} {\bibinfo {author} {\bibfnamefont {X.}~\bibnamefont
  {Yang}}, \bibinfo {author} {\bibfnamefont {Y.}~\bibnamefont {Liu}}, \bibinfo
  {author} {\bibfnamefont {Z.}~\bibnamefont {Wang}}, \bibinfo {author}
  {\bibfnamefont {Y.}~\bibnamefont {Zheng}}, \ and\ \bibinfo {author}
  {\bibfnamefont {Z.-a.}\ \bibnamefont {Xu}},\ }\href@noop {} {\bibfield
  {journal} {\bibinfo  {journal} {arXiv preprint arXiv:1506.03190}\ } (\bibinfo
  {year} {2015}{\natexlab{b}})}\BibitemShut {NoStop}%
\bibitem [{\citenamefont {Ali}\ \emph {et~al.}(2014{\natexlab{b}})\citenamefont
  {Ali}, \citenamefont {Xiong}, \citenamefont {Flynn}, \citenamefont {Tao},
  \citenamefont {Gibson}, \citenamefont {Schoop}, \citenamefont {Liang},
  \citenamefont {Haldolaarachchige}, \citenamefont {Hirschberger},
  \citenamefont {Ong} \emph {et~al.}}]{ali2014large}%
  \BibitemOpen
  \bibfield  {author} {\bibinfo {author} {\bibfnamefont {M.}~\bibnamefont
  {Ali}}, \bibinfo {author} {\bibfnamefont {J.}~\bibnamefont {Xiong}}, \bibinfo
  {author} {\bibfnamefont {S.}~\bibnamefont {Flynn}}, \bibinfo {author}
  {\bibfnamefont {J.}~\bibnamefont {Tao}}, \bibinfo {author} {\bibfnamefont
  {Q.}~\bibnamefont {Gibson}}, \bibinfo {author} {\bibfnamefont
  {L.}~\bibnamefont {Schoop}}, \bibinfo {author} {\bibfnamefont
  {T.}~\bibnamefont {Liang}}, \bibinfo {author} {\bibfnamefont
  {N.}~\bibnamefont {Haldolaarachchige}}, \bibinfo {author} {\bibfnamefont
  {M.}~\bibnamefont {Hirschberger}}, \bibinfo {author} {\bibfnamefont
  {N.}~\bibnamefont {Ong}},  \emph {et~al.},\ }\href@noop {} {\bibfield
  {journal} {\bibinfo  {journal} {Nature}\ }\textbf {\bibinfo {volume} {514}},\
  \bibinfo {pages} {205} (\bibinfo {year} {2014}{\natexlab{b}})}\BibitemShut
  {NoStop}%
\bibitem [{\citenamefont {Shekhar}\ \emph {et~al.}(2015)\citenamefont
  {Shekhar}, \citenamefont {Nayak}, \citenamefont {Sun}, \citenamefont
  {Schmidt}, \citenamefont {Nicklas}, \citenamefont {Leermakers}, \citenamefont
  {Zeitler}, \citenamefont {Skourski}, \citenamefont {Wosnitza}, \citenamefont
  {Liu} \emph {et~al.}}]{shekhar2015extremely}%
  \BibitemOpen
  \bibfield  {author} {\bibinfo {author} {\bibfnamefont {C.}~\bibnamefont
  {Shekhar}}, \bibinfo {author} {\bibfnamefont {A.~K.}\ \bibnamefont {Nayak}},
  \bibinfo {author} {\bibfnamefont {Y.}~\bibnamefont {Sun}}, \bibinfo {author}
  {\bibfnamefont {M.}~\bibnamefont {Schmidt}}, \bibinfo {author} {\bibfnamefont
  {M.}~\bibnamefont {Nicklas}}, \bibinfo {author} {\bibfnamefont
  {I.}~\bibnamefont {Leermakers}}, \bibinfo {author} {\bibfnamefont
  {U.}~\bibnamefont {Zeitler}}, \bibinfo {author} {\bibfnamefont
  {Y.}~\bibnamefont {Skourski}}, \bibinfo {author} {\bibfnamefont
  {J.}~\bibnamefont {Wosnitza}}, \bibinfo {author} {\bibfnamefont
  {Z.}~\bibnamefont {Liu}},  \emph {et~al.},\ }\href@noop {} {\bibfield
  {journal} {\bibinfo  {journal} {Nature Physics}\ }\textbf {\bibinfo {volume}
  {11}},\ \bibinfo {pages} {645} (\bibinfo {year} {2015})}\BibitemShut
  {NoStop}%
\bibitem [{\citenamefont {Bzdusek}\ \emph {et~al.}(2016)\citenamefont
  {Bzdusek}, \citenamefont {Wu}, \citenamefont {Ruegg}, \citenamefont
  {Sigrist}, \citenamefont {Soluyanov} \emph {et~al.}}]{bzdusek2016nodal}%
  \BibitemOpen
  \bibfield  {author} {\bibinfo {author} {\bibfnamefont {T.}~\bibnamefont
  {Bzdusek}}, \bibinfo {author} {\bibfnamefont {Q.}~\bibnamefont {Wu}},
  \bibinfo {author} {\bibfnamefont {A.}~\bibnamefont {Ruegg}}, \bibinfo
  {author} {\bibfnamefont {M.}~\bibnamefont {Sigrist}}, \bibinfo {author}
  {\bibfnamefont {A.}~\bibnamefont {Soluyanov}},  \emph {et~al.},\ }\href@noop
  {} {\bibfield  {journal} {\bibinfo  {journal} {Nature}\ }\textbf {\bibinfo
  {volume} {538}},\ \bibinfo {pages} {75} (\bibinfo {year} {2016})}\BibitemShut
  {NoStop}%
\bibitem [{\citenamefont {Weng}\ \emph
  {et~al.}(2015{\natexlab{b}})\citenamefont {Weng}, \citenamefont {Liang},
  \citenamefont {Xu}, \citenamefont {Yu}, \citenamefont {Fang}, \citenamefont
  {Dai},\ and\ \citenamefont {Kawazoe}}]{weng2015topological}%
  \BibitemOpen
  \bibfield  {author} {\bibinfo {author} {\bibfnamefont {H.}~\bibnamefont
  {Weng}}, \bibinfo {author} {\bibfnamefont {Y.}~\bibnamefont {Liang}},
  \bibinfo {author} {\bibfnamefont {Q.}~\bibnamefont {Xu}}, \bibinfo {author}
  {\bibfnamefont {R.}~\bibnamefont {Yu}}, \bibinfo {author} {\bibfnamefont
  {Z.}~\bibnamefont {Fang}}, \bibinfo {author} {\bibfnamefont {X.}~\bibnamefont
  {Dai}}, \ and\ \bibinfo {author} {\bibfnamefont {Y.}~\bibnamefont
  {Kawazoe}},\ }\href@noop {} {\bibfield  {journal} {\bibinfo  {journal}
  {Physical Review B}\ }\textbf {\bibinfo {volume} {92}},\ \bibinfo {pages}
  {045108} (\bibinfo {year} {2015}{\natexlab{b}})}\BibitemShut {NoStop}%
\bibitem [{\citenamefont {Burkov}\ \emph {et~al.}(2011)\citenamefont {Burkov},
  \citenamefont {Hook},\ and\ \citenamefont {Balents}}]{burkov2011topological}%
  \BibitemOpen
  \bibfield  {author} {\bibinfo {author} {\bibfnamefont {A.}~\bibnamefont
  {Burkov}}, \bibinfo {author} {\bibfnamefont {M.}~\bibnamefont {Hook}}, \ and\
  \bibinfo {author} {\bibfnamefont {L.}~\bibnamefont {Balents}},\ }\href@noop
  {} {\bibfield  {journal} {\bibinfo  {journal} {Physical Review B}\ }\textbf
  {\bibinfo {volume} {84}},\ \bibinfo {pages} {235126} (\bibinfo {year}
  {2011})}\BibitemShut {NoStop}%
\bibitem [{\citenamefont {Heikkil{\"a}}\ and\ \citenamefont
  {Volovik}(2011)}]{heikkila2011dimensional}%
  \BibitemOpen
  \bibfield  {author} {\bibinfo {author} {\bibfnamefont {T.~T.}\ \bibnamefont
  {Heikkil{\"a}}}\ and\ \bibinfo {author} {\bibfnamefont {G.~E.}\ \bibnamefont
  {Volovik}},\ }\href@noop {} {\bibfield  {journal} {\bibinfo  {journal} {JETP
  letters}\ }\textbf {\bibinfo {volume} {93}},\ \bibinfo {pages} {59} (\bibinfo
  {year} {2011})}\BibitemShut {NoStop}%
\bibitem [{\citenamefont {Kopnin}\ \emph {et~al.}(2011)\citenamefont {Kopnin},
  \citenamefont {Heikkil{\"a}},\ and\ \citenamefont
  {Volovik}}]{kopnin2011high}%
  \BibitemOpen
  \bibfield  {author} {\bibinfo {author} {\bibfnamefont {N.}~\bibnamefont
  {Kopnin}}, \bibinfo {author} {\bibfnamefont {T.}~\bibnamefont
  {Heikkil{\"a}}}, \ and\ \bibinfo {author} {\bibfnamefont {G.}~\bibnamefont
  {Volovik}},\ }\href@noop {} {\bibfield  {journal} {\bibinfo  {journal}
  {Physical Review B}\ }\textbf {\bibinfo {volume} {83}},\ \bibinfo {pages}
  {220503} (\bibinfo {year} {2011})}\BibitemShut {NoStop}%
\bibitem [{\citenamefont {Heikkil{\"a}}\ \emph {et~al.}(2011)\citenamefont
  {Heikkil{\"a}}, \citenamefont {Kopnin},\ and\ \citenamefont
  {Volovik}}]{heikkila2011flat}%
  \BibitemOpen
  \bibfield  {author} {\bibinfo {author} {\bibfnamefont {T.~T.}\ \bibnamefont
  {Heikkil{\"a}}}, \bibinfo {author} {\bibfnamefont {N.~B.}\ \bibnamefont
  {Kopnin}}, \ and\ \bibinfo {author} {\bibfnamefont {G.~E.}\ \bibnamefont
  {Volovik}},\ }\href@noop {} {\bibfield  {journal} {\bibinfo  {journal} {JETP
  letters}\ }\textbf {\bibinfo {volume} {94}},\ \bibinfo {pages} {233}
  (\bibinfo {year} {2011})}\BibitemShut {NoStop}%
\bibitem [{\citenamefont {Huh}\ \emph {et~al.}(2016)\citenamefont {Huh},
  \citenamefont {Moon},\ and\ \citenamefont {Kim}}]{huh2016long}%
  \BibitemOpen
  \bibfield  {author} {\bibinfo {author} {\bibfnamefont {Y.}~\bibnamefont
  {Huh}}, \bibinfo {author} {\bibfnamefont {E.-G.}\ \bibnamefont {Moon}}, \
  and\ \bibinfo {author} {\bibfnamefont {Y.~B.}\ \bibnamefont {Kim}},\
  }\href@noop {} {\bibfield  {journal} {\bibinfo  {journal} {Physical Review
  B}\ }\textbf {\bibinfo {volume} {93}},\ \bibinfo {pages} {035138} (\bibinfo
  {year} {2016})}\BibitemShut {NoStop}%
\bibitem [{\citenamefont {Rhim}\ and\ \citenamefont
  {Kim}(2015)}]{rhim2015landau}%
  \BibitemOpen
  \bibfield  {author} {\bibinfo {author} {\bibfnamefont {J.-W.}\ \bibnamefont
  {Rhim}}\ and\ \bibinfo {author} {\bibfnamefont {Y.~B.}\ \bibnamefont {Kim}},\
  }\href@noop {} {\bibfield  {journal} {\bibinfo  {journal} {Physical Review
  B}\ }\textbf {\bibinfo {volume} {92}},\ \bibinfo {pages} {045126} (\bibinfo
  {year} {2015})}\BibitemShut {NoStop}%
\bibitem [{\citenamefont {Ramamurthy}\ and\ \citenamefont
  {Hughes}(2015)}]{ramamurthy2015quasi}%
  \BibitemOpen
  \bibfield  {author} {\bibinfo {author} {\bibfnamefont {S.~T.}\ \bibnamefont
  {Ramamurthy}}\ and\ \bibinfo {author} {\bibfnamefont {T.~L.}\ \bibnamefont
  {Hughes}},\ }\href@noop {} {\bibfield  {journal} {\bibinfo  {journal} {arXiv
  preprint arXiv:1508.01205}\ } (\bibinfo {year} {2015})}\BibitemShut {NoStop}%
\bibitem [{\citenamefont {Ebihara}\ \emph {et~al.}(2016)\citenamefont
  {Ebihara}, \citenamefont {Fukushima},\ and\ \citenamefont
  {Oka}}]{ebihara2016chiral}%
  \BibitemOpen
  \bibfield  {author} {\bibinfo {author} {\bibfnamefont {S.}~\bibnamefont
  {Ebihara}}, \bibinfo {author} {\bibfnamefont {K.}~\bibnamefont {Fukushima}},
  \ and\ \bibinfo {author} {\bibfnamefont {T.}~\bibnamefont {Oka}},\
  }\href@noop {} {\bibfield  {journal} {\bibinfo  {journal} {Physical Review
  B}\ }\textbf {\bibinfo {volume} {93}},\ \bibinfo {pages} {155107} (\bibinfo
  {year} {2016})}\BibitemShut {NoStop}%
\bibitem [{\citenamefont {Chan}\ \emph
  {et~al.}(2016{\natexlab{a}})\citenamefont {Chan}, \citenamefont {Lee},
  \citenamefont {Burch}, \citenamefont {Han},\ and\ \citenamefont
  {Ran}}]{chan2016chiral}%
  \BibitemOpen
  \bibfield  {author} {\bibinfo {author} {\bibfnamefont {C.-K.}\ \bibnamefont
  {Chan}}, \bibinfo {author} {\bibfnamefont {P.~A.}\ \bibnamefont {Lee}},
  \bibinfo {author} {\bibfnamefont {K.~S.}\ \bibnamefont {Burch}}, \bibinfo
  {author} {\bibfnamefont {J.~H.}\ \bibnamefont {Han}}, \ and\ \bibinfo
  {author} {\bibfnamefont {Y.}~\bibnamefont {Ran}},\ }\href@noop {} {\bibfield
  {journal} {\bibinfo  {journal} {Physical review letters}\ }\textbf {\bibinfo
  {volume} {116}},\ \bibinfo {pages} {026805} (\bibinfo {year}
  {2016}{\natexlab{a}})}\BibitemShut {NoStop}%
\bibitem [{\citenamefont {Taguchi}\ \emph
  {et~al.}(2016{\natexlab{a}})\citenamefont {Taguchi}, \citenamefont {Imaeda},
  \citenamefont {Sato},\ and\ \citenamefont
  {Tanaka}}]{taguchi2016photovoltaic}%
  \BibitemOpen
  \bibfield  {author} {\bibinfo {author} {\bibfnamefont {K.}~\bibnamefont
  {Taguchi}}, \bibinfo {author} {\bibfnamefont {T.}~\bibnamefont {Imaeda}},
  \bibinfo {author} {\bibfnamefont {M.}~\bibnamefont {Sato}}, \ and\ \bibinfo
  {author} {\bibfnamefont {Y.}~\bibnamefont {Tanaka}},\ }\href@noop {}
  {\bibfield  {journal} {\bibinfo  {journal} {Physical Review B}\ }\textbf
  {\bibinfo {volume} {93}},\ \bibinfo {pages} {201202} (\bibinfo {year}
  {2016}{\natexlab{a}})}\BibitemShut {NoStop}%
\bibitem [{\citenamefont {Takahashi}\ \emph {et~al.}(2017)\citenamefont
  {Takahashi}, \citenamefont {Hirayama},\ and\ \citenamefont
  {Murakami}}]{takahashi2017topological}%
  \BibitemOpen
  \bibfield  {author} {\bibinfo {author} {\bibfnamefont {R.}~\bibnamefont
  {Takahashi}}, \bibinfo {author} {\bibfnamefont {M.}~\bibnamefont {Hirayama}},
  \ and\ \bibinfo {author} {\bibfnamefont {S.}~\bibnamefont {Murakami}},\
  }\href@noop {} {\bibfield  {journal} {\bibinfo  {journal} {arXiv preprint
  arXiv:1704.02151}\ } (\bibinfo {year} {2017})}\BibitemShut {NoStop}%
\bibitem [{\citenamefont {Fang}\ \emph {et~al.}(2015)\citenamefont {Fang},
  \citenamefont {Chen}, \citenamefont {Kee},\ and\ \citenamefont
  {Fu}}]{fang2015topological}%
  \BibitemOpen
  \bibfield  {author} {\bibinfo {author} {\bibfnamefont {C.}~\bibnamefont
  {Fang}}, \bibinfo {author} {\bibfnamefont {Y.}~\bibnamefont {Chen}}, \bibinfo
  {author} {\bibfnamefont {H.-Y.}\ \bibnamefont {Kee}}, \ and\ \bibinfo
  {author} {\bibfnamefont {L.}~\bibnamefont {Fu}},\ }\href@noop {} {\bibfield
  {journal} {\bibinfo  {journal} {Physical Review B}\ }\textbf {\bibinfo
  {volume} {92}},\ \bibinfo {pages} {081201} (\bibinfo {year}
  {2015})}\BibitemShut {NoStop}%
\bibitem [{\citenamefont {Fang}\ \emph {et~al.}(2016)\citenamefont {Fang},
  \citenamefont {Weng}, \citenamefont {Dai},\ and\ \citenamefont
  {Fang}}]{fang2016topological}%
  \BibitemOpen
  \bibfield  {author} {\bibinfo {author} {\bibfnamefont {C.}~\bibnamefont
  {Fang}}, \bibinfo {author} {\bibfnamefont {H.}~\bibnamefont {Weng}}, \bibinfo
  {author} {\bibfnamefont {X.}~\bibnamefont {Dai}}, \ and\ \bibinfo {author}
  {\bibfnamefont {Z.}~\bibnamefont {Fang}},\ }\href@noop {} {\bibfield
  {journal} {\bibinfo  {journal} {Chinese Physics B}\ }\textbf {\bibinfo
  {volume} {25}},\ \bibinfo {pages} {117106} (\bibinfo {year}
  {2016})}\BibitemShut {NoStop}%
\bibitem [{\citenamefont {Yan}\ and\ \citenamefont
  {Felser}(2017)}]{yan2017topological}%
  \BibitemOpen
  \bibfield  {author} {\bibinfo {author} {\bibfnamefont {B.}~\bibnamefont
  {Yan}}\ and\ \bibinfo {author} {\bibfnamefont {C.}~\bibnamefont {Felser}},\
  }\href@noop {} {\bibfield  {journal} {\bibinfo  {journal} {Annual Review of
  Condensed Matter Physics}\ }\textbf {\bibinfo {volume} {8}},\ \bibinfo
  {pages} {337} (\bibinfo {year} {2017})}\BibitemShut {NoStop}%
\bibitem [{\citenamefont {Jia}\ \emph {et~al.}(2016)\citenamefont {Jia},
  \citenamefont {Xu},\ and\ \citenamefont {Hasan}}]{jia2016weyl}%
  \BibitemOpen
  \bibfield  {author} {\bibinfo {author} {\bibfnamefont {S.}~\bibnamefont
  {Jia}}, \bibinfo {author} {\bibfnamefont {S.-Y.}\ \bibnamefont {Xu}}, \ and\
  \bibinfo {author} {\bibfnamefont {M.~Z.}\ \bibnamefont {Hasan}},\ }\href@noop
  {} {\bibfield  {journal} {\bibinfo  {journal} {Nature materials}\ }\textbf
  {\bibinfo {volume} {15}},\ \bibinfo {pages} {1140} (\bibinfo {year}
  {2016})}\BibitemShut {NoStop}%
\bibitem [{\citenamefont {Wang}\ \emph {et~al.}(2017)\citenamefont {Wang},
  \citenamefont {Lin}, \citenamefont {Wang}, \citenamefont {Yu},\ and\
  \citenamefont {Liao}}]{wang2017quantum}%
  \BibitemOpen
  \bibfield  {author} {\bibinfo {author} {\bibfnamefont {S.}~\bibnamefont
  {Wang}}, \bibinfo {author} {\bibfnamefont {B.-C.}\ \bibnamefont {Lin}},
  \bibinfo {author} {\bibfnamefont {A.-Q.}\ \bibnamefont {Wang}}, \bibinfo
  {author} {\bibfnamefont {D.-P.}\ \bibnamefont {Yu}}, \ and\ \bibinfo {author}
  {\bibfnamefont {Z.-M.}\ \bibnamefont {Liao}},\ }\href@noop {} {\bibfield
  {journal} {\bibinfo  {journal} {Advances in Physics: X}\ }\textbf {\bibinfo
  {volume} {2}},\ \bibinfo {pages} {518} (\bibinfo {year} {2017})}\BibitemShut
  {NoStop}%
\bibitem [{\citenamefont {Yu}\ \emph {et~al.}(2017{\natexlab{a}})\citenamefont
  {Yu}, \citenamefont {Fang}, \citenamefont {Dai},\ and\ \citenamefont
  {Weng}}]{yu2017topological}%
  \BibitemOpen
  \bibfield  {author} {\bibinfo {author} {\bibfnamefont {R.}~\bibnamefont
  {Yu}}, \bibinfo {author} {\bibfnamefont {Z.}~\bibnamefont {Fang}}, \bibinfo
  {author} {\bibfnamefont {X.}~\bibnamefont {Dai}}, \ and\ \bibinfo {author}
  {\bibfnamefont {H.}~\bibnamefont {Weng}},\ }\href@noop {} {\bibfield
  {journal} {\bibinfo  {journal} {Frontiers of Physics}\ }\textbf {\bibinfo
  {volume} {12}},\ \bibinfo {pages} {127202} (\bibinfo {year}
  {2017}{\natexlab{a}})}\BibitemShut {NoStop}%
\bibitem [{\citenamefont {Weng}\ \emph {et~al.}(2016)\citenamefont {Weng},
  \citenamefont {Dai},\ and\ \citenamefont {Fang}}]{weng2016topological}%
  \BibitemOpen
  \bibfield  {author} {\bibinfo {author} {\bibfnamefont {H.}~\bibnamefont
  {Weng}}, \bibinfo {author} {\bibfnamefont {X.}~\bibnamefont {Dai}}, \ and\
  \bibinfo {author} {\bibfnamefont {Z.}~\bibnamefont {Fang}},\ }\href@noop {}
  {\bibfield  {journal} {\bibinfo  {journal} {J. Phys.: Condens. Matter}\
  }\textbf {\bibinfo {volume} {28}},\ \bibinfo {pages} {303001} (\bibinfo
  {year} {2016})}\BibitemShut {NoStop}%
\bibitem [{\citenamefont {Bian}\ \emph
  {et~al.}(2016{\natexlab{a}})\citenamefont {Bian}, \citenamefont {Chang},
  \citenamefont {Sankar}, \citenamefont {Xu}, \citenamefont {Zheng},
  \citenamefont {Neupert}, \citenamefont {Chiu}, \citenamefont {Huang},
  \citenamefont {Chang}, \citenamefont {Belopolski} \emph
  {et~al.}}]{bian2016topological}%
  \BibitemOpen
  \bibfield  {author} {\bibinfo {author} {\bibfnamefont {G.}~\bibnamefont
  {Bian}}, \bibinfo {author} {\bibfnamefont {T.-R.}\ \bibnamefont {Chang}},
  \bibinfo {author} {\bibfnamefont {R.}~\bibnamefont {Sankar}}, \bibinfo
  {author} {\bibfnamefont {S.-Y.}\ \bibnamefont {Xu}}, \bibinfo {author}
  {\bibfnamefont {H.}~\bibnamefont {Zheng}}, \bibinfo {author} {\bibfnamefont
  {T.}~\bibnamefont {Neupert}}, \bibinfo {author} {\bibfnamefont {C.-K.}\
  \bibnamefont {Chiu}}, \bibinfo {author} {\bibfnamefont {S.-M.}\ \bibnamefont
  {Huang}}, \bibinfo {author} {\bibfnamefont {G.}~\bibnamefont {Chang}},
  \bibinfo {author} {\bibfnamefont {I.}~\bibnamefont {Belopolski}},  \emph
  {et~al.},\ }\href@noop {} {\bibfield  {journal} {\bibinfo  {journal} {Nature
  communications}\ }\textbf {\bibinfo {volume} {7}},\ \bibinfo {pages} {10556}
  (\bibinfo {year} {2016}{\natexlab{a}})}\BibitemShut {NoStop}%
\bibitem [{\citenamefont {Schoop}\ \emph {et~al.}(2016)\citenamefont {Schoop},
  \citenamefont {Ali}, \citenamefont {Stra{\ss}er}, \citenamefont {Topp},
  \citenamefont {Varykhalov}, \citenamefont {Marchenko}, \citenamefont
  {Duppel}, \citenamefont {Parkin}, \citenamefont {Lotsch},\ and\ \citenamefont
  {Ast}}]{schoop2016dirac}%
  \BibitemOpen
  \bibfield  {author} {\bibinfo {author} {\bibfnamefont {L.~M.}\ \bibnamefont
  {Schoop}}, \bibinfo {author} {\bibfnamefont {M.~N.}\ \bibnamefont {Ali}},
  \bibinfo {author} {\bibfnamefont {C.}~\bibnamefont {Stra{\ss}er}}, \bibinfo
  {author} {\bibfnamefont {A.}~\bibnamefont {Topp}}, \bibinfo {author}
  {\bibfnamefont {A.}~\bibnamefont {Varykhalov}}, \bibinfo {author}
  {\bibfnamefont {D.}~\bibnamefont {Marchenko}}, \bibinfo {author}
  {\bibfnamefont {V.}~\bibnamefont {Duppel}}, \bibinfo {author} {\bibfnamefont
  {S.~S.}\ \bibnamefont {Parkin}}, \bibinfo {author} {\bibfnamefont {B.~V.}\
  \bibnamefont {Lotsch}}, \ and\ \bibinfo {author} {\bibfnamefont {C.~R.}\
  \bibnamefont {Ast}},\ }\href@noop {} {\bibfield  {journal} {\bibinfo
  {journal} {Nature communications}\ }\textbf {\bibinfo {volume} {7}},\
  \bibinfo {pages} {11696} (\bibinfo {year} {2016})}\BibitemShut {NoStop}%
\bibitem [{\citenamefont {Sun}\ \emph {et~al.}(2017)\citenamefont {Sun},
  \citenamefont {Zhang}, \citenamefont {Liu}, \citenamefont {Felser},\ and\
  \citenamefont {Yan}}]{sun2017dirac}%
  \BibitemOpen
  \bibfield  {author} {\bibinfo {author} {\bibfnamefont {Y.}~\bibnamefont
  {Sun}}, \bibinfo {author} {\bibfnamefont {Y.}~\bibnamefont {Zhang}}, \bibinfo
  {author} {\bibfnamefont {C.-X.}\ \bibnamefont {Liu}}, \bibinfo {author}
  {\bibfnamefont {C.}~\bibnamefont {Felser}}, \ and\ \bibinfo {author}
  {\bibfnamefont {B.}~\bibnamefont {Yan}},\ }\href@noop {} {\bibfield
  {journal} {\bibinfo  {journal} {arXiv preprint arXiv:1701.09089}\ } (\bibinfo
  {year} {2017})}\BibitemShut {NoStop}%
\bibitem [{\citenamefont {Liu}\ \emph {et~al.}(2016{\natexlab{b}})\citenamefont
  {Liu}, \citenamefont {Li}, \citenamefont {Li}, \citenamefont {Liu},
  \citenamefont {Li}, \citenamefont {Yang}, \citenamefont {Yao}, \citenamefont
  {Fan}, \citenamefont {Wan}, \citenamefont {Wang} \emph
  {et~al.}}]{liu2016direct}%
  \BibitemOpen
  \bibfield  {author} {\bibinfo {author} {\bibfnamefont {Z.}~\bibnamefont
  {Liu}}, \bibinfo {author} {\bibfnamefont {M.}~\bibnamefont {Li}}, \bibinfo
  {author} {\bibfnamefont {Q.}~\bibnamefont {Li}}, \bibinfo {author}
  {\bibfnamefont {J.}~\bibnamefont {Liu}}, \bibinfo {author} {\bibfnamefont
  {W.}~\bibnamefont {Li}}, \bibinfo {author} {\bibfnamefont {H.}~\bibnamefont
  {Yang}}, \bibinfo {author} {\bibfnamefont {Q.}~\bibnamefont {Yao}}, \bibinfo
  {author} {\bibfnamefont {C.}~\bibnamefont {Fan}}, \bibinfo {author}
  {\bibfnamefont {X.}~\bibnamefont {Wan}}, \bibinfo {author} {\bibfnamefont
  {Z.}~\bibnamefont {Wang}},  \emph {et~al.},\ }\href@noop {} {\bibfield
  {journal} {\bibinfo  {journal} {Scientific reports}\ }\textbf {\bibinfo
  {volume} {6}},\ \bibinfo {pages} {30309} (\bibinfo {year}
  {2016}{\natexlab{b}})}\BibitemShut {NoStop}%
\bibitem [{\citenamefont {Neupane}\ \emph {et~al.}(2016)\citenamefont
  {Neupane}, \citenamefont {Belopolski}, \citenamefont {Hosen}, \citenamefont
  {Sanchez}, \citenamefont {Sankar}, \citenamefont {Szlawska}, \citenamefont
  {Xu}, \citenamefont {Dimitri}, \citenamefont {Dhakal}, \citenamefont
  {Maldonado} \emph {et~al.}}]{neupane2016observation}%
  \BibitemOpen
  \bibfield  {author} {\bibinfo {author} {\bibfnamefont {M.}~\bibnamefont
  {Neupane}}, \bibinfo {author} {\bibfnamefont {I.}~\bibnamefont {Belopolski}},
  \bibinfo {author} {\bibfnamefont {M.~M.}\ \bibnamefont {Hosen}}, \bibinfo
  {author} {\bibfnamefont {D.~S.}\ \bibnamefont {Sanchez}}, \bibinfo {author}
  {\bibfnamefont {R.}~\bibnamefont {Sankar}}, \bibinfo {author} {\bibfnamefont
  {M.}~\bibnamefont {Szlawska}}, \bibinfo {author} {\bibfnamefont {S.-Y.}\
  \bibnamefont {Xu}}, \bibinfo {author} {\bibfnamefont {K.}~\bibnamefont
  {Dimitri}}, \bibinfo {author} {\bibfnamefont {N.}~\bibnamefont {Dhakal}},
  \bibinfo {author} {\bibfnamefont {P.}~\bibnamefont {Maldonado}},  \emph
  {et~al.},\ }\href@noop {} {\bibfield  {journal} {\bibinfo  {journal}
  {Physical Review B}\ }\textbf {\bibinfo {volume} {93}},\ \bibinfo {pages}
  {201104} (\bibinfo {year} {2016})}\BibitemShut {NoStop}%
\bibitem [{\citenamefont {Chen}\ \emph {et~al.}(2017)\citenamefont {Chen},
  \citenamefont {Xu}, \citenamefont {Jiang}, \citenamefont {Wu}, \citenamefont
  {Qi}, \citenamefont {Yang}, \citenamefont {Wang}, \citenamefont {Sun},
  \citenamefont {Schr{\"o}ter}, \citenamefont {Yang} \emph
  {et~al.}}]{chen2017dirac}%
  \BibitemOpen
  \bibfield  {author} {\bibinfo {author} {\bibfnamefont {C.}~\bibnamefont
  {Chen}}, \bibinfo {author} {\bibfnamefont {X.}~\bibnamefont {Xu}}, \bibinfo
  {author} {\bibfnamefont {J.}~\bibnamefont {Jiang}}, \bibinfo {author}
  {\bibfnamefont {S.-C.}\ \bibnamefont {Wu}}, \bibinfo {author} {\bibfnamefont
  {Y.}~\bibnamefont {Qi}}, \bibinfo {author} {\bibfnamefont {L.}~\bibnamefont
  {Yang}}, \bibinfo {author} {\bibfnamefont {M.}~\bibnamefont {Wang}}, \bibinfo
  {author} {\bibfnamefont {Y.}~\bibnamefont {Sun}}, \bibinfo {author}
  {\bibfnamefont {N.}~\bibnamefont {Schr{\"o}ter}}, \bibinfo {author}
  {\bibfnamefont {H.}~\bibnamefont {Yang}},  \emph {et~al.},\ }\href@noop {}
  {\bibfield  {journal} {\bibinfo  {journal} {Physical Review B}\ }\textbf
  {\bibinfo {volume} {95}},\ \bibinfo {pages} {125126} (\bibinfo {year}
  {2017})}\BibitemShut {NoStop}%
\bibitem [{\citenamefont {Ali}\ \emph {et~al.}(2014{\natexlab{c}})\citenamefont
  {Ali}, \citenamefont {Gibson}, \citenamefont {Klimczuk},\ and\ \citenamefont
  {Cava}}]{ali2014noncentrosymmetric}%
  \BibitemOpen
  \bibfield  {author} {\bibinfo {author} {\bibfnamefont {M.~N.}\ \bibnamefont
  {Ali}}, \bibinfo {author} {\bibfnamefont {Q.~D.}\ \bibnamefont {Gibson}},
  \bibinfo {author} {\bibfnamefont {T.}~\bibnamefont {Klimczuk}}, \ and\
  \bibinfo {author} {\bibfnamefont {R.}~\bibnamefont {Cava}},\ }\href@noop {}
  {\bibfield  {journal} {\bibinfo  {journal} {Physical Review B}\ }\textbf
  {\bibinfo {volume} {89}},\ \bibinfo {pages} {020505} (\bibinfo {year}
  {2014}{\natexlab{c}})}\BibitemShut {NoStop}%
\bibitem [{\citenamefont {Guan}\ \emph {et~al.}(2016)\citenamefont {Guan},
  \citenamefont {Chen}, \citenamefont {Chu}, \citenamefont {Sankar},
  \citenamefont {Chou}, \citenamefont {Jeng}, \citenamefont {Chang},\ and\
  \citenamefont {Chuang}}]{guan2016superconducting}%
  \BibitemOpen
  \bibfield  {author} {\bibinfo {author} {\bibfnamefont {S.-Y.}\ \bibnamefont
  {Guan}}, \bibinfo {author} {\bibfnamefont {P.-J.}\ \bibnamefont {Chen}},
  \bibinfo {author} {\bibfnamefont {M.-W.}\ \bibnamefont {Chu}}, \bibinfo
  {author} {\bibfnamefont {R.}~\bibnamefont {Sankar}}, \bibinfo {author}
  {\bibfnamefont {F.}~\bibnamefont {Chou}}, \bibinfo {author} {\bibfnamefont
  {H.-T.}\ \bibnamefont {Jeng}}, \bibinfo {author} {\bibfnamefont {C.-S.}\
  \bibnamefont {Chang}}, \ and\ \bibinfo {author} {\bibfnamefont {T.-M.}\
  \bibnamefont {Chuang}},\ }\href@noop {} {\bibfield  {journal} {\bibinfo
  {journal} {Science Advances}\ }\textbf {\bibinfo {volume} {2}},\ \bibinfo
  {pages} {e1600894} (\bibinfo {year} {2016})}\BibitemShut {NoStop}%
\bibitem [{\citenamefont {Chang}\ \emph {et~al.}(2016)\citenamefont {Chang},
  \citenamefont {Chen}, \citenamefont {Bian}, \citenamefont {Huang},
  \citenamefont {Zheng}, \citenamefont {Neupert}, \citenamefont {Sankar},
  \citenamefont {Xu}, \citenamefont {Belopolski}, \citenamefont {Chang} \emph
  {et~al.}}]{chang2016topological}%
  \BibitemOpen
  \bibfield  {author} {\bibinfo {author} {\bibfnamefont {T.-R.}\ \bibnamefont
  {Chang}}, \bibinfo {author} {\bibfnamefont {P.-J.}\ \bibnamefont {Chen}},
  \bibinfo {author} {\bibfnamefont {G.}~\bibnamefont {Bian}}, \bibinfo {author}
  {\bibfnamefont {S.-M.}\ \bibnamefont {Huang}}, \bibinfo {author}
  {\bibfnamefont {H.}~\bibnamefont {Zheng}}, \bibinfo {author} {\bibfnamefont
  {T.}~\bibnamefont {Neupert}}, \bibinfo {author} {\bibfnamefont
  {R.}~\bibnamefont {Sankar}}, \bibinfo {author} {\bibfnamefont {S.-Y.}\
  \bibnamefont {Xu}}, \bibinfo {author} {\bibfnamefont {I.}~\bibnamefont
  {Belopolski}}, \bibinfo {author} {\bibfnamefont {G.}~\bibnamefont {Chang}},
  \emph {et~al.},\ }\href@noop {} {\bibfield  {journal} {\bibinfo  {journal}
  {Physical Review B}\ }\textbf {\bibinfo {volume} {93}},\ \bibinfo {pages}
  {245130} (\bibinfo {year} {2016})}\BibitemShut {NoStop}%
\bibitem [{\citenamefont {Bian}\ \emph
  {et~al.}(2016{\natexlab{b}})\citenamefont {Bian}, \citenamefont {Chang},
  \citenamefont {Zheng}, \citenamefont {Velury}, \citenamefont {Xu},
  \citenamefont {Neupert}, \citenamefont {Chiu}, \citenamefont {Huang},
  \citenamefont {Sanchez}, \citenamefont {Belopolski} \emph
  {et~al.}}]{bian2016drumhead}%
  \BibitemOpen
  \bibfield  {author} {\bibinfo {author} {\bibfnamefont {G.}~\bibnamefont
  {Bian}}, \bibinfo {author} {\bibfnamefont {T.-R.}\ \bibnamefont {Chang}},
  \bibinfo {author} {\bibfnamefont {H.}~\bibnamefont {Zheng}}, \bibinfo
  {author} {\bibfnamefont {S.}~\bibnamefont {Velury}}, \bibinfo {author}
  {\bibfnamefont {S.-Y.}\ \bibnamefont {Xu}}, \bibinfo {author} {\bibfnamefont
  {T.}~\bibnamefont {Neupert}}, \bibinfo {author} {\bibfnamefont {C.-K.}\
  \bibnamefont {Chiu}}, \bibinfo {author} {\bibfnamefont {S.-M.}\ \bibnamefont
  {Huang}}, \bibinfo {author} {\bibfnamefont {D.~S.}\ \bibnamefont {Sanchez}},
  \bibinfo {author} {\bibfnamefont {I.}~\bibnamefont {Belopolski}},  \emph
  {et~al.},\ }\href@noop {} {\bibfield  {journal} {\bibinfo  {journal}
  {Physical Review B}\ }\textbf {\bibinfo {volume} {93}},\ \bibinfo {pages}
  {121113} (\bibinfo {year} {2016}{\natexlab{b}})}\BibitemShut {NoStop}%
\bibitem [{\citenamefont {Kim}\ \emph {et~al.}(2015)\citenamefont {Kim},
  \citenamefont {Wieder}, \citenamefont {Kane},\ and\ \citenamefont
  {Rappe}}]{kim2015dirac}%
  \BibitemOpen
  \bibfield  {author} {\bibinfo {author} {\bibfnamefont {Y.}~\bibnamefont
  {Kim}}, \bibinfo {author} {\bibfnamefont {B.~J.}\ \bibnamefont {Wieder}},
  \bibinfo {author} {\bibfnamefont {C.}~\bibnamefont {Kane}}, \ and\ \bibinfo
  {author} {\bibfnamefont {A.~M.}\ \bibnamefont {Rappe}},\ }\href@noop {}
  {\bibfield  {journal} {\bibinfo  {journal} {Physical review letters}\
  }\textbf {\bibinfo {volume} {115}},\ \bibinfo {pages} {036806} (\bibinfo
  {year} {2015})}\BibitemShut {NoStop}%
\bibitem [{\citenamefont {Yu}\ \emph {et~al.}(2015)\citenamefont {Yu},
  \citenamefont {Weng}, \citenamefont {Fang}, \citenamefont {Dai},\ and\
  \citenamefont {Hu}}]{yu2015topological}%
  \BibitemOpen
  \bibfield  {author} {\bibinfo {author} {\bibfnamefont {R.}~\bibnamefont
  {Yu}}, \bibinfo {author} {\bibfnamefont {H.}~\bibnamefont {Weng}}, \bibinfo
  {author} {\bibfnamefont {Z.}~\bibnamefont {Fang}}, \bibinfo {author}
  {\bibfnamefont {X.}~\bibnamefont {Dai}}, \ and\ \bibinfo {author}
  {\bibfnamefont {X.}~\bibnamefont {Hu}},\ }\href@noop {} {\bibfield  {journal}
  {\bibinfo  {journal} {Physical review letters}\ }\textbf {\bibinfo {volume}
  {115}},\ \bibinfo {pages} {036807} (\bibinfo {year} {2015})}\BibitemShut
  {NoStop}%
\bibitem [{\citenamefont {Yu}\ \emph {et~al.}(2017{\natexlab{b}})\citenamefont
  {Yu}, \citenamefont {Wu}, \citenamefont {Fang},\ and\ \citenamefont
  {Weng}}]{yu2017nodal}%
  \BibitemOpen
  \bibfield  {author} {\bibinfo {author} {\bibfnamefont {R.}~\bibnamefont
  {Yu}}, \bibinfo {author} {\bibfnamefont {Q.}~\bibnamefont {Wu}}, \bibinfo
  {author} {\bibfnamefont {Z.}~\bibnamefont {Fang}}, \ and\ \bibinfo {author}
  {\bibfnamefont {H.}~\bibnamefont {Weng}},\ }\href@noop {} {\bibfield
  {journal} {\bibinfo  {journal} {arXiv preprint arXiv:1701.08502}\ } (\bibinfo
  {year} {2017}{\natexlab{b}})}\BibitemShut {NoStop}%
\bibitem [{\citenamefont {Gibson}\ \emph {et~al.}(2015)\citenamefont {Gibson},
  \citenamefont {Schoop}, \citenamefont {Muechler}, \citenamefont {Xie},
  \citenamefont {Hirschberger}, \citenamefont {Ong}, \citenamefont {Car},\ and\
  \citenamefont {Cava}}]{gibson2015three}%
  \BibitemOpen
  \bibfield  {author} {\bibinfo {author} {\bibfnamefont {Q.}~\bibnamefont
  {Gibson}}, \bibinfo {author} {\bibfnamefont {L.}~\bibnamefont {Schoop}},
  \bibinfo {author} {\bibfnamefont {L.}~\bibnamefont {Muechler}}, \bibinfo
  {author} {\bibfnamefont {L.}~\bibnamefont {Xie}}, \bibinfo {author}
  {\bibfnamefont {M.}~\bibnamefont {Hirschberger}}, \bibinfo {author}
  {\bibfnamefont {N.}~\bibnamefont {Ong}}, \bibinfo {author} {\bibfnamefont
  {R.}~\bibnamefont {Car}}, \ and\ \bibinfo {author} {\bibfnamefont
  {R.}~\bibnamefont {Cava}},\ }\href@noop {} {\bibfield  {journal} {\bibinfo
  {journal} {Physical Review B}\ }\textbf {\bibinfo {volume} {91}},\ \bibinfo
  {pages} {205128} (\bibinfo {year} {2015})}\BibitemShut {NoStop}%
\bibitem [{\citenamefont {Young}\ and\ \citenamefont
  {Kane}(2015)}]{young2015dirac}%
  \BibitemOpen
  \bibfield  {author} {\bibinfo {author} {\bibfnamefont {S.~M.}\ \bibnamefont
  {Young}}\ and\ \bibinfo {author} {\bibfnamefont {C.~L.}\ \bibnamefont
  {Kane}},\ }\href@noop {} {\bibfield  {journal} {\bibinfo  {journal} {Physical
  review letters}\ }\textbf {\bibinfo {volume} {115}},\ \bibinfo {pages}
  {126803} (\bibinfo {year} {2015})}\BibitemShut {NoStop}%
\bibitem [{\citenamefont {Murakami}(2017)}]{lecturenotes}%
  \BibitemOpen
  \bibfield  {author} {\bibinfo {author} {\bibfnamefont {S.}~\bibnamefont
  {Murakami}},\ }\href@noop {} {\ \textbf {\bibinfo {volume} {139}} (\bibinfo
  {year} {2017})}\BibitemShut {NoStop}%
\bibitem [{\citenamefont {Chan}\ \emph
  {et~al.}(2016{\natexlab{b}})\citenamefont {Chan}, \citenamefont {Chiu},
  \citenamefont {Chou},\ and\ \citenamefont {Schnyder}}]{chan20163}%
  \BibitemOpen
  \bibfield  {author} {\bibinfo {author} {\bibfnamefont {Y.-H.}\ \bibnamefont
  {Chan}}, \bibinfo {author} {\bibfnamefont {C.-K.}\ \bibnamefont {Chiu}},
  \bibinfo {author} {\bibfnamefont {M.}~\bibnamefont {Chou}}, \ and\ \bibinfo
  {author} {\bibfnamefont {A.~P.}\ \bibnamefont {Schnyder}},\ }\href@noop {}
  {\bibfield  {journal} {\bibinfo  {journal} {Physical Review B}\ }\textbf
  {\bibinfo {volume} {93}},\ \bibinfo {pages} {205132} (\bibinfo {year}
  {2016}{\natexlab{b}})}\BibitemShut {NoStop}%
\bibitem [{\citenamefont {Hoffmann}(1987)}]{hoffmann1987chemistry}%
  \BibitemOpen
  \bibfield  {author} {\bibinfo {author} {\bibfnamefont {R.}~\bibnamefont
  {Hoffmann}},\ }\href@noop {} {\bibfield  {journal} {\bibinfo  {journal}
  {Angewandte Chemie International Edition in English}\ }\textbf {\bibinfo
  {volume} {26}},\ \bibinfo {pages} {846} (\bibinfo {year} {1987})}\BibitemShut
  {NoStop}%
\bibitem [{\citenamefont {Tremel}\ and\ \citenamefont
  {Hoffmann}(1987)}]{tremel1987square}%
  \BibitemOpen
  \bibfield  {author} {\bibinfo {author} {\bibfnamefont {W.}~\bibnamefont
  {Tremel}}\ and\ \bibinfo {author} {\bibfnamefont {R.}~\bibnamefont
  {Hoffmann}},\ }\href@noop {} {\bibfield  {journal} {\bibinfo  {journal} {J.
  Am. Chem. Soc}\ }\textbf {\bibinfo {volume} {109}},\ \bibinfo {pages} {124}
  (\bibinfo {year} {1987})}\BibitemShut {NoStop}%
\bibitem [{\citenamefont {Du}\ \emph {et~al.}(2016{\natexlab{b}})\citenamefont
  {Du}, \citenamefont {Tang}, \citenamefont {Wang}, \citenamefont {Sheng},
  \citenamefont {Kan}, \citenamefont {Duan}, \citenamefont {Savrasov},\ and\
  \citenamefont {Wan}}]{du2016cate}%
  \BibitemOpen
  \bibfield  {author} {\bibinfo {author} {\bibfnamefont {Y.}~\bibnamefont
  {Du}}, \bibinfo {author} {\bibfnamefont {F.}~\bibnamefont {Tang}}, \bibinfo
  {author} {\bibfnamefont {D.}~\bibnamefont {Wang}}, \bibinfo {author}
  {\bibfnamefont {L.}~\bibnamefont {Sheng}}, \bibinfo {author} {\bibfnamefont
  {E.-j.}\ \bibnamefont {Kan}}, \bibinfo {author} {\bibfnamefont {C.-G.}\
  \bibnamefont {Duan}}, \bibinfo {author} {\bibfnamefont {S.~Y.}\ \bibnamefont
  {Savrasov}}, \ and\ \bibinfo {author} {\bibfnamefont {X.}~\bibnamefont
  {Wan}},\ }\href@noop {} {\bibfield  {journal} {\bibinfo  {journal} {arXiv
  preprint arXiv:1605.07998}\ } (\bibinfo {year}
  {2016}{\natexlab{b}})}\BibitemShut {NoStop}%
\bibitem [{\citenamefont {Zeng}\ \emph {et~al.}(2015)\citenamefont {Zeng},
  \citenamefont {Fang}, \citenamefont {Chang}, \citenamefont {Chen},
  \citenamefont {Hsieh}, \citenamefont {Bansil}, \citenamefont {Lin},\ and\
  \citenamefont {Fu}}]{zeng2015topological}%
  \BibitemOpen
  \bibfield  {author} {\bibinfo {author} {\bibfnamefont {M.}~\bibnamefont
  {Zeng}}, \bibinfo {author} {\bibfnamefont {C.}~\bibnamefont {Fang}}, \bibinfo
  {author} {\bibfnamefont {G.}~\bibnamefont {Chang}}, \bibinfo {author}
  {\bibfnamefont {Y.-A.}\ \bibnamefont {Chen}}, \bibinfo {author}
  {\bibfnamefont {T.}~\bibnamefont {Hsieh}}, \bibinfo {author} {\bibfnamefont
  {A.}~\bibnamefont {Bansil}}, \bibinfo {author} {\bibfnamefont
  {H.}~\bibnamefont {Lin}}, \ and\ \bibinfo {author} {\bibfnamefont
  {L.}~\bibnamefont {Fu}},\ }\href@noop {} {\bibfield  {journal} {\bibinfo
  {journal} {arXiv preprint arXiv:1504.03492}\ } (\bibinfo {year}
  {2015})}\BibitemShut {NoStop}%
\bibitem [{\citenamefont {Xie}\ \emph {et~al.}(2015)\citenamefont {Xie},
  \citenamefont {Schoop}, \citenamefont {Seibel}, \citenamefont {Gibson},
  \citenamefont {Xie},\ and\ \citenamefont {Cava}}]{xie2015new}%
  \BibitemOpen
  \bibfield  {author} {\bibinfo {author} {\bibfnamefont {L.~S.}\ \bibnamefont
  {Xie}}, \bibinfo {author} {\bibfnamefont {L.~M.}\ \bibnamefont {Schoop}},
  \bibinfo {author} {\bibfnamefont {E.~M.}\ \bibnamefont {Seibel}}, \bibinfo
  {author} {\bibfnamefont {Q.~D.}\ \bibnamefont {Gibson}}, \bibinfo {author}
  {\bibfnamefont {W.}~\bibnamefont {Xie}}, \ and\ \bibinfo {author}
  {\bibfnamefont {R.~J.}\ \bibnamefont {Cava}},\ }\href@noop {} {\bibfield
  {journal} {\bibinfo  {journal} {Apl Materials}\ }\textbf {\bibinfo {volume}
  {3}},\ \bibinfo {pages} {083602} (\bibinfo {year} {2015})}\BibitemShut
  {NoStop}%
\bibitem [{\citenamefont {Xu}\ \emph {et~al.}(2017)\citenamefont {Xu},
  \citenamefont {Yu}, \citenamefont {Fang}, \citenamefont {Dai},\ and\
  \citenamefont {Weng}}]{xu2017topological}%
  \BibitemOpen
  \bibfield  {author} {\bibinfo {author} {\bibfnamefont {Q.}~\bibnamefont
  {Xu}}, \bibinfo {author} {\bibfnamefont {R.}~\bibnamefont {Yu}}, \bibinfo
  {author} {\bibfnamefont {Z.}~\bibnamefont {Fang}}, \bibinfo {author}
  {\bibfnamefont {X.}~\bibnamefont {Dai}}, \ and\ \bibinfo {author}
  {\bibfnamefont {H.}~\bibnamefont {Weng}},\ }\href@noop {} {\bibfield
  {journal} {\bibinfo  {journal} {Physical Review B}\ }\textbf {\bibinfo
  {volume} {95}},\ \bibinfo {pages} {045136} (\bibinfo {year}
  {2017})}\BibitemShut {NoStop}%
\bibitem [{\citenamefont {Huang}\ \emph
  {et~al.}(2016{\natexlab{b}})\citenamefont {Huang}, \citenamefont {Liu},
  \citenamefont {Vanderbilt},\ and\ \citenamefont
  {Duan}}]{huang2016topological}%
  \BibitemOpen
  \bibfield  {author} {\bibinfo {author} {\bibfnamefont {H.}~\bibnamefont
  {Huang}}, \bibinfo {author} {\bibfnamefont {J.}~\bibnamefont {Liu}}, \bibinfo
  {author} {\bibfnamefont {D.}~\bibnamefont {Vanderbilt}}, \ and\ \bibinfo
  {author} {\bibfnamefont {W.}~\bibnamefont {Duan}},\ }\href@noop {} {\bibfield
   {journal} {\bibinfo  {journal} {Physical Review B}\ }\textbf {\bibinfo
  {volume} {93}},\ \bibinfo {pages} {201114} (\bibinfo {year}
  {2016}{\natexlab{b}})}\BibitemShut {NoStop}%
\bibitem [{\citenamefont {Nayak}\ \emph {et~al.}(2017)\citenamefont {Nayak},
  \citenamefont {Wu}, \citenamefont {Kumar}, \citenamefont {Shekhar},
  \citenamefont {Singh}, \citenamefont {Fink}, \citenamefont {Rienks},
  \citenamefont {Fecher}, \citenamefont {Parkin}, \citenamefont {Yan} \emph
  {et~al.}}]{nayak2017multiple}%
  \BibitemOpen
  \bibfield  {author} {\bibinfo {author} {\bibfnamefont {J.}~\bibnamefont
  {Nayak}}, \bibinfo {author} {\bibfnamefont {S.-C.}\ \bibnamefont {Wu}},
  \bibinfo {author} {\bibfnamefont {N.}~\bibnamefont {Kumar}}, \bibinfo
  {author} {\bibfnamefont {C.}~\bibnamefont {Shekhar}}, \bibinfo {author}
  {\bibfnamefont {S.}~\bibnamefont {Singh}}, \bibinfo {author} {\bibfnamefont
  {J.}~\bibnamefont {Fink}}, \bibinfo {author} {\bibfnamefont {E.~E.}\
  \bibnamefont {Rienks}}, \bibinfo {author} {\bibfnamefont {G.~H.}\
  \bibnamefont {Fecher}}, \bibinfo {author} {\bibfnamefont {S.~S.}\
  \bibnamefont {Parkin}}, \bibinfo {author} {\bibfnamefont {B.}~\bibnamefont
  {Yan}},  \emph {et~al.},\ }\href@noop {} {\bibfield  {journal} {\bibinfo
  {journal} {Nature communications}\ }\textbf {\bibinfo {volume} {8}} (\bibinfo
  {year} {2017})}\BibitemShut {NoStop}%
\bibitem [{\citenamefont {Feng}\ \emph {et~al.}(2017)\citenamefont {Feng},
  \citenamefont {Yue}, \citenamefont {Song}, \citenamefont {Wu},\ and\
  \citenamefont {Wen}}]{feng2017topological}%
  \BibitemOpen
  \bibfield  {author} {\bibinfo {author} {\bibfnamefont {X.}~\bibnamefont
  {Feng}}, \bibinfo {author} {\bibfnamefont {C.}~\bibnamefont {Yue}}, \bibinfo
  {author} {\bibfnamefont {Z.}~\bibnamefont {Song}}, \bibinfo {author}
  {\bibfnamefont {Q.}~\bibnamefont {Wu}}, \ and\ \bibinfo {author}
  {\bibfnamefont {B.}~\bibnamefont {Wen}},\ }\href@noop {} {\bibfield
  {journal} {\bibinfo  {journal} {arXiv preprint arXiv:1705.00511}\ } (\bibinfo
  {year} {2017})}\BibitemShut {NoStop}%
\bibitem [{\citenamefont {Kumar}\ \emph {et~al.}(2012)\citenamefont {Kumar},
  \citenamefont {Mishra}, \citenamefont {Sharma}, \citenamefont {Sharma},
  \citenamefont {Lowther}, \citenamefont {Vyas},\ and\ \citenamefont
  {Sharma}}]{kumar2012electronic}%
  \BibitemOpen
  \bibfield  {author} {\bibinfo {author} {\bibfnamefont {R.}~\bibnamefont
  {Kumar}}, \bibinfo {author} {\bibfnamefont {M.}~\bibnamefont {Mishra}},
  \bibinfo {author} {\bibfnamefont {B.}~\bibnamefont {Sharma}}, \bibinfo
  {author} {\bibfnamefont {V.}~\bibnamefont {Sharma}}, \bibinfo {author}
  {\bibfnamefont {J.}~\bibnamefont {Lowther}}, \bibinfo {author} {\bibfnamefont
  {V.}~\bibnamefont {Vyas}}, \ and\ \bibinfo {author} {\bibfnamefont
  {G.}~\bibnamefont {Sharma}},\ }\href@noop {} {\bibfield  {journal} {\bibinfo
  {journal} {Computational Materials Science}\ }\textbf {\bibinfo {volume}
  {61}},\ \bibinfo {pages} {150} (\bibinfo {year} {2012})}\BibitemShut
  {NoStop}%
\bibitem [{\citenamefont {Zhang}\ \emph
  {et~al.}(2017{\natexlab{b}})\citenamefont {Zhang}, \citenamefont {Yu},
  \citenamefont {Sheng}, \citenamefont {Yang},\ and\ \citenamefont
  {Yang}}]{zhang2017coexistence}%
  \BibitemOpen
  \bibfield  {author} {\bibinfo {author} {\bibfnamefont {X.}~\bibnamefont
  {Zhang}}, \bibinfo {author} {\bibfnamefont {Z.-M.}\ \bibnamefont {Yu}},
  \bibinfo {author} {\bibfnamefont {X.-L.}\ \bibnamefont {Sheng}}, \bibinfo
  {author} {\bibfnamefont {H.~Y.}\ \bibnamefont {Yang}}, \ and\ \bibinfo
  {author} {\bibfnamefont {S.~A.}\ \bibnamefont {Yang}},\ }\href@noop {}
  {\bibfield  {journal} {\bibinfo  {journal} {Physical Review B}\ }\textbf
  {\bibinfo {volume} {95}},\ \bibinfo {pages} {235116} (\bibinfo {year}
  {2017}{\natexlab{b}})}\BibitemShut {NoStop}%
\bibitem [{\citenamefont {Mullen}\ \emph {et~al.}(2015)\citenamefont {Mullen},
  \citenamefont {Uchoa},\ and\ \citenamefont {Glatzhofer}}]{mullen2015line}%
  \BibitemOpen
  \bibfield  {author} {\bibinfo {author} {\bibfnamefont {K.}~\bibnamefont
  {Mullen}}, \bibinfo {author} {\bibfnamefont {B.}~\bibnamefont {Uchoa}}, \
  and\ \bibinfo {author} {\bibfnamefont {D.~T.}\ \bibnamefont {Glatzhofer}},\
  }\href@noop {} {\bibfield  {journal} {\bibinfo  {journal} {Physical review
  letters}\ }\textbf {\bibinfo {volume} {115}},\ \bibinfo {pages} {026403}
  (\bibinfo {year} {2015})}\BibitemShut {NoStop}%
\bibitem [{\citenamefont {Feng}\ \emph {et~al.}(2016)\citenamefont {Feng},
  \citenamefont {Fu}, \citenamefont {Kasamatsu}, \citenamefont {Ito},
  \citenamefont {Cheng}, \citenamefont {Liu}, \citenamefont {Mahatha},
  \citenamefont {Sheverdyaeva}, \citenamefont {Moras}, \citenamefont {Arita}
  \emph {et~al.}}]{feng2016discovery}%
  \BibitemOpen
  \bibfield  {author} {\bibinfo {author} {\bibfnamefont {B.}~\bibnamefont
  {Feng}}, \bibinfo {author} {\bibfnamefont {B.}~\bibnamefont {Fu}}, \bibinfo
  {author} {\bibfnamefont {S.}~\bibnamefont {Kasamatsu}}, \bibinfo {author}
  {\bibfnamefont {S.}~\bibnamefont {Ito}}, \bibinfo {author} {\bibfnamefont
  {P.}~\bibnamefont {Cheng}}, \bibinfo {author} {\bibfnamefont {C.-C.}\
  \bibnamefont {Liu}}, \bibinfo {author} {\bibfnamefont {S.~K.}\ \bibnamefont
  {Mahatha}}, \bibinfo {author} {\bibfnamefont {P.}~\bibnamefont
  {Sheverdyaeva}}, \bibinfo {author} {\bibfnamefont {P.}~\bibnamefont {Moras}},
  \bibinfo {author} {\bibfnamefont {M.}~\bibnamefont {Arita}},  \emph
  {et~al.},\ }\href@noop {} {\bibfield  {journal} {\bibinfo  {journal} {arXiv
  preprint arXiv:1611.09578}\ } (\bibinfo {year} {2016})}\BibitemShut {NoStop}%
\bibitem [{\citenamefont {Lu}\ \emph {et~al.}(2016)\citenamefont {Lu},
  \citenamefont {Luo}, \citenamefont {Li}, \citenamefont {Yang}, \citenamefont
  {Cao}, \citenamefont {Gong},\ and\ \citenamefont {Xiang}}]{lu2016two}%
  \BibitemOpen
  \bibfield  {author} {\bibinfo {author} {\bibfnamefont {J.}~\bibnamefont
  {Lu}}, \bibinfo {author} {\bibfnamefont {W.}~\bibnamefont {Luo}}, \bibinfo
  {author} {\bibfnamefont {X.}~\bibnamefont {Li}}, \bibinfo {author}
  {\bibfnamefont {S.}~\bibnamefont {Yang}}, \bibinfo {author} {\bibfnamefont
  {J.}~\bibnamefont {Cao}}, \bibinfo {author} {\bibfnamefont {X.}~\bibnamefont
  {Gong}}, \ and\ \bibinfo {author} {\bibfnamefont {H.}~\bibnamefont {Xiang}},\
  }\href@noop {} {\bibfield  {journal} {\bibinfo  {journal} {arXiv preprint
  arXiv:1603.04596}\ } (\bibinfo {year} {2016})}\BibitemShut {NoStop}%
\bibitem [{\citenamefont {Tafti}\ \emph
  {et~al.}(2016{\natexlab{a}})\citenamefont {Tafti}, \citenamefont {Gibson},
  \citenamefont {Kushwaha}, \citenamefont {Krizan}, \citenamefont
  {Haldolaarachchige},\ and\ \citenamefont {Cava}}]{tafti2016temperature}%
  \BibitemOpen
  \bibfield  {author} {\bibinfo {author} {\bibfnamefont {F.~F.}\ \bibnamefont
  {Tafti}}, \bibinfo {author} {\bibfnamefont {Q.}~\bibnamefont {Gibson}},
  \bibinfo {author} {\bibfnamefont {S.}~\bibnamefont {Kushwaha}}, \bibinfo
  {author} {\bibfnamefont {J.~W.}\ \bibnamefont {Krizan}}, \bibinfo {author}
  {\bibfnamefont {N.}~\bibnamefont {Haldolaarachchige}}, \ and\ \bibinfo
  {author} {\bibfnamefont {R.~J.}\ \bibnamefont {Cava}},\ }\href@noop {}
  {\bibfield  {journal} {\bibinfo  {journal} {Proceedings of the National
  Academy of Sciences}\ }\textbf {\bibinfo {volume} {113}},\ \bibinfo {pages}
  {E3475} (\bibinfo {year} {2016}{\natexlab{a}})}\BibitemShut {NoStop}%
\bibitem [{\citenamefont {Tafti}\ \emph
  {et~al.}(2016{\natexlab{b}})\citenamefont {Tafti}, \citenamefont {Gibson},
  \citenamefont {Kushwaha}, \citenamefont {Haldolaarachchige},\ and\
  \citenamefont {Cava}}]{tafti2016resistivity}%
  \BibitemOpen
  \bibfield  {author} {\bibinfo {author} {\bibfnamefont {F.}~\bibnamefont
  {Tafti}}, \bibinfo {author} {\bibfnamefont {Q.}~\bibnamefont {Gibson}},
  \bibinfo {author} {\bibfnamefont {S.}~\bibnamefont {Kushwaha}}, \bibinfo
  {author} {\bibfnamefont {N.}~\bibnamefont {Haldolaarachchige}}, \ and\
  \bibinfo {author} {\bibfnamefont {R.}~\bibnamefont {Cava}},\ }\href@noop {}
  {\bibfield  {journal} {\bibinfo  {journal} {Nature Physics}\ }\textbf
  {\bibinfo {volume} {12}},\ \bibinfo {pages} {272} (\bibinfo {year}
  {2016}{\natexlab{b}})}\BibitemShut {NoStop}%
\bibitem [{\citenamefont {Kumar}\ \emph
  {et~al.}(2016{\natexlab{a}})\citenamefont {Kumar}, \citenamefont {Shekhar},
  \citenamefont {Wu}, \citenamefont {Leermakers}, \citenamefont {Young},
  \citenamefont {Zeitler}, \citenamefont {Yan},\ and\ \citenamefont
  {Felser}}]{kumar2016observation}%
  \BibitemOpen
  \bibfield  {author} {\bibinfo {author} {\bibfnamefont {N.}~\bibnamefont
  {Kumar}}, \bibinfo {author} {\bibfnamefont {C.}~\bibnamefont {Shekhar}},
  \bibinfo {author} {\bibfnamefont {S.-C.}\ \bibnamefont {Wu}}, \bibinfo
  {author} {\bibfnamefont {I.}~\bibnamefont {Leermakers}}, \bibinfo {author}
  {\bibfnamefont {O.}~\bibnamefont {Young}}, \bibinfo {author} {\bibfnamefont
  {U.}~\bibnamefont {Zeitler}}, \bibinfo {author} {\bibfnamefont
  {B.}~\bibnamefont {Yan}}, \ and\ \bibinfo {author} {\bibfnamefont
  {C.}~\bibnamefont {Felser}},\ }\href@noop {} {\bibfield  {journal} {\bibinfo
  {journal} {Physical Review B}\ }\textbf {\bibinfo {volume} {93}},\ \bibinfo
  {pages} {241106} (\bibinfo {year} {2016}{\natexlab{a}})}\BibitemShut
  {NoStop}%
\bibitem [{\citenamefont {Zeng}\ \emph {et~al.}(2016)\citenamefont {Zeng},
  \citenamefont {Lou}, \citenamefont {Wu}, \citenamefont {Xu}, \citenamefont
  {Guo}, \citenamefont {Kong}, \citenamefont {Zhong}, \citenamefont {Ma},
  \citenamefont {Fu}, \citenamefont {Richard} \emph
  {et~al.}}]{zeng2016compensated}%
  \BibitemOpen
  \bibfield  {author} {\bibinfo {author} {\bibfnamefont {L.-K.}\ \bibnamefont
  {Zeng}}, \bibinfo {author} {\bibfnamefont {R.}~\bibnamefont {Lou}}, \bibinfo
  {author} {\bibfnamefont {D.-S.}\ \bibnamefont {Wu}}, \bibinfo {author}
  {\bibfnamefont {Q.}~\bibnamefont {Xu}}, \bibinfo {author} {\bibfnamefont
  {P.-J.}\ \bibnamefont {Guo}}, \bibinfo {author} {\bibfnamefont {L.-Y.}\
  \bibnamefont {Kong}}, \bibinfo {author} {\bibfnamefont {Y.-G.}\ \bibnamefont
  {Zhong}}, \bibinfo {author} {\bibfnamefont {J.-Z.}\ \bibnamefont {Ma}},
  \bibinfo {author} {\bibfnamefont {B.-B.}\ \bibnamefont {Fu}}, \bibinfo
  {author} {\bibfnamefont {P.}~\bibnamefont {Richard}},  \emph {et~al.},\
  }\href@noop {} {\bibfield  {journal} {\bibinfo  {journal} {Physical review
  letters}\ }\textbf {\bibinfo {volume} {117}},\ \bibinfo {pages} {127204}
  (\bibinfo {year} {2016})}\BibitemShut {NoStop}%
\bibitem [{\citenamefont {Wu}\ \emph {et~al.}(2016)\citenamefont {Wu},
  \citenamefont {Kong}, \citenamefont {Wang}, \citenamefont {Johnson},
  \citenamefont {Mou}, \citenamefont {Huang}, \citenamefont {Schrunk},
  \citenamefont {Bud'ko}, \citenamefont {Canfield},\ and\ \citenamefont
  {Kaminski}}]{wu2016asymmetric}%
  \BibitemOpen
  \bibfield  {author} {\bibinfo {author} {\bibfnamefont {Y.}~\bibnamefont
  {Wu}}, \bibinfo {author} {\bibfnamefont {T.}~\bibnamefont {Kong}}, \bibinfo
  {author} {\bibfnamefont {L.-L.}\ \bibnamefont {Wang}}, \bibinfo {author}
  {\bibfnamefont {D.~D.}\ \bibnamefont {Johnson}}, \bibinfo {author}
  {\bibfnamefont {D.}~\bibnamefont {Mou}}, \bibinfo {author} {\bibfnamefont
  {L.}~\bibnamefont {Huang}}, \bibinfo {author} {\bibfnamefont
  {B.}~\bibnamefont {Schrunk}}, \bibinfo {author} {\bibfnamefont {S.~L.}\
  \bibnamefont {Bud'ko}}, \bibinfo {author} {\bibfnamefont {P.~C.}\
  \bibnamefont {Canfield}}, \ and\ \bibinfo {author} {\bibfnamefont
  {A.}~\bibnamefont {Kaminski}},\ }\href@noop {} {\bibfield  {journal}
  {\bibinfo  {journal} {Physical Review B}\ }\textbf {\bibinfo {volume} {94}},\
  \bibinfo {pages} {081108} (\bibinfo {year} {2016})}\BibitemShut {NoStop}%
\bibitem [{\citenamefont {Wa{\'s}kowska}\ \emph {et~al.}(2011)\citenamefont
  {Wa{\'s}kowska}, \citenamefont {Gerward}, \citenamefont {Olsen},
  \citenamefont {Babu}, \citenamefont {Vaitheeswaran}, \citenamefont
  {Kanchana}, \citenamefont {Svane}, \citenamefont {Filipov}, \citenamefont
  {Levchenko},\ and\ \citenamefont {Lyaschenko}}]{waskowska2011thermoelastic}%
  \BibitemOpen
  \bibfield  {author} {\bibinfo {author} {\bibfnamefont {A.}~\bibnamefont
  {Wa{\'s}kowska}}, \bibinfo {author} {\bibfnamefont {L.}~\bibnamefont
  {Gerward}}, \bibinfo {author} {\bibfnamefont {J.~S.}\ \bibnamefont {Olsen}},
  \bibinfo {author} {\bibfnamefont {K.~R.}\ \bibnamefont {Babu}}, \bibinfo
  {author} {\bibfnamefont {G.}~\bibnamefont {Vaitheeswaran}}, \bibinfo {author}
  {\bibfnamefont {V.}~\bibnamefont {Kanchana}}, \bibinfo {author}
  {\bibfnamefont {A.}~\bibnamefont {Svane}}, \bibinfo {author} {\bibfnamefont
  {V.}~\bibnamefont {Filipov}}, \bibinfo {author} {\bibfnamefont
  {G.}~\bibnamefont {Levchenko}}, \ and\ \bibinfo {author} {\bibfnamefont
  {A.}~\bibnamefont {Lyaschenko}},\ }\href@noop {} {\bibfield  {journal}
  {\bibinfo  {journal} {Acta Materialia}\ }\textbf {\bibinfo {volume} {59}},\
  \bibinfo {pages} {4886} (\bibinfo {year} {2011})}\BibitemShut {NoStop}%
\bibitem [{\citenamefont {Okamoto}\ \emph {et~al.}(2010)\citenamefont
  {Okamoto}, \citenamefont {Kusakari}, \citenamefont {Tanaka}, \citenamefont
  {Inui},\ and\ \citenamefont {Otani}}]{okamoto2010anisotropic}%
  \BibitemOpen
  \bibfield  {author} {\bibinfo {author} {\bibfnamefont {N.~L.}\ \bibnamefont
  {Okamoto}}, \bibinfo {author} {\bibfnamefont {M.}~\bibnamefont {Kusakari}},
  \bibinfo {author} {\bibfnamefont {K.}~\bibnamefont {Tanaka}}, \bibinfo
  {author} {\bibfnamefont {H.}~\bibnamefont {Inui}}, \ and\ \bibinfo {author}
  {\bibfnamefont {S.}~\bibnamefont {Otani}},\ }\href@noop {} {\bibfield
  {journal} {\bibinfo  {journal} {Acta Materialia}\ }\textbf {\bibinfo {volume}
  {58}},\ \bibinfo {pages} {76} (\bibinfo {year} {2010})}\BibitemShut {NoStop}%
\bibitem [{\citenamefont {Ezawa}(2016)}]{ezawa2016loop}%
  \BibitemOpen
  \bibfield  {author} {\bibinfo {author} {\bibfnamefont {M.}~\bibnamefont
  {Ezawa}},\ }\href@noop {} {\bibfield  {journal} {\bibinfo  {journal}
  {Physical review letters}\ }\textbf {\bibinfo {volume} {116}},\ \bibinfo
  {pages} {127202} (\bibinfo {year} {2016})}\BibitemShut {NoStop}%
\bibitem [{\citenamefont {Lee}\ \emph {et~al.}(2014{\natexlab{a}})\citenamefont
  {Lee}, \citenamefont {Bhattacharjee}, \citenamefont {Hwang}, \citenamefont
  {Kim}, \citenamefont {Jin},\ and\ \citenamefont {Kim}}]{lee2014topological}%
  \BibitemOpen
  \bibfield  {author} {\bibinfo {author} {\bibfnamefont {E.~K.-H.}\
  \bibnamefont {Lee}}, \bibinfo {author} {\bibfnamefont {S.}~\bibnamefont
  {Bhattacharjee}}, \bibinfo {author} {\bibfnamefont {K.}~\bibnamefont
  {Hwang}}, \bibinfo {author} {\bibfnamefont {H.-S.}\ \bibnamefont {Kim}},
  \bibinfo {author} {\bibfnamefont {H.}~\bibnamefont {Jin}}, \ and\ \bibinfo
  {author} {\bibfnamefont {Y.~B.}\ \bibnamefont {Kim}},\ }\href@noop {}
  {\bibfield  {journal} {\bibinfo  {journal} {Physical Review B}\ }\textbf
  {\bibinfo {volume} {89}},\ \bibinfo {pages} {205132} (\bibinfo {year}
  {2014}{\natexlab{a}})}\BibitemShut {NoStop}%
\bibitem [{\citenamefont {Takayama}\ \emph {et~al.}(2015)\citenamefont
  {Takayama}, \citenamefont {Kato}, \citenamefont {Dinnebier}, \citenamefont
  {Nuss}, \citenamefont {Kono}, \citenamefont {Veiga}, \citenamefont {Fabbris},
  \citenamefont {Haskel},\ and\ \citenamefont
  {Takagi}}]{takayama2015hyperhoneycomb}%
  \BibitemOpen
  \bibfield  {author} {\bibinfo {author} {\bibfnamefont {T.}~\bibnamefont
  {Takayama}}, \bibinfo {author} {\bibfnamefont {A.}~\bibnamefont {Kato}},
  \bibinfo {author} {\bibfnamefont {R.}~\bibnamefont {Dinnebier}}, \bibinfo
  {author} {\bibfnamefont {J.}~\bibnamefont {Nuss}}, \bibinfo {author}
  {\bibfnamefont {H.}~\bibnamefont {Kono}}, \bibinfo {author} {\bibfnamefont
  {L.}~\bibnamefont {Veiga}}, \bibinfo {author} {\bibfnamefont
  {G.}~\bibnamefont {Fabbris}}, \bibinfo {author} {\bibfnamefont
  {D.}~\bibnamefont {Haskel}}, \ and\ \bibinfo {author} {\bibfnamefont
  {H.}~\bibnamefont {Takagi}},\ }\href@noop {} {\bibfield  {journal} {\bibinfo
  {journal} {Physical review letters}\ }\textbf {\bibinfo {volume} {114}},\
  \bibinfo {pages} {077202} (\bibinfo {year} {2015})}\BibitemShut {NoStop}%
\bibitem [{\citenamefont {Biffin}\ \emph {et~al.}(2014)\citenamefont {Biffin},
  \citenamefont {Johnson}, \citenamefont {Choi}, \citenamefont {Freund},
  \citenamefont {Manni}, \citenamefont {Bombardi}, \citenamefont {Manuel},
  \citenamefont {Gegenwart},\ and\ \citenamefont
  {Coldea}}]{biffin2014unconventional}%
  \BibitemOpen
  \bibfield  {author} {\bibinfo {author} {\bibfnamefont {A.}~\bibnamefont
  {Biffin}}, \bibinfo {author} {\bibfnamefont {R.}~\bibnamefont {Johnson}},
  \bibinfo {author} {\bibfnamefont {S.}~\bibnamefont {Choi}}, \bibinfo {author}
  {\bibfnamefont {F.}~\bibnamefont {Freund}}, \bibinfo {author} {\bibfnamefont
  {S.}~\bibnamefont {Manni}}, \bibinfo {author} {\bibfnamefont
  {A.}~\bibnamefont {Bombardi}}, \bibinfo {author} {\bibfnamefont
  {P.}~\bibnamefont {Manuel}}, \bibinfo {author} {\bibfnamefont
  {P.}~\bibnamefont {Gegenwart}}, \ and\ \bibinfo {author} {\bibfnamefont
  {R.}~\bibnamefont {Coldea}},\ }\href@noop {} {\bibfield  {journal} {\bibinfo
  {journal} {Physical Review B}\ }\textbf {\bibinfo {volume} {90}},\ \bibinfo
  {pages} {205116} (\bibinfo {year} {2014})}\BibitemShut {NoStop}%
\bibitem [{\citenamefont {Modic}\ \emph {et~al.}(2014)\citenamefont {Modic},
  \citenamefont {Smidt}, \citenamefont {Kimchi}, \citenamefont {Breznay},
  \citenamefont {Biffin}, \citenamefont {Choi}, \citenamefont {Johnson},
  \citenamefont {Coldea}, \citenamefont {Watkins-Curry}, \citenamefont
  {McCandless} \emph {et~al.}}]{modic2014new}%
  \BibitemOpen
  \bibfield  {author} {\bibinfo {author} {\bibfnamefont {K.~A.}\ \bibnamefont
  {Modic}}, \bibinfo {author} {\bibfnamefont {T.~E.}\ \bibnamefont {Smidt}},
  \bibinfo {author} {\bibfnamefont {I.}~\bibnamefont {Kimchi}}, \bibinfo
  {author} {\bibfnamefont {N.~P.}\ \bibnamefont {Breznay}}, \bibinfo {author}
  {\bibfnamefont {A.}~\bibnamefont {Biffin}}, \bibinfo {author} {\bibfnamefont
  {S.}~\bibnamefont {Choi}}, \bibinfo {author} {\bibfnamefont {R.~D.}\
  \bibnamefont {Johnson}}, \bibinfo {author} {\bibfnamefont {R.}~\bibnamefont
  {Coldea}}, \bibinfo {author} {\bibfnamefont {P.}~\bibnamefont
  {Watkins-Curry}}, \bibinfo {author} {\bibfnamefont {G.~T.}\ \bibnamefont
  {McCandless}},  \emph {et~al.},\ }\href@noop {} {\bibfield  {journal}
  {\bibinfo  {journal} {arXiv preprint arXiv:1402.3254}\ } (\bibinfo {year}
  {2014})}\BibitemShut {NoStop}%
\bibitem [{\citenamefont {Lee}\ \emph {et~al.}(2014{\natexlab{b}})\citenamefont
  {Lee}, \citenamefont {Schaffer}, \citenamefont {Bhattacharjee},\ and\
  \citenamefont {Kim}}]{lee2014heisenberg}%
  \BibitemOpen
  \bibfield  {author} {\bibinfo {author} {\bibfnamefont {E.~K.-H.}\
  \bibnamefont {Lee}}, \bibinfo {author} {\bibfnamefont {R.}~\bibnamefont
  {Schaffer}}, \bibinfo {author} {\bibfnamefont {S.}~\bibnamefont
  {Bhattacharjee}}, \ and\ \bibinfo {author} {\bibfnamefont {Y.~B.}\
  \bibnamefont {Kim}},\ }\href@noop {} {\bibfield  {journal} {\bibinfo
  {journal} {Physical Review B}\ }\textbf {\bibinfo {volume} {89}},\ \bibinfo
  {pages} {045117} (\bibinfo {year} {2014}{\natexlab{b}})}\BibitemShut
  {NoStop}%
\bibitem [{\citenamefont {Kimchi}\ \emph {et~al.}(2014)\citenamefont {Kimchi},
  \citenamefont {Analytis},\ and\ \citenamefont
  {Vishwanath}}]{kimchi2014three}%
  \BibitemOpen
  \bibfield  {author} {\bibinfo {author} {\bibfnamefont {I.}~\bibnamefont
  {Kimchi}}, \bibinfo {author} {\bibfnamefont {J.~G.}\ \bibnamefont
  {Analytis}}, \ and\ \bibinfo {author} {\bibfnamefont {A.}~\bibnamefont
  {Vishwanath}},\ }\href@noop {} {\bibfield  {journal} {\bibinfo  {journal}
  {Physical Review B}\ }\textbf {\bibinfo {volume} {90}},\ \bibinfo {pages}
  {205126} (\bibinfo {year} {2014})}\BibitemShut {NoStop}%
\bibitem [{\citenamefont {Lee}\ \emph {et~al.}(2014{\natexlab{c}})\citenamefont
  {Lee}, \citenamefont {Lee}, \citenamefont {Paramekanti},\ and\ \citenamefont
  {Kim}}]{lee2014order}%
  \BibitemOpen
  \bibfield  {author} {\bibinfo {author} {\bibfnamefont {S.}~\bibnamefont
  {Lee}}, \bibinfo {author} {\bibfnamefont {E.~K.-H.}\ \bibnamefont {Lee}},
  \bibinfo {author} {\bibfnamefont {A.}~\bibnamefont {Paramekanti}}, \ and\
  \bibinfo {author} {\bibfnamefont {Y.~B.}\ \bibnamefont {Kim}},\ }\href@noop
  {} {\bibfield  {journal} {\bibinfo  {journal} {Physical Review B}\ }\textbf
  {\bibinfo {volume} {89}},\ \bibinfo {pages} {014424} (\bibinfo {year}
  {2014}{\natexlab{c}})}\BibitemShut {NoStop}%
\bibitem [{\citenamefont {Pang}\ \emph {et~al.}(2016)\citenamefont {Pang},
  \citenamefont {Smidman}, \citenamefont {Zhao}, \citenamefont {Wang},
  \citenamefont {Weng}, \citenamefont {Che}, \citenamefont {Chen},
  \citenamefont {Lu}, \citenamefont {Chen},\ and\ \citenamefont
  {Yuan}}]{pang2016nodeless}%
  \BibitemOpen
  \bibfield  {author} {\bibinfo {author} {\bibfnamefont {G.}~\bibnamefont
  {Pang}}, \bibinfo {author} {\bibfnamefont {M.}~\bibnamefont {Smidman}},
  \bibinfo {author} {\bibfnamefont {L.}~\bibnamefont {Zhao}}, \bibinfo {author}
  {\bibfnamefont {Y.}~\bibnamefont {Wang}}, \bibinfo {author} {\bibfnamefont
  {Z.}~\bibnamefont {Weng}}, \bibinfo {author} {\bibfnamefont {L.}~\bibnamefont
  {Che}}, \bibinfo {author} {\bibfnamefont {Y.}~\bibnamefont {Chen}}, \bibinfo
  {author} {\bibfnamefont {X.}~\bibnamefont {Lu}}, \bibinfo {author}
  {\bibfnamefont {G.}~\bibnamefont {Chen}}, \ and\ \bibinfo {author}
  {\bibfnamefont {H.}~\bibnamefont {Yuan}},\ }\href@noop {} {\bibfield
  {journal} {\bibinfo  {journal} {Physical Review B}\ }\textbf {\bibinfo
  {volume} {93}},\ \bibinfo {pages} {060506} (\bibinfo {year}
  {2016})}\BibitemShut {NoStop}%
\bibitem [{\citenamefont {Zhang}\ \emph
  {et~al.}(2016{\natexlab{a}})\citenamefont {Zhang}, \citenamefont {Yuan},
  \citenamefont {Bian}, \citenamefont {Xu}, \citenamefont {Zhang},
  \citenamefont {Hasan},\ and\ \citenamefont {Jia}}]{zhang2016superconducting}%
  \BibitemOpen
  \bibfield  {author} {\bibinfo {author} {\bibfnamefont {C.-L.}\ \bibnamefont
  {Zhang}}, \bibinfo {author} {\bibfnamefont {Z.}~\bibnamefont {Yuan}},
  \bibinfo {author} {\bibfnamefont {G.}~\bibnamefont {Bian}}, \bibinfo {author}
  {\bibfnamefont {S.-Y.}\ \bibnamefont {Xu}}, \bibinfo {author} {\bibfnamefont
  {X.}~\bibnamefont {Zhang}}, \bibinfo {author} {\bibfnamefont {M.~Z.}\
  \bibnamefont {Hasan}}, \ and\ \bibinfo {author} {\bibfnamefont
  {S.}~\bibnamefont {Jia}},\ }\href@noop {} {\bibfield  {journal} {\bibinfo
  {journal} {Physical Review B}\ }\textbf {\bibinfo {volume} {93}},\ \bibinfo
  {pages} {054520} (\bibinfo {year} {2016}{\natexlab{a}})}\BibitemShut
  {NoStop}%
\bibitem [{\citenamefont {Sankar}\ \emph {et~al.}(2017)\citenamefont {Sankar},
  \citenamefont {Rao}, \citenamefont {Muthuselvam}, \citenamefont {Chang},
  \citenamefont {Jeng}, \citenamefont {Murugan}, \citenamefont {Lee},\ and\
  \citenamefont {Chou}}]{sankar2017anisotropic}%
  \BibitemOpen
  \bibfield  {author} {\bibinfo {author} {\bibfnamefont {R.}~\bibnamefont
  {Sankar}}, \bibinfo {author} {\bibfnamefont {G.~N.}\ \bibnamefont {Rao}},
  \bibinfo {author} {\bibfnamefont {I.~P.}\ \bibnamefont {Muthuselvam}},
  \bibinfo {author} {\bibfnamefont {T.-R.}\ \bibnamefont {Chang}}, \bibinfo
  {author} {\bibfnamefont {H.}~\bibnamefont {Jeng}}, \bibinfo {author}
  {\bibfnamefont {G.~S.}\ \bibnamefont {Murugan}}, \bibinfo {author}
  {\bibfnamefont {W.-L.}\ \bibnamefont {Lee}}, \ and\ \bibinfo {author}
  {\bibfnamefont {F.}~\bibnamefont {Chou}},\ }\href@noop {} {\bibfield
  {journal} {\bibinfo  {journal} {Journal of Physics: Condensed Matter}\
  }\textbf {\bibinfo {volume} {29}},\ \bibinfo {pages} {095601} (\bibinfo
  {year} {2017})}\BibitemShut {NoStop}%
\bibitem [{\citenamefont {Belopolski}\ \emph {et~al.}(2016)\citenamefont
  {Belopolski}, \citenamefont {Xu}, \citenamefont {Sanchez}, \citenamefont
  {Chang}, \citenamefont {Guo}, \citenamefont {Neupane}, \citenamefont {Zheng},
  \citenamefont {Lee}, \citenamefont {Huang}, \citenamefont {Bian} \emph
  {et~al.}}]{belopolski2016criteria}%
  \BibitemOpen
  \bibfield  {author} {\bibinfo {author} {\bibfnamefont {I.}~\bibnamefont
  {Belopolski}}, \bibinfo {author} {\bibfnamefont {S.-Y.}\ \bibnamefont {Xu}},
  \bibinfo {author} {\bibfnamefont {D.~S.}\ \bibnamefont {Sanchez}}, \bibinfo
  {author} {\bibfnamefont {G.}~\bibnamefont {Chang}}, \bibinfo {author}
  {\bibfnamefont {C.}~\bibnamefont {Guo}}, \bibinfo {author} {\bibfnamefont
  {M.}~\bibnamefont {Neupane}}, \bibinfo {author} {\bibfnamefont
  {H.}~\bibnamefont {Zheng}}, \bibinfo {author} {\bibfnamefont {C.-C.}\
  \bibnamefont {Lee}}, \bibinfo {author} {\bibfnamefont {S.-M.}\ \bibnamefont
  {Huang}}, \bibinfo {author} {\bibfnamefont {G.}~\bibnamefont {Bian}},  \emph
  {et~al.},\ }\href@noop {} {\bibfield  {journal} {\bibinfo  {journal}
  {Physical review letters}\ }\textbf {\bibinfo {volume} {116}},\ \bibinfo
  {pages} {066802} (\bibinfo {year} {2016})}\BibitemShut {NoStop}%
\bibitem [{\citenamefont {Inoue}\ \emph {et~al.}(2016)\citenamefont {Inoue},
  \citenamefont {Gyenis}, \citenamefont {Wang}, \citenamefont {Li},
  \citenamefont {Oh}, \citenamefont {Jiang}, \citenamefont {Ni}, \citenamefont
  {Bernevig},\ and\ \citenamefont {Yazdani}}]{inoue2016quasiparticle}%
  \BibitemOpen
  \bibfield  {author} {\bibinfo {author} {\bibfnamefont {H.}~\bibnamefont
  {Inoue}}, \bibinfo {author} {\bibfnamefont {A.}~\bibnamefont {Gyenis}},
  \bibinfo {author} {\bibfnamefont {Z.}~\bibnamefont {Wang}}, \bibinfo {author}
  {\bibfnamefont {J.}~\bibnamefont {Li}}, \bibinfo {author} {\bibfnamefont
  {S.~W.}\ \bibnamefont {Oh}}, \bibinfo {author} {\bibfnamefont
  {S.}~\bibnamefont {Jiang}}, \bibinfo {author} {\bibfnamefont
  {N.}~\bibnamefont {Ni}}, \bibinfo {author} {\bibfnamefont {B.~A.}\
  \bibnamefont {Bernevig}}, \ and\ \bibinfo {author} {\bibfnamefont
  {A.}~\bibnamefont {Yazdani}},\ }\href@noop {} {\bibfield  {journal} {\bibinfo
   {journal} {Science}\ }\textbf {\bibinfo {volume} {351}},\ \bibinfo {pages}
  {1184} (\bibinfo {year} {2016})}\BibitemShut {NoStop}%
\bibitem [{\citenamefont {Batabyal}\ \emph {et~al.}(2016)\citenamefont
  {Batabyal}, \citenamefont {Morali}, \citenamefont {Avraham}, \citenamefont
  {Sun}, \citenamefont {Schmidt}, \citenamefont {Felser}, \citenamefont
  {Stern}, \citenamefont {Yan},\ and\ \citenamefont
  {Beidenkopf}}]{batabyal2016visualizing}%
  \BibitemOpen
  \bibfield  {author} {\bibinfo {author} {\bibfnamefont {R.}~\bibnamefont
  {Batabyal}}, \bibinfo {author} {\bibfnamefont {N.}~\bibnamefont {Morali}},
  \bibinfo {author} {\bibfnamefont {N.}~\bibnamefont {Avraham}}, \bibinfo
  {author} {\bibfnamefont {Y.}~\bibnamefont {Sun}}, \bibinfo {author}
  {\bibfnamefont {M.}~\bibnamefont {Schmidt}}, \bibinfo {author} {\bibfnamefont
  {C.}~\bibnamefont {Felser}}, \bibinfo {author} {\bibfnamefont
  {A.}~\bibnamefont {Stern}}, \bibinfo {author} {\bibfnamefont
  {B.}~\bibnamefont {Yan}}, \ and\ \bibinfo {author} {\bibfnamefont
  {H.}~\bibnamefont {Beidenkopf}},\ }\href@noop {} {\bibfield  {journal}
  {\bibinfo  {journal} {Science advances}\ }\textbf {\bibinfo {volume} {2}},\
  \bibinfo {pages} {e1600709} (\bibinfo {year} {2016})}\BibitemShut {NoStop}%
\bibitem [{\citenamefont {Ghimire}\ \emph {et~al.}(2015)\citenamefont
  {Ghimire}, \citenamefont {Luo}, \citenamefont {Neupane}, \citenamefont
  {Williams}, \citenamefont {Bauer},\ and\ \citenamefont
  {Ronning}}]{ghimire2015magnetotransport}%
  \BibitemOpen
  \bibfield  {author} {\bibinfo {author} {\bibfnamefont {N.~J.}\ \bibnamefont
  {Ghimire}}, \bibinfo {author} {\bibfnamefont {Y.}~\bibnamefont {Luo}},
  \bibinfo {author} {\bibfnamefont {M.}~\bibnamefont {Neupane}}, \bibinfo
  {author} {\bibfnamefont {D.}~\bibnamefont {Williams}}, \bibinfo {author}
  {\bibfnamefont {E.}~\bibnamefont {Bauer}}, \ and\ \bibinfo {author}
  {\bibfnamefont {F.}~\bibnamefont {Ronning}},\ }\href@noop {} {\bibfield
  {journal} {\bibinfo  {journal} {Journal of Physics: Condensed Matter}\
  }\textbf {\bibinfo {volume} {27}},\ \bibinfo {pages} {152201} (\bibinfo
  {year} {2015})}\BibitemShut {NoStop}%
\bibitem [{\citenamefont {Wang}\ \emph
  {et~al.}(2016{\natexlab{b}})\citenamefont {Wang}, \citenamefont {Zheng},
  \citenamefont {Shen}, \citenamefont {Lu}, \citenamefont {Fang}, \citenamefont
  {Sheng}, \citenamefont {Zhou}, \citenamefont {Yang}, \citenamefont {Li},
  \citenamefont {Feng} \emph {et~al.}}]{wang2016helicity}%
  \BibitemOpen
  \bibfield  {author} {\bibinfo {author} {\bibfnamefont {Z.}~\bibnamefont
  {Wang}}, \bibinfo {author} {\bibfnamefont {Y.}~\bibnamefont {Zheng}},
  \bibinfo {author} {\bibfnamefont {Z.}~\bibnamefont {Shen}}, \bibinfo {author}
  {\bibfnamefont {Y.}~\bibnamefont {Lu}}, \bibinfo {author} {\bibfnamefont
  {H.}~\bibnamefont {Fang}}, \bibinfo {author} {\bibfnamefont {F.}~\bibnamefont
  {Sheng}}, \bibinfo {author} {\bibfnamefont {Y.}~\bibnamefont {Zhou}},
  \bibinfo {author} {\bibfnamefont {X.}~\bibnamefont {Yang}}, \bibinfo {author}
  {\bibfnamefont {Y.}~\bibnamefont {Li}}, \bibinfo {author} {\bibfnamefont
  {C.}~\bibnamefont {Feng}},  \emph {et~al.},\ }\href@noop {} {\bibfield
  {journal} {\bibinfo  {journal} {Physical Review B}\ }\textbf {\bibinfo
  {volume} {93}},\ \bibinfo {pages} {121112} (\bibinfo {year}
  {2016}{\natexlab{b}})}\BibitemShut {NoStop}%
\bibitem [{\citenamefont {Zhang}\ \emph
  {et~al.}(2016{\natexlab{b}})\citenamefont {Zhang}, \citenamefont {Xu},
  \citenamefont {Belopolski}, \citenamefont {Yuan}, \citenamefont {Lin},
  \citenamefont {Tong}, \citenamefont {Bian}, \citenamefont {Alidoust},
  \citenamefont {Lee}, \citenamefont {Huang} \emph
  {et~al.}}]{zhang2016signatures}%
  \BibitemOpen
  \bibfield  {author} {\bibinfo {author} {\bibfnamefont {C.-L.}\ \bibnamefont
  {Zhang}}, \bibinfo {author} {\bibfnamefont {S.-Y.}\ \bibnamefont {Xu}},
  \bibinfo {author} {\bibfnamefont {I.}~\bibnamefont {Belopolski}}, \bibinfo
  {author} {\bibfnamefont {Z.}~\bibnamefont {Yuan}}, \bibinfo {author}
  {\bibfnamefont {Z.}~\bibnamefont {Lin}}, \bibinfo {author} {\bibfnamefont
  {B.}~\bibnamefont {Tong}}, \bibinfo {author} {\bibfnamefont {G.}~\bibnamefont
  {Bian}}, \bibinfo {author} {\bibfnamefont {N.}~\bibnamefont {Alidoust}},
  \bibinfo {author} {\bibfnamefont {C.-C.}\ \bibnamefont {Lee}}, \bibinfo
  {author} {\bibfnamefont {S.-M.}\ \bibnamefont {Huang}},  \emph {et~al.},\
  }\href@noop {} {\bibfield  {journal} {\bibinfo  {journal} {Nature
  communications}\ }\textbf {\bibinfo {volume} {7}} (\bibinfo {year}
  {2016}{\natexlab{b}})}\BibitemShut {NoStop}%
\bibitem [{\citenamefont {Luo}\ \emph {et~al.}(2015)\citenamefont {Luo},
  \citenamefont {Ghimire}, \citenamefont {Wartenbe}, \citenamefont {Choi},
  \citenamefont {Neupane}, \citenamefont {McDonald}, \citenamefont {Bauer},
  \citenamefont {Zhu}, \citenamefont {Thompson},\ and\ \citenamefont
  {Ronning}}]{luo2015electron}%
  \BibitemOpen
  \bibfield  {author} {\bibinfo {author} {\bibfnamefont {Y.}~\bibnamefont
  {Luo}}, \bibinfo {author} {\bibfnamefont {N.}~\bibnamefont {Ghimire}},
  \bibinfo {author} {\bibfnamefont {M.}~\bibnamefont {Wartenbe}}, \bibinfo
  {author} {\bibfnamefont {H.}~\bibnamefont {Choi}}, \bibinfo {author}
  {\bibfnamefont {M.}~\bibnamefont {Neupane}}, \bibinfo {author} {\bibfnamefont
  {R.}~\bibnamefont {McDonald}}, \bibinfo {author} {\bibfnamefont
  {E.}~\bibnamefont {Bauer}}, \bibinfo {author} {\bibfnamefont
  {J.}~\bibnamefont {Zhu}}, \bibinfo {author} {\bibfnamefont {J.}~\bibnamefont
  {Thompson}}, \ and\ \bibinfo {author} {\bibfnamefont {F.}~\bibnamefont
  {Ronning}},\ }\href@noop {} {\bibfield  {journal} {\bibinfo  {journal}
  {Physical Review B}\ }\textbf {\bibinfo {volume} {92}},\ \bibinfo {pages}
  {205134} (\bibinfo {year} {2015})}\BibitemShut {NoStop}%
\bibitem [{\citenamefont {Moll}\ \emph {et~al.}(2016)\citenamefont {Moll},
  \citenamefont {Potter}, \citenamefont {Nair}, \citenamefont {Ramshaw},
  \citenamefont {Modic}, \citenamefont {Riggs}, \citenamefont {Zeng},
  \citenamefont {Ghimire}, \citenamefont {Bauer}, \citenamefont {Kealhofer}
  \emph {et~al.}}]{moll2016magnetic}%
  \BibitemOpen
  \bibfield  {author} {\bibinfo {author} {\bibfnamefont {P.~J.}\ \bibnamefont
  {Moll}}, \bibinfo {author} {\bibfnamefont {A.~C.}\ \bibnamefont {Potter}},
  \bibinfo {author} {\bibfnamefont {N.~L.}\ \bibnamefont {Nair}}, \bibinfo
  {author} {\bibfnamefont {B.}~\bibnamefont {Ramshaw}}, \bibinfo {author}
  {\bibfnamefont {K.}~\bibnamefont {Modic}}, \bibinfo {author} {\bibfnamefont
  {S.}~\bibnamefont {Riggs}}, \bibinfo {author} {\bibfnamefont
  {B.}~\bibnamefont {Zeng}}, \bibinfo {author} {\bibfnamefont {N.~J.}\
  \bibnamefont {Ghimire}}, \bibinfo {author} {\bibfnamefont {E.~D.}\
  \bibnamefont {Bauer}}, \bibinfo {author} {\bibfnamefont {R.}~\bibnamefont
  {Kealhofer}},  \emph {et~al.},\ }\href@noop {} {\bibfield  {journal}
  {\bibinfo  {journal} {Nature communications}\ }\textbf {\bibinfo {volume}
  {7}} (\bibinfo {year} {2016})}\BibitemShut {NoStop}%
\bibitem [{\citenamefont {Yamakage}\ \emph {et~al.}(2015)\citenamefont
  {Yamakage}, \citenamefont {Yamakawa}, \citenamefont {Tanaka},\ and\
  \citenamefont {Okamoto}}]{yamakage2015line}%
  \BibitemOpen
  \bibfield  {author} {\bibinfo {author} {\bibfnamefont {A.}~\bibnamefont
  {Yamakage}}, \bibinfo {author} {\bibfnamefont {Y.}~\bibnamefont {Yamakawa}},
  \bibinfo {author} {\bibfnamefont {Y.}~\bibnamefont {Tanaka}}, \ and\ \bibinfo
  {author} {\bibfnamefont {Y.}~\bibnamefont {Okamoto}},\ }\href@noop {}
  {\bibfield  {journal} {\bibinfo  {journal} {Journal of the Physical Society
  of Japan}\ }\textbf {\bibinfo {volume} {85}},\ \bibinfo {pages} {013708}
  (\bibinfo {year} {2015})}\BibitemShut {NoStop}%
\bibitem [{\citenamefont {Takane}\ \emph {et~al.}(2016)\citenamefont {Takane},
  \citenamefont {Wang}, \citenamefont {Souma}, \citenamefont {Nakayama},
  \citenamefont {Trang}, \citenamefont {Sato}, \citenamefont {Takahashi},\ and\
  \citenamefont {Ando}}]{takane2016dirac}%
  \BibitemOpen
  \bibfield  {author} {\bibinfo {author} {\bibfnamefont {D.}~\bibnamefont
  {Takane}}, \bibinfo {author} {\bibfnamefont {Z.}~\bibnamefont {Wang}},
  \bibinfo {author} {\bibfnamefont {S.}~\bibnamefont {Souma}}, \bibinfo
  {author} {\bibfnamefont {K.}~\bibnamefont {Nakayama}}, \bibinfo {author}
  {\bibfnamefont {C.}~\bibnamefont {Trang}}, \bibinfo {author} {\bibfnamefont
  {T.}~\bibnamefont {Sato}}, \bibinfo {author} {\bibfnamefont {T.}~\bibnamefont
  {Takahashi}}, \ and\ \bibinfo {author} {\bibfnamefont {Y.}~\bibnamefont
  {Ando}},\ }\href@noop {} {\bibfield  {journal} {\bibinfo  {journal} {Physical
  Review B}\ }\textbf {\bibinfo {volume} {94}},\ \bibinfo {pages} {121108}
  (\bibinfo {year} {2016})}\BibitemShut {NoStop}%
\bibitem [{\citenamefont {Ali}\ \emph {et~al.}(2016)\citenamefont {Ali},
  \citenamefont {Schoop}, \citenamefont {Garg}, \citenamefont {Lippmann},
  \citenamefont {Lara}, \citenamefont {Lotsch},\ and\ \citenamefont
  {Parkin}}]{ali2016butterfly}%
  \BibitemOpen
  \bibfield  {author} {\bibinfo {author} {\bibfnamefont {M.~N.}\ \bibnamefont
  {Ali}}, \bibinfo {author} {\bibfnamefont {L.~M.}\ \bibnamefont {Schoop}},
  \bibinfo {author} {\bibfnamefont {C.}~\bibnamefont {Garg}}, \bibinfo {author}
  {\bibfnamefont {J.~M.}\ \bibnamefont {Lippmann}}, \bibinfo {author}
  {\bibfnamefont {E.}~\bibnamefont {Lara}}, \bibinfo {author} {\bibfnamefont
  {B.}~\bibnamefont {Lotsch}}, \ and\ \bibinfo {author} {\bibfnamefont
  {S.}~\bibnamefont {Parkin}},\ }\href@noop {} {\bibfield  {journal} {\bibinfo
  {journal} {arXiv preprint arXiv:1603.09318}\ } (\bibinfo {year}
  {2016})}\BibitemShut {NoStop}%
\bibitem [{\citenamefont {Wang}\ \emph
  {et~al.}(2016{\natexlab{c}})\citenamefont {Wang}, \citenamefont {Pan},
  \citenamefont {Gao}, \citenamefont {Yu}, \citenamefont {Jiang}, \citenamefont
  {Zhang}, \citenamefont {Zuo}, \citenamefont {Zhang}, \citenamefont {Wei},
  \citenamefont {Niu} \emph {et~al.}}]{wang2016evidence}%
  \BibitemOpen
  \bibfield  {author} {\bibinfo {author} {\bibfnamefont {X.}~\bibnamefont
  {Wang}}, \bibinfo {author} {\bibfnamefont {X.}~\bibnamefont {Pan}}, \bibinfo
  {author} {\bibfnamefont {M.}~\bibnamefont {Gao}}, \bibinfo {author}
  {\bibfnamefont {J.}~\bibnamefont {Yu}}, \bibinfo {author} {\bibfnamefont
  {J.}~\bibnamefont {Jiang}}, \bibinfo {author} {\bibfnamefont
  {J.}~\bibnamefont {Zhang}}, \bibinfo {author} {\bibfnamefont
  {H.}~\bibnamefont {Zuo}}, \bibinfo {author} {\bibfnamefont {M.}~\bibnamefont
  {Zhang}}, \bibinfo {author} {\bibfnamefont {Z.}~\bibnamefont {Wei}}, \bibinfo
  {author} {\bibfnamefont {W.}~\bibnamefont {Niu}},  \emph {et~al.},\
  }\href@noop {} {\bibfield  {journal} {\bibinfo  {journal} {arXiv preprint
  arXiv:1604.00108}\ } (\bibinfo {year} {2016}{\natexlab{c}})}\BibitemShut
  {NoStop}%
\bibitem [{\citenamefont {Singha}\ \emph {et~al.}(2016)\citenamefont {Singha},
  \citenamefont {Pariari}, \citenamefont {Satpati},\ and\ \citenamefont
  {Mandal}}]{singha2016titanic}%
  \BibitemOpen
  \bibfield  {author} {\bibinfo {author} {\bibfnamefont {R.}~\bibnamefont
  {Singha}}, \bibinfo {author} {\bibfnamefont {A.}~\bibnamefont {Pariari}},
  \bibinfo {author} {\bibfnamefont {B.}~\bibnamefont {Satpati}}, \ and\
  \bibinfo {author} {\bibfnamefont {P.}~\bibnamefont {Mandal}},\ }\href@noop {}
  {\bibfield  {journal} {\bibinfo  {journal} {arXiv preprint arXiv:1602.01993}\
  } (\bibinfo {year} {2016})}\BibitemShut {NoStop}%
\bibitem [{\citenamefont {Hu}\ \emph {et~al.}(2016{\natexlab{a}})\citenamefont
  {Hu}, \citenamefont {Tang}, \citenamefont {Liu}, \citenamefont {Zhu},
  \citenamefont {Wei},\ and\ \citenamefont {Mao}}]{hu2016evidence1}%
  \BibitemOpen
  \bibfield  {author} {\bibinfo {author} {\bibfnamefont {J.}~\bibnamefont
  {Hu}}, \bibinfo {author} {\bibfnamefont {Z.}~\bibnamefont {Tang}}, \bibinfo
  {author} {\bibfnamefont {J.}~\bibnamefont {Liu}}, \bibinfo {author}
  {\bibfnamefont {Y.}~\bibnamefont {Zhu}}, \bibinfo {author} {\bibfnamefont
  {J.}~\bibnamefont {Wei}}, \ and\ \bibinfo {author} {\bibfnamefont
  {Z.}~\bibnamefont {Mao}},\ }\href@noop {} {\bibfield  {journal} {\bibinfo
  {journal} {arXiv preprint arXiv:1604.01567}\ } (\bibinfo {year}
  {2016}{\natexlab{a}})}\BibitemShut {NoStop}%
\bibitem [{\citenamefont {Kumar}\ \emph
  {et~al.}(2016{\natexlab{b}})\citenamefont {Kumar}, \citenamefont {Manna},
  \citenamefont {Qi}, \citenamefont {Wu}, \citenamefont {Wang}, \citenamefont
  {Yan}, \citenamefont {Felser},\ and\ \citenamefont
  {Shekhar}}]{kumar2016unusual}%
  \BibitemOpen
  \bibfield  {author} {\bibinfo {author} {\bibfnamefont {N.}~\bibnamefont
  {Kumar}}, \bibinfo {author} {\bibfnamefont {K.}~\bibnamefont {Manna}},
  \bibinfo {author} {\bibfnamefont {Y.}~\bibnamefont {Qi}}, \bibinfo {author}
  {\bibfnamefont {S.-C.}\ \bibnamefont {Wu}}, \bibinfo {author} {\bibfnamefont
  {L.}~\bibnamefont {Wang}}, \bibinfo {author} {\bibfnamefont {B.}~\bibnamefont
  {Yan}}, \bibinfo {author} {\bibfnamefont {C.}~\bibnamefont {Felser}}, \ and\
  \bibinfo {author} {\bibfnamefont {C.}~\bibnamefont {Shekhar}},\ }\href@noop
  {} {\bibfield  {journal} {\bibinfo  {journal} {arXiv preprint
  arXiv:1612.05176}\ } (\bibinfo {year} {2016}{\natexlab{b}})}\BibitemShut
  {NoStop}%
\bibitem [{\citenamefont {Topp}\ \emph {et~al.}(2016)\citenamefont {Topp},
  \citenamefont {Lippmann}, \citenamefont {Varykhalov}, \citenamefont {Duppel},
  \citenamefont {Lotsch}, \citenamefont {Ast},\ and\ \citenamefont
  {Schoop}}]{topp2016non}%
  \BibitemOpen
  \bibfield  {author} {\bibinfo {author} {\bibfnamefont {A.}~\bibnamefont
  {Topp}}, \bibinfo {author} {\bibfnamefont {J.~M.}\ \bibnamefont {Lippmann}},
  \bibinfo {author} {\bibfnamefont {A.}~\bibnamefont {Varykhalov}}, \bibinfo
  {author} {\bibfnamefont {V.}~\bibnamefont {Duppel}}, \bibinfo {author}
  {\bibfnamefont {B.~V.}\ \bibnamefont {Lotsch}}, \bibinfo {author}
  {\bibfnamefont {C.~R.}\ \bibnamefont {Ast}}, \ and\ \bibinfo {author}
  {\bibfnamefont {L.~M.}\ \bibnamefont {Schoop}},\ }\href@noop {} {\bibfield
  {journal} {\bibinfo  {journal} {New Journal of Physics}\ }\textbf {\bibinfo
  {volume} {18}},\ \bibinfo {pages} {125014} (\bibinfo {year}
  {2016})}\BibitemShut {NoStop}%
\bibitem [{\citenamefont {Lv}\ \emph {et~al.}(2016)\citenamefont {Lv},
  \citenamefont {Zhang}, \citenamefont {Li}, \citenamefont {Yao}, \citenamefont
  {Chen}, \citenamefont {Zhou}, \citenamefont {Zhang}, \citenamefont {Lu},\
  and\ \citenamefont {Chen}}]{lv2016extremely}%
  \BibitemOpen
  \bibfield  {author} {\bibinfo {author} {\bibfnamefont {Y.-Y.}\ \bibnamefont
  {Lv}}, \bibinfo {author} {\bibfnamefont {B.-B.}\ \bibnamefont {Zhang}},
  \bibinfo {author} {\bibfnamefont {X.}~\bibnamefont {Li}}, \bibinfo {author}
  {\bibfnamefont {S.-H.}\ \bibnamefont {Yao}}, \bibinfo {author} {\bibfnamefont
  {Y.}~\bibnamefont {Chen}}, \bibinfo {author} {\bibfnamefont {J.}~\bibnamefont
  {Zhou}}, \bibinfo {author} {\bibfnamefont {S.-T.}\ \bibnamefont {Zhang}},
  \bibinfo {author} {\bibfnamefont {M.-H.}\ \bibnamefont {Lu}}, \ and\ \bibinfo
  {author} {\bibfnamefont {Y.-F.}\ \bibnamefont {Chen}},\ }\href@noop {}
  {\bibfield  {journal} {\bibinfo  {journal} {Applied Physics Letters}\
  }\textbf {\bibinfo {volume} {108}},\ \bibinfo {pages} {244101} (\bibinfo
  {year} {2016})}\BibitemShut {NoStop}%
\bibitem [{\citenamefont {Nie}\ \emph {et~al.}(2015)\citenamefont {Nie},
  \citenamefont {King}, \citenamefont {Kim}, \citenamefont {Uchida},
  \citenamefont {Wei}, \citenamefont {Faeth}, \citenamefont {Ruf},
  \citenamefont {Ruff}, \citenamefont {Xie}, \citenamefont {Pan} \emph
  {et~al.}}]{nie2015interplay}%
  \BibitemOpen
  \bibfield  {author} {\bibinfo {author} {\bibfnamefont {Y.}~\bibnamefont
  {Nie}}, \bibinfo {author} {\bibfnamefont {P.}~\bibnamefont {King}}, \bibinfo
  {author} {\bibfnamefont {C.}~\bibnamefont {Kim}}, \bibinfo {author}
  {\bibfnamefont {M.}~\bibnamefont {Uchida}}, \bibinfo {author} {\bibfnamefont
  {H.}~\bibnamefont {Wei}}, \bibinfo {author} {\bibfnamefont {B.}~\bibnamefont
  {Faeth}}, \bibinfo {author} {\bibfnamefont {J.}~\bibnamefont {Ruf}}, \bibinfo
  {author} {\bibfnamefont {J.}~\bibnamefont {Ruff}}, \bibinfo {author}
  {\bibfnamefont {L.}~\bibnamefont {Xie}}, \bibinfo {author} {\bibfnamefont
  {X.}~\bibnamefont {Pan}},  \emph {et~al.},\ }\href@noop {} {\bibfield
  {journal} {\bibinfo  {journal} {Physical review letters}\ }\textbf {\bibinfo
  {volume} {114}},\ \bibinfo {pages} {016401} (\bibinfo {year}
  {2015})}\BibitemShut {NoStop}%
\bibitem [{\citenamefont {Lou}\ \emph {et~al.}(2016)\citenamefont {Lou},
  \citenamefont {Ma}, \citenamefont {Xu}, \citenamefont {Fu}, \citenamefont
  {Kong}, \citenamefont {Shi}, \citenamefont {Richard}, \citenamefont {Weng},
  \citenamefont {Fang}, \citenamefont {Sun} \emph {et~al.}}]{lou2016emergence}%
  \BibitemOpen
  \bibfield  {author} {\bibinfo {author} {\bibfnamefont {R.}~\bibnamefont
  {Lou}}, \bibinfo {author} {\bibfnamefont {J.-Z.}\ \bibnamefont {Ma}},
  \bibinfo {author} {\bibfnamefont {Q.-N.}\ \bibnamefont {Xu}}, \bibinfo
  {author} {\bibfnamefont {B.-B.}\ \bibnamefont {Fu}}, \bibinfo {author}
  {\bibfnamefont {L.-Y.}\ \bibnamefont {Kong}}, \bibinfo {author}
  {\bibfnamefont {Y.-G.}\ \bibnamefont {Shi}}, \bibinfo {author} {\bibfnamefont
  {P.}~\bibnamefont {Richard}}, \bibinfo {author} {\bibfnamefont {H.-M.}\
  \bibnamefont {Weng}}, \bibinfo {author} {\bibfnamefont {Z.}~\bibnamefont
  {Fang}}, \bibinfo {author} {\bibfnamefont {S.-S.}\ \bibnamefont {Sun}},
  \emph {et~al.},\ }\href@noop {} {\bibfield  {journal} {\bibinfo  {journal}
  {Physical Review B}\ }\textbf {\bibinfo {volume} {93}},\ \bibinfo {pages}
  {241104} (\bibinfo {year} {2016})}\BibitemShut {NoStop}%
\bibitem [{\citenamefont {Hu}\ \emph {et~al.}(2016{\natexlab{b}})\citenamefont
  {Hu}, \citenamefont {Tang}, \citenamefont {Liu}, \citenamefont {Liu},
  \citenamefont {Zhu}, \citenamefont {Graf}, \citenamefont {Myhro},
  \citenamefont {Tran}, \citenamefont {Lau}, \citenamefont {Wei} \emph
  {et~al.}}]{hu2016evidence}%
  \BibitemOpen
  \bibfield  {author} {\bibinfo {author} {\bibfnamefont {J.}~\bibnamefont
  {Hu}}, \bibinfo {author} {\bibfnamefont {Z.}~\bibnamefont {Tang}}, \bibinfo
  {author} {\bibfnamefont {J.}~\bibnamefont {Liu}}, \bibinfo {author}
  {\bibfnamefont {X.}~\bibnamefont {Liu}}, \bibinfo {author} {\bibfnamefont
  {Y.}~\bibnamefont {Zhu}}, \bibinfo {author} {\bibfnamefont {D.}~\bibnamefont
  {Graf}}, \bibinfo {author} {\bibfnamefont {K.}~\bibnamefont {Myhro}},
  \bibinfo {author} {\bibfnamefont {S.}~\bibnamefont {Tran}}, \bibinfo {author}
  {\bibfnamefont {C.~N.}\ \bibnamefont {Lau}}, \bibinfo {author} {\bibfnamefont
  {J.}~\bibnamefont {Wei}},  \emph {et~al.},\ }\href@noop {} {\bibfield
  {journal} {\bibinfo  {journal} {Physical Review Letters}\ }\textbf {\bibinfo
  {volume} {117}},\ \bibinfo {pages} {016602} (\bibinfo {year}
  {2016}{\natexlab{b}})}\BibitemShut {NoStop}%
\bibitem [{\citenamefont {Xu}\ \emph {et~al.}(2015{\natexlab{c}})\citenamefont
  {Xu}, \citenamefont {Song}, \citenamefont {Nie}, \citenamefont {Weng},
  \citenamefont {Fang},\ and\ \citenamefont {Dai}}]{xu2015two}%
  \BibitemOpen
  \bibfield  {author} {\bibinfo {author} {\bibfnamefont {Q.}~\bibnamefont
  {Xu}}, \bibinfo {author} {\bibfnamefont {Z.}~\bibnamefont {Song}}, \bibinfo
  {author} {\bibfnamefont {S.}~\bibnamefont {Nie}}, \bibinfo {author}
  {\bibfnamefont {H.}~\bibnamefont {Weng}}, \bibinfo {author} {\bibfnamefont
  {Z.}~\bibnamefont {Fang}}, \ and\ \bibinfo {author} {\bibfnamefont
  {X.}~\bibnamefont {Dai}},\ }\href@noop {} {\bibfield  {journal} {\bibinfo
  {journal} {Physical Review B}\ }\textbf {\bibinfo {volume} {92}},\ \bibinfo
  {pages} {205310} (\bibinfo {year} {2015}{\natexlab{c}})}\BibitemShut
  {NoStop}%
\bibitem [{\citenamefont {Chen}\ \emph {et~al.}(2015)\citenamefont {Chen},
  \citenamefont {Lu},\ and\ \citenamefont {Kee}}]{chen2015topological}%
  \BibitemOpen
  \bibfield  {author} {\bibinfo {author} {\bibfnamefont {Y.}~\bibnamefont
  {Chen}}, \bibinfo {author} {\bibfnamefont {Y.-M.}\ \bibnamefont {Lu}}, \ and\
  \bibinfo {author} {\bibfnamefont {H.-Y.}\ \bibnamefont {Kee}},\ }\href@noop
  {} {\bibfield  {journal} {\bibinfo  {journal} {Nature communications}\
  }\textbf {\bibinfo {volume} {6}},\ \bibinfo {pages} {6593} (\bibinfo {year}
  {2015})}\BibitemShut {NoStop}%
\bibitem [{\citenamefont {Zeb}\ and\ \citenamefont
  {Kee}(2012)}]{zeb2012interplay}%
  \BibitemOpen
  \bibfield  {author} {\bibinfo {author} {\bibfnamefont {M.~A.}\ \bibnamefont
  {Zeb}}\ and\ \bibinfo {author} {\bibfnamefont {H.-Y.}\ \bibnamefont {Kee}},\
  }\href@noop {} {\bibfield  {journal} {\bibinfo  {journal} {Physical Review
  B}\ }\textbf {\bibinfo {volume} {86}},\ \bibinfo {pages} {085149} (\bibinfo
  {year} {2012})}\BibitemShut {NoStop}%
\bibitem [{\citenamefont {Carter}\ \emph {et~al.}(2012)\citenamefont {Carter},
  \citenamefont {Shankar}, \citenamefont {Zeb},\ and\ \citenamefont
  {Kee}}]{carter2012semimetal}%
  \BibitemOpen
  \bibfield  {author} {\bibinfo {author} {\bibfnamefont {J.-M.}\ \bibnamefont
  {Carter}}, \bibinfo {author} {\bibfnamefont {V.~V.}\ \bibnamefont {Shankar}},
  \bibinfo {author} {\bibfnamefont {M.~A.}\ \bibnamefont {Zeb}}, \ and\
  \bibinfo {author} {\bibfnamefont {H.-Y.}\ \bibnamefont {Kee}},\ }\href@noop
  {} {\bibfield  {journal} {\bibinfo  {journal} {Physical Review B}\ }\textbf
  {\bibinfo {volume} {85}},\ \bibinfo {pages} {115105} (\bibinfo {year}
  {2012})}\BibitemShut {NoStop}%
\bibitem [{\citenamefont {Liu}\ \emph {et~al.}(2016{\natexlab{c}})\citenamefont
  {Liu}, \citenamefont {Kriegner}, \citenamefont {Horak}, \citenamefont
  {Puggioni}, \citenamefont {Serrao}, \citenamefont {Chen}, \citenamefont {Yi},
  \citenamefont {Frontera}, \citenamefont {Holy}, \citenamefont {Vishwanath}
  \emph {et~al.}}]{liu2016strain}%
  \BibitemOpen
  \bibfield  {author} {\bibinfo {author} {\bibfnamefont {J.}~\bibnamefont
  {Liu}}, \bibinfo {author} {\bibfnamefont {D.}~\bibnamefont {Kriegner}},
  \bibinfo {author} {\bibfnamefont {L.}~\bibnamefont {Horak}}, \bibinfo
  {author} {\bibfnamefont {D.}~\bibnamefont {Puggioni}}, \bibinfo {author}
  {\bibfnamefont {C.~R.}\ \bibnamefont {Serrao}}, \bibinfo {author}
  {\bibfnamefont {R.}~\bibnamefont {Chen}}, \bibinfo {author} {\bibfnamefont
  {D.}~\bibnamefont {Yi}}, \bibinfo {author} {\bibfnamefont {C.}~\bibnamefont
  {Frontera}}, \bibinfo {author} {\bibfnamefont {V.}~\bibnamefont {Holy}},
  \bibinfo {author} {\bibfnamefont {A.}~\bibnamefont {Vishwanath}},  \emph
  {et~al.},\ }\href@noop {} {\bibfield  {journal} {\bibinfo  {journal}
  {Physical Review B}\ }\textbf {\bibinfo {volume} {93}},\ \bibinfo {pages}
  {085118} (\bibinfo {year} {2016}{\natexlab{c}})}\BibitemShut {NoStop}%
\bibitem [{\citenamefont {Fujioka}\ \emph {et~al.}(2017)\citenamefont
  {Fujioka}, \citenamefont {Okawa}, \citenamefont {Yamamoto},\ and\
  \citenamefont {Tokura}}]{fujioka2017correlated}%
  \BibitemOpen
  \bibfield  {author} {\bibinfo {author} {\bibfnamefont {J.}~\bibnamefont
  {Fujioka}}, \bibinfo {author} {\bibfnamefont {T.}~\bibnamefont {Okawa}},
  \bibinfo {author} {\bibfnamefont {A.}~\bibnamefont {Yamamoto}}, \ and\
  \bibinfo {author} {\bibfnamefont {Y.}~\bibnamefont {Tokura}},\ }\href@noop {}
  {\bibfield  {journal} {\bibinfo  {journal} {Physical Review B}\ }\textbf
  {\bibinfo {volume} {95}},\ \bibinfo {pages} {121102} (\bibinfo {year}
  {2017})}\BibitemShut {NoStop}%
\bibitem [{\citenamefont {Zhang}\ \emph {et~al.}(2014)\citenamefont {Zhang},
  \citenamefont {Chen}, \citenamefont {Zhang}, \citenamefont {Zhou},
  \citenamefont {Zhang}, \citenamefont {Gu}, \citenamefont {Yao},\ and\
  \citenamefont {Chen}}]{zhang2014sensitively}%
  \BibitemOpen
  \bibfield  {author} {\bibinfo {author} {\bibfnamefont {L.}~\bibnamefont
  {Zhang}}, \bibinfo {author} {\bibfnamefont {Y.}~\bibnamefont {Chen}},
  \bibinfo {author} {\bibfnamefont {B.}~\bibnamefont {Zhang}}, \bibinfo
  {author} {\bibfnamefont {J.}~\bibnamefont {Zhou}}, \bibinfo {author}
  {\bibfnamefont {S.}~\bibnamefont {Zhang}}, \bibinfo {author} {\bibfnamefont
  {Z.}~\bibnamefont {Gu}}, \bibinfo {author} {\bibfnamefont {S.}~\bibnamefont
  {Yao}}, \ and\ \bibinfo {author} {\bibfnamefont {Y.}~\bibnamefont {Chen}},\
  }\href@noop {} {\bibfield  {journal} {\bibinfo  {journal} {Journal of the
  Physical Society of Japan}\ }\textbf {\bibinfo {volume} {83}},\ \bibinfo
  {pages} {054707} (\bibinfo {year} {2014})}\BibitemShut {NoStop}%
\bibitem [{\citenamefont {Gruenewald}\ \emph {et~al.}(2014)\citenamefont
  {Gruenewald}, \citenamefont {Nichols}, \citenamefont {Terzic}, \citenamefont
  {Cao}, \citenamefont {Brill},\ and\ \citenamefont
  {Seo}}]{gruenewald2014compressive}%
  \BibitemOpen
  \bibfield  {author} {\bibinfo {author} {\bibfnamefont {J.~H.}\ \bibnamefont
  {Gruenewald}}, \bibinfo {author} {\bibfnamefont {J.}~\bibnamefont {Nichols}},
  \bibinfo {author} {\bibfnamefont {J.}~\bibnamefont {Terzic}}, \bibinfo
  {author} {\bibfnamefont {G.}~\bibnamefont {Cao}}, \bibinfo {author}
  {\bibfnamefont {J.~W.}\ \bibnamefont {Brill}}, \ and\ \bibinfo {author}
  {\bibfnamefont {S.~S.~A.}\ \bibnamefont {Seo}},\ }\href@noop {} {\bibfield
  {journal} {\bibinfo  {journal} {Journal of Materials Research}\ }\textbf
  {\bibinfo {volume} {29}},\ \bibinfo {pages} {2491} (\bibinfo {year}
  {2014})}\BibitemShut {NoStop}%
\bibitem [{\citenamefont {Biswas}\ \emph {et~al.}(2014)\citenamefont {Biswas},
  \citenamefont {Kim},\ and\ \citenamefont {Jeong}}]{biswas2014metal}%
  \BibitemOpen
  \bibfield  {author} {\bibinfo {author} {\bibfnamefont {A.}~\bibnamefont
  {Biswas}}, \bibinfo {author} {\bibfnamefont {K.-S.}\ \bibnamefont {Kim}}, \
  and\ \bibinfo {author} {\bibfnamefont {Y.~H.}\ \bibnamefont {Jeong}},\
  }\href@noop {} {\bibfield  {journal} {\bibinfo  {journal} {Journal of Applied
  Physics}\ }\textbf {\bibinfo {volume} {116}},\ \bibinfo {pages} {213704}
  (\bibinfo {year} {2014})}\BibitemShut {NoStop}%
\bibitem [{\citenamefont {Schnyder}\ \emph {et~al.}(2008)\citenamefont
  {Schnyder}, \citenamefont {Ryu}, \citenamefont {Furusaki},\ and\
  \citenamefont {Ludwig}}]{schnyder2008classification}%
  \BibitemOpen
  \bibfield  {author} {\bibinfo {author} {\bibfnamefont {A.~P.}\ \bibnamefont
  {Schnyder}}, \bibinfo {author} {\bibfnamefont {S.}~\bibnamefont {Ryu}},
  \bibinfo {author} {\bibfnamefont {A.}~\bibnamefont {Furusaki}}, \ and\
  \bibinfo {author} {\bibfnamefont {A.~W.}\ \bibnamefont {Ludwig}},\
  }\href@noop {} {\bibfield  {journal} {\bibinfo  {journal} {Physical Review
  B}\ }\textbf {\bibinfo {volume} {78}},\ \bibinfo {pages} {195125} (\bibinfo
  {year} {2008})}\BibitemShut {NoStop}%
\bibitem [{\citenamefont {Lin}\ \emph {et~al.}(2004)\citenamefont {Lin},
  \citenamefont {Huang}, \citenamefont {Lin}, \citenamefont {Lee},
  \citenamefont {Liu}, \citenamefont {Zhang}, \citenamefont {Chen},\ and\
  \citenamefont {Huang}}]{lin2004low}%
  \BibitemOpen
  \bibfield  {author} {\bibinfo {author} {\bibfnamefont {J.}~\bibnamefont
  {Lin}}, \bibinfo {author} {\bibfnamefont {S.}~\bibnamefont {Huang}}, \bibinfo
  {author} {\bibfnamefont {Y.}~\bibnamefont {Lin}}, \bibinfo {author}
  {\bibfnamefont {T.}~\bibnamefont {Lee}}, \bibinfo {author} {\bibfnamefont
  {H.}~\bibnamefont {Liu}}, \bibinfo {author} {\bibfnamefont {X.}~\bibnamefont
  {Zhang}}, \bibinfo {author} {\bibfnamefont {R.}~\bibnamefont {Chen}}, \ and\
  \bibinfo {author} {\bibfnamefont {Y.}~\bibnamefont {Huang}},\ }\href@noop {}
  {\bibfield  {journal} {\bibinfo  {journal} {Journal of Physics: Condensed
  Matter}\ }\textbf {\bibinfo {volume} {16}},\ \bibinfo {pages} {8035}
  (\bibinfo {year} {2004})}\BibitemShut {NoStop}%
\bibitem [{\citenamefont {Ryden}\ \emph {et~al.}(1972)\citenamefont {Ryden},
  \citenamefont {Reed},\ and\ \citenamefont {Greiner}}]{ryden1972high}%
  \BibitemOpen
  \bibfield  {author} {\bibinfo {author} {\bibfnamefont {W.}~\bibnamefont
  {Ryden}}, \bibinfo {author} {\bibfnamefont {W.}~\bibnamefont {Reed}}, \ and\
  \bibinfo {author} {\bibfnamefont {E.}~\bibnamefont {Greiner}},\ }\href@noop
  {} {\bibfield  {journal} {\bibinfo  {journal} {Physical Review B}\ }\textbf
  {\bibinfo {volume} {6}},\ \bibinfo {pages} {2089} (\bibinfo {year}
  {1972})}\BibitemShut {NoStop}%
\bibitem [{\citenamefont {Patschke}\ \emph {et~al.}(1999)\citenamefont
  {Patschke}, \citenamefont {Brazis}, \citenamefont {Kannewurf},\ and\
  \citenamefont {Kanatzidis}}]{patschke1999cu}%
  \BibitemOpen
  \bibfield  {author} {\bibinfo {author} {\bibfnamefont {R.}~\bibnamefont
  {Patschke}}, \bibinfo {author} {\bibfnamefont {P.}~\bibnamefont {Brazis}},
  \bibinfo {author} {\bibfnamefont {C.~R.}\ \bibnamefont {Kannewurf}}, \ and\
  \bibinfo {author} {\bibfnamefont {M.~G.}\ \bibnamefont {Kanatzidis}},\
  }\href@noop {} {\bibfield  {journal} {\bibinfo  {journal} {Journal of
  Materials Chemistry}\ }\textbf {\bibinfo {volume} {9}},\ \bibinfo {pages}
  {2293} (\bibinfo {year} {1999})}\BibitemShut {NoStop}%
\bibitem [{\citenamefont {Liu}\ and\ \citenamefont
  {Balents}(2017)}]{liu2017correlation}%
  \BibitemOpen
  \bibfield  {author} {\bibinfo {author} {\bibfnamefont {J.}~\bibnamefont
  {Liu}}\ and\ \bibinfo {author} {\bibfnamefont {L.}~\bibnamefont {Balents}},\
  }\href@noop {} {\bibfield  {journal} {\bibinfo  {journal} {Physical Review
  B}\ }\textbf {\bibinfo {volume} {95}},\ \bibinfo {pages} {075426} (\bibinfo
  {year} {2017})}\BibitemShut {NoStop}%
\bibitem [{\citenamefont {Chan}\ \emph {et~al.}(2017)\citenamefont {Chan},
  \citenamefont {Lindner}, \citenamefont {Refael},\ and\ \citenamefont
  {Lee}}]{chan2017photocurrents}%
  \BibitemOpen
  \bibfield  {author} {\bibinfo {author} {\bibfnamefont {C.-K.}\ \bibnamefont
  {Chan}}, \bibinfo {author} {\bibfnamefont {N.~H.}\ \bibnamefont {Lindner}},
  \bibinfo {author} {\bibfnamefont {G.}~\bibnamefont {Refael}}, \ and\ \bibinfo
  {author} {\bibfnamefont {P.~A.}\ \bibnamefont {Lee}},\ }\href@noop {}
  {\bibfield  {journal} {\bibinfo  {journal} {Physical Review B}\ }\textbf
  {\bibinfo {volume} {95}},\ \bibinfo {pages} {041104} (\bibinfo {year}
  {2017})}\BibitemShut {NoStop}%
\bibitem [{\citenamefont {Taguchi}\ \emph
  {et~al.}(2016{\natexlab{b}})\citenamefont {Taguchi}, \citenamefont {Xu},
  \citenamefont {Yamakage},\ and\ \citenamefont
  {Law}}]{taguchi2016photovoltaic1}%
  \BibitemOpen
  \bibfield  {author} {\bibinfo {author} {\bibfnamefont {K.}~\bibnamefont
  {Taguchi}}, \bibinfo {author} {\bibfnamefont {D.-H.}\ \bibnamefont {Xu}},
  \bibinfo {author} {\bibfnamefont {A.}~\bibnamefont {Yamakage}}, \ and\
  \bibinfo {author} {\bibfnamefont {K.}~\bibnamefont {Law}},\ }\href@noop {}
  {\bibfield  {journal} {\bibinfo  {journal} {Physical Review B}\ }\textbf
  {\bibinfo {volume} {94}},\ \bibinfo {pages} {155206} (\bibinfo {year}
  {2016}{\natexlab{b}})}\BibitemShut {NoStop}%
\bibitem [{\citenamefont {Yan}\ and\ \citenamefont
  {Wang}(2016)}]{yan2016tunable}%
  \BibitemOpen
  \bibfield  {author} {\bibinfo {author} {\bibfnamefont {Z.}~\bibnamefont
  {Yan}}\ and\ \bibinfo {author} {\bibfnamefont {Z.}~\bibnamefont {Wang}},\
  }\href@noop {} {\bibfield  {journal} {\bibinfo  {journal} {Physical review
  letters}\ }\textbf {\bibinfo {volume} {117}},\ \bibinfo {pages} {087402}
  (\bibinfo {year} {2016})}\BibitemShut {NoStop}%
\bibitem [{\citenamefont {Ezawa}(2017)}]{ezawa2017photoinduced}%
  \BibitemOpen
  \bibfield  {author} {\bibinfo {author} {\bibfnamefont {M.}~\bibnamefont
  {Ezawa}},\ }\href@noop {} {\bibfield  {journal} {\bibinfo  {journal} {arXiv
  preprint arXiv:1705.02140}\ } (\bibinfo {year} {2017})}\BibitemShut {NoStop}%
  
\end{thebibliography}

%

\end{document}